\documentclass[11pt,preprint]{aastex}

\usepackage{amsmath}
\usepackage{amsfonts}
\usepackage{amssymb}
\usepackage{moreverb}
\usepackage{subfig}
\usepackage{graphicx}
\renewcommand{\vec}[1]{\mbox{\boldmath $#1$}}

\begin{document}

\title{Modeling of gyrosynchrotron radio emission pulsations produced by MHD loop oscillations in solar flares}
\author{George Mossessian\altaffilmark{1,2} \& Gregory D. Fleishman\altaffilmark{1,3}}
\altaffiltext{1}{Center For Solar-Terrestrial Research, New Jersey Institute of Technology, Newark, NJ 07102}
\altaffiltext{2}{Department of Physics, University of Illinois at Urbana-Champaign, Urbana, IL 61801}
\altaffiltext{3}{Ioffe Institute, St. Petersburg 194021, Russia}

\begin{abstract}A quantitative study of the observable radio signatures of the sausage, kink, and torsional MHD oscillation  modes  in flaring coronal loops is performed. Considering first non-zero order effect of these various MHD oscillation modes on the radio source parameters such as magnetic field, line of sight, plasma density and temperature, electron distribution function, and the source dimensions, we compute time dependent radio emission (spectra and light curves). The radio light curves (of both flux density and degree of polarization) at all considered radio frequencies are than quantified in both time domain (via computation of the full modulation amplitude as a function of frequency) and in Fourier domain (oscillation spectra, phases, and partial modulation amplitude) to form the signatures specific to a particular oscillation mode and/or   source parameter regime. We found that the parameter regime and the involved MHD mode can indeed be distinguished using the quantitative measures derived in the modeling. We apply the developed approach to analyze radio burst recorded by Owens Valley Solar Array and report possible detection of the sausage mode oscillation in one (partly occulted) flare and kink or torsional oscillations in another flare.

\end{abstract}

\section{Introduction}
Oscillations and quasi-periodic pulsations (QPPs) in radio emission from solar flares have been observed for half a century \citep{Young61}. Pulsation phenomena can be formally categorized by wavelength of the emission and by frequency/period of the QPPs. For example, at meter  and decimeter wavelengths, rapid QPPs occur with periods of tens to hundreds of milliseconds. Typically, these QPPs have rapidly varying amplitudes, variable periods, and narrow bandwidths $\delta f/f\lesssim 1$, \citep[e.g.][]{Young61, Zlobec_etal_1987,
Stepanov_Yurovsky_1990, Kurths_etal_1991, Yurovsky_1991, Fleishman94,
Aschwanden_etal_1995, Fl_etal_2002, PRL, Magdalenic_etal_2002,
Benz_etal_2005, Meszarosova_etal_2005, Chen_Yan_2007, Benz_etal_2011, Chen_etal_2011}.
Longer-period QPPs, with periods $\tau\gtrsim10$~s  are observed in microwaves; such QPPs are often accompanied by QPPs in hard X-rays \citep[HXR;
e.g.,][] {Parks_Winckler_1969, Janssens_etal_1973,
Zaitsev_Stepanov_1982b, Kane_etal_1983, Nakajima_etal_1983,
Asai_etal_2001, Grechnev_etal_2003, Nakariakov_etal_2003,
Stepanov_etal_2004, Melnikov_etal_2005, Fl_etal_2008}.

Numerous models have been proposed to account for the various types of oscillations, QPPs, and related phenomena observed at radio wavelengths \citep[e.g.,][and references therein]{Aschwanden87, Nakariakov_etal_2003, Fl_etal_2008}. Generally speaking, for radio emissions produced by coherent radiation processes (such as fast millisecond pulsations), QPPs are believed to result from nonlinear, self-organizing wave-wave or wave-particle interactions \citep{Zaitsev_Stepanov_1975b,
Meerson_etal_1978, Bardakov_Stepanov_1979, Aschwanden88,Fleishman94,Korsakov98}. In contrast, for radio emission produced by incoherent gyrosynchrotron (GS) radiation from energetic electrons, QPPs are believed to result from modulation of the source parameters (such as the energetic electron distribution, the magnetic field strength, the line of sight, etc.) via magnetohydrodynamic (MHD) oscillations (kink, sausage, or torsional modes) \citep{Aschwanden04} and/or modulation of electron distribution function via time-dependent acceleration and injection. Most of the reported so far QPPs of GS microwave emission \citep[e.g.,][]{Kane_etal_1983, Nakajima_etal_1983, Fl_etal_2008} are consistent with modulation of fast electron acceleration/injection, while there is no compelling case in which modulation of GS emission by an MHD oscillation mode would be firmly confirmed. As has been shown by \citet{Fl_etal_2008}, a well spectrally resolved radio spectrum is crucial to distinguish between MHD oscillations and modulation of fast electron injection/acceleration. Another component needed to firmly detect a particular MHD oscillation mode from the observed radio data is detailed quantitative modeling of observable measures, such as oscillation phase or modulation power, of the GS flaring emission modulated by one or another MHD mode, which is yet unavailable.

This paper attempts to partly remedy the situation by analyzing  dependence of GS radiation on modulation of the source parameters by various MHD modes, namely,
large-scale kink, sausage, and torsional oscillation modes, while we do not consider a special case of the longitudinal slow mode (essentially, the sound wave) where only the plasma density oscillates since this mode can only affect low-frequency radiation from a dense plasma, which has already been studied by \citet{Nakariakov_Melnikov_2006}. For each mode, time-domain and Fourier analyses are performed on the radio emission and polarization to characterize the properties of the pulsations, and the results are evaluated to identity a signature by which each oscillation mode can be distinguished in observational data. In contrast to approximate sausage-mode modulation modeling performed by  \citet{Fl_etal_2008} based on relatively crude Dulk-Marsh approximation, in this study the microwave (GS plus free-free) emission is modeled using precise fast GS codes recently developed by \cite{Fl_Kuzn_2010}.

The results of the modeling are then used to search for the corresponding QPP signatures in the total power microwave database available from the Owens Valley Solar Array (OVSA). Specifically, we looked through well-calibrated data recorded during 2001-2002 when microwave emission from 412 solar flares was recorded \citep{Nita_etal_2004}.
We found the the vast majority of the radio bursts does not display unambiguous QPPs in the dynamic spectra; only about 10\% of all events show time variations at the light curves that can be classified as QPPs. From them we selected a few events in which the presence of QPPs is apparent from their dynamic spectra; for those selected events
we computed all possible quantitative measures to be compared with the corresponding measures derived from the modeling in the attempt to detect an indication of one or another MHD loop oscillation mode. In most of the case we were unable to firmly identify the genuine cause of the QPPs; however, in two best cases, presented in the paper, the observed QPPs indeed could be produced by an MHD mode: the sausage mode in an (occulted) flare of 28 August 2002 21:41 UT and kink or torsional mode in a flare of 20 July 2002 21:27 UT. 
We note that some MHD oscillation modes may or may not produce  corresponding oscillations in HXR and/or soft X-rays (SXR), which we also tested for the mentioned events using HXR data available from the Reuven Ramaty High Energy Solar Spectroscopic Imager (RHESSI) and SXR data from the Geostationary Operational Environmental Satellites (GOES). 

\section{Modeling of GS emission modulated by MHD modes}
\label{procedure}

\subsection{MHD loop oscillation modes}	
\label{S_MHD_modes}

	The different modes of MHD loop oscillations are modeled by oscillating perturbations of magnetic field vector and further calculating the linear effect of these oscillations on relevant parameters affecting the GS radiation. In particular, the magnetic flux conservation affects cross-sectional area and volume of the loop. Then, assumed particle number conservation has an effect on the number density. The first adiabatic invariant results in oscillations of the particle energy.
Oscillations of the magnetic vector direction implies oscillations of the viewing angle between the magnetic field lines and the line of sight.

For the sake of simplicity, the flaring coronal loop is approximated to be a cylinder as in \citet{Zaitsev_Stepanov_1982a, Aschwanden06}. The  oscillation modes of a finite cylinder must necessarily display both time and spatial dependence; however, the latter strongly complicates the entire treatment as it implies building a nonuniform 3D model and solving the radiation transfer equation along many lines of sight \citep[see, e.g.,][]{Kuznetsov_etal_2011}, which is far beyond our intention to address a spatially uniform problem. Thus, for the sake of simplicity, we entirely discard any dependence of the oscillations on the spatial coordinates $s$, $r$, and $\varphi$ and retain only time dependence of the MHD modes; 
so for the case of the time-oscillating cross-sectional area and surface area, both the side and base of the cylinder were considered as affecting GS radiation by taking the visible area to be

\begin{equation}
\label{Eq_Area}
A(t)\approx S(t)\sin(\theta(t))+\frac{\pi}{4} L^2(t)\cos(\theta(t))
\end{equation}
where $S(t)$ is the cylinder side area, $L(t)$ is the cylinder base diameter, and $\theta(t)$ is the viewing angle between the cylinder axes and the line of sight. Accordingly, the source depth $d$ is the smallest of $S(t)/L(t)$ and
\begin{equation}
\label{Eq_Depth}
d(t)\approx  L(t)/\sin(\theta(t)).
\end{equation}
		
	
The large-scale (global) {\it sausage mode} is modelled as an oscillation of the magnetic field parallel to the coronal loop:
\begin{equation}
\label{Eq_Bz_sausage}
B_z(t)=B_0\left(1+a\sin(\omega t+\phi)\right).
\end{equation}
The first adiabatic invariant implies corresponding oscillations of the transverse particle momentum $p_\perp$. Without energy losses and  angular scattering of the  particles, these $p_\perp$ oscillations imply variations of both particle energy and pitch-angle, i.e., variation of electron energy and angular distributions. If, however, electron isotropization due to angular scattering occurs faster than the oscillation period then the transverse energy will be redistributed between the transverse and parallel motions to retain the angular distribution isotropic independently on the oscillation phase still conserving the total energy transmitted from the oscillation to the particles. This energy conservation can, however, break down if the processes of the energy loss/gain are important. For example, the energy can decrease due to thermal conduction or radiative losses or increase due to a flare energy release including the Alfv\'en pump effect driven by the very same MHD oscillations. Below we adopt that a balance between the energy gain and loss results in a gradual variation of the plasma mean (over the MHD period) temperature on top of which the MHD-wave induced oscillations are superimposed; thus, the temperature oscillates as
\begin{equation}
\label{Eq_T_sausage}
T(t)=T(0)\left(\frac{2}{3}\frac{B}{B_0}+\frac{1}{3}\right)
\end{equation}
around a mean level, $T(0)$, which, perhaps, itself smoothly varies with time.
The upper and lower energy cutoffs ($E_{\min}$, $E_{\max}$) \citep{Fl_Kuzn_2010} for accelerated electrons fluctuate by the same law. By flux conservation, this induces an oscillation of the cylinder cross-sectional area directly proportional to $(B/B_0)^{-1}$ and, therefore, the cylinder radius $\sim(B/B_0)^{-1/2}$, and then in number density of both thermal and nonthermal particles  $\sim B/B_0$. The viewing angle is not affected in the sausage mode. Thus, the source visible area oscillates according to Eqn~(\ref{Eq_Area}) due to oscillations of the cylinder dimensions, while the viewing angle remains constant and does affect the oscillations.
	
	In the {\it kink mode}, the oscillation of the magnetic field was taken as
\begin{equation}
\label{Eq_kink_m_B_osc}
\vec{B}_\perp(t)=\vec{B}_{\perp0}\sin(\omega t+\phi)
\end{equation}
where $\vec{B}_\perp$ is the magnetic field in the direction of the kink perturbation. The kink mode is known to be weakly compressible, to the order of $r/\lambda$, where $\lambda$ is the oscillation wavelength, so there are some variations of the density and the cross-sectional area  in this wave mode. However, the adopted spatially uniform oscillations imply $r/\lambda\ll 1$, which allows discarding these variations, so we  assume a constant cross-sectional area of the cylinder in the kink oscillations; however,  visible source area (\ref{Eq_Area}) and its depth (\ref{Eq_Depth}) are affected by the magnetic field oscillations described by Eqn~(\ref{Eq_kink_m_B_osc}) due to the viewing angle variations. Thus, the main linear effect in these oscillations is produced by the viewing angle, while the magnetic field value experience only a second order oscillations. If the unit vector toward the observer is $\vec{I}=\left(I_\|,I_\perp\right)$, then
\begin{equation}
\cos\theta(t)=\frac{\vec{I}\cdot\vec{B}}{|\vec{B}|}=\cos\theta_0+a\sin(\omega t+\phi) \sin\theta_0 \cos\varphi,
\end{equation}	
where $a=B_{\perp0}/B_0$ and  $\varphi$ is the angle between the direction of oscillation of $\vec{B}_\perp$ and the line of sight.
Therefore, there are two main extreme cases to be considered here, $\cos\varphi= 1$ and  $\cos\varphi = 0$. In the first case,
the radiation variation arises mostly from the oscillation of the viewing angle. In the second
case, however, the viewing angle will not oscillate, and the radiation oscillation will arise from a
second-order oscillation of the magnetic field strength. Apparently, all intermediate cases, $0<\cos\varphi< 1$ are also possible; the results will be between the mentioned extreme cases.

Similarly, the {\it torsional mode} only affects the pitch angle of the magnetic field in the linear approximation. The magnetic field oscillates as
\begin{equation}
B_\varphi(t)=B_{\varphi0}\sin(\omega t+\phi)
\end{equation}
so only the viewing angle oscillates as \citep{Tapping83}
\begin{equation}
\cos\theta(t)=\cos\theta_0\cos\left(\psi(t)\right) - \sin\theta_0\sin \left(\psi(t)\right),
\end{equation}
where $\psi(t)\equiv a\sin\omega t$ is the pitch-angle of the magnetic field. In contrast to the kink mode, there no oscillations of the source depth and visible area in the torsional mode, because oscillations of the azimuth component $B_\varphi$ do not imply spatial variation of the cylinder as a whole. 

These oscillation modes have been studied and described extensively by \citet{Rosenberg70,Roberts83,Aschwanden87,Aschwanden99, Nakariakov_etal_2003,Grechnev_etal_2003}. 
The kink and sausage modes both have fast and slow MHD wave versions. For the fast sausage mode, the observed period is defined by the cylinder length $l$ and the wave phase speed $v_{ph}$
\begin{equation}\tau_{\text{sausage}}\approx \frac{2l}{v_{ph}}<120\frac{r_8\sqrt{n_{10}}}{B}\text{seconds} \end{equation}
where $r_8$ is the tube radius in $10^8$ cm, $n_{10}$ is the tube density in $10^{10}$ cm$^{-3}$, and $B$ the magnetic field in gauss. The fast kink mode will have a period of
\begin{equation}\tau_{\text{kink}}\approx100\frac{L_8\sqrt{n_{10}}}{jB}\text{seconds}\end{equation}
where $j-1$ is the number of nodes and $L_8$ is the length of the loop in $10^8$ cm. The slow wave versions both have the same period,
\begin{equation}\tau_{\text{slow}}\approx 13\frac{L_8}{\sqrt{T_6}}\text{seconds}\end{equation} where $T_6$ is the temperature measured in $10^6$ K. The sausage mode also has an unstable propagating version with the period that is 0.6 times less fast than the fast sausage mode \citep{Grechnev_etal_2003}, while the torsional mode has a period
\begin{equation}
\tau_{\text{torsional}}\approx100\frac{L_8\sqrt{n_{10}}}{mB}\text{seconds},
\end{equation}
where $m$ is the number of windings along the loop.


\subsection{Characterization of radio oscillations}

The time-oscillating  radio flux and polarization modulated by a given MHD oscillation mode was straightforwardly simulated using the fast GS codes: in each instance we computed an array of the source parameters specific to this particular mode, see \S~\ref{S_MHD_modes}, and sent it to GS code, which returned the instant radio spectrum and polarization at the selected spectral range, typically, from 1 to 100 GHz. Repeating this procedure many times during the oscillation period we, thus, formed the oscillating radio emission representing the subject of further analysis of the QPPs. 

As has been already said, QPPs can be characterized in both time and Fourier domains \citep{Fl_etal_2008}.
In the time domain, the modulation amplitude $m(f)$ for each radio frequency can be calculated as
\begin{equation}
\label{Eq_full_m}
m(f) = \left( \frac{1}{T}\int_0^T S^2(f,t)\;dt\right)^{1/2} \end{equation}
where
\begin{equation} S(f,t)=\frac{F(f,t)-\left<F(f,t)\right>}{\left<F(f,t)\right>} \end{equation}
is the normalized modulation of the signal, $f$ is the frequency in GHz, $t$ is the time in seconds, and $F(f,t)$ is the total flux density at frequency $f$ \citep{Fl_etal_2008}. This value is a measure of the variation of emission with respect to the running average. The same measure can also be calculated for the absolute value of the degree of polarization of the emission.

	To find the  oscillation periods of the modeled radio light curves, we perform a FFT of the GS data across all the emission frequencies. 
To smooth out the result of the FFT, an array of zero-valued points was added to the end of the data \citep[sf.,][]{PRL}. Since only one oscillation period was simulated, the Fourier coefficients of the MHD oscillations form a narrow peak in the FFT spectrum describing the simulated oscillation period. The radio light curves, however, represent a complicated nonlinear response on the MHD oscillations and so, their Fourier spectra can strongly deviate from the original single narrow peak. 
	
At each significant Fourier peak, the phase of the pulsations relative to the oscillations at a reference frequency $f_{ref}$, that is selected somewhere in the optically thin region, 
is calculated as
\begin{equation}\varphi(f,\nu)=2\tan^{-1}\frac{\text{Im}\,\left\{ \Phi(f,\nu)\right\}}{\text{Re}\,\left\{\Phi(f,\nu)\right\}}\end{equation}
where
\begin{equation}\Phi(f,\nu)=\sqrt{\frac{\mathcal{F}(f_{ref},\nu)\cdot\mathcal{F}^*(f,\nu)}{|\mathcal{F}(f_{ref},\nu)|\cdot|\mathcal{F}(f,\nu)| }}\end{equation}
and $\mathcal{F}(f,\nu)$ is the coefficient of the complex FFT array at emission frequency $f$ corresponding to a pulsation frequency $\nu$, and $\mathcal{F}^*(f,\nu)$ is the complex conjugate of this coefficient.

	In addition to the above, the Fourier analysis allows for consideration of the partial modulation of the signal by each pulsating component at a Fourier spectrum. Analysis of partial modulation is necessary when there is more than one significant Fourier peak in the power spectrum. Furthermore, different peaks may be the results of different physical processes, or of different modes of MHD oscillations, and a quantitative analysis of partial modulation can help discern these different causes. The partial modulation amplitude $m_p(f)$ of radiation from a limited region of Fourier harmonics from $n_1$ to $n_2$ covering each main Fourier peak is defined as follows \citep{Fl_etal_2008}:
\begin{equation}
\label{Eq_partial_m}
m_p(f)=\left[ 2 \frac{N_{tot}}{N_i}\sum_{n=n_1}^{n_2}\left| S(f, \nu(n))\right|^2\right]^{1/2},\end{equation}
where $N_i$ is the total number of the measurements in the time domain and $N_{tot}$ is the total number of the oscillation frequencies $\nu(n)$ where the Fourier amplitudes are determined, and the factor of 2 arises from the fact that each partial oscillation consists of two Fourier peaks, one at positive frequency $\nu$ and one at the corresponding $-\nu$. It is worthwhile to note that adding up all the available Fourier components from $n=0$ to $n=N_{tot}-1$ yields the total modulation amplitude, which is equivalent to that given by Eqn~(\ref{Eq_full_m}) because of familiar Parceval's identity.


Microwave emission from solar flares differs depending on the ambient plasma density, magnetic field, fast electron energy or angular distributions. Below we consider a few different parameter combinations covering main distinct regimes of the GS emission in flares. For each considered parameter regime we simulate the time domain response of the corresponding radio source on all mentioned MHD loop oscillations as described above. Specifically, to characterize radio response on the MHD oscillations and determine corresponding GS emission signatures, the following measures were computed on both the flux density and polarization:
\begin{enumerate}
	\item the full modulation amplitude as a function of emission frequency $f$;
	\item the Fourier spectra of the emission at each emission frequency $f$;
	\item the relative phase of oscillations at Fourier peaks in the power spectrum  as a function of emission frequency $f$;
	\item the partial modulation amplitudes as a function of emission frequency $f$  for each significant peak in the Fourier spectrum.
\end{enumerate}

\section{Modeling results}

\begin{figure}[ht!]
	\centering
    \subfloat[Dynamic spectrum, flux]{\includegraphics[width=0.5\linewidth,height =0.001\linewidth, bb=0 0 1 1]{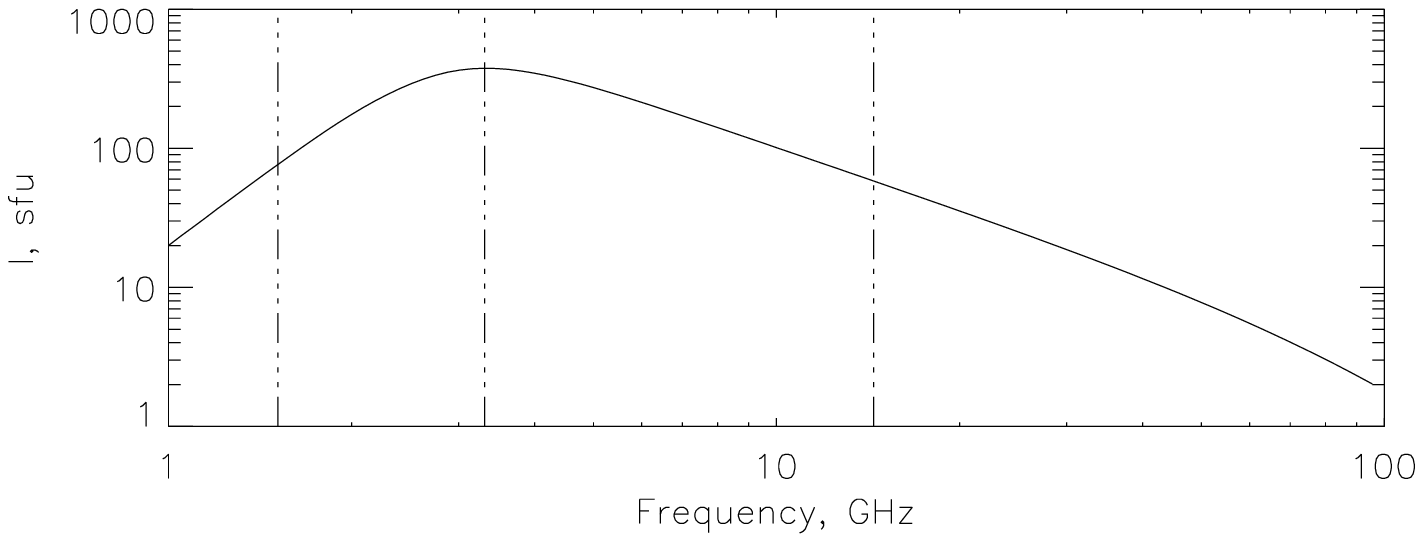}}
	\subfloat[Dynamic spectrum, pol.]{\includegraphics[width=0.5\linewidth,height=0.001\linewidth, bb=0 0 1 1]{figs/sausage/sausage_f.eps}}
	\vspace{-8mm}

	\subfloat[Default flux]{\includegraphics[width=0.45\linewidth]{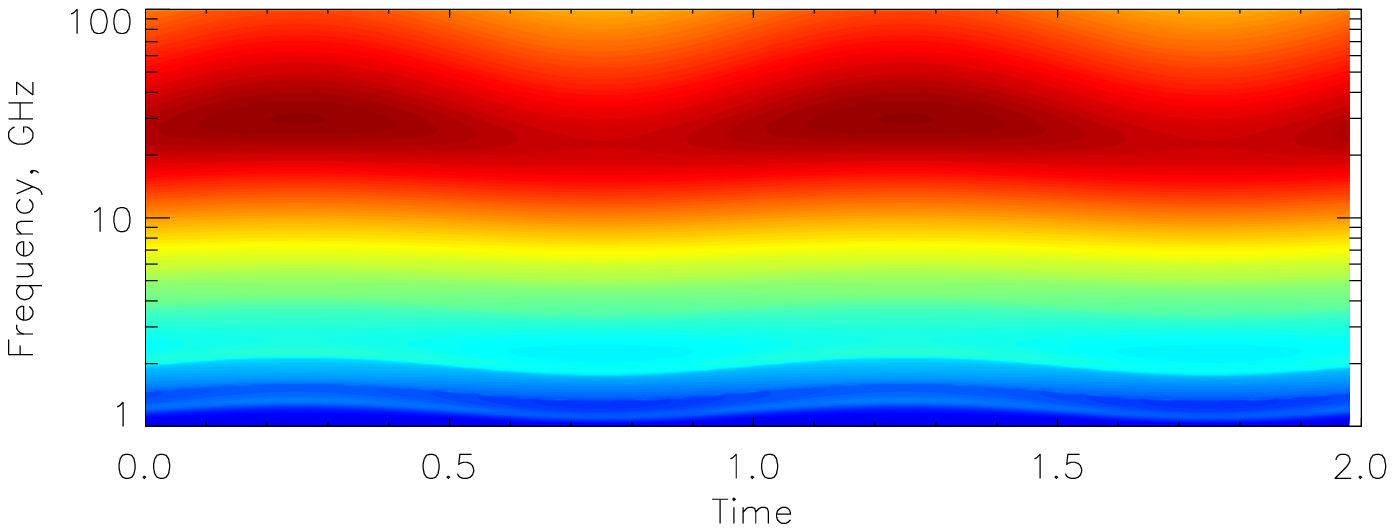}}
    \subfloat[Default pol.]{\includegraphics[width=0.45\linewidth]{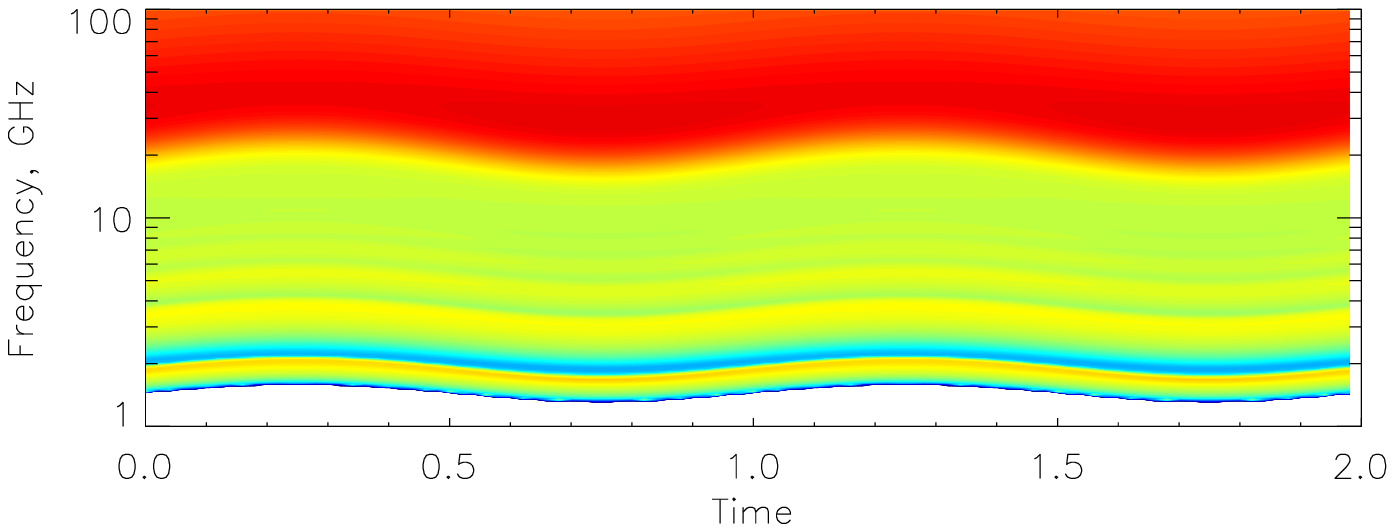}}
	\vspace{-8mm}

	\subfloat[One pixel; flux]{\includegraphics[width=0.45\linewidth]{figs/sausage/sausage_f.eps}}
	\subfloat[One pixel; pol.]{\includegraphics[width=0.45\linewidth]{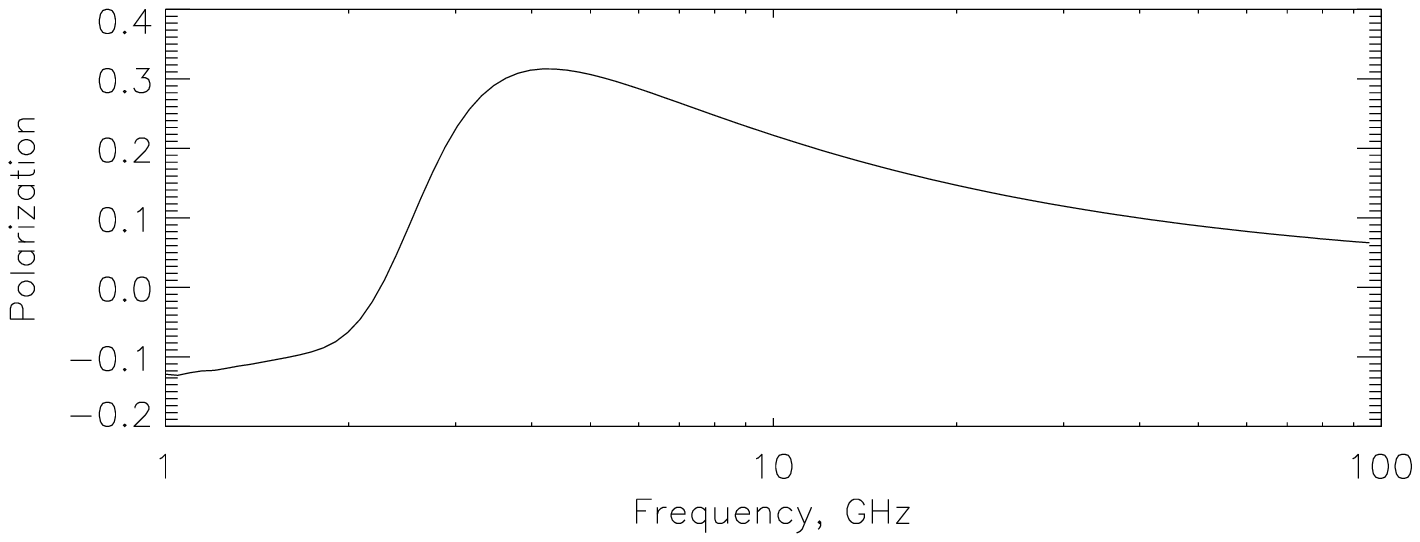}}
	\vspace{-8mm}

	\subfloat[Loss-cone; flux]{\includegraphics[width=0.45\linewidth]{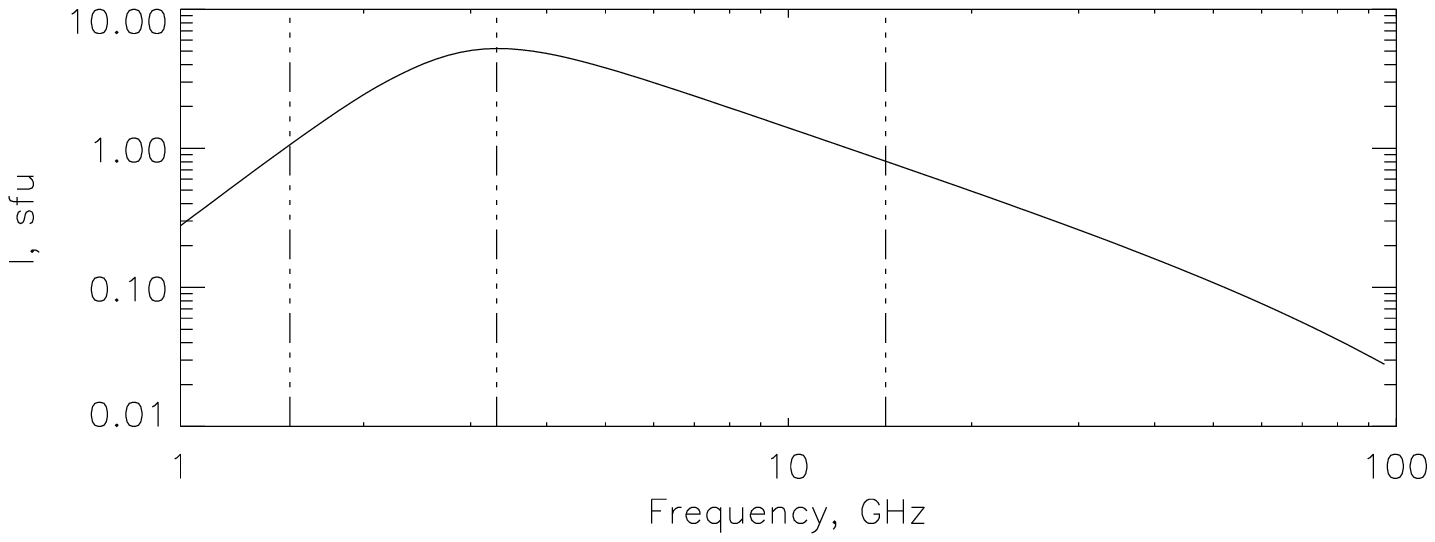}}
	\subfloat[Loss-cone; pol.]{\includegraphics[width=0.45\linewidth]{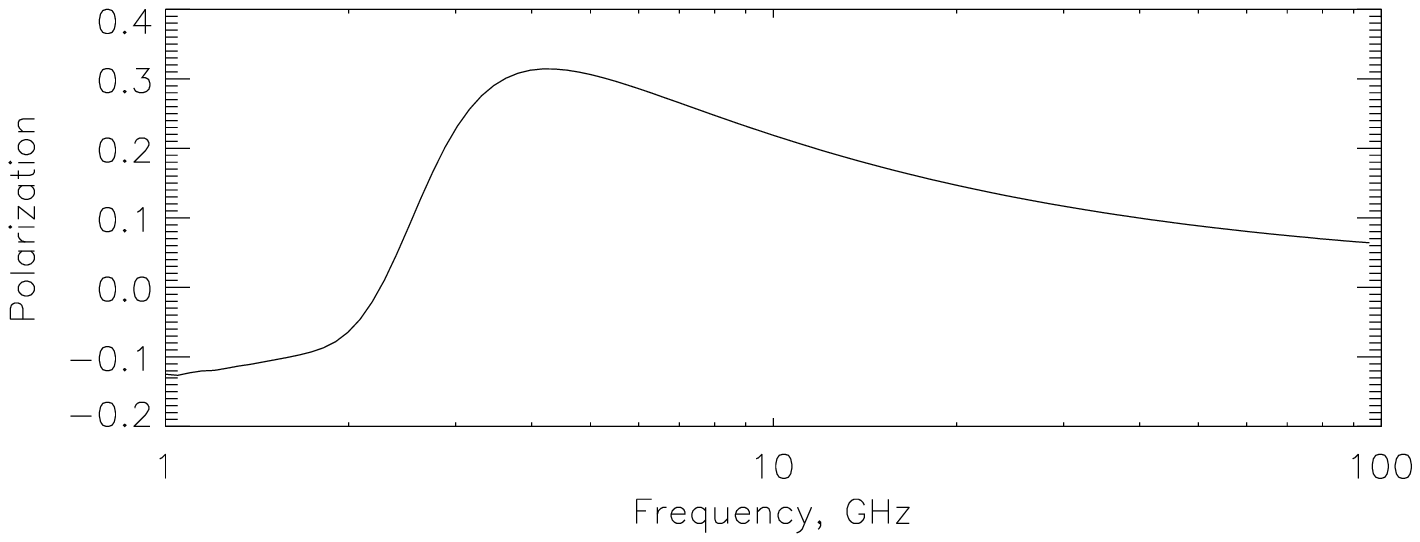}}
	\vspace{-8mm}

	\subfloat[Razin Effect; flux]{\includegraphics[width=0.45\linewidth]{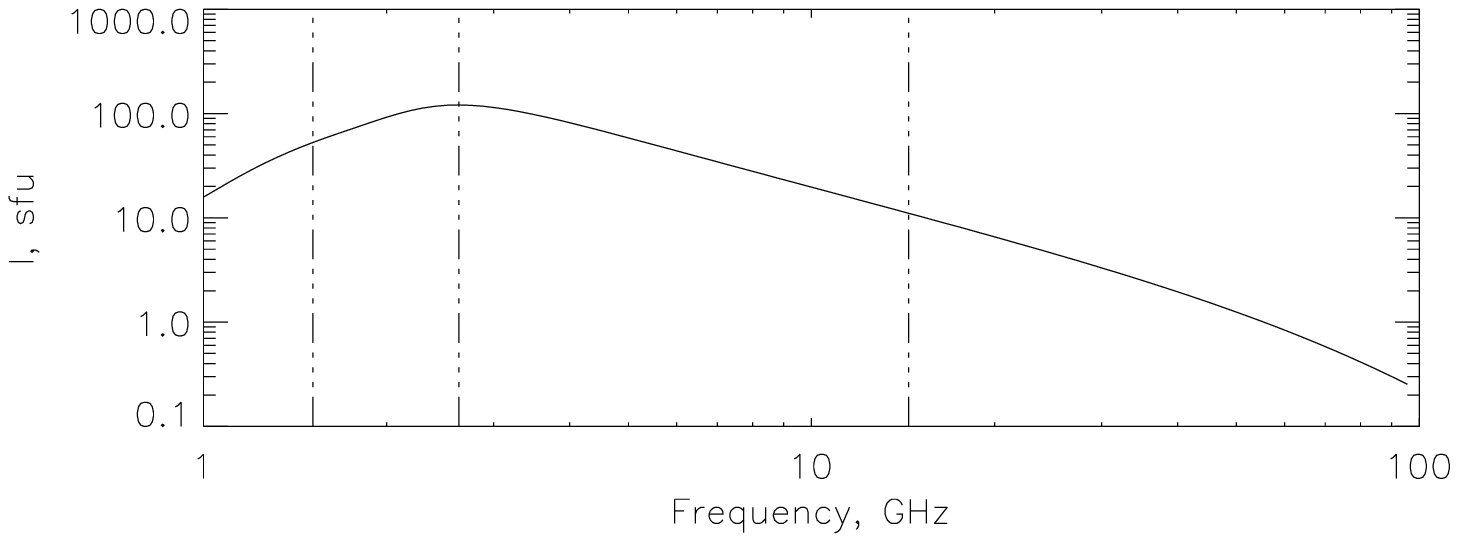}}
	\subfloat[Razin Effect; pol.]{\includegraphics[width=0.45\linewidth]{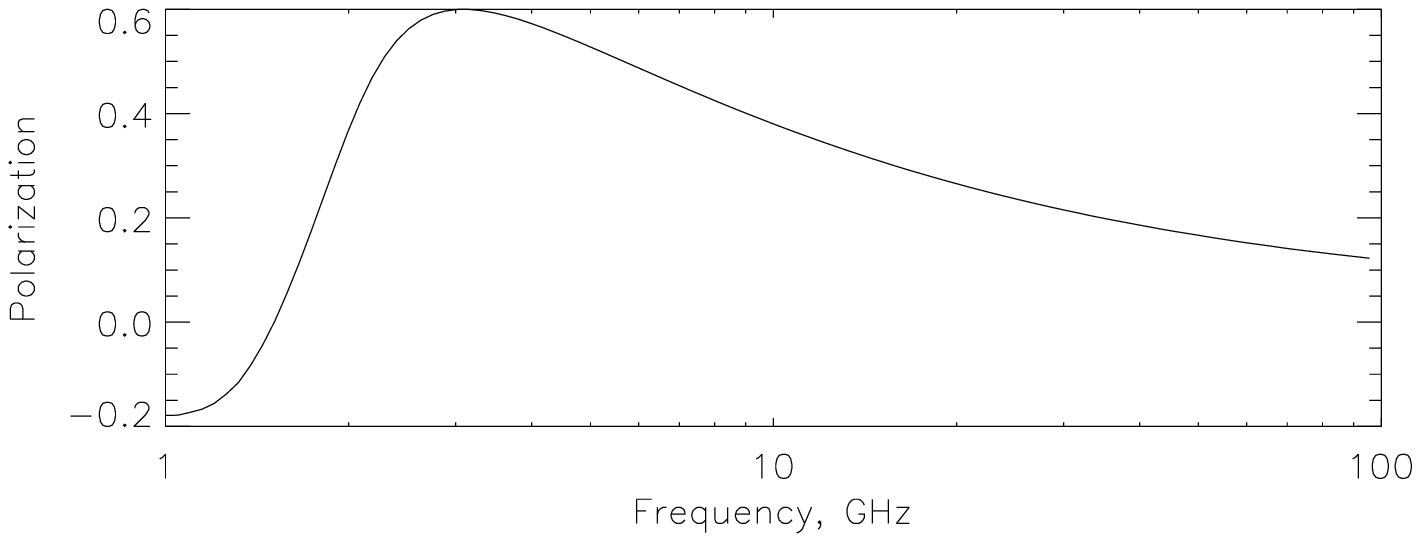}}
	\vspace{-8mm}

	\subfloat[High field; flux; $\theta=80^\circ$]{\includegraphics[width=0.45\linewidth]{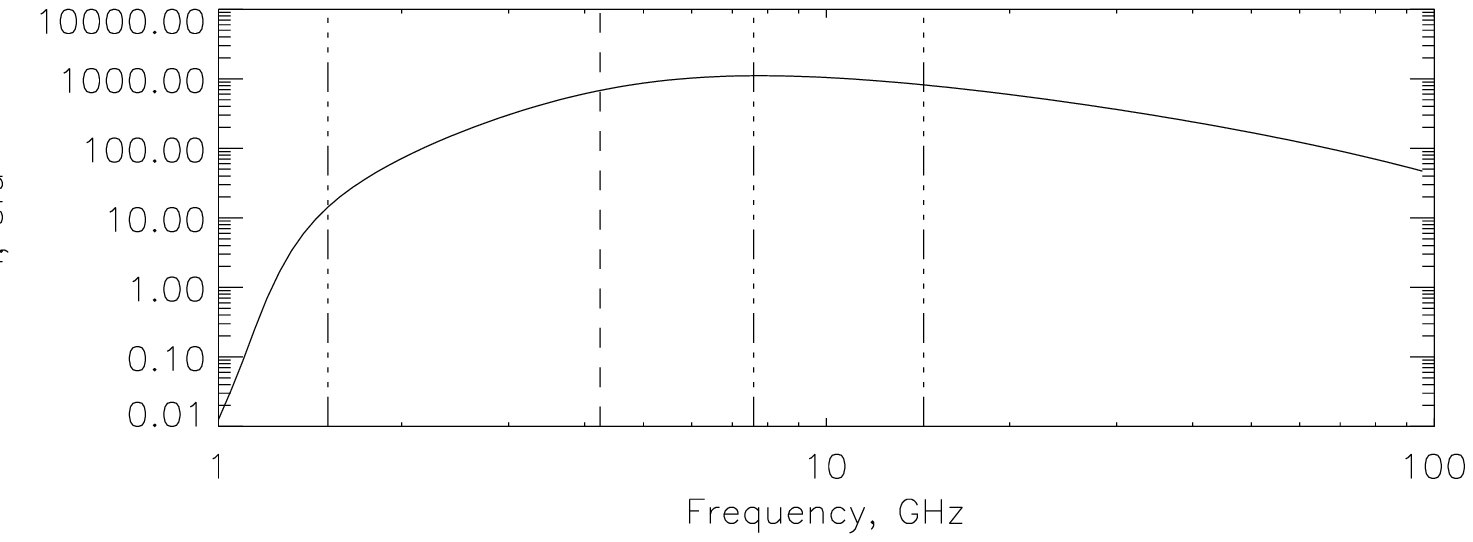}}
	\subfloat[High field; pol.; $\theta=80^\circ$]{\includegraphics[width=0.45\linewidth]{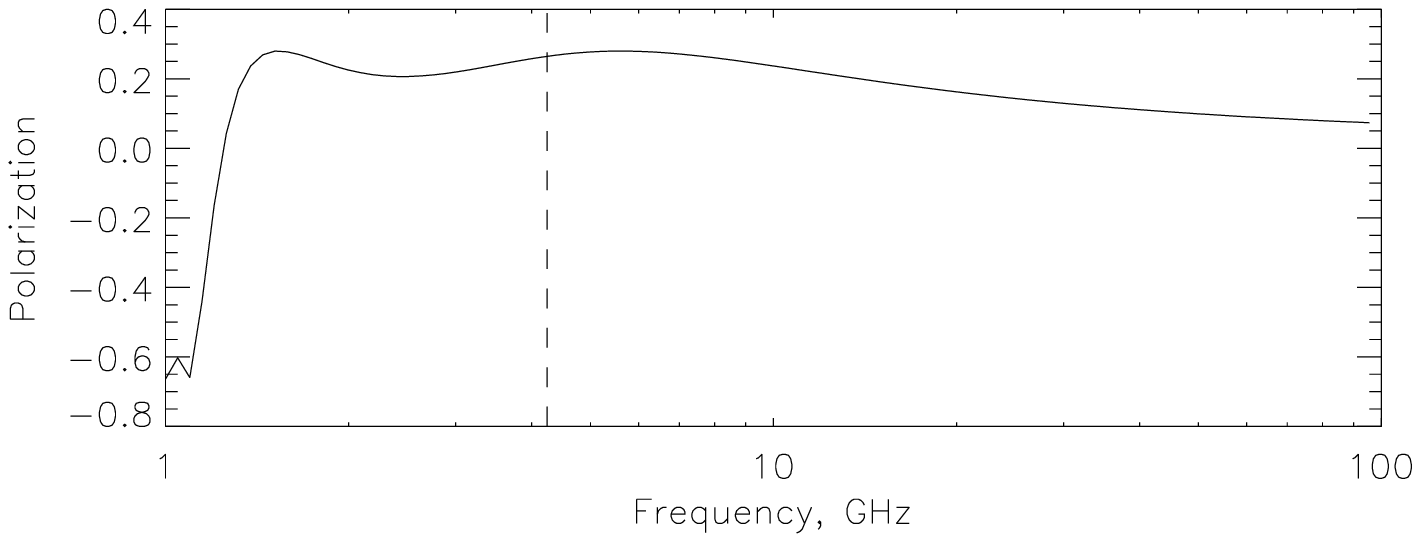}}
	\vspace{-8mm}
%
%
	
	\subfloat{\includegraphics[width=0.45\linewidth]{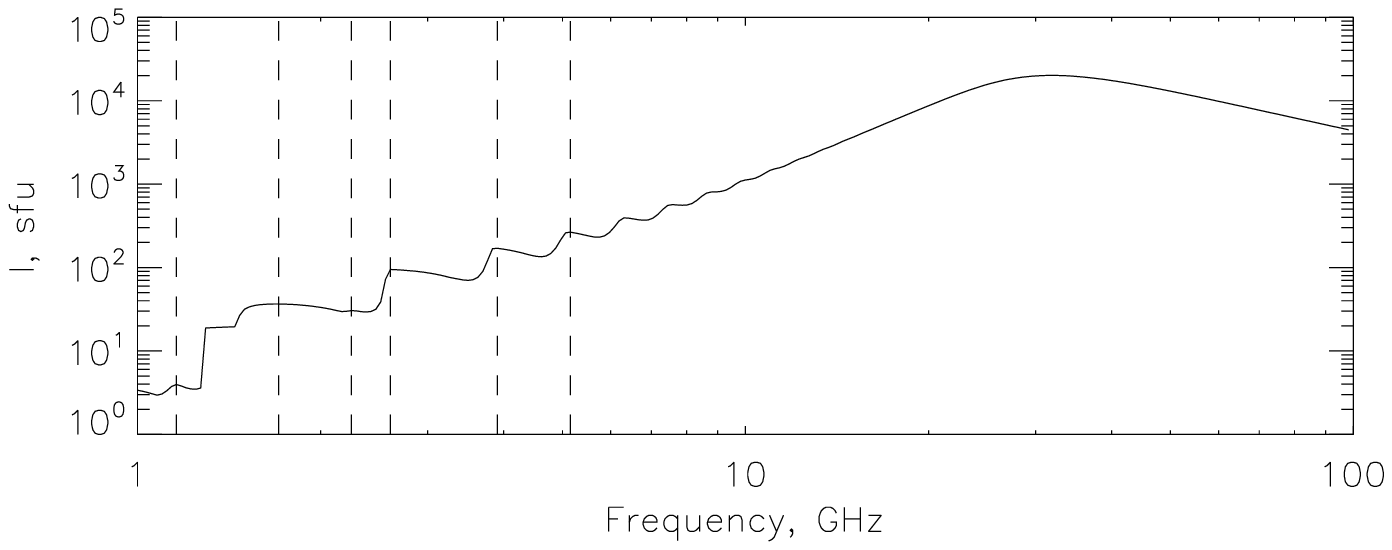}}
	\subfloat{\includegraphics[width=0.45\linewidth]{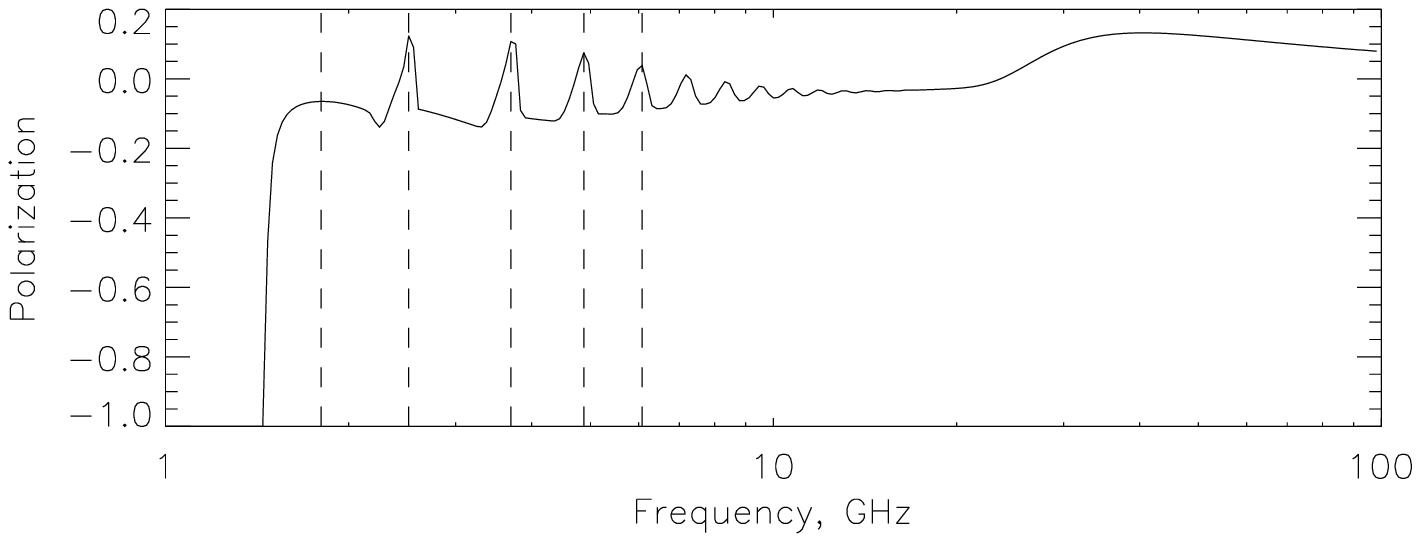}}

	\caption{Flux density (left column) and polarization (right column) for, from top to bottom, examples of dynamic spectra with the low gyroharmonics present, and spectra for default parameters, single pixel emission, anisotropic pitch-angle distribution, Razin effect parameters (dashed line marks Razin cutoff frequency), and low-frequency spectrum peaks.}
	\label{spectra_all}
\end{figure}

We considered a number of distinct parameter regimes forming the microwave emission including the cases of the spectral peak formation by either optical thickness effect (low-density case) or Razin-effect (high-density case); then, we also analyzed the case of strong magnetic field when distinct gyroharmonics appear at the low-frequency part ($1-10$~GHz) of the microwave spectrum. Furthermore, we considered different angular distributions of the radiating fast electrons, namely, the isotropic and anisotropic loss-cone distributions; note, that the loss-cone boundary itself can oscillate in response to the MHD modulation. Finally, we distinguished cases when the total power data are analyzed (e.g., oscillations of the source area contribute to the overall modulation) and when we deal with a spatially resolved data from a fixed single pixel (whose area is by definition constant).  The implied big number of the corresponding models, see overview of them in Figure~\ref{spectra_all}, results apparently in a large number of illustrating plots; most of which are collected in the on-line supplement to the paper. In contrast, in the main part of the paper we consider in more detail only a limited number of the cases needed for direct comparison with  either available simplified modeling \citep{Fl_etal_2008} or with observational data analyzed below in the paper. We adopt the following default parameters:
the cylinder side area $S_0=2.5\times10^{18}$~cm$^2$, the cylinder depth $L_0=2.5\times10^{8}$~cm, the background temperature
$T_0=2\times 10^7$~K, the thermal number density $n_0=10^9$~cm$^{-3}$, the nonthermal number density
$n_b=0.02\times n_0$, the magnetic field
$B_0= 50$~G, the viewing angle $\theta_0=45^\circ$, the electron energy spectral index $\delta=3.5$ for $0.1<E<10$~MeV, the angular distribution is isotropic.

\begin{figure}[htp!]
	\centering	
	\subfloat[1.5 GHz]{\includegraphics[width=0.3\linewidth]{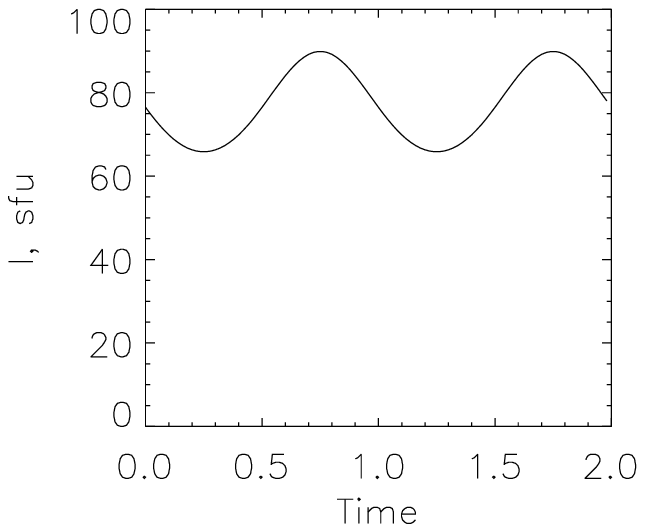}}
	\subfloat[2.2 GHz]{\includegraphics[width=0.3\linewidth]{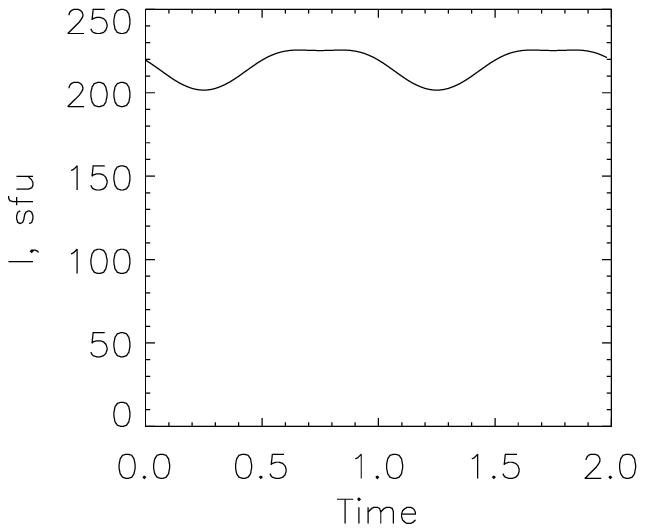}}
	\subfloat[13.8 GHz]{\includegraphics[width=0.3\linewidth]{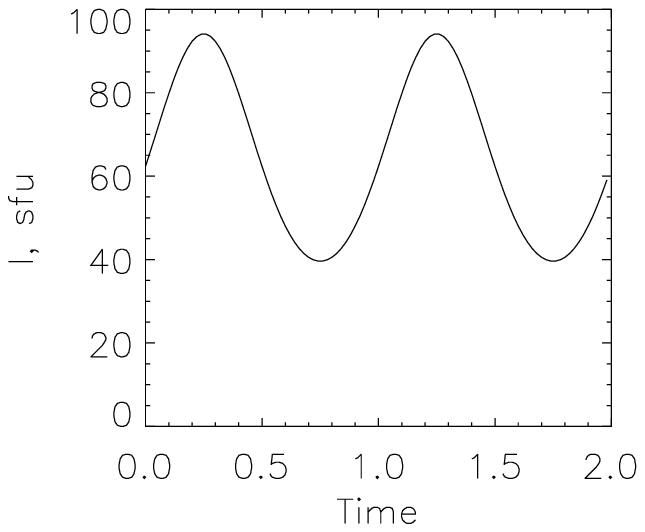}}
	\caption{Sausage mode, default parameters, oscillation of flux at indicated frequencies. Note that (i) the low- and high- frequency light curves are out of phase and (ii) the oscillating component deviates strongly from a sinusoidal around the spectrum peak frequency, e.g., 2.2~GHz. }
	\label{sausage_default_osc_f}

\end{figure}
\begin{figure}[hbp!]
	\centering
	\subfloat[1.5 GHz]{\includegraphics[width=0.3\textwidth]{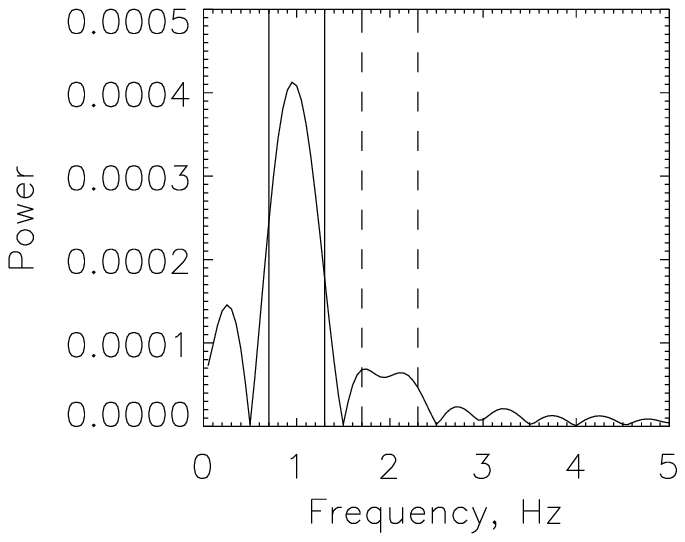}}
    \subfloat[2.2 GHz]{\includegraphics[width=0.3\textwidth]{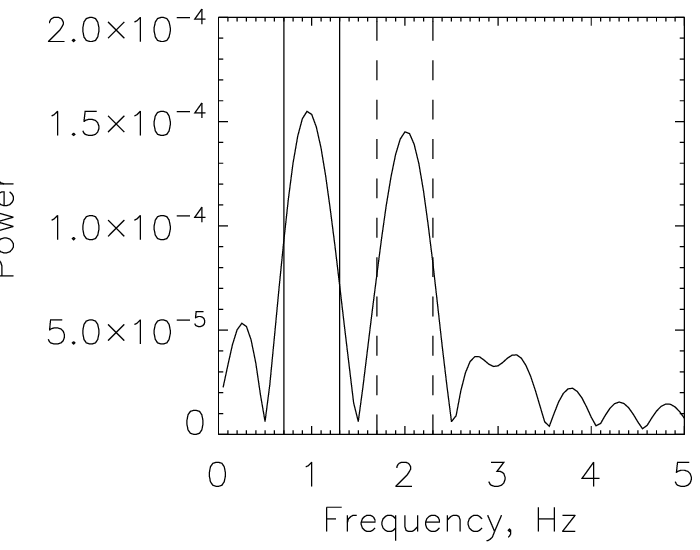}}
	\subfloat[13.8 GHz]{\includegraphics[width=0.3\textwidth]{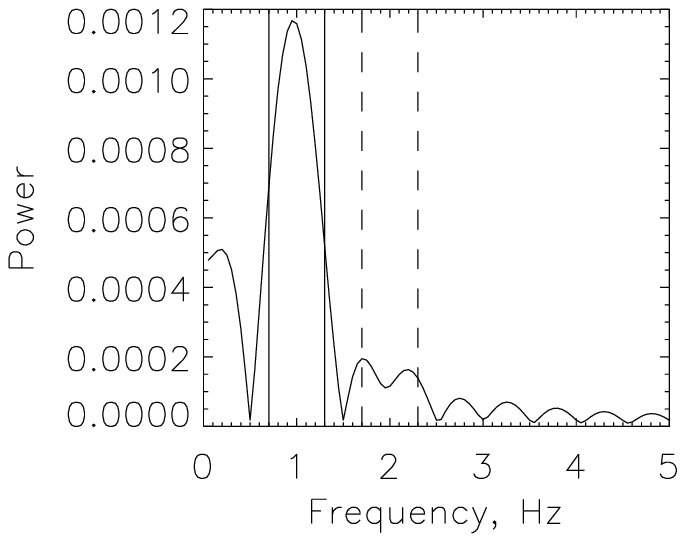}}
	\vspace{-5mm}
	\subfloat[Phase]{\includegraphics[width=0.3\textwidth]{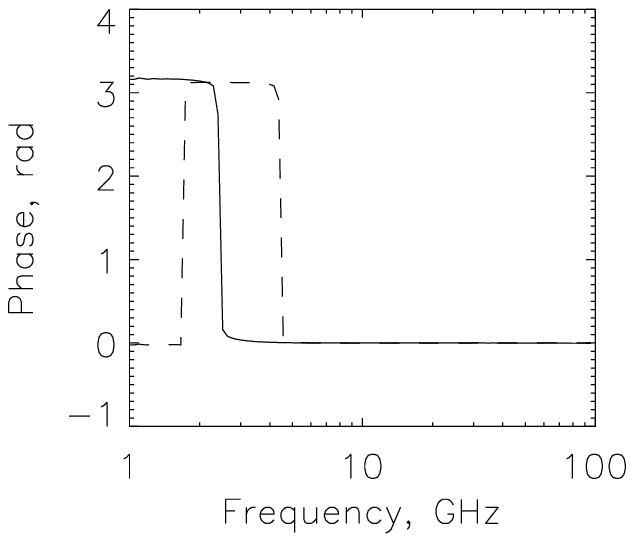}}
	\subfloat[Partial modulation]{\includegraphics[width=0.3\textwidth]{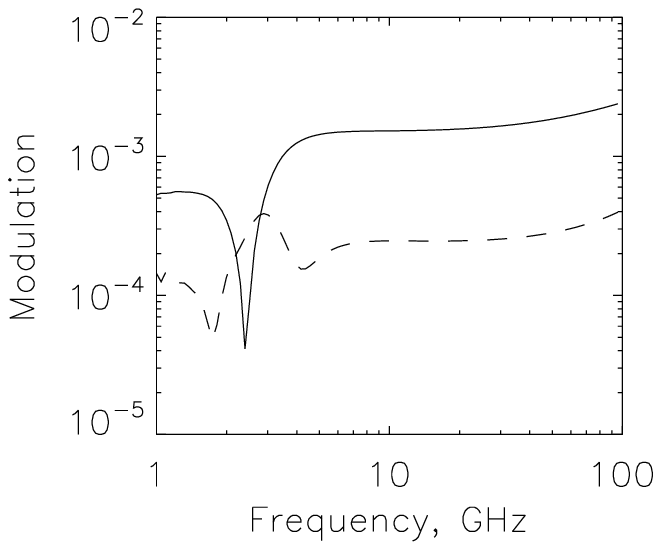}}
	\subfloat[Full modulation]{\includegraphics[width=0.3\textwidth]{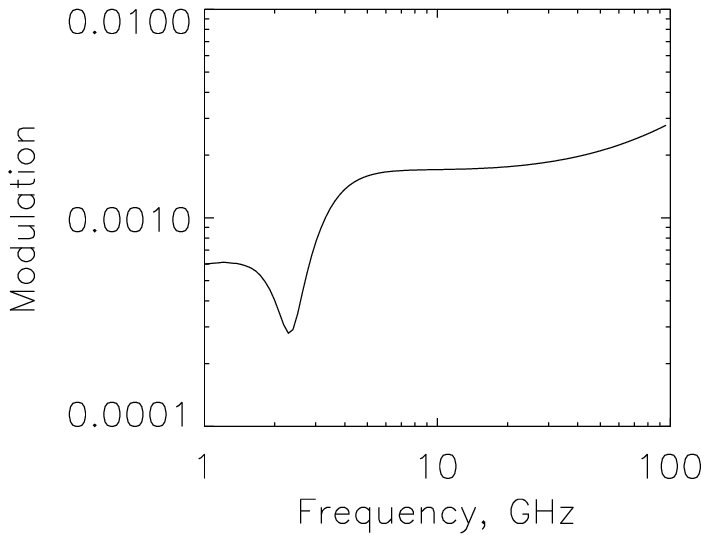}}

	\caption{Sausage mode, default parameters, oscillations of the flux density. {\bf(a)-(c)} Fourier spectra for the emission frequencies indicated, vertical lines mark off the  peak regions considered for partial modulation according to Eqn~(\ref{Eq_partial_m}); note enhanced second harmonics at 2.2~GHz. {\bf(d)} Relative phase of flux, solid line is the fundamental harmonics (1~Hz), dashed line is the second harmonics  (2~Hz); note the $\pi$ phase change around the spectrum peak frequency for both oscillation harmonics. {\bf(e)} Partial modulation amplitude for the indicated range of Fourier coefficients for two harmonics: the solid (dashed) line represents integration between the solid (dashed) vertical lines in panels {\bf(a)-(c)}.  {\bf(f)} Full modulation amplitude for the oscillating component; it has a deep minimum around the spectrum peak frequency likewise the partial modulation amplitude of the first harmonics; in contrast, the second harmonics modulation amplitude has a maximum at this range, see panel {\bf(e)}.}
	\label{sausage_default_osc_f_analysis}

\end{figure}

We start from the \emph{sausage mode} oscillations of a radio source with default-value parameters where the radio spectrum peak, see Figure~\ref{spectra_all}c, is formed by the optical thickness effect like in \citet{Fl_etal_2008}, while the Razin-effect is unimportant. Figure~\ref{sausage_default_osc_f} displays the oscillating components of the radio emission at three characteristic frequencies selected in the optically thick, spectrum peak, and optically thin regions of the spectrum, respectively. One can easily notice that although the modulation resembles a sinusoidal signal at both thick and thin frequencies, it departs from the sinusoidal greatly around the spectrum peak, Figure~\ref{sausage_default_osc_f}b, which means that even small-amplitude sinusoidal MHD oscillations can result in a nonlinear response producing higher harmonics in the radio light curves. This conclusion is explicitly confirmed by the Fourier spectra of these same oscillating components given in Figure~\ref{sausage_default_osc_f_analysis}a-c. Indeed, the first Fourier harmonics dominates the power spectra at thick (1.5 GHz) and thin (13.8 GHz) frequency, while the second harmonics dominates the power spectrum around the radio spectrum peak (2.2 GHz). The bottom raw of the same Figure displays results of the Fourier spectra analysis: the relative phase, partial, and full modulation amplitudes as functions of the radio frequency.

\begin{figure}[htp!]
	\centering	
	\subfloat[1.5 GHz]{\includegraphics[width=0.3\linewidth]{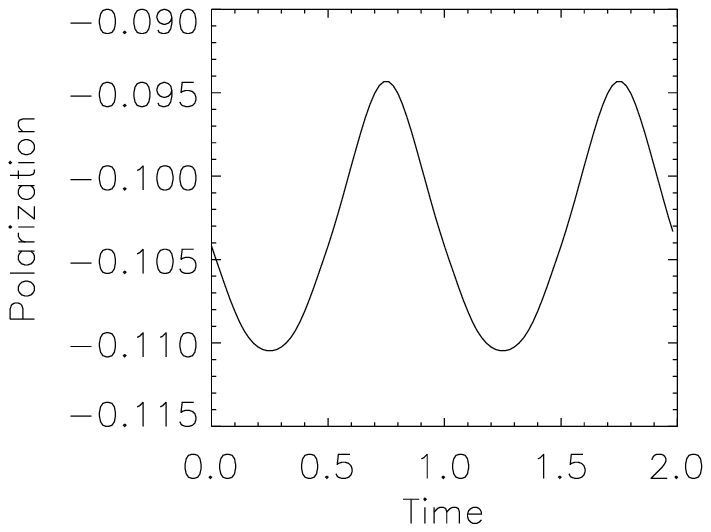}}
	\subfloat[2.2 GHz]{\includegraphics[width=0.3\linewidth]{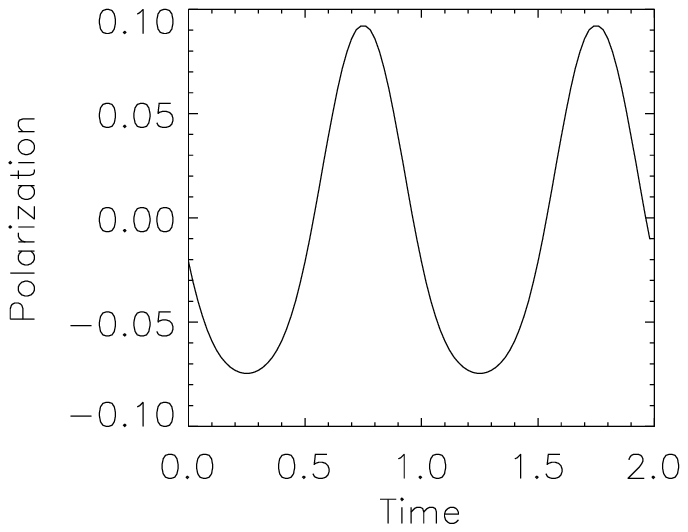}}
	\subfloat[13.8 GHz]{\includegraphics[width=0.3\linewidth]{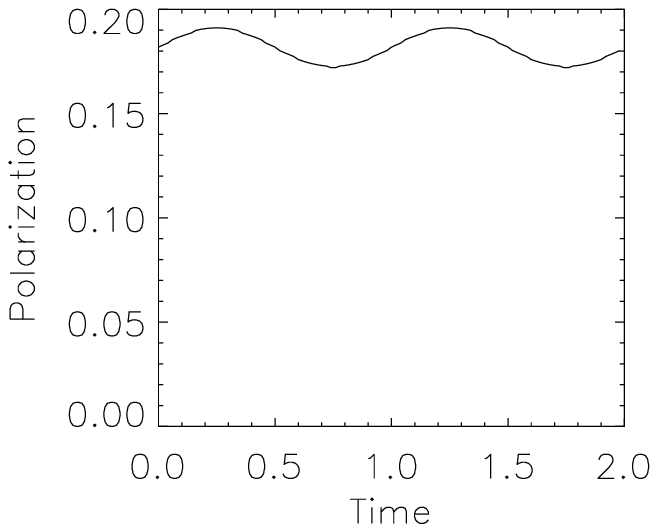}}
	\caption{Sausage mode, default parameters, oscillation of polarization at indicated frequencies. The oscillating components at low frequencies are $\pi$ out of phase relative to the high frequencies. The high- (low-) frequency polarization is in phase with the high- (low-) frequency flux density, see Figure~\ref{sausage_default_osc_f}.}
	\label{sausage_default_osc_p}
\end{figure}

\begin{figure}[hbp!]
	\centering
	\subfloat[1.5 GHz]{\includegraphics[width=0.3\textwidth]{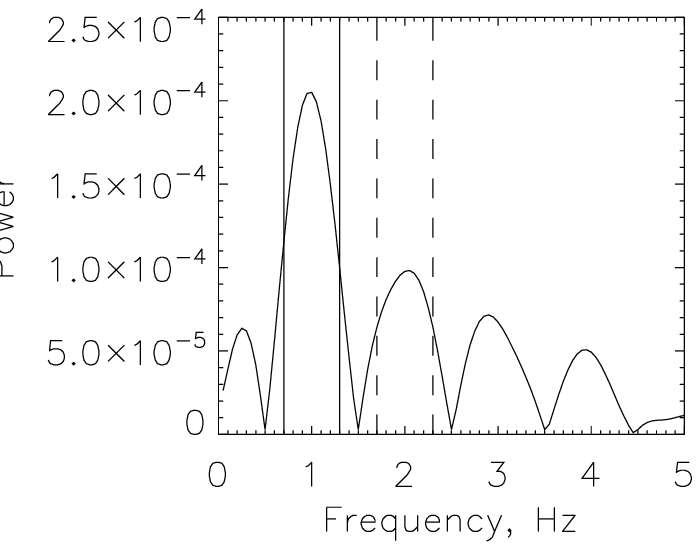}}
	\subfloat[2.2 GHz]{\includegraphics[width=0.3\textwidth]{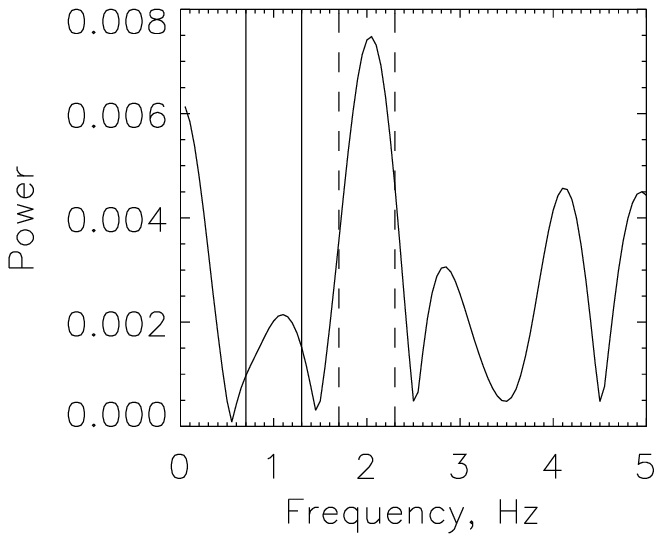}}
	\subfloat[13.8 GHz]{\includegraphics[width=0.3\textwidth]{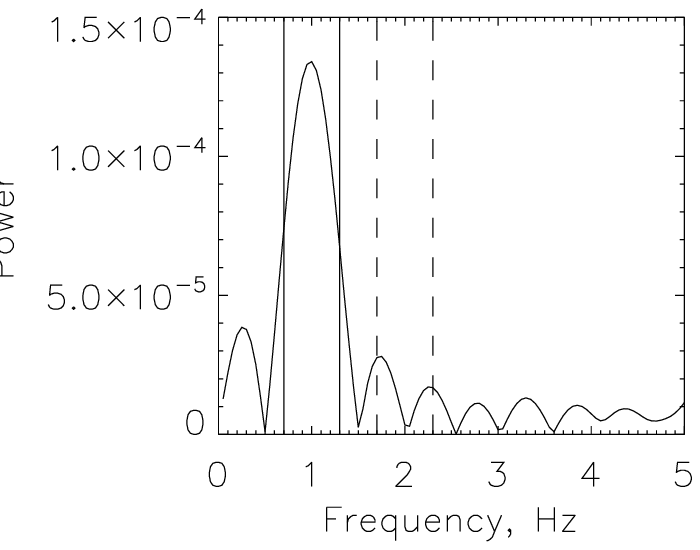}}
	\vspace{-5mm}
	\subfloat[Phase]{\includegraphics[width=0.3\textwidth]{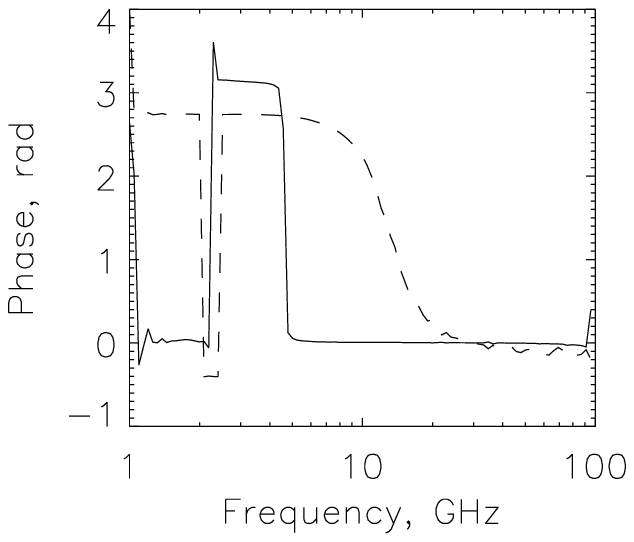}}
	\subfloat[Partial modulation]{\includegraphics[width=0.3\textwidth]{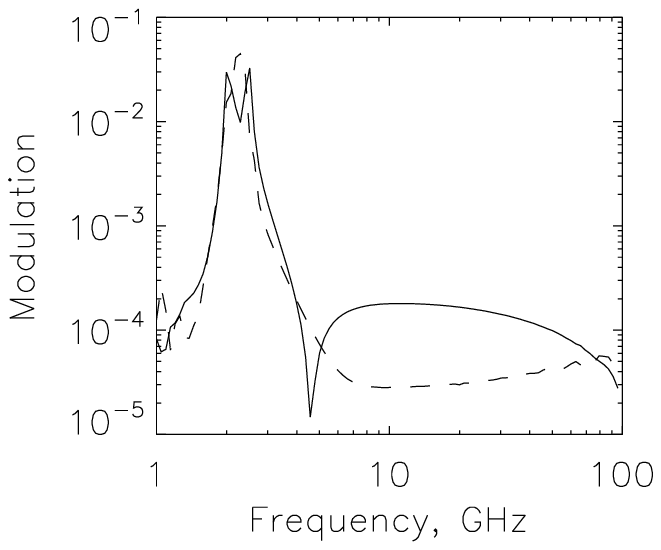}}
	\subfloat[Full modulation]{\includegraphics[width=0.3\textwidth]{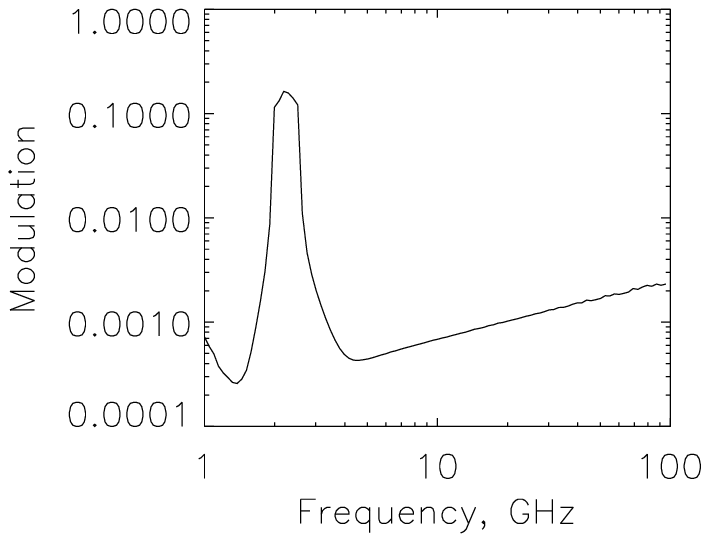}}
	\caption{Sausage mode, default parameters, oscillations of polarization. {\bf(a)-(c)} Fourier spectra for the emission frequencies indicated. Vertical lines mark off peaks considered for partial modulation, like in Figure~\ref{sausage_default_osc_f_analysis}. {\bf(d)} Relative phase of flux. Solid line is the fundamental harmonic, dashed line is the second harmonic. {\bf(e)} Partial modulation amplitude for the indicated range of Fourier coefficients for both harmonics. {\bf(f)} Full modulation amplitude for both harmonics. The modulation amplitudes reach the peak values around the spectrum peak frequency, where the sense of polarization changes from X-mode (at high frequencies) to O-mode (at low frequencies).}
	\label{sausage_default_osc_p_analysis}
\end{figure}

Apparently, the results are  consistent with those of the simplified modeling reported by \citet{Fl_etal_2008}: the oscillation phase changes by $\pi$ around the peak frequency, while the modulation amplitude displays a curve with a minimum around the frequency of the spectrum peak and with modulation at the optically thin part exceeding by a few times the modulation at the optically thick part. Thus, the performed here more advanced modeling confirms conclusion of \citet{Fl_etal_2008} that the QPPs analyzed by them cannot be caused by an MHD oscillation mode. Then, Figures~\ref{sausage_default_osc_p} and \ref{sausage_default_osc_p_analysis} display the same parameters but computed for the absolute value of the degree of polarization, which can be used for the QPP analysis if the polarization measurements are available.

The partial modulation amplitude of the degree of polarization comes to a peak at the  frequency range where the degree of polarization changes the sign and therefore it is around zero, while decrease strongly at low and high frequencies. At low and high frequencies respectively, the flux is in phase with the degree of polarization, Figures~\ref{sausage_default_osc_f} and \ref{sausage_default_osc_p}.

Another kind of observations, which can help in identifying the cause of QPPs, is the imaging spectroscopy data taken with spatial resolution. In particular, the detailed shapes of the modulation amplitudes for the total power data and single-pixel view are different from each other, see on-line supplement, so their comparison, when available, will yield an estimate of the source visible area pulsations, which is highly valuable in identifying the oscillation mode. The shape of the modulation amplitude curve is also different for the loss-cone distribution of the fast electrons compared with the isotropic case,  see on-line supplement; however, the difference is not that big to allow unambiguous distinction between the isotropic and anisotropic electron distributions at the current level of the available radio data.

The presence of low-frequency peaks in the emission is an indicator of high magnetic field. In emission from a pulsating source, these peaks will oscillate as well, Figure~\ref{spectra_all}a,~b. As there is a phase shift associated with each local peak frequency and also amplitude of the base of each such peak, the Fourier spectra in the region of the low-frequency peaks do not exhibit consistent Fourier peaks, see on-line supplement, which makes both the partial modulation amplitude and pulsation phase fluctuating at those low frequencies. The phase repeatedly changes by $\pi$ around the gyroharmonics peaks.  The full modulation amplitude naturally exhibits peaks at each low-frequency emission peak, while for high frequencies behaves similarly to the case of the default parameters; these low-frequency behavior makes the case of the strong magnetic field easily  distinguishable from the default case without gyroharmonics contribution.  The degree of polarization of the emission behaves in a similar way, see on-line supplement for more detail. 

\begin{figure}[hbp!]
	\centering
	\subfloat[1.5 GHz]{\includegraphics[width=0.3\textwidth]{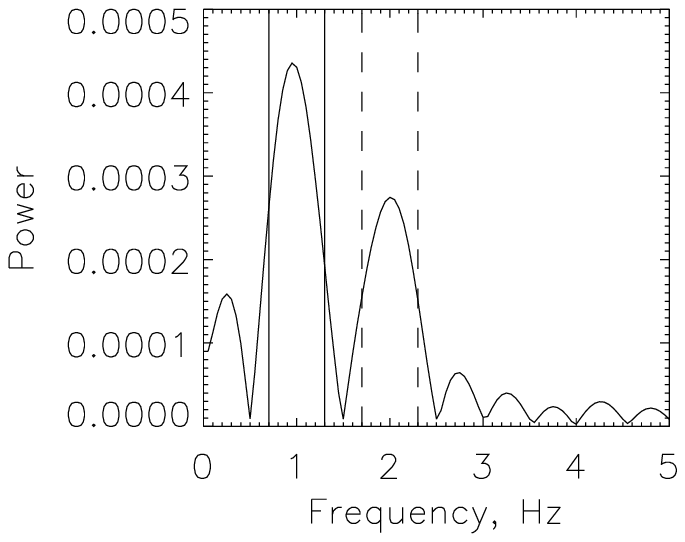}}
	\subfloat[6.6 GHz]{\includegraphics[width=0.3\textwidth]{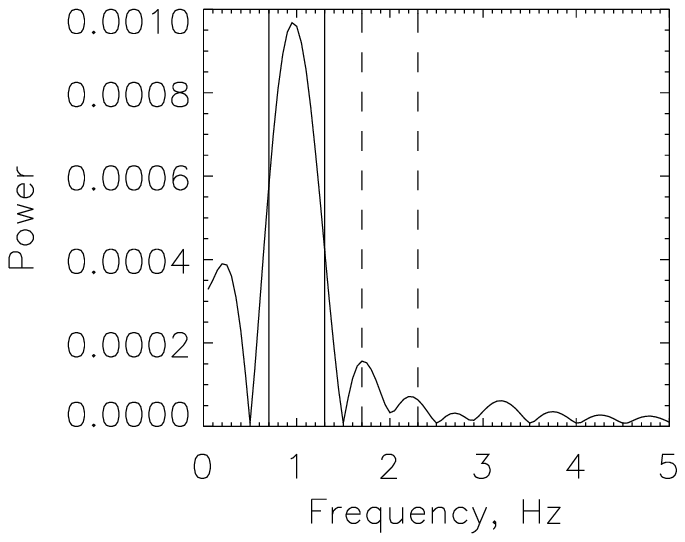}}
	\subfloat[13.8 GHz]{\includegraphics[width=0.3\textwidth]{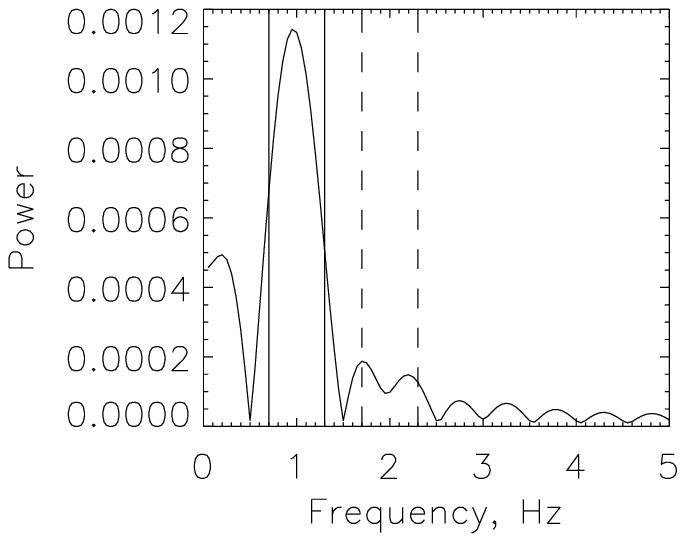}}
	
	\subfloat[\vspace{-0mm}Phase]{\includegraphics[width=0.3\textwidth]{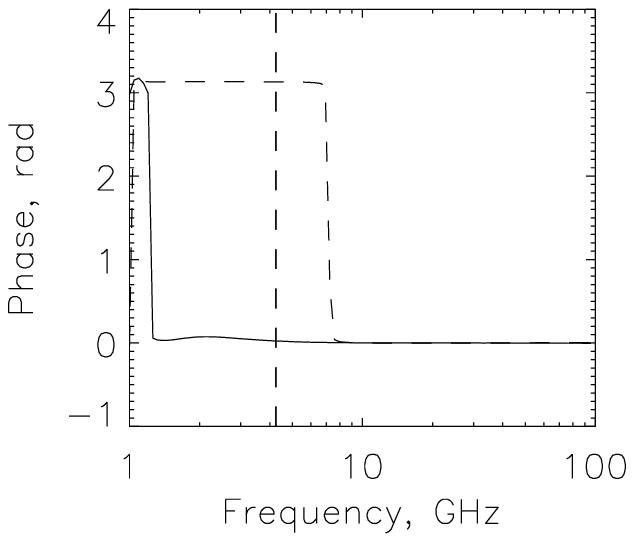}}
	\subfloat[Partial modulation]{\includegraphics[width=0.3\textwidth]{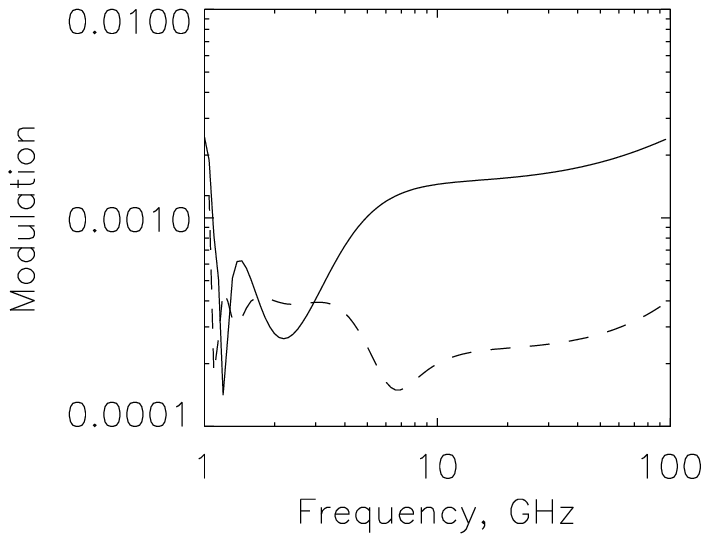}}
	\subfloat[Full modulation]{\includegraphics[width=0.3\textwidth]{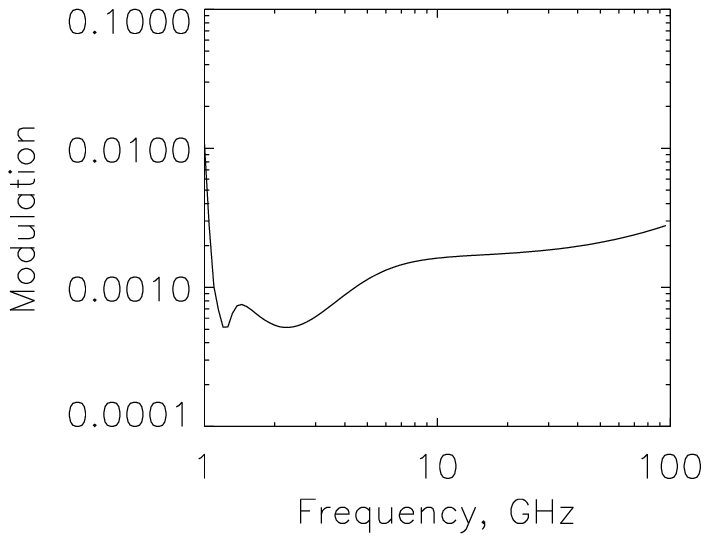}}
	
	\subfloat[1.5 GHz]{\includegraphics[width=0.3\textwidth]{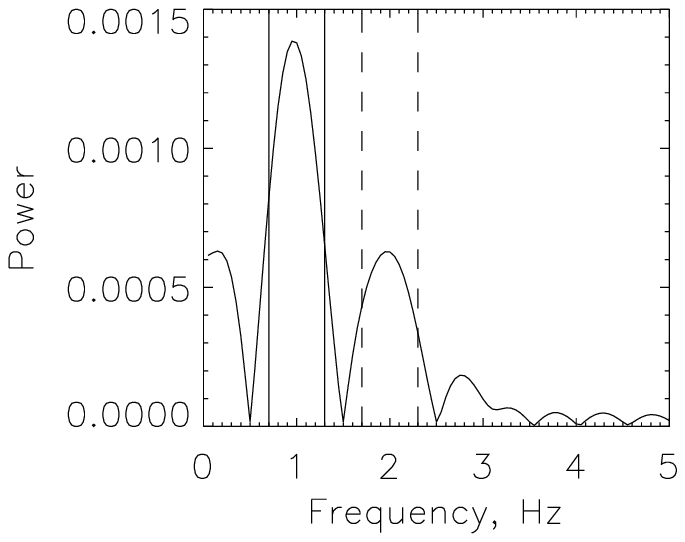}}
	\subfloat[6.6 GHz]{\includegraphics[width=0.3\textwidth]{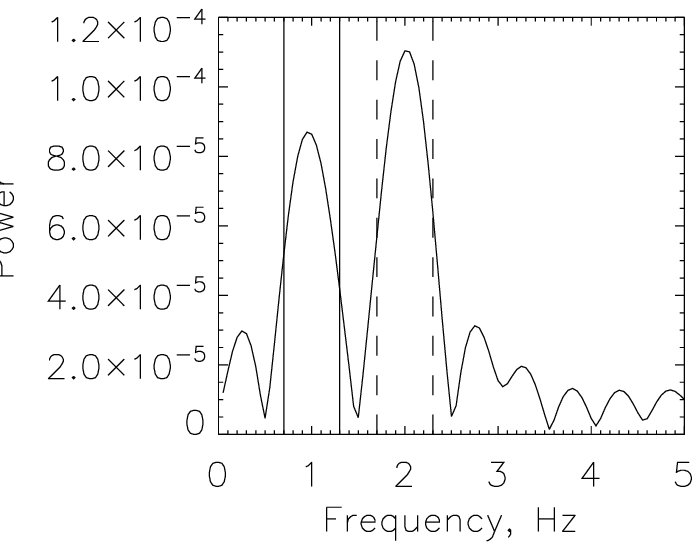}}
	\subfloat[13.8 GHz]{\includegraphics[width=0.3\textwidth]{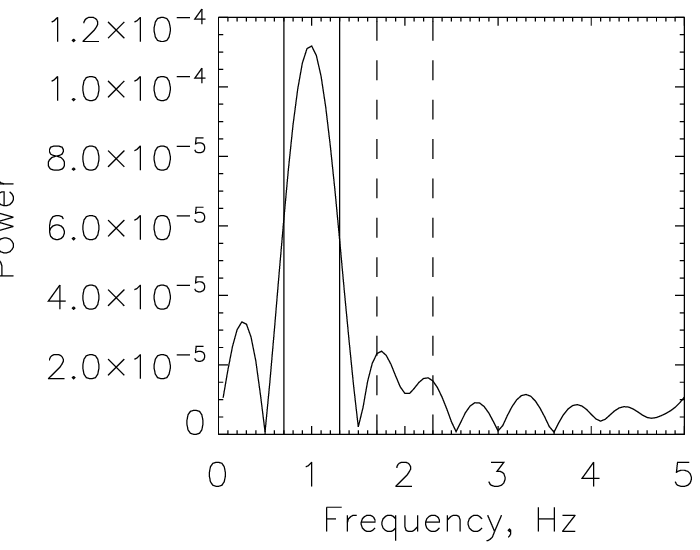}}
	
	\subfloat[Phase]{\includegraphics[width=0.3\textwidth]{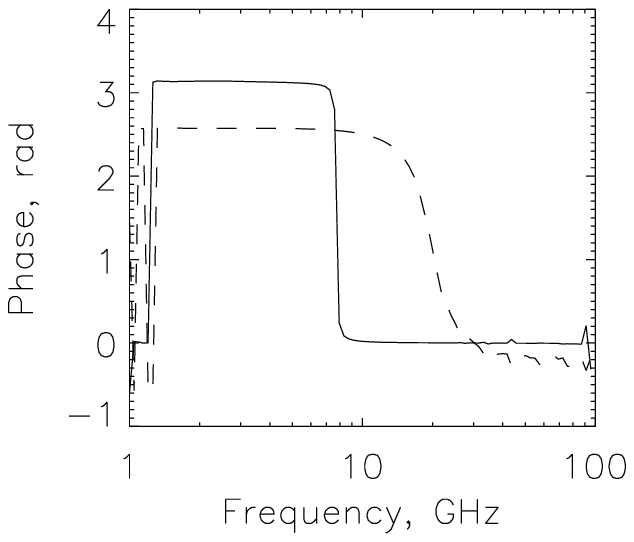}}
	\subfloat[Partial modulation]{\includegraphics[width=0.3\textwidth]{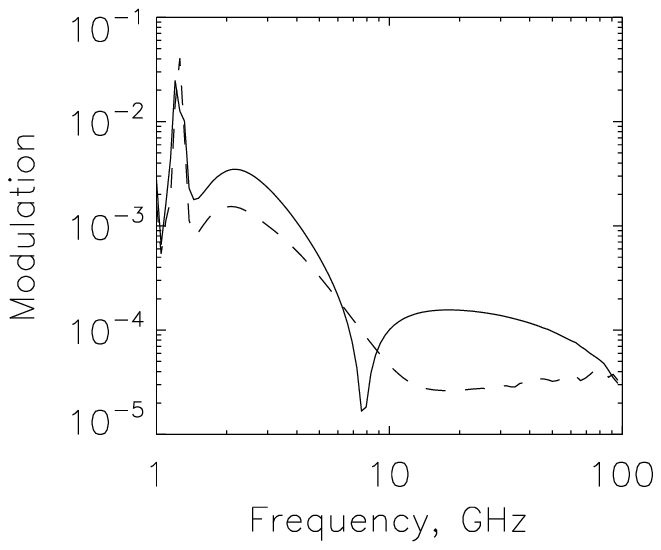}}
	\subfloat[Full modulation]{\includegraphics[width=0.3\textwidth]{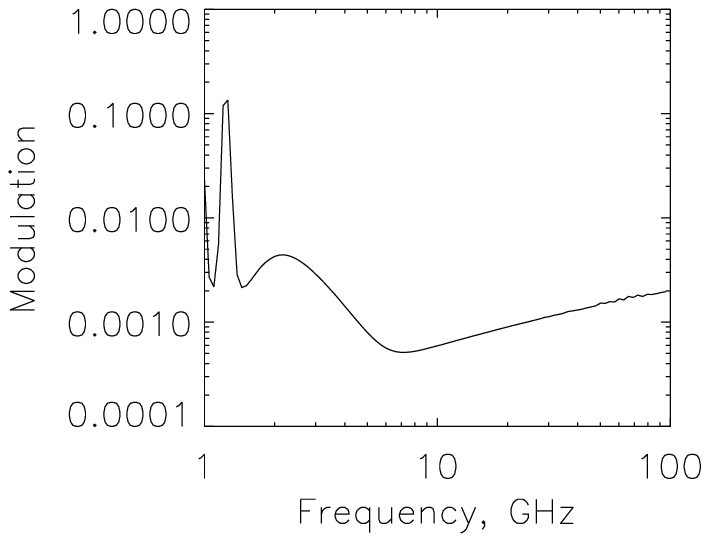}}

	\caption{Sausage mode, Razin effect, oscillations of flux and polarization: {\bf(a)-(f)} same as in Figures~\ref{sausage_default_osc_f_analysis} and {\bf(g)-(l)} same as \ref{sausage_default_osc_p_analysis}.  }
	\label{sausage_razin_all}
\end{figure}

The sausage mode modulation of a dense radio source, $n_0=10^{10}$~cm$^{-3}$, where the Razin effect is important, is clearly distinguishable from the case without the Razin effect even though the effect of the optical depth can still be important just around the spectral peak as in 
Figure~\ref{spectra_all}g, where the Razin cutoff frequency  \citep[e.g.,][]{Simon69}
\begin{equation}
\nu_R\approx 15\frac{n_e}{B\sin\theta}
\end{equation}
is marked by a vertical dashed line slightly below the spectrum peak frequency.
The relative phase of the flux displays no phase shift across the entire radio spectrum, i.e., oscillation of both low- and high- frequency light curves occur strictly in phase. The modulation amplitude differs from the case of the default parameters; although it still rises at the high frequencies it does not display any flat region at the low frequencies, see 
Figure~\ref{sausage_razin_all}.

The \emph{kink mode}, which we consider here only for the line-of-sight oscillations, $\cos\varphi=1$, in a dense plasma volume  (see on-line supplement for all other parameter regimes), is easily distinguishable from the sausage mode. Indeed, unlike the sausage mode, here the flux and polarization at a given frequency are $\pi$ out of phase with each other. The relative phase of the flux light curves is constant across the entire spectrum like for the sausage mode oscillations in a dense source. However, dependence of the modulation amplitude on frequency is strongly different from that for the sausage mode. Now the modulation amplitude is large at low frequencies, then decreases and remains relatively constant at high frequencies.

The behavior of the kink mode oscillations is particularly interesting in case of the default parameters, see on-line supplement. The explicit effect of the viewing angle oscillations and corresponding source visible area oscillations are counter directed at the optically thick part of the spectrum, which has two observable consequences: (i) there is no phase shift across the entire spectrum (unlike the sausage mode, default parameters) and (ii) the modulation amplitude decreases toward low frequencies, while for the sausage mode case the modulation amplitude reaches a minimum around the radio flux spectrum peak. Remarkably, that with spatially resolved observations, the flux density from a fixed single pixel is unaffected by the visible area oscillations; as a result
the kink mode oscillations observed from a single pixel do produce a $\pi$ phase shift in both flux and polarization in the region of the spectrum peak; the shape of the modulation amplitude function is also significantly different from that from the total power data, see on-line supplement. Thus, comparison of these measures for the total power and spatially resolved data is highly informative to distinguish the kink mode oscillations of the flaring loop. Note, that the oscillations of the source visible area has no effect on the optically thin part of the spectrum; in particular, it is the case for the entire spectrum when the Razin effect forms the spectrum peak, see above.
	
	
	
The \emph{torsional mode}, like the kink mode, is to the first order an oscillation of the viewing angle, although, unlike the kink mode, the source visible area and depth are not affected by the oscillations. For this reason, the torsional mode oscillations are distinct from the kink mode in case of the default parameters, while it is undistinguishable from the line-of-sight kink-mode oscillations at the optically thin part of the radio spectrum, in particular, in the Razin effect regime; see on-line supplement.

	
\section{Search for MHD oscillations in flaring microwave bursts}

The most detailed database of the solar microwave bursts recorded with reasonably high spectral resolution (40 frequencies across the range of $1-18$~GHz) was obtained by OVSA during solar cycle \# 23. \citet{Nita_etal_2004} presented a thorough statistical analysis of a subset of the entire database composed of 412 radio bursts recorded during 2001--2002. Given the subset is sufficiently large and the corresponding total power data are well calibrated, we restrict our search of the MHD oscillation signatures to this subset; we note, however, that a number of radio bursts with QPPs suitable for a case study analysis was recorded after 2002, see example studied in detail by \citet{Fl_etal_2008}.

To identify the events candidates for further quantitative analysis we manually search throughout the dynamic spectra and light curves of all 412 bursts and picked up those displaying QPPs. Surprisingly, the total number of the events-candidates is very small: less than 10 events (i.e., less than 2.5\%) demonstrate apparent QPPs visible in the dynamic spectrum; about 40 events ($\sim10\%$) display light curves with QPPs at a few frequencies.

We performed analysis of the QPPs in those events demonstrating pulsations in the dynamic spectra because this means stronger signal modulation, which is apparently more suitable for the quantitative study, while retain all other (potentially pulsating) events for further analysis. However, in most of the events selected for our study we were unable to firmly determine the cause of the observed QPPs. Below we describe  two  events in which detection of MHD oscillation signatures is likely,  August 28, 2002 21:41 UT and July 20, 2002 21:27 UT. 

\subsection{August 28, 2002}

\begin{figure}[hp!]
	\centering
	\includegraphics{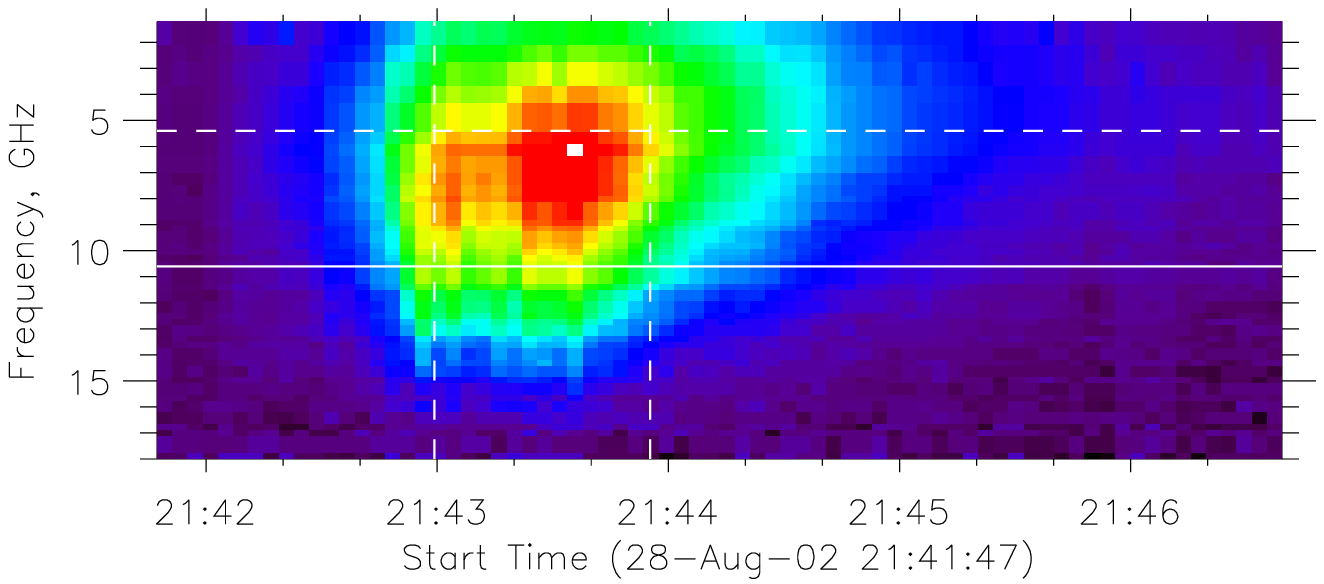}
	\includegraphics{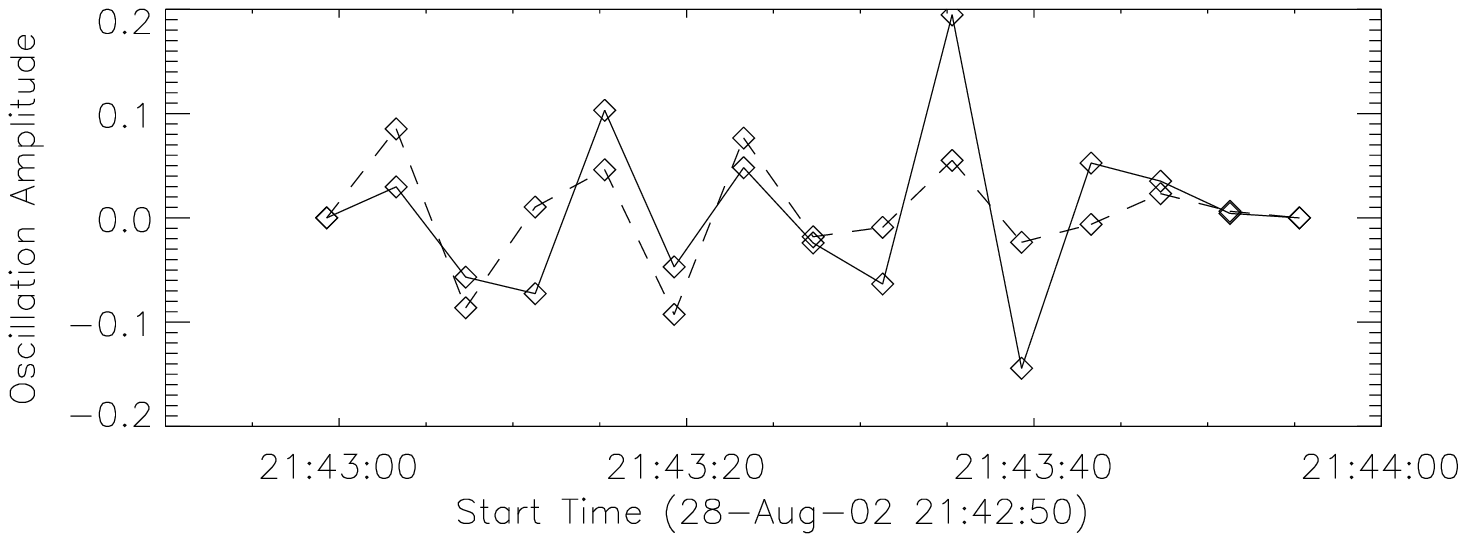}
	\caption{August 28, 2002. Top: Dynamic spectrum of the flux density normalized by the absolute peak flux of 212 sfu. The vertical stripe-like features in the dynamic spectrum are indicative of prominent QPPs in the event. Vertical dashed lines indicate examined region.  Bottom: oscillating components at frequencies 5.4 (dashed) and 10.6~GHz (solid), shown in the top panel by the dashed and solid lines, respectively. The low- and high- frequency oscillations occur roughly in phase.}
	\label{20020828_spectrum}
\end{figure}

The event of August 28, 2002 21:41 UT, Figure~\ref{20020828_spectrum}, is particularly interesting because it was reported among other partly occulted events \citep{Krucker_Lin_2008}. The fact that a part of the flaring loop is occulted is helpful for further analysis as effect of the loop nonuniformity on the radio emission is reduced (no contribution from the occulted part is recorded). This implies that a uniform source model adopted for our analysis is better justified for interpretation of the occulted flares compared with an average flare. Nevertheless, even for this favorable geometry the radio spectrum of this flare is noticeably broader than a typical spectrum from a uniform source indicating that the source inhomogeneity still plays some role here.

\begin{figure}[hp!]
	\centering
	\subfloat[3.4 GHz]{\includegraphics[width=0.35\linewidth]{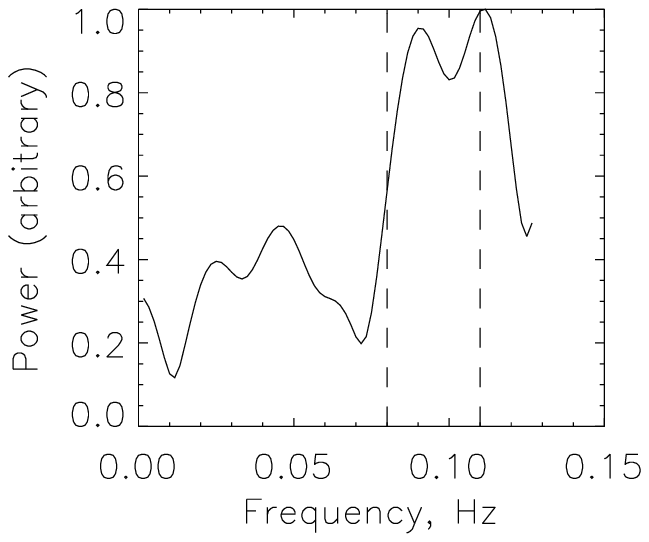}}
	\subfloat[5.4 GHz]{\includegraphics[width=0.35\linewidth]{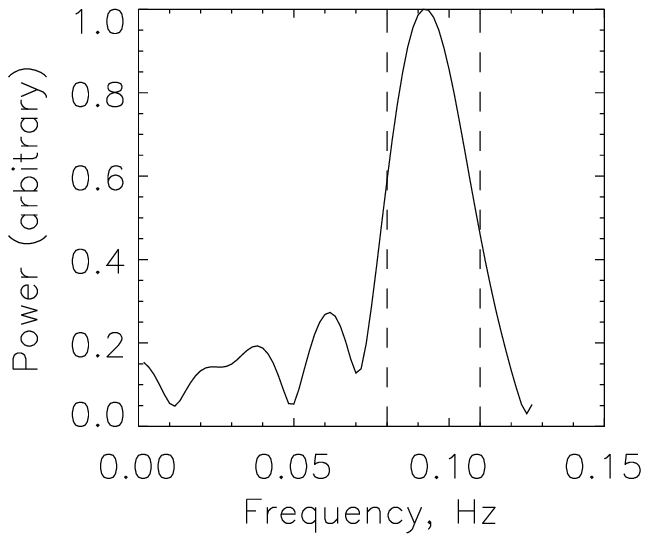}}
	\subfloat[10.6 GHz]{\includegraphics[width=0.35\linewidth]{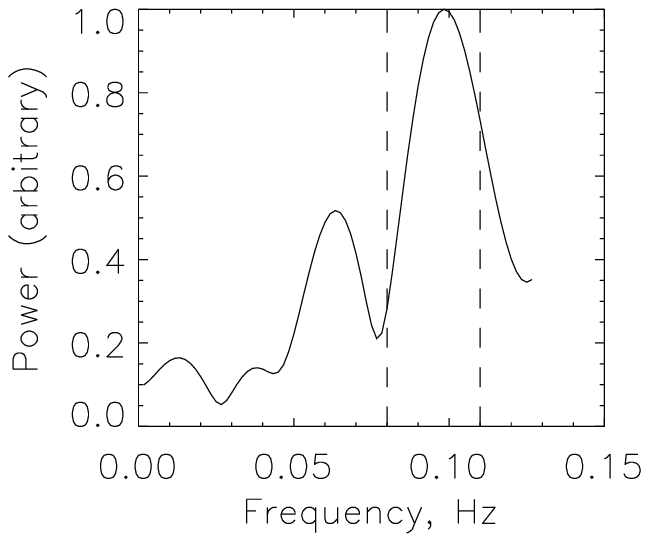}}
	\caption{August 28, 2002. Fourier spectra for indicated frequencies. Dashed lines indicate peak regions considered for further partial modulation calculation, see Figure~\ref{20020828_analysis} below.}
	\label{20020828_fourier}
\end{figure}

\begin{figure}[hp!]
	\centering
	\subfloat[Phase]{\includegraphics[width=0.35\linewidth]{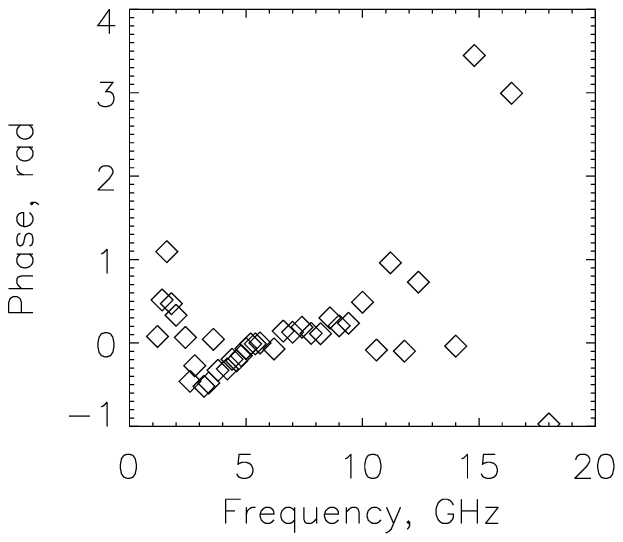}}
	\subfloat[Partial modulation]{\includegraphics[width=0.35\linewidth]{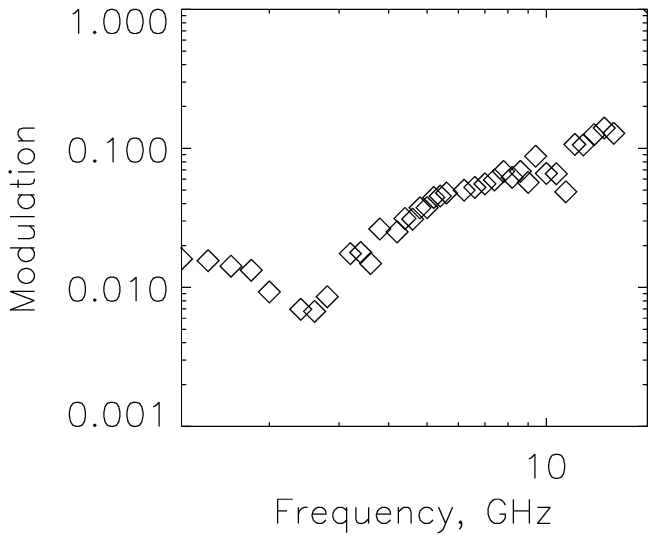}}
	\subfloat[Full modulation]{\includegraphics[width=0.35\linewidth]{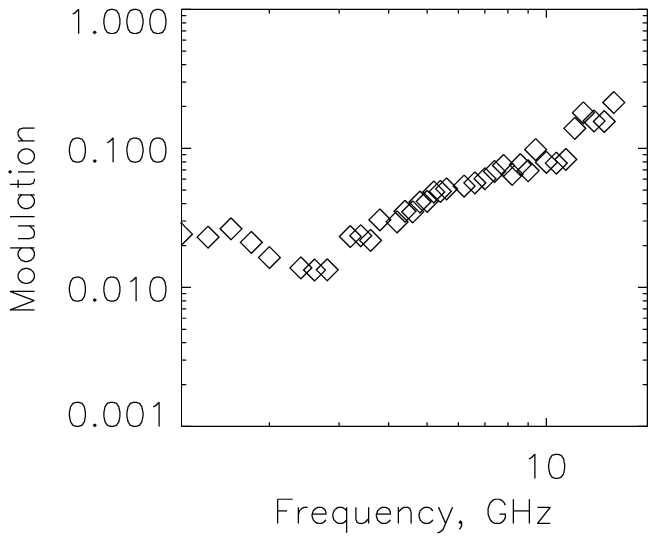}}
	\caption{August 28, 2002. \textbf{(a)}: Relative phase of flux oscillations at 0.09~Hz. \textbf{(b)}: Partial modulation amplitude at the harmonic range selected for integration by the dashed lines in Figure~\ref{20020828_fourier}. \textbf{(c)}: Full modulation amplitude during the time segment selected for the analysis; see dashed vertical lines in Figure~\ref{20020828_spectrum}, top.}
	\label{20020828_analysis}
\end{figure}

The Fourier spectra of the oscillating component of the radio light curves display consistent peaks around 11~s throughout the entire radio spectrum, Figure~\ref{20020828_fourier}, although their shapes are not strictly identical to each other. The phase of the pulsations, however, displays neither exact constancy nor distinct $\pi$ shift around the radio spectrum peak, Figure~\ref{20020828_analysis}a. On the other hand, we note that the scatter of individual data points, e.g., at frequencies 10 to 15 GHz, is of the same order of magnitude as the range of overall phase variation over all frequencies. Thus, given overall weakness of the oscillating component, $\pm(10-20)$~sfu, implying relatively large error in the phase determination, we conclude that the phase does not display significant variations over the spectral range considered. As has been shown in \S~\ref{procedure}, the phase constancy across the spectrum is indicative of the Razin-effect importance in forming the radio emission, which means a relatively dense flaring loop involved in the QPPs or the kink mode oscillations, which produce constant phase in both high- and low- density cases.

To distinguish between possible MHD modes capable of producing the observed QPPs, we address the modulation amplitude as a function of emission frequency, Figure~\ref{20020828_analysis}b,~c. We note that both partial and full modulation amplitudes have similar shape, which is consistent with that produced by the sausage mode oscillations but deviates from those produced by the kink (for either high- or low- density cases)  or torsional modes. We conclude that the kink and torsional modes are irrelevant for the QPPs from the event studied, while the sausage mode oscillations is still a plausible candidate.

\begin{figure}
	\centering
	\includegraphics[width=0.6\linewidth]{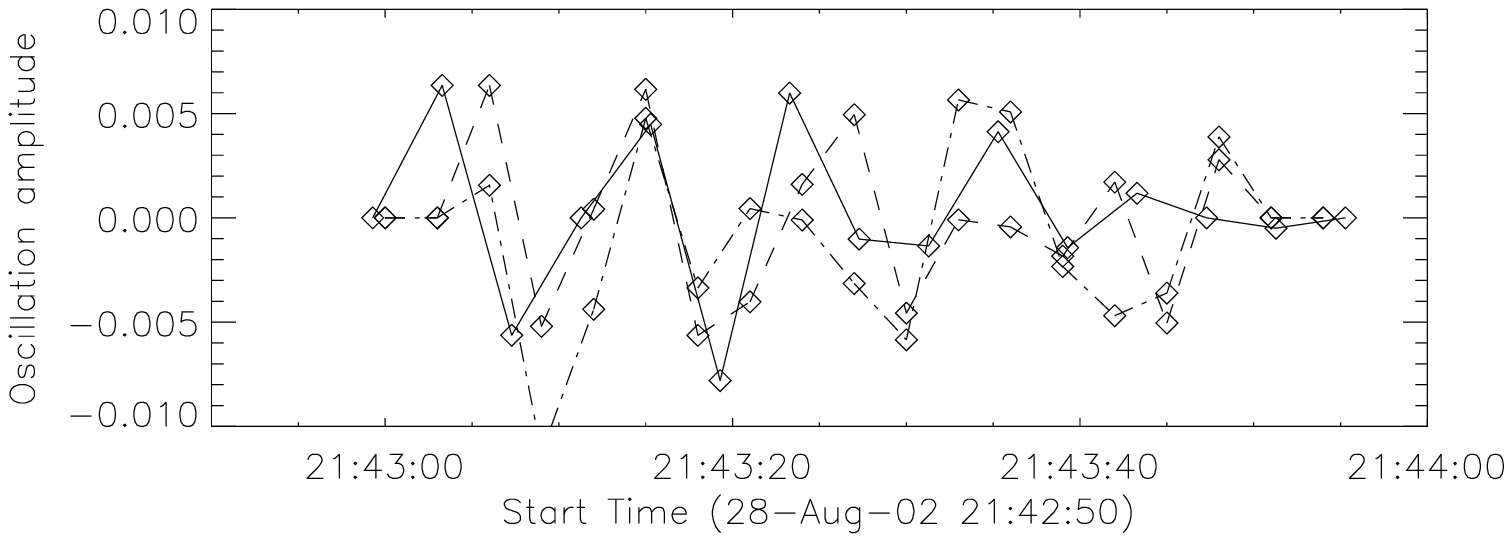}{a}
	\includegraphics[width=0.6\linewidth]{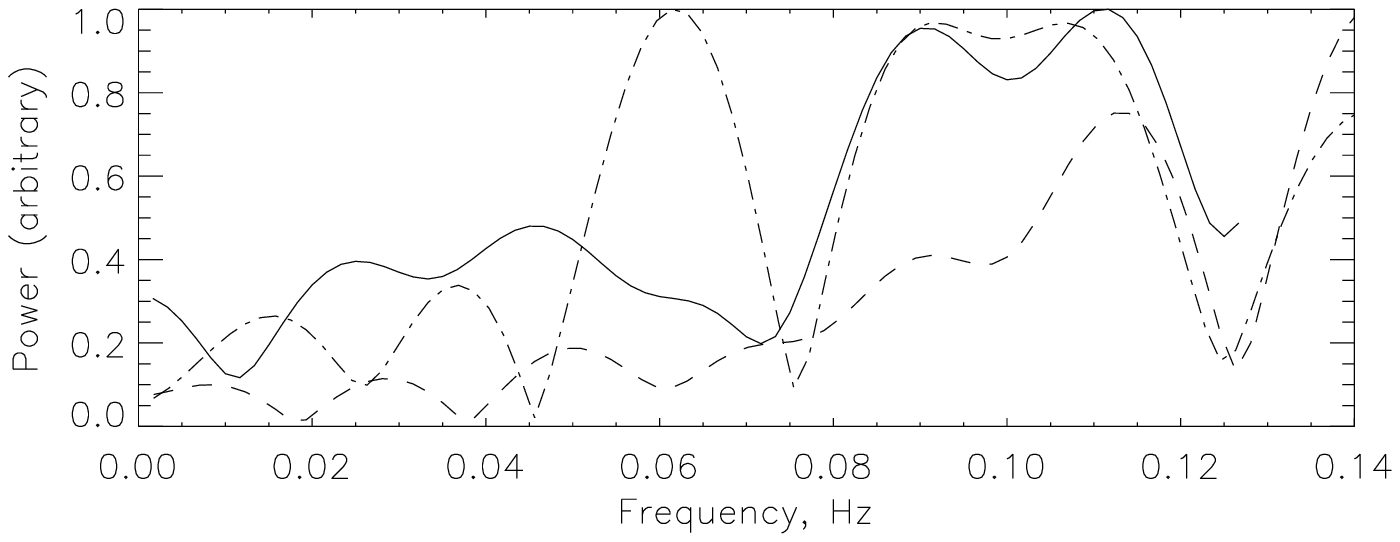}{b}
	\includegraphics[width=0.6\linewidth]{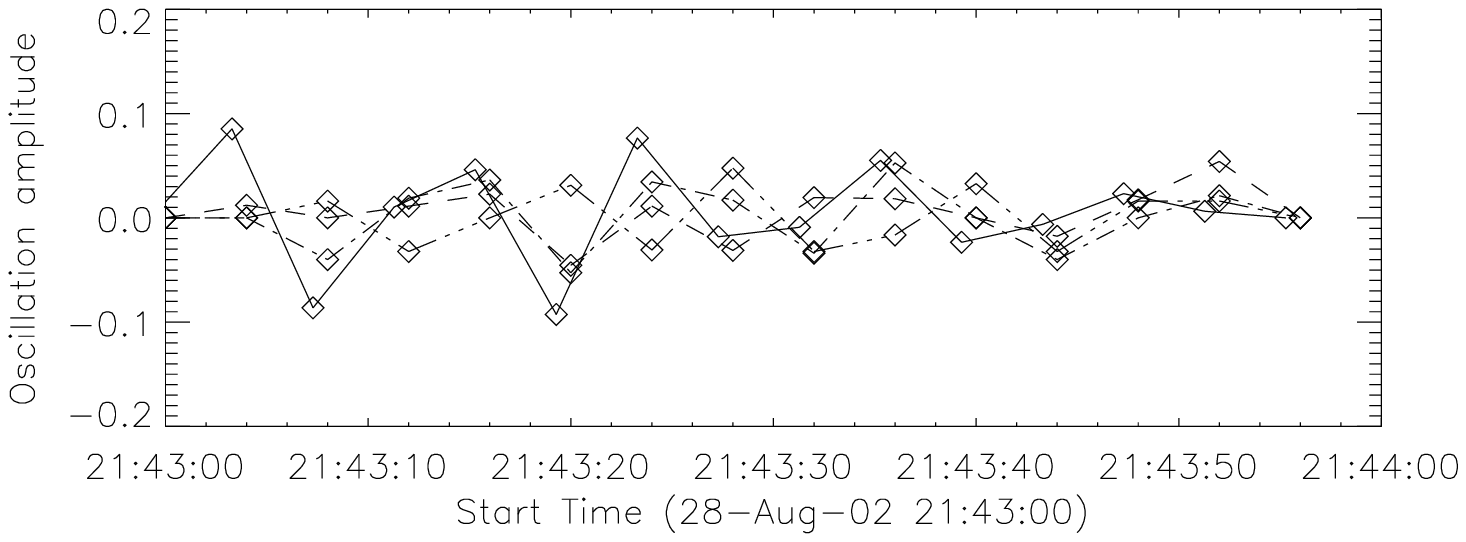}{c}
	\includegraphics[width=0.6\linewidth]{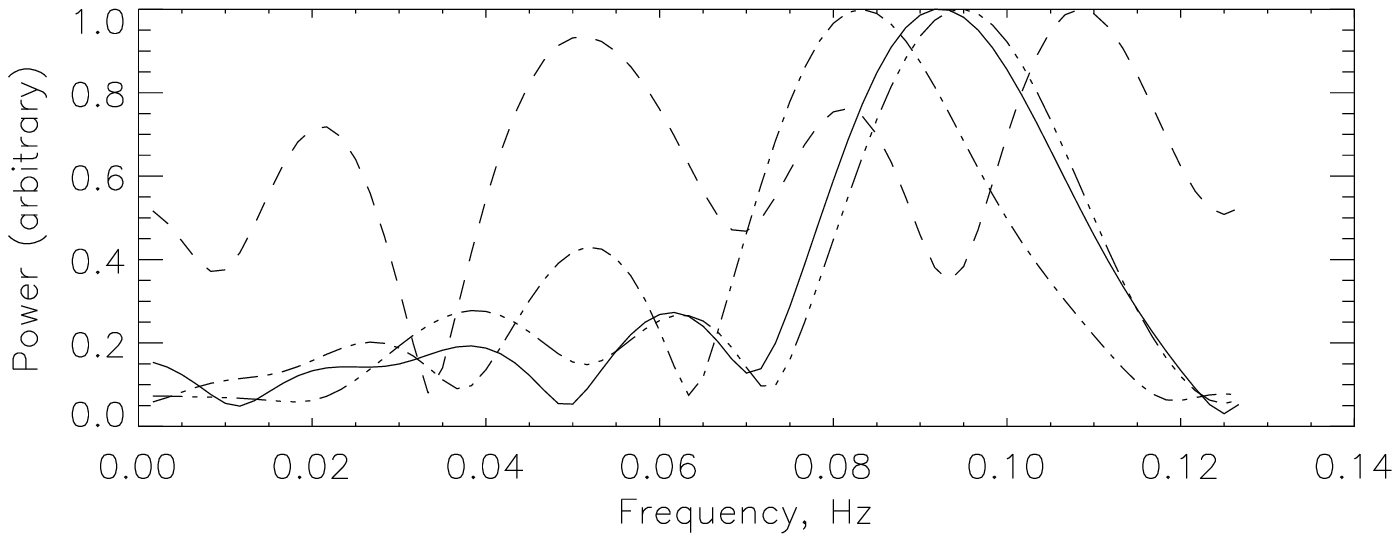}{d}
	\caption{August 28, 2002. \textbf{(a)}:  Dashed and dash-dotted lines show the oscillating components of the low- and high- energy GOES channels, respectively; the solid line shows the radio light curve at 3.4~GHz for comparison. \textbf{(b)}: the corresponding Fourier spectra for the considered oscillating components. Note overall similarity of the broad double peaks around 0.1~Hz in the Fourier spectra of both SXR channels and the radio light curve.
\textbf{(c)}: Dashed, single-dotted, and triple-dotted lines show oscillation components for the RHESSI channels 12-25 keV, 25-50 keV and 50-100 keV respectively. The solid line represents the radio emission at 10.6~GHz, which is roughly $\pi$ out of phase compared with the highest RHESSI channel, 50-100~keV. \textbf{(d)}: Fourier spectra of the above channels. Note that the solid (radio) Fourier spectrum is almost identical to the triple-dotted (highest energy RHESSI channel) Fourier spectrum.}
	\label{20020828_goes} 
\end{figure}

To further evaluate this hypothesis we recall that sausage mode oscillations stimulate corresponding oscillations of the plasma density, temperature, and fast electron distribution function, which must result in QPPs in both SXR and HXR emission. To test this expectation we use the data available from GOES and RHESSI, respectively.
Figure~\ref{20020828_goes}a displays the oscillating components of two GOES SXR channels superimposed on a optically thin radio oscillating component; these light curves, although not identical, do display an overall consistent behavior, which is further quantified by the corresponding Fourier spectra given in Figure~\ref{20020828_goes}b. The Fourier spectra of the GOES channels do display broad peaks roughly coinciding with the radio peak around 10~s period (note a double peak structures in both 3.4~GHz and GOES spectra around $\nu=0.1$~Hz). However, the spectra do not coincide precisely with each other and the high GOES channel displays an additional peak around $\tau=17$~s not present in two other light curves.

Figures~\ref{20020828_goes}c,~d display oscillation amplitudes and Fourier spectra for the RHESSI data. Likewise the SXR, the HXR oscillations show overall similarity with the radio QPPs with the highest energy channel, 50--100~keV, being roughly $\pi$ out of phase with the radio light curve and lower-energy RHESSI channels. As for the Fourier spectra, that for the highest RHESSI channel  coincides almost identically with the Fourier spectrum of radio oscillations at a high frequency, 10.6~GHz, Figure~\ref{20020828_goes}d, while two other, lower-energy channels, 12--25 \& 25--50~keV, display a peak at somewhat lower frequency, $\nu\approx0.08-0.085$~Hz and a secondary peak at $\nu\approx0.05$~Hz. We suppose that oscillations of thermal plasma may contribute to this additional peak and if so, it must correspond to the peak in the GOES Fourier spectrum at $\nu\approx0.06$~Hz, which gives us a rough estimate of the QPP frequency determination accuracy of $\delta\nu\approx0.01$~Hz. With this accuracy, the Fourier peaks at $\nu\approx0.09$~Hz and $\nu\approx0.08$~Hz are identical, which confirms a common nature of the radio and HXR pulsations, while the thermal plasma (seen via GOES and lower-energy RHESSI channels) displays some additional QPPs ($\tau\sim17$~s) not seen in the microwave or HXR emission.

We conclude that the sausage mode oscillations of a dense flaring loop could be the cause of the observed QPPs in the August 28, 2002 21:41 UT event, although the presented evidence is not fully unambiguous. In contrast, a model of QPPs produced by pulsating acceleration/injection of the fast electrons is less consistent with the data because this model does not predict QPPs of the SHR emission. We note that there is no radio polarization data for this event, which, if present, might further help in identifying the relevant oscillation mode.


\subsection{July 20, 2002}

\begin{figure}[hp!]
	\centering
    \includegraphics{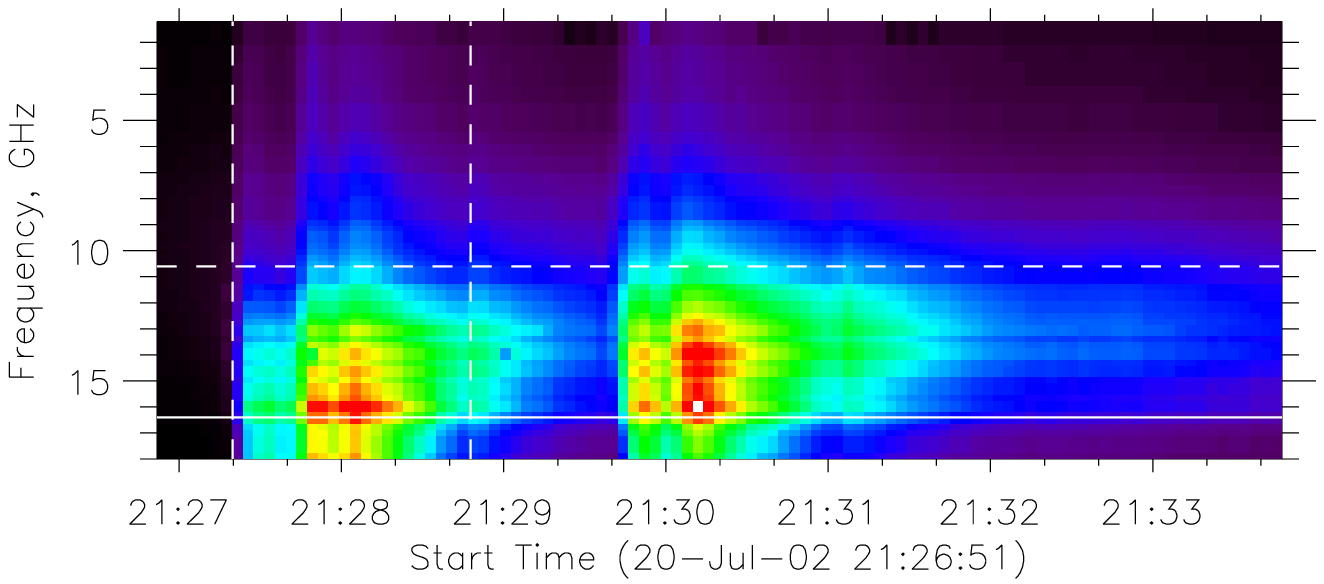}
	\includegraphics{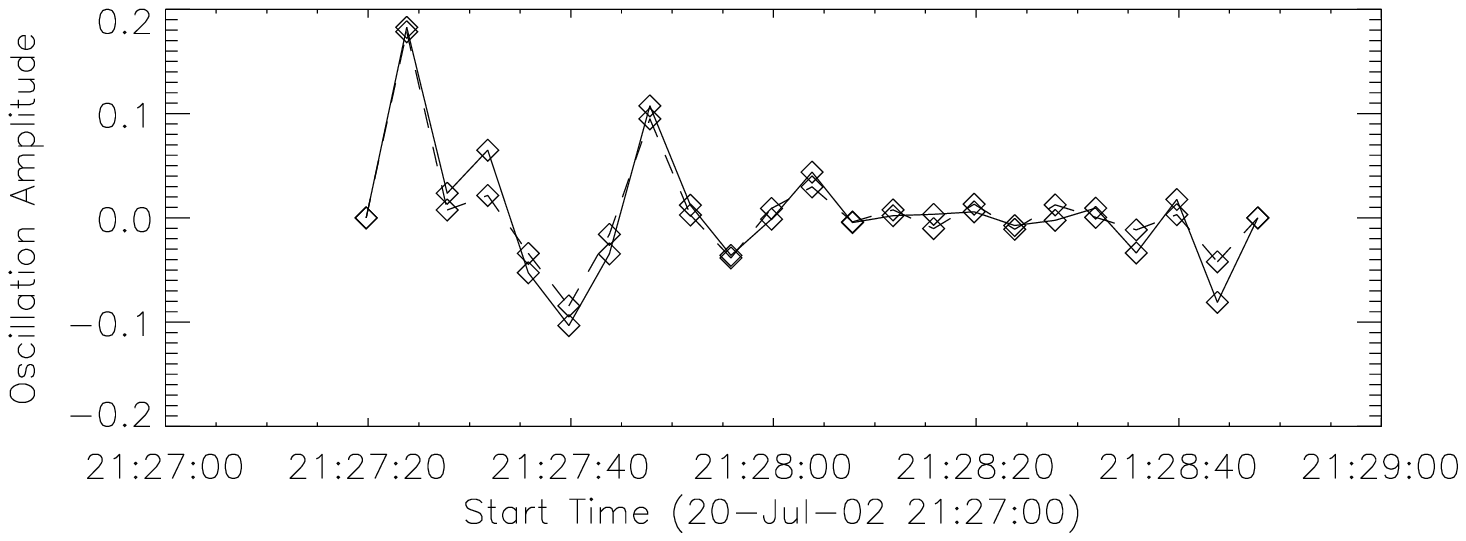}
	\caption{July 20, 2002. Top: Dynamic spectrum of the  flux density  normalized by the absolute peak flux of 9858 sfu. Vertical dashed lines indicate examined region.  Bottom: Oscillating components for two radio frequencies, 10.6 and 16.4~GHz, shown by the dashed and solid lines, respectively.}
	\label{20020720_spectrum}
\end{figure}

\begin{figure}[hp!]
	\centering
	\subfloat[3.4 GHz]{\includegraphics[width=0.35\linewidth]{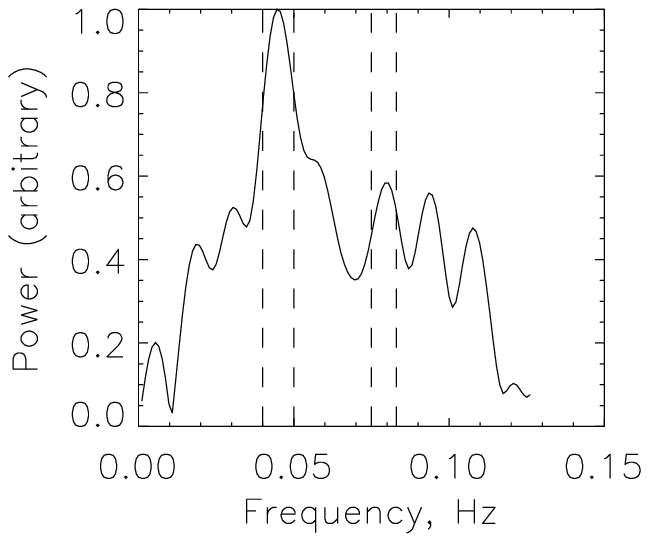}}
	\subfloat[6.6 GHz]{\includegraphics[width=0.35\linewidth]{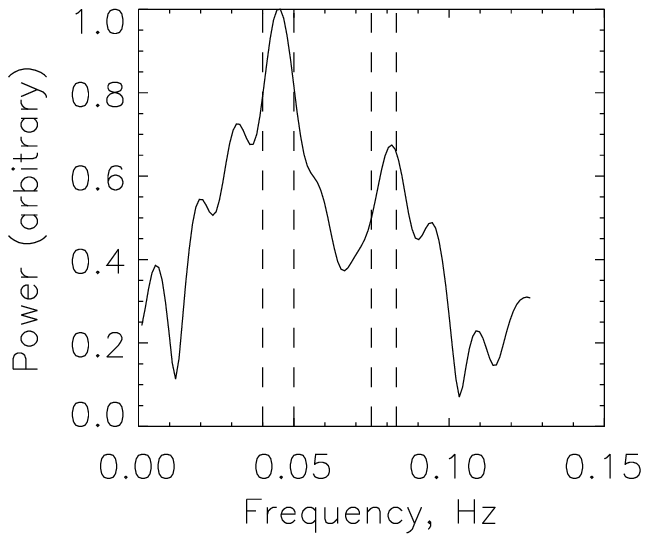}}
    \subfloat[10.6 GHz]{\includegraphics[width=0.35\linewidth]{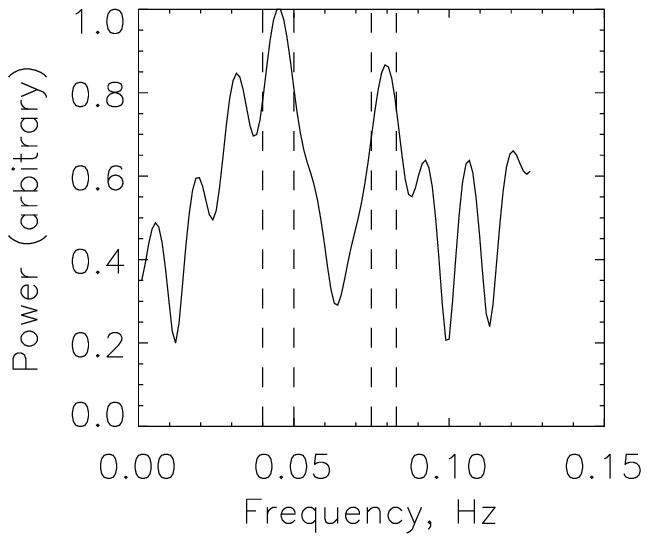}}
	\caption{July 20, 2002. Fourier spectra for indicated frequencies. Vertical dashed lines indicate the peak regions considered for calculation of partial modulation amplitudes.}
	\label{20020720_fourier}
\end{figure}

\begin{figure}[hp!]
	\centering
	\subfloat[Phase 22~s]{\includegraphics[width=0.4\linewidth]{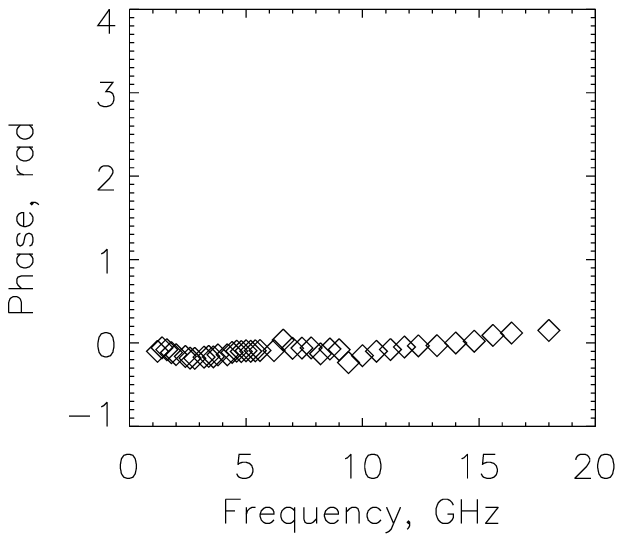}}
	\subfloat[Partial modulation 22~s]{\includegraphics[width=0.4\linewidth]{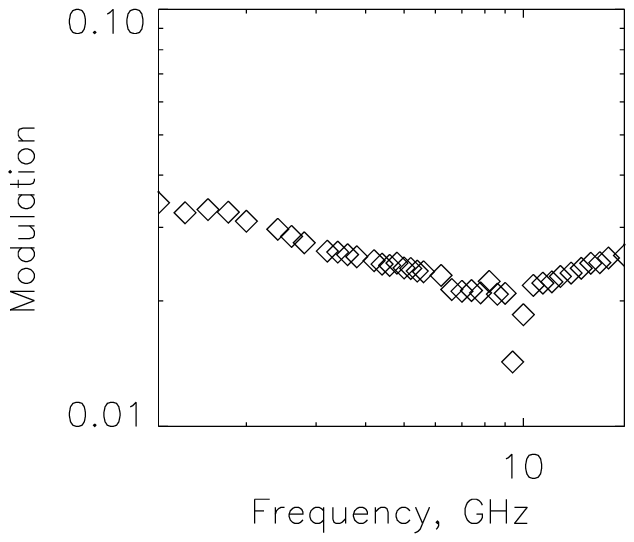}}
	\vspace{-7mm}
	\subfloat[Phase 13~s]{\includegraphics[width=0.4\linewidth]{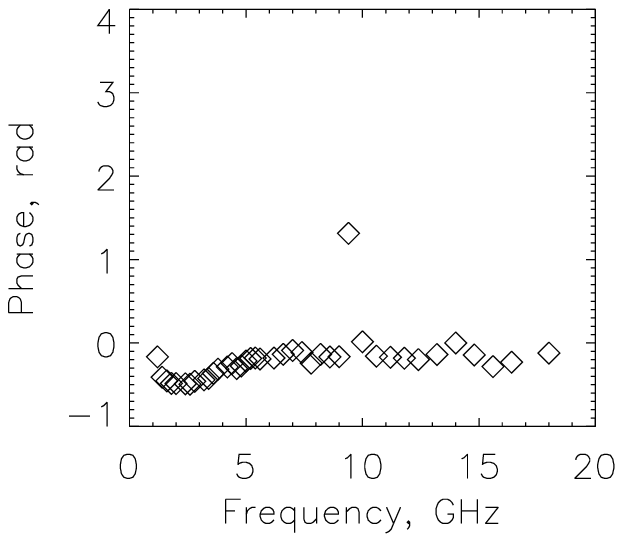}}
	\subfloat[Partial modulation 13~s]{\includegraphics[width=0.4\linewidth]{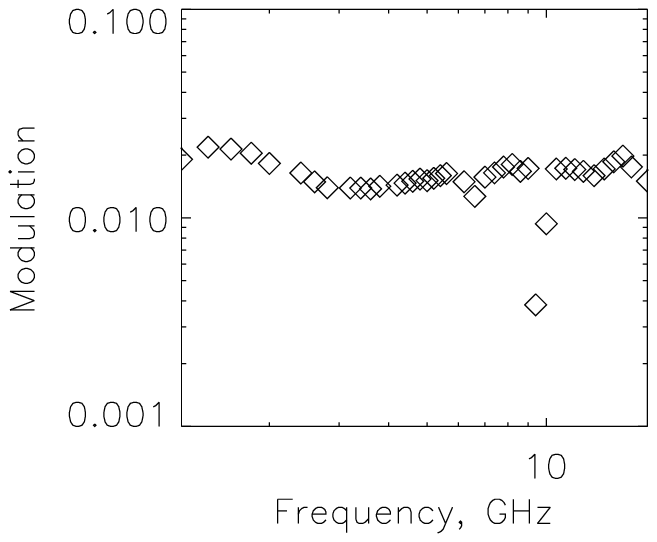}}
	\vspace{-5mm}
	\subfloat[Full modulation]{\includegraphics[width=0.45\linewidth]{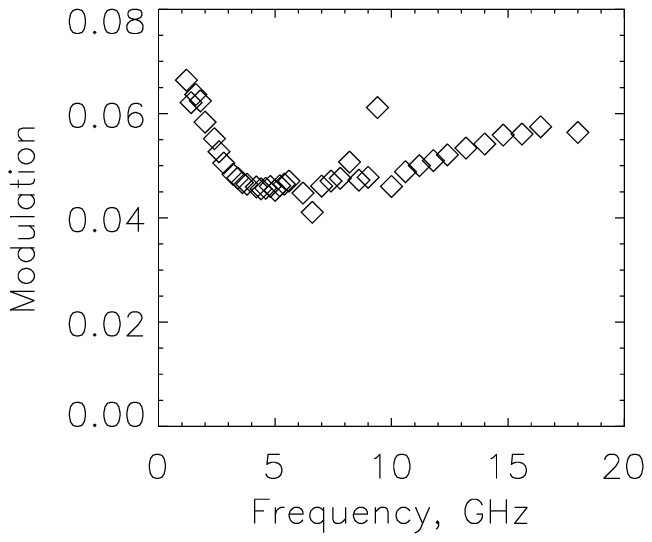}}
	\caption{July 20, 2002. Top: Phase and partial modulation amplitude for the Fourier peak at $\sim22$~s ($\nu=0.045$~Hz). Middle: Phase and partial modulation amplitutde for the Fourier peak at $\sim13$~s ($\nu=0.077$~Hz). Bottom: Full modulation amplitude shown here in linear scale giving  better view of the modulation amplitude variation with frequency. General similarity  of the partial modulation amplitudes for the two Fourier peaks and of the full modulation amplitude is suggestive of these two oscillations to be driven by the same mechanism.}
	\label{20020720_analysis}
\end{figure}

The event of July 20, 2002  21:27 UT, Figure~\ref{20020720_spectrum}, is different from the event of August 28, 2002 in many respects. It is a usual, not occulted, flare with relatively large radio flux up to 10$^4$~sfu at the absolute peak implying higher accuracy of the derived measures characterizing QPPs.
The Fourier spectra display two  consistent peaks across the entire radio spectrum, one at $\sim22$~s ($\nu\approx0.045$~Hz) and the other at $\sim13$~s ($\nu\approx0.08$~Hz), Figure~\ref{20020720_fourier}. The relative phases of the QPPs at both oscillations are almost constant across the radio spectrum,  Figure~\ref{20020720_analysis}a,~c, again indicative of the Razin-effect importance in the radio spectrum formation implying relatively high plasma density in the flaring source.    The partial modulation amplitude retains a roughly similar concave-up shape in both oscillations, Figure~\ref{20020720_analysis}b,~d, same as the full modulation amplitude Figure~\ref{20020720_analysis}e. Perhaps, this suggests that the $\sim13$~s oscillation is a higher harmonic in the Fourier decomposition of the main $\sim22$~s oscillation. Comparison of the observed and model modulation amplitudes does not favor the sausage mode oscillations in this case, while consistent with either torsional or line-of-sight kink mode oscillations, which are very hard to distinguish from each other with the developed approach.

\begin{figure}
	\centering
	\includegraphics{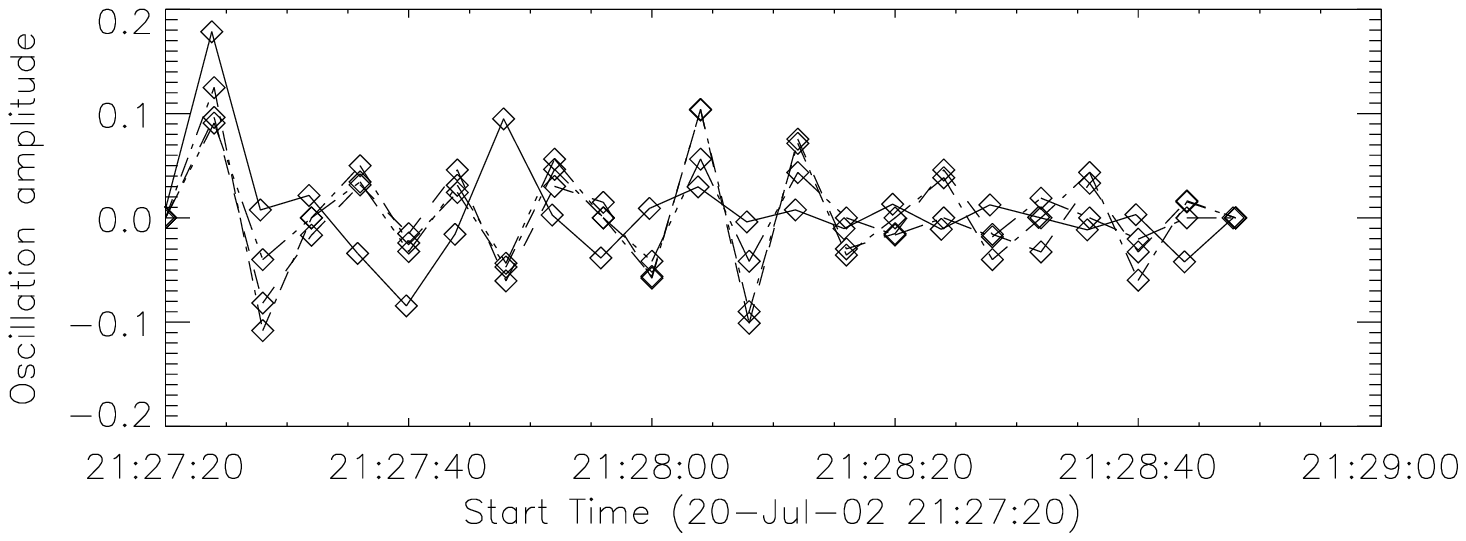}
	\includegraphics{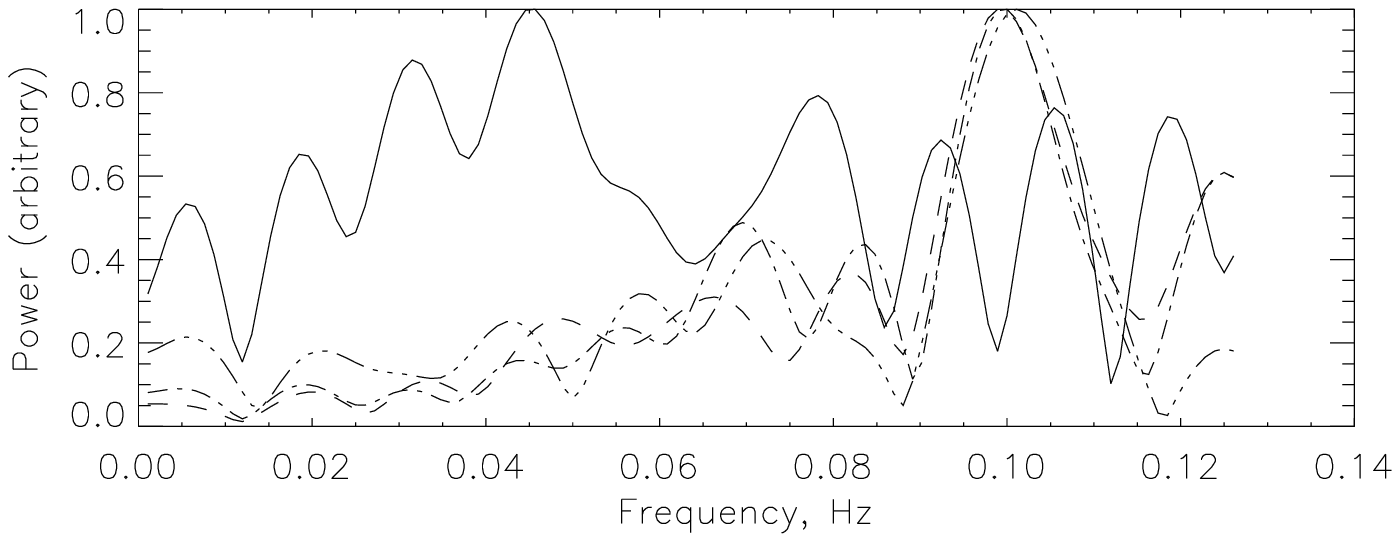}
	\caption{July 20, 2002. Top: Dashed, single-dotted, and triple-dotted lines show oscillation components for the RHESSI channels 12-25 keV, 25-50 keV, and 50-100 keV respectively. The solid line represents the radio emission at 16.4~GHz. Bottom: Fourier spectra of the above channels. }
	\label{20020720_rhessi}
\end{figure}

Those 'transverse' MHD oscillations do not involve pulsations of the loop cross-sectional area or volume and, accordingly, on the plasma density, temperature, and fast electron distribution. Therefore, we do not expect any SXR or HXR QPPs coherent with the radio QPPs. Indeed, no relevant oscillations is detected in the GOES light curves (not shown here), both of which are very smooth in this event. The HXR emission observed with RHESSI does display fluctuations with some peaks in the light curves being in phase with the radio light curves, while others not in phase or opposite phase. As a result, the main significant Fourier peak of HXR oscillations, being consistent at different HXR channels, $\nu\approx0.1$~Hz, does not coincide with any of the significant Fourier peak of the radio data, Figure~\ref{20020720_rhessi}, suggesting that there are oscillations in the HXR emission, but likely from a different cause. 
We conclude that kink or torsional oscillations of a dense flaring loop is a possible cause of the radio QPPs observed from the event of July 20, 2002, while the data is more difficult to reconcile with the idea of fast electron QPPs within a simple single-loop source model adopted in this paper: the fast electron pulsations would imply coherent in-phase radio and HXR QPPs, like in \citet{Fl_etal_2008}, which is not observed in the event considered here. We cannot exclude, however, a possibility of a more complex source geometry in which less straightforward relationships between radio and X-ray QPPs are well possible. To address this question, a detailed imaging spectroscopy data is needed at a currently unavailable level.

\section{Discussion}

We have described a detailed modeling of the MHD loop oscillation effect on the GS microwave emission from solar flares. Unlike most of other studies concentrating on observed QPP periods and their possible correspondence to one or another MHD mode, we address the question how the oscillation phase and corresponding modulation amplitude behave as a function of the radio frequency, given that these QPP measures were shown to be highly efficient to identify the true cause of QPPs \citep{Fl_etal_2008}.
We found that the QPP relative phase behavior is distinctly different for the cases of relatively strong magnetic field when the phase repeatedly changes at each of the local peaks at low gyroharmonics  and of  either
dense or tenuous flaring loops  with a lower field, displaying no or $\pi$ phase shift around the (single) spectral peak, respectively. Furthermore, the sausage mode oscillations are distinguishable from the kink and torsional mode oscillations using the frequency behavior of the modulation amplitude.
The kink and torsional modes are distinguishable from each other in the case of tenuous source, when oscillations of the source visible area dominate the kink mode effect on the radio QPPs in the optically thick region.
However, the kink and torsional mode oscillations in a dense loop are indistinguishable from each other as far as linear order effect (which is identical for these modes) dominates the radio emission response on these MHD oscillations.

We have used the modeling results to analyze the OVSA microwave burst database and search for the expected MHD oscillation signatures in some bursts displaying QPPs. Although we did not find many promising events-candidates, we did identified two reasonably good examples in which detection of one or another MHD oscillation mode is likely. Specifically, in one of them the radio QPPs might be caused by the sausage mode, while in the other one by either kink or torsional mode. Comparison of the radio QPPs with SXR and HXR QPPs favors the MHD loop oscillations for these two events compared with the fast electron injection/acceleration oscillations detected in another event using a similar analysis \citep{Fl_etal_2008}. However, this conclusion must be taken with some caution (given incompleteness and insufficient accuracy of the data) and must be further evaluated and tested on other events with radio QPPs; we plan to perform a more comprehensive statistical study based on full available OVSA database elsewhere including analysis of the polarization data when available, because as evident from the above modeling, such a study would benefit greatly from  accurate polarization data.

We anticipate that similar data analysis will be even more important for soon-to-be-available imaging spectroscopy data after on-going OVSA expansion (EOVSA) has been completed. In particular, all EOVSA antennas will be equipped with dual-polarization feeds for simultaneous right-hand and left-hand circular polarization measurements. Furthermore, the expansion will greatly improve the image quality and  the frequency resolution, which are all needed to perform a more meaningful analysis and yield more reliable  conclusions on the cause of radio QPPs in each particular case.

\acknowledgements
This work was supported in part by NSF grants AGS-0961867, AST-0908344, and NASA grant NNX10AF27G to New Jersey
Institute of Technology, and by the RFBR grants 09-02-00226, 09-02-00624, and 11-02-91175. The authors are sincerely thankful to our colleagues, Professors Dale Gary and Gelu Nita for their valuable discussions on the subject and persistent help.
GM is grateful to  Center for Solar-Terrestrial Research at the New Jersey Institute of Technology, where he had a summer studentship under NSF REU Supplement funding to grant  ATM0707319, and is also indebted to Professor Lance Cooper and Celia Elliott of the University of Illinois at Urbana-Champaign for their guidance in writing and their editing.

\bibliographystyle{apj}
\bibliography{main,pulsations,fleishman}

\clearpage

\clearpage

\clearpage

\clearpage

\begin{figure}[hbp!]
	\centering
	\subfloat[1.5 GHz]{\includegraphics[width=0.3\textwidth]{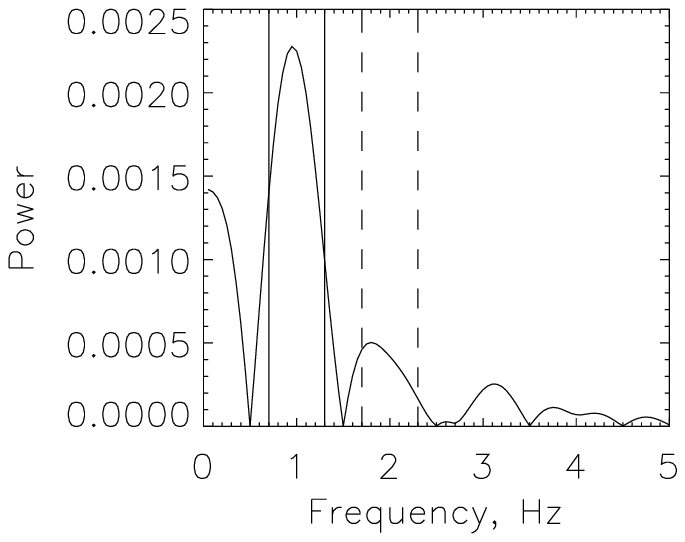}}
	\subfloat[6.6 GHz]{\includegraphics[width=0.3\textwidth]{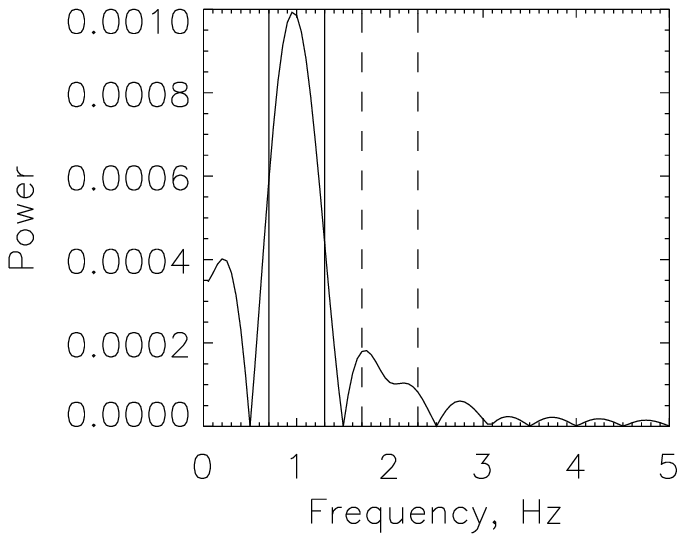}}
	\subfloat[13.8 GHz]{\includegraphics[width=0.3\textwidth]{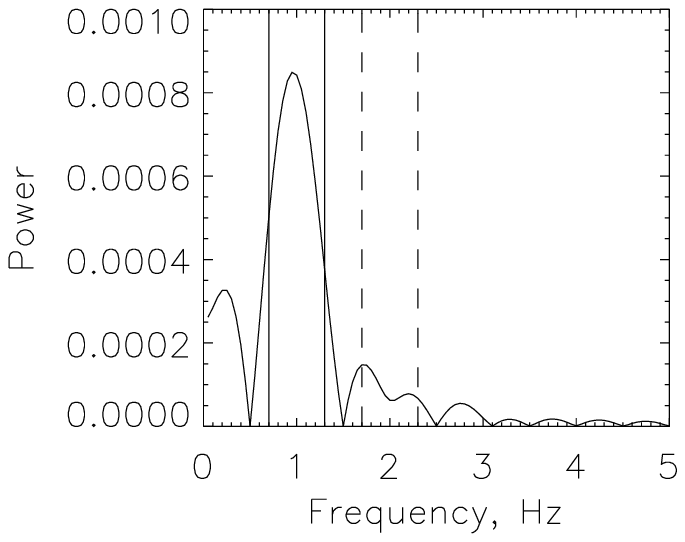}}
	
	\subfloat[\vspace{-0mm}Phase]{\includegraphics[width=0.3\textwidth]{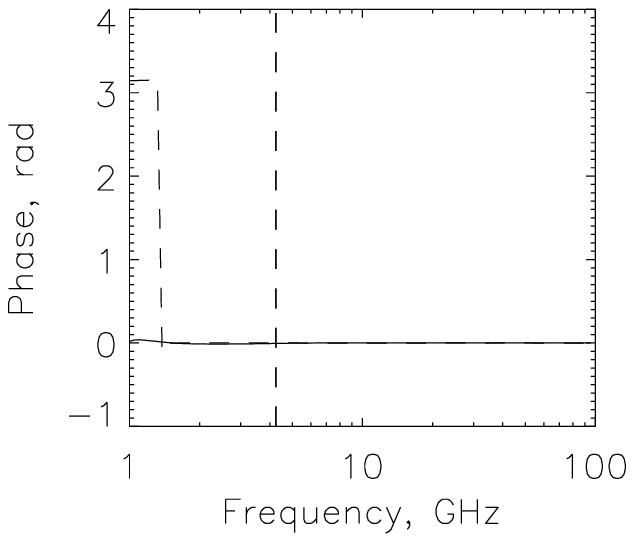}}
	\subfloat[Partial modulation]{\includegraphics[width=0.3\textwidth]{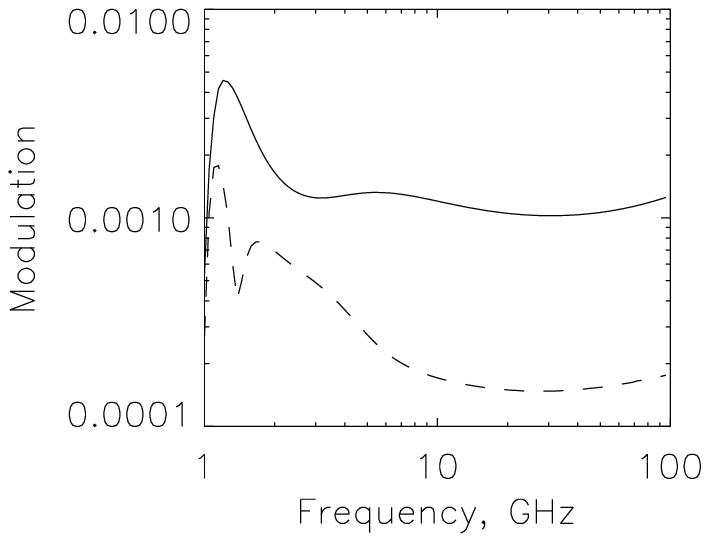}}
	\subfloat[Full modulation]{\includegraphics[width=0.3\textwidth]{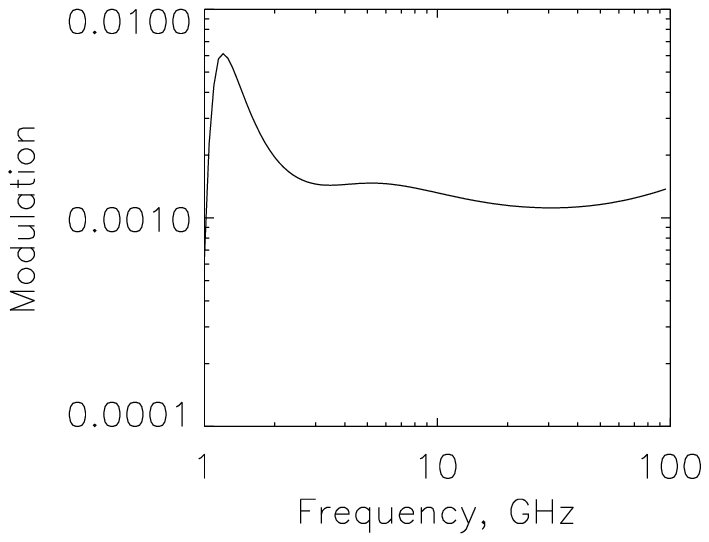}}
	
	\subfloat[1.5 GHz]{\includegraphics[width=0.3\textwidth]{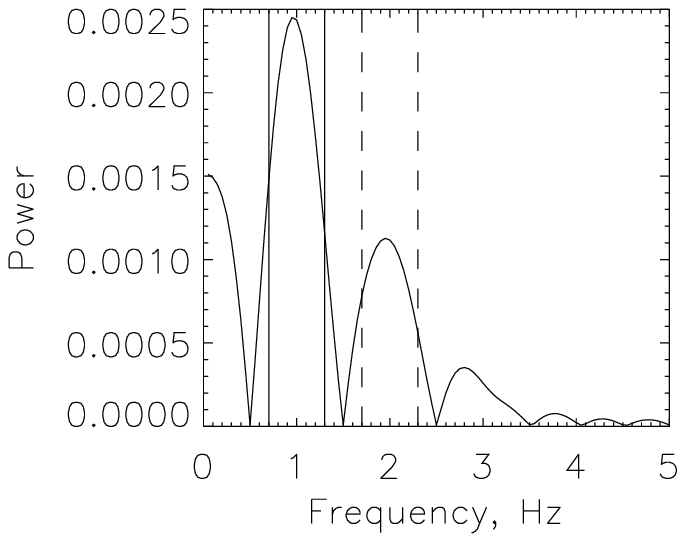}}
	\subfloat[6.6 GHz]{\includegraphics[width=0.3\textwidth]{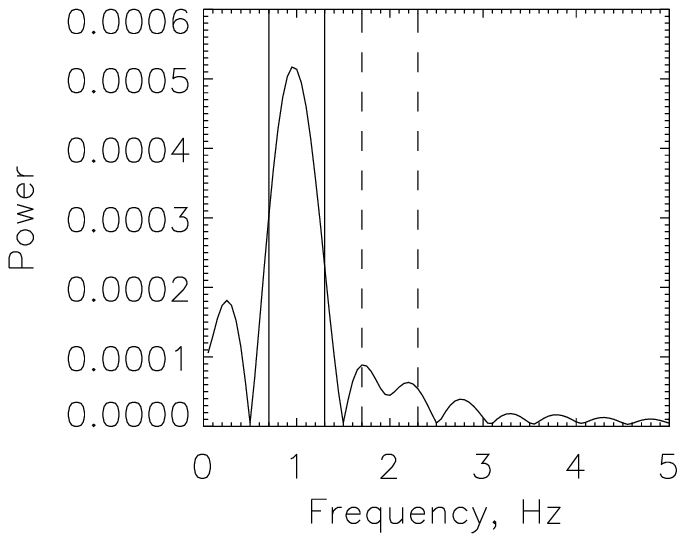}}
	\subfloat[13.8 GHz]{\includegraphics[width=0.3\textwidth]{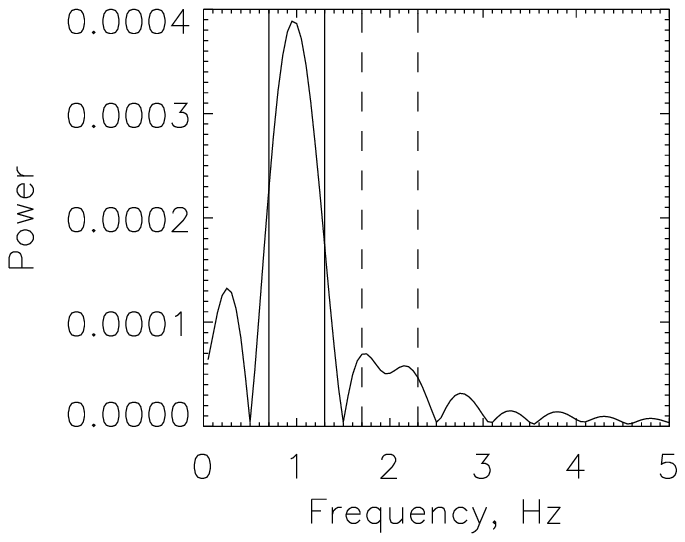}}
	
	\subfloat[Phase]{\includegraphics[width=0.3\textwidth]{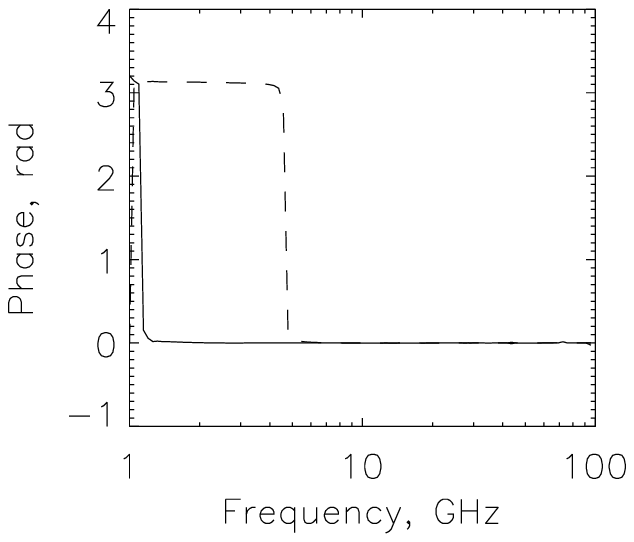}}
	\subfloat[Partial modulation]{\includegraphics[width=0.3\textwidth]{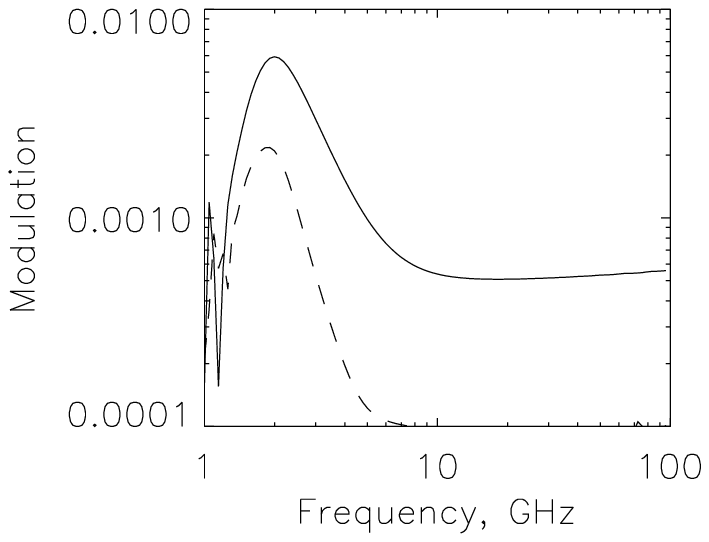}}
	\subfloat[Full modulation]{\includegraphics[width=0.3\textwidth]{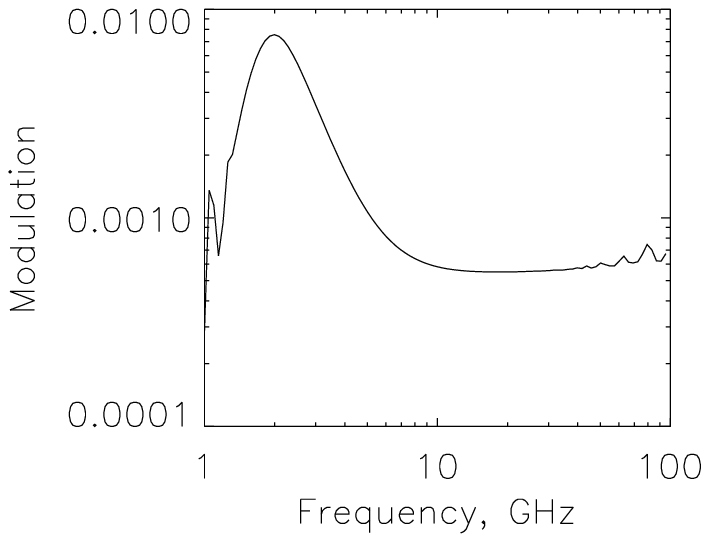}}

	\caption{Kink mode, Razin effect, oscillations of flux and polarization; same as in Figure~\ref{sausage_razin_all}. 
}
	\label{kink_razin_all}
\end{figure}

\clearpage

\newpage

\appendix
\section*{On-line Album of quasiperiodic pulsations (QPP) of flaring radio emission driven by  MHD loop oscillations }
%
%

Like in the main part of the paper, we adopt the following default parameters:
the cylinder side area $S_0=2.5\times10^{18}$~cm$^2$, the cylinder depth $L_0=2.5\times10^{8}$~cm, the background temperature
$T_0=2\times 10^7$~K, the thermal number density $n_0=10^9$~cm$^{-3}$, the nonthermal number density
$n_b=0.02\times n_0$, the magnetic field
$B_0= 50$~G, the viewing angle $\theta_0=45^\circ$, the electron energy spectral index $\delta=3.5$ for $0.1<E<10$~MeV, the angular distribution is isotropic, unless otherwise is explicitly stated. For the Razin-effect parameter regime the thermal number density is $n_0=10^{10}$~cm$^{-3}$. For the anisotropic distribution of the fast electrons the loss-cone angle $\theta=60^\circ$. For the high magnetic field case the magnetic field $B_0= 500$~G.

\section{Sausage mode}

\begin{figure}[hp!]
	\centering	
	\subfloat[1.5 GHz]{\includegraphics[width=0.3\textwidth]{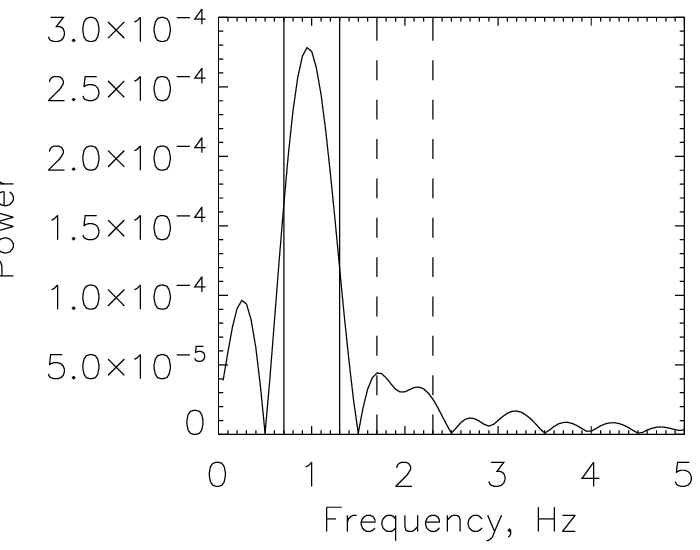}}
	\subfloat[2.2 GHz]{\includegraphics[width=0.3\textwidth]{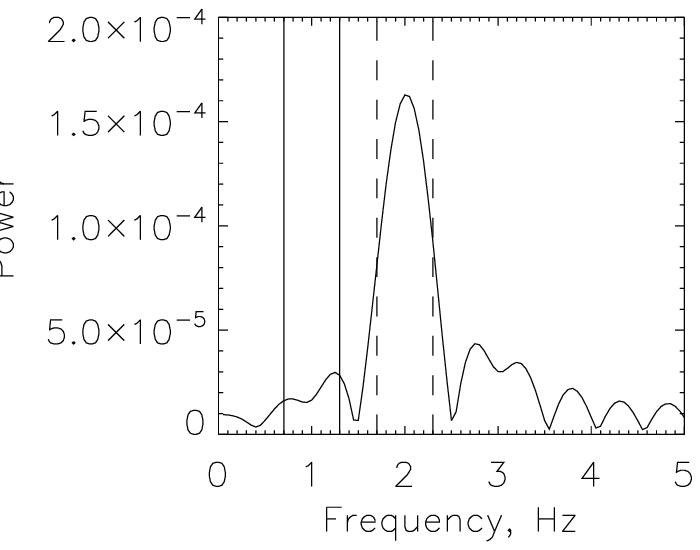}}
	\subfloat[13.8 GHz]{\includegraphics[width=0.3\textwidth]{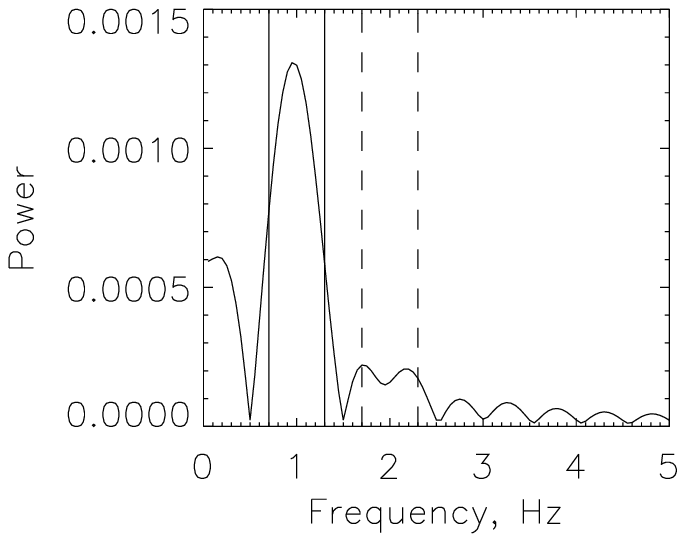}}
\vspace{-4mm}
	\subfloat[Phase]{\includegraphics[width=0.3\textwidth]{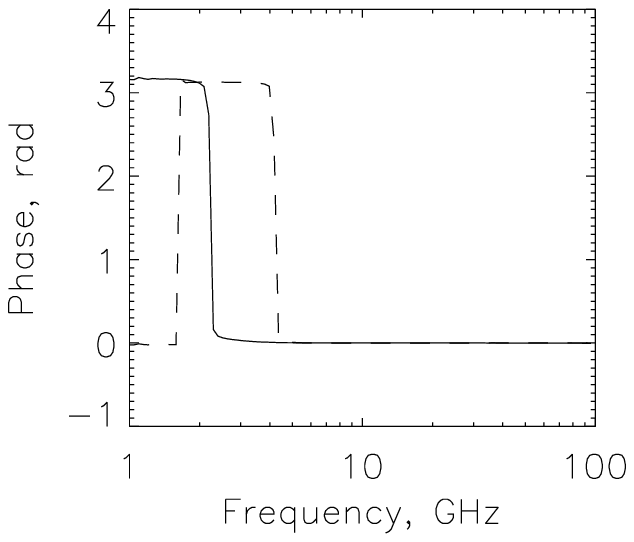}}
	\subfloat[Partial modulation]{\includegraphics[width=0.3\textwidth]{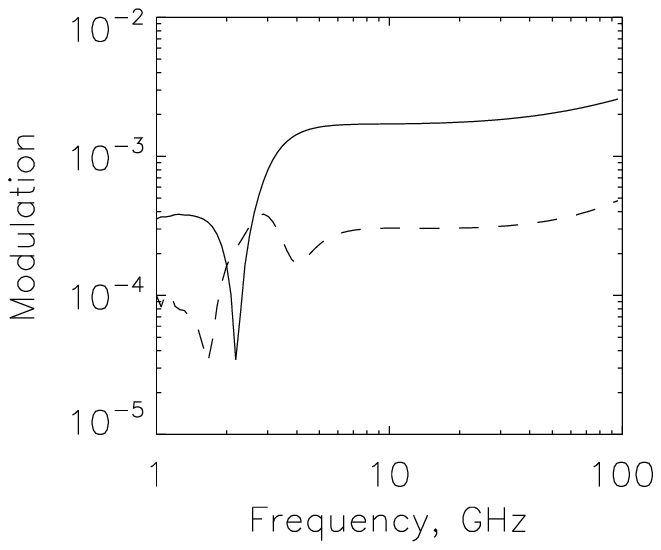}}
	\subfloat[Full modulation]{\includegraphics[width=0.3\textwidth]{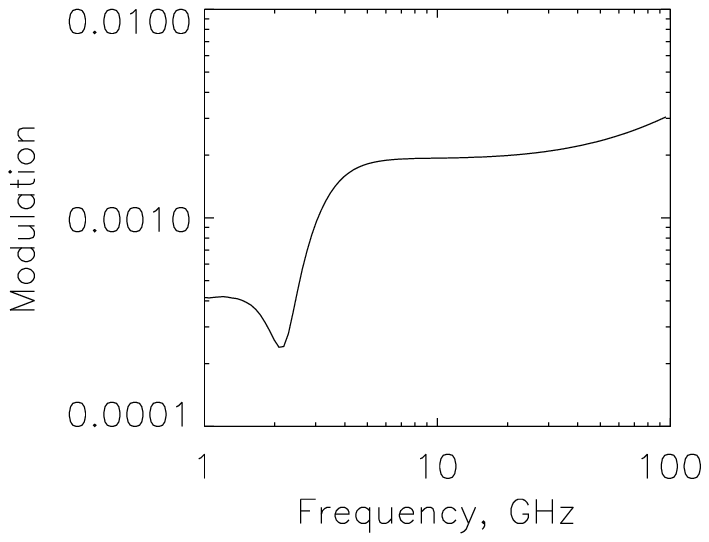}}
\vspace{-4mm}
	\subfloat[1.5 GHz]{\includegraphics[width=0.3\textwidth]{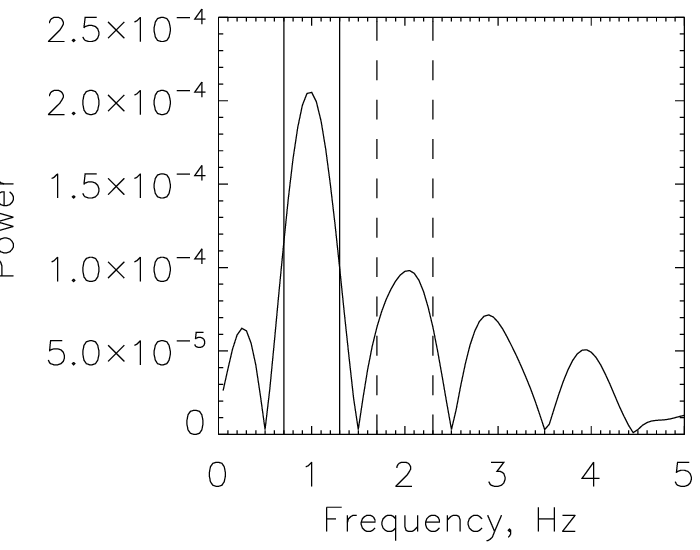}}
	\subfloat[2.2 GHz]{\includegraphics[width=0.3\textwidth]{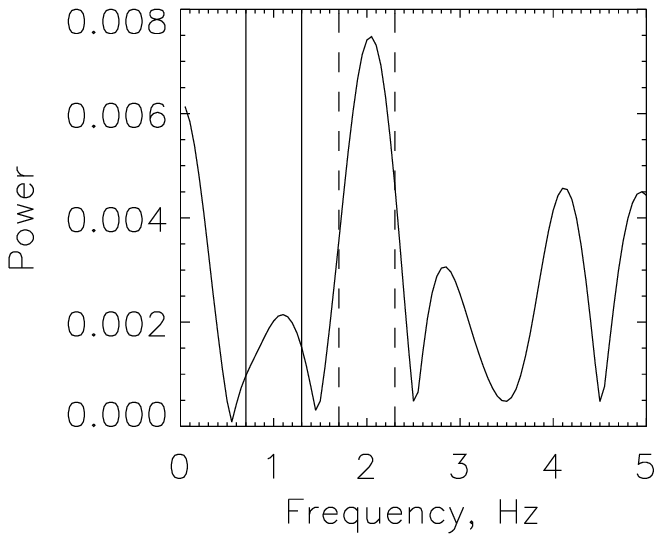}}
	\subfloat[13.8 GHz]{\includegraphics[width=0.3\textwidth]{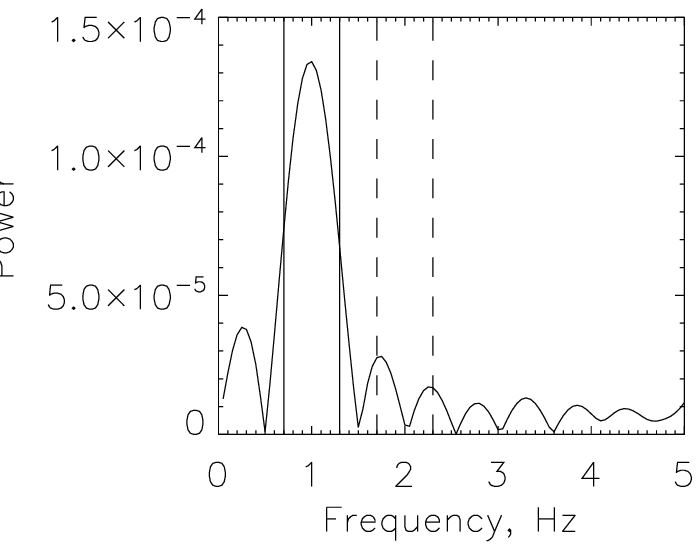}}
\vspace{-4mm}
	\subfloat[Phase]{\includegraphics[width=0.3\textwidth]{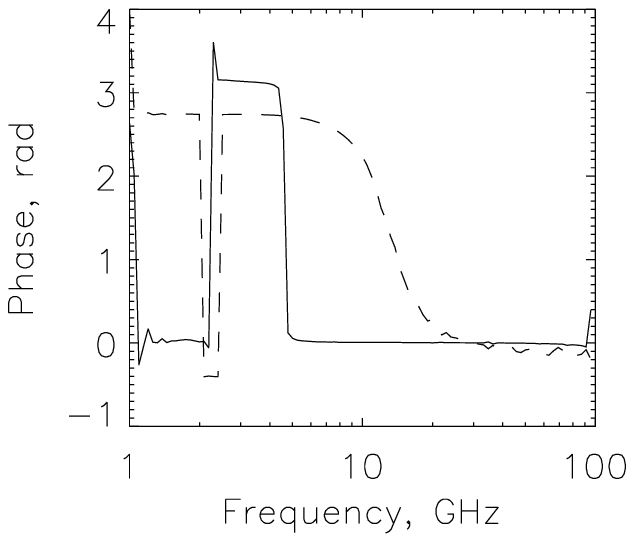}}
	\subfloat[Partial modulation]{\includegraphics[width=0.3\textwidth]{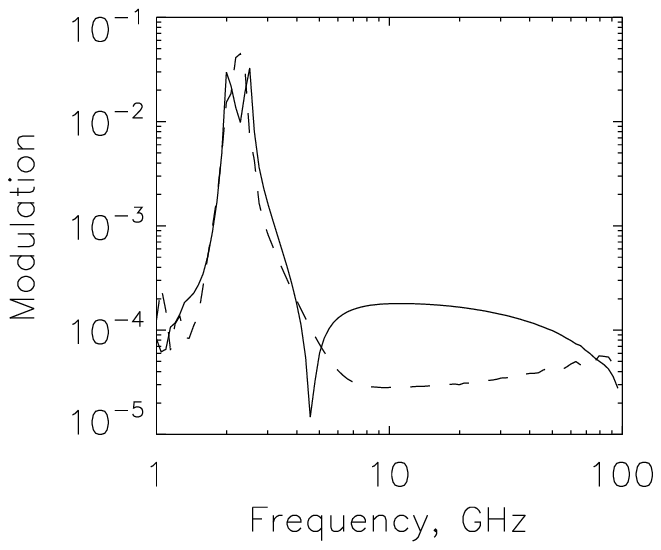}}
	\subfloat[Full modulation]{\includegraphics[width=0.3\textwidth]{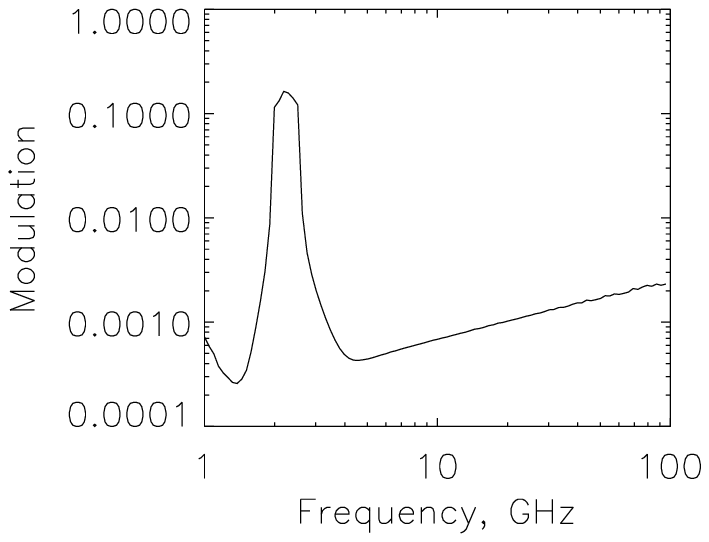}}
	\caption{Sausage mode, default parameters, single pixel view. {\bf(a)-(c)} Fourier spectra of the flux density oscillating components for the emission frequencies indicated. Vertical lines mark off peaks considered for partial modulation. {\bf(d)} Relative phase of flux. Solid line is the fundamental harmonics (1~Hz), dashed line is the second harmonics (2~Hz). {\bf(e)} Partial modulation amplitude for the indicated range of Fourier coefficients for two harmonics: the solid (dashed) line represents integration between the solid (dashed) vertical lines in panels {\bf(a)-(c)}. {\bf(f)} Full modulation amplitude. {\bf(g)-(l)} Same for polarization.}
	\label{sausage_pixel_all}
\end{figure}

\begin{figure}[hbp!]
	\centering
	\subfloat[1.5 GHz]{\includegraphics[width=0.3\textwidth]{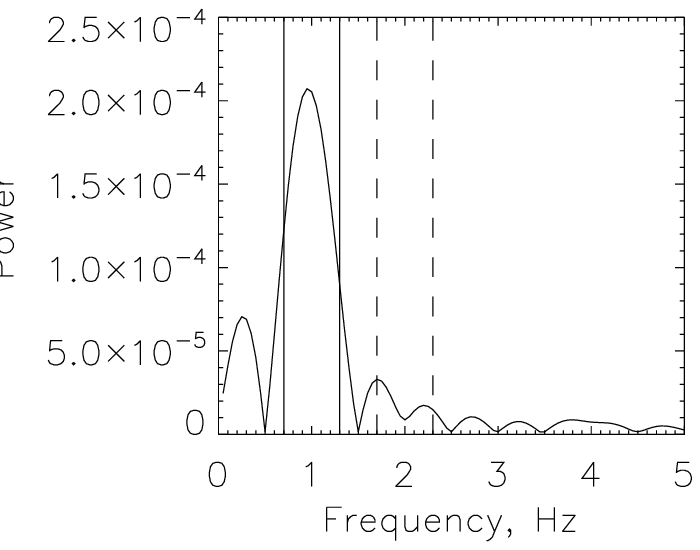}}
	\subfloat[2.2 GHz]{\includegraphics[width=0.3\textwidth]{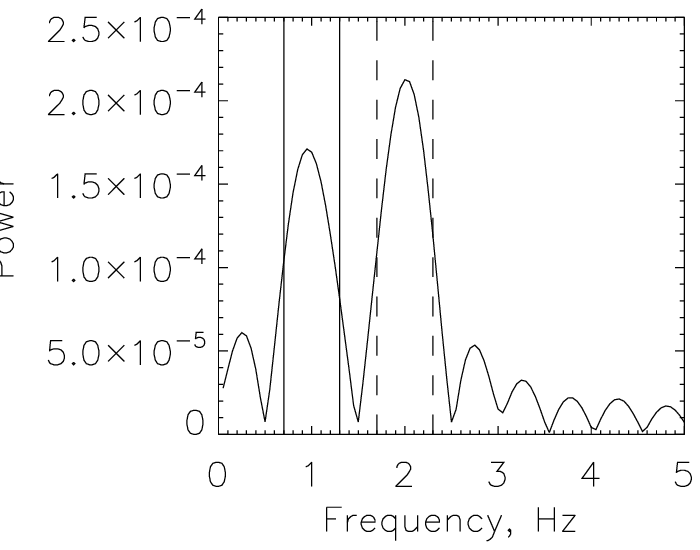}}
	\subfloat[13.8 GHz]{\includegraphics[width=0.3\textwidth]{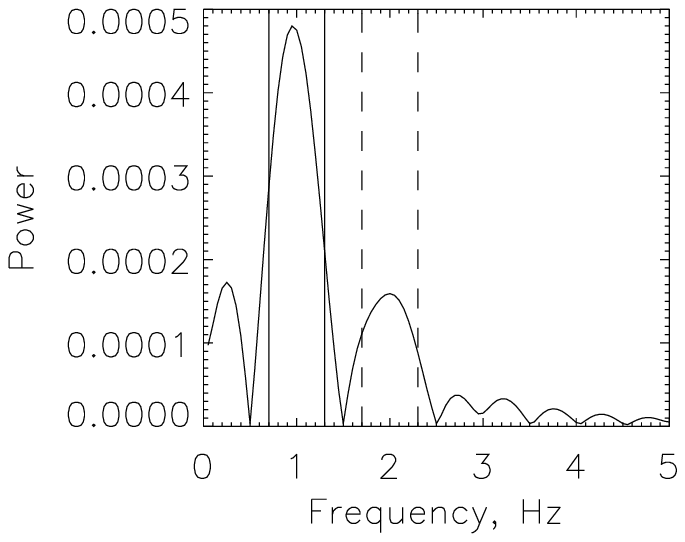}}
	
	\subfloat[\vspace{-0mm}Phase]{\includegraphics[width=0.3\textwidth]{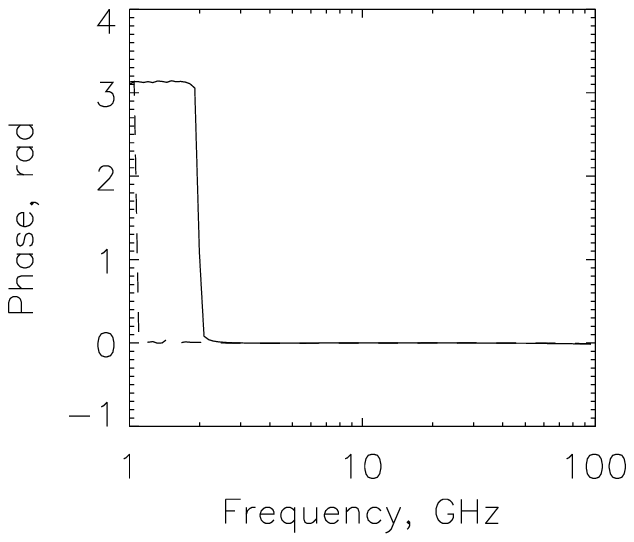}}
	\subfloat[Partial modulation]{\includegraphics[width=0.3\textwidth]{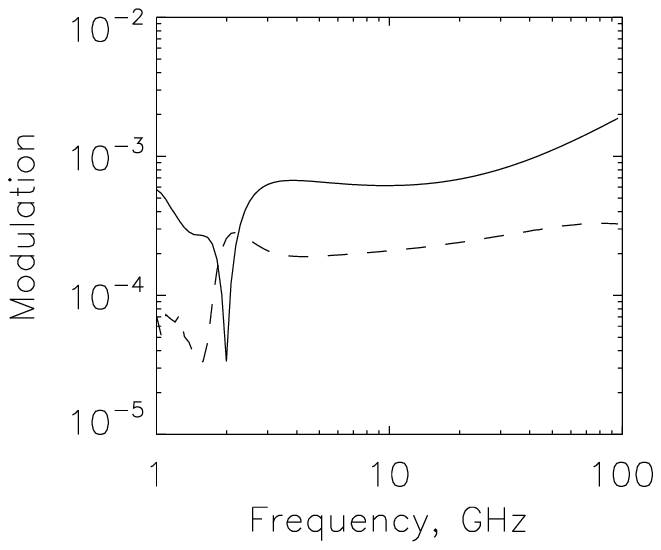}}
	\subfloat[Full modulation]{\includegraphics[width=0.3\textwidth]{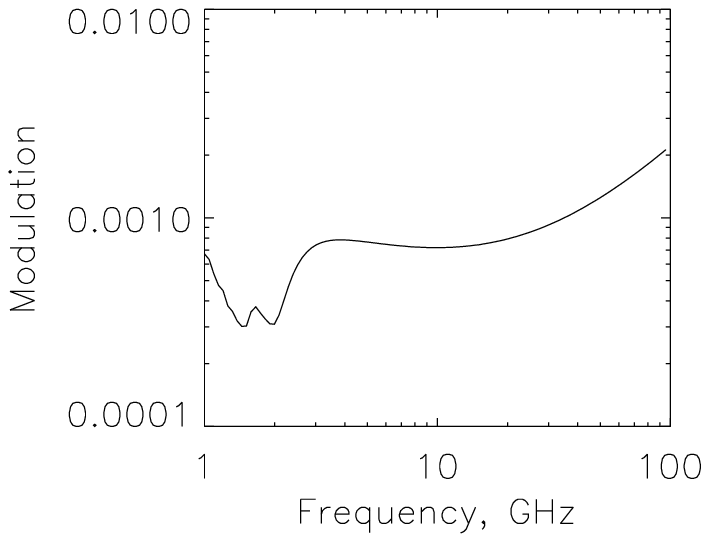}}
	
	\subfloat[1.5 GHz]{\includegraphics[width=0.3\textwidth]{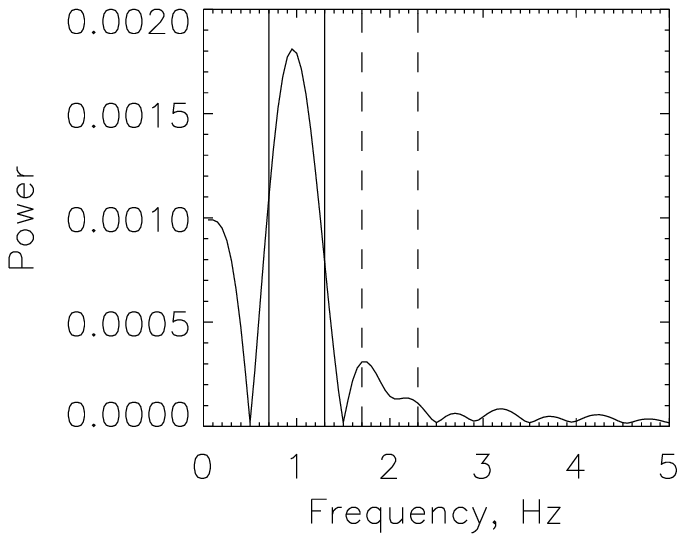}}
	\subfloat[2.2 GHz]{\includegraphics[width=0.3\textwidth]{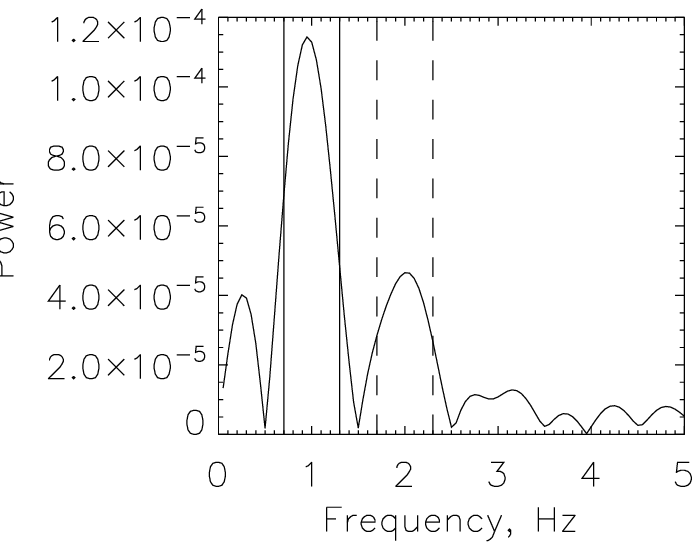}}
	\subfloat[13.8 GHz]{\includegraphics[width=0.3\textwidth]{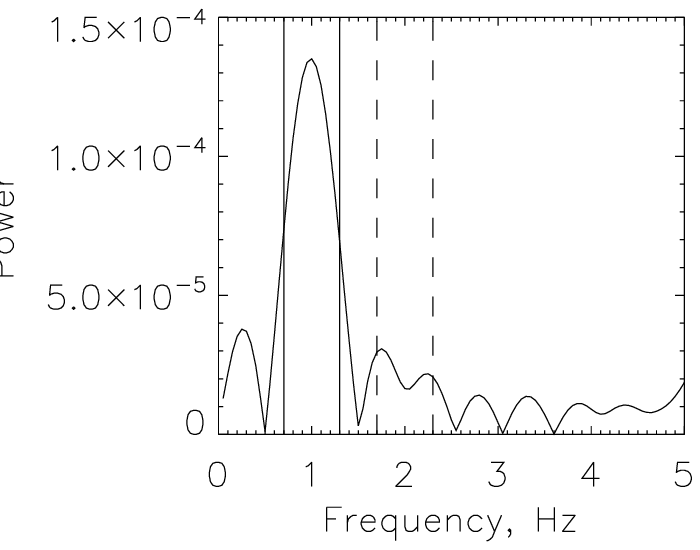}}
	
	\subfloat[Phase]{\includegraphics[width=0.3\textwidth]{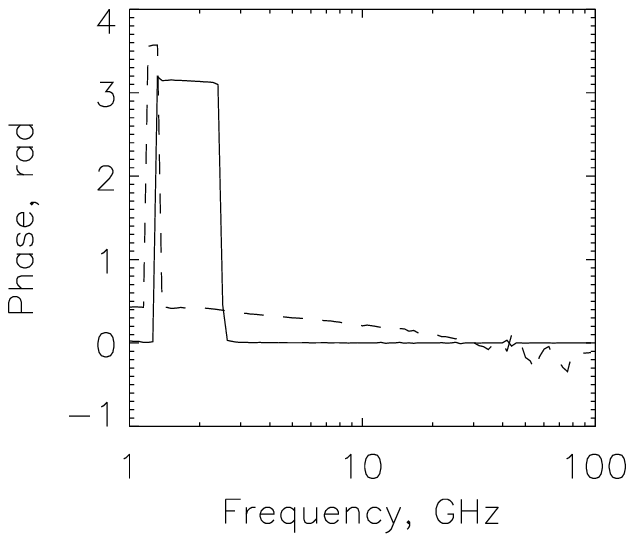}}
	\subfloat[Partial modulation]{\includegraphics[width=0.3\textwidth]{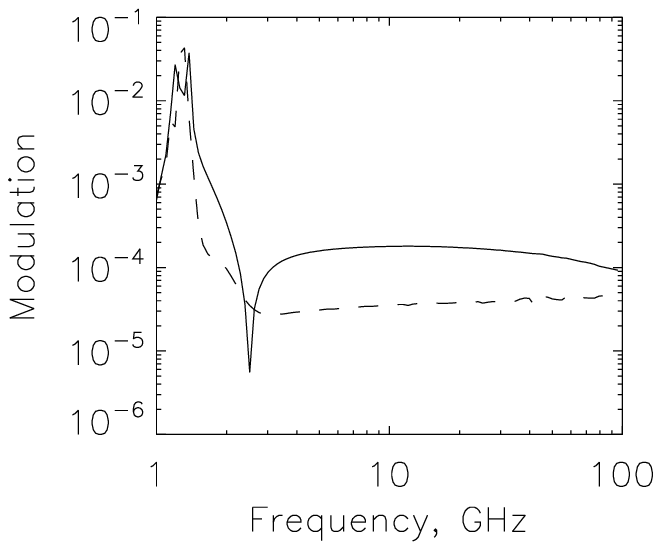}}
	\subfloat[Full modulation]{\includegraphics[width=0.3\textwidth]{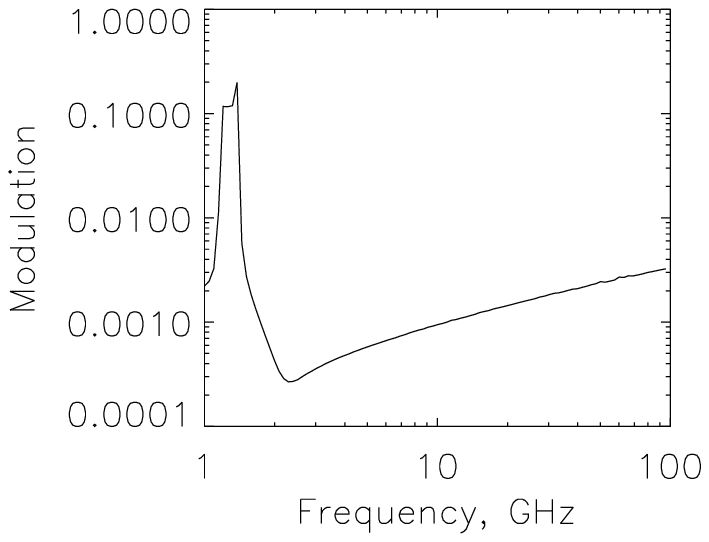}}

	\caption{Same as in Figure~\ref{sausage_pixel_all} for  sausage mode, default parameters but with anisotropic loss-cone pitch-angle distribution; $\theta_c=60^\circ$.  }
	\label{sausage_aniso_all}
\end{figure}

In addition to the plots for the default-value parameters and Razin-effect regime presented in the paper itself, we give here corresponding plots for the single pixel view when the visible area $A$  remains fixed, while all other parameters  oscillate as before, Figure~\ref{sausage_pixel_all}, as well as the case of anisotropic loss-cone distribution of fast electrons, when the loss-cone boundary oscillates in addition to all other source parameters Figure~\ref{sausage_aniso_all}.
	
\begin{figure}[htp!]
	\centering
	\includegraphics{figs/sausage/lowfreq/sausage_lowfreq_f_spectrum.eps}
	\includegraphics{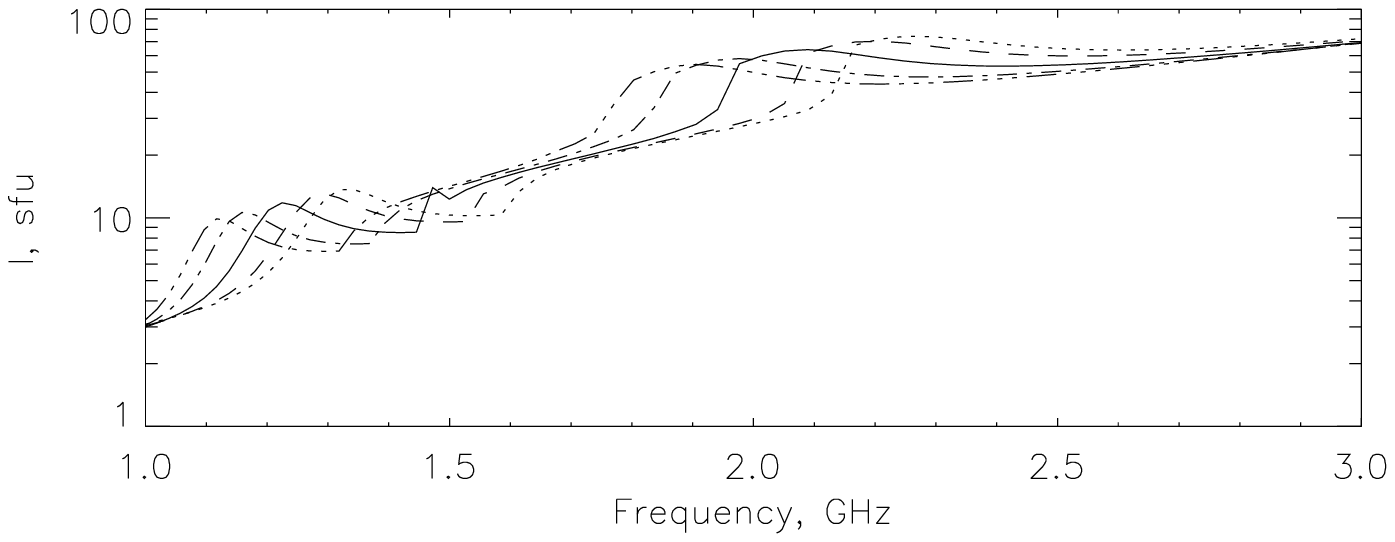}
	\includegraphics{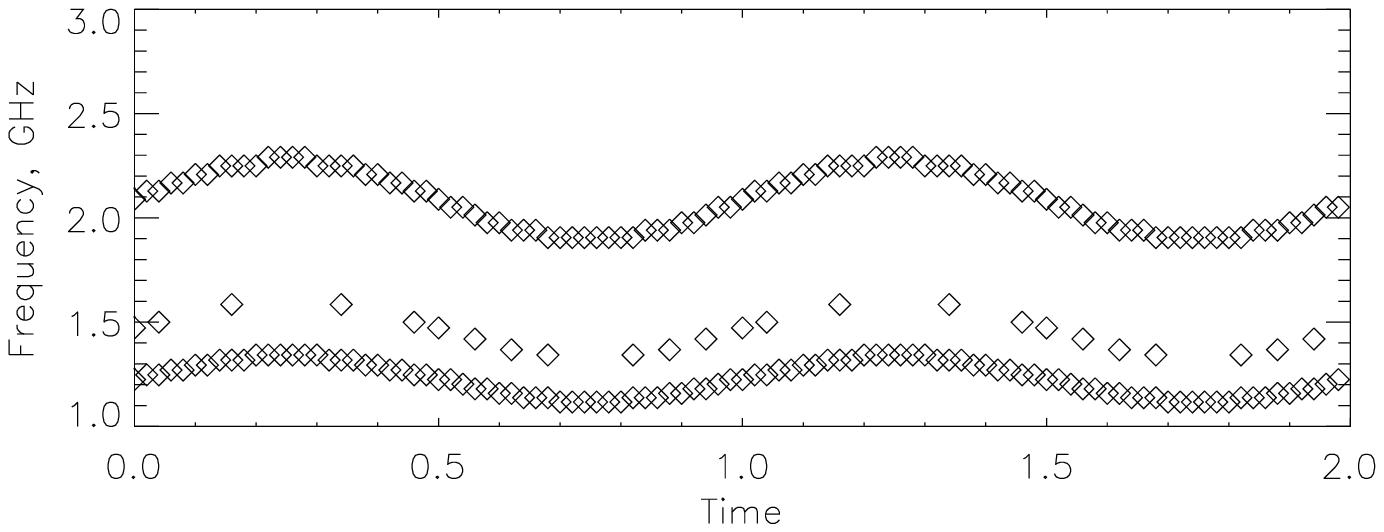}
	\caption{Sausage mode, high magnetic field, the viewing angle $\theta=45^\circ$:  low frequency gyroharmonics are present in the spectrum. Top: Dynamic spectrum of simulation -- oscillations of peaks are clearly visible. Middle: Locations of low-frequency peaks over one period of oscillation. Bottom: Variation of  the peak positions over one period of oscillation.}
	\label{sausage_lowfreq_f_light}
\end{figure}

\begin{figure}[htp!]
	\centering
	\subfloat[1.6 GHz]{\includegraphics[width=0.5\linewidth]{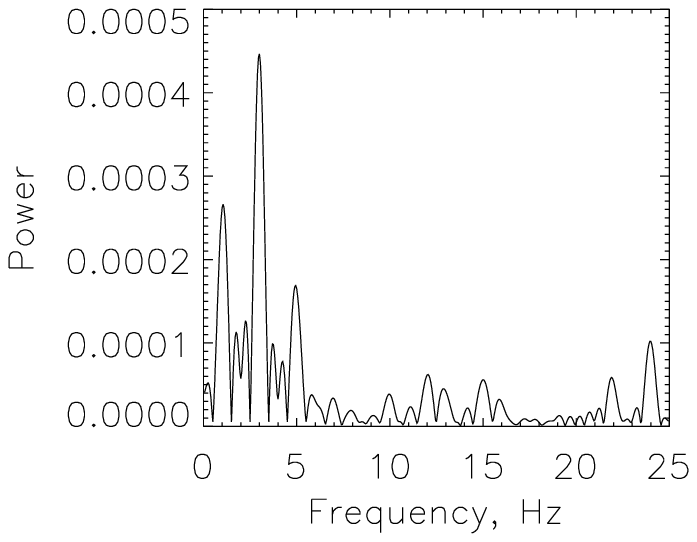}}
	\subfloat[1.7 GHz]{\includegraphics[width=0.5\linewidth]{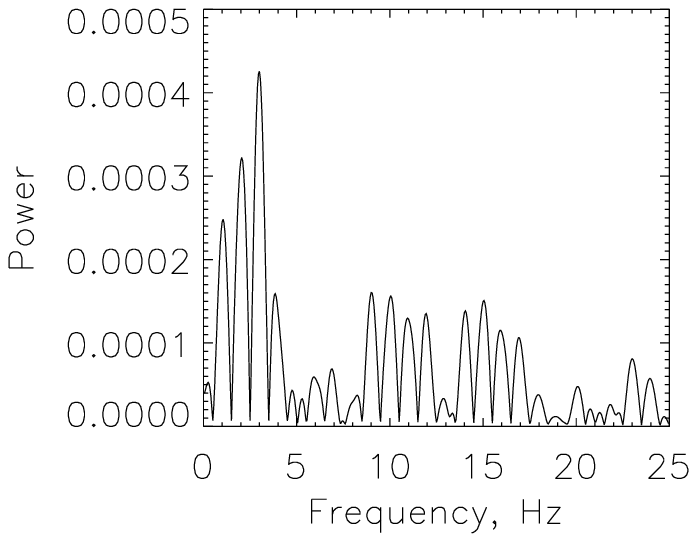}}
	\vspace{-8mm}
	\subfloat[1.9 GHz]{\includegraphics[width=0.5\linewidth]{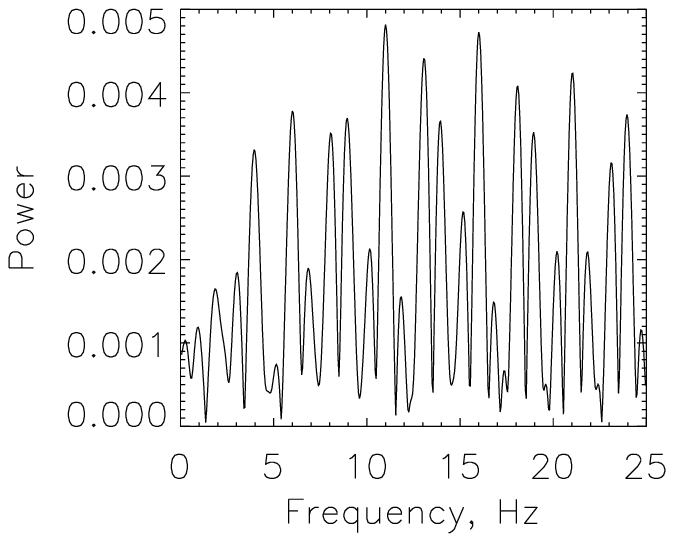}}
	\subfloat[2.1 GHz]{\includegraphics[width=0.5\linewidth]{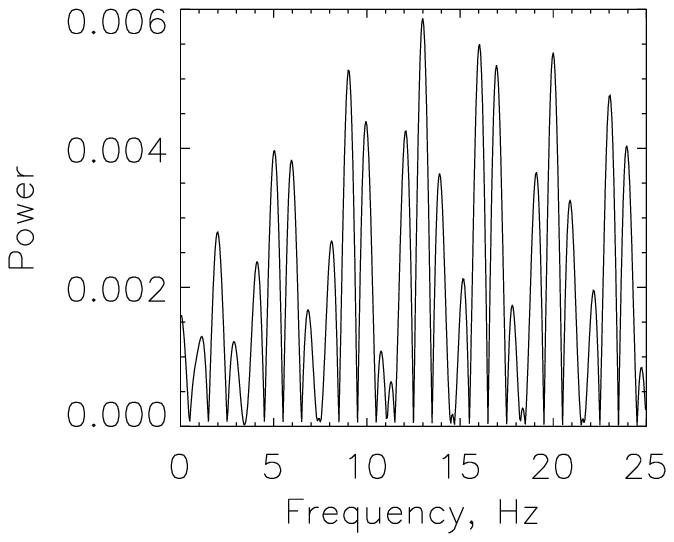}}
	\vspace{-8mm}
	\subfloat[2.3 GHz]{\includegraphics[width=0.5\linewidth]{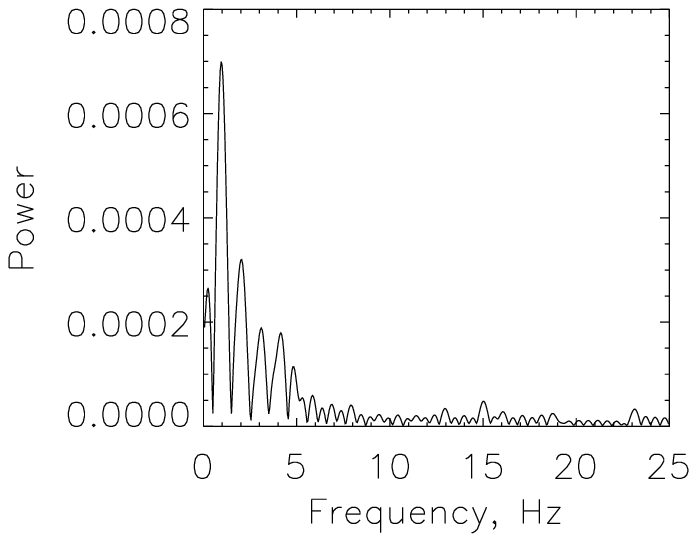}}
	\subfloat[2.5 GHz]{\includegraphics[width=0.5\linewidth]{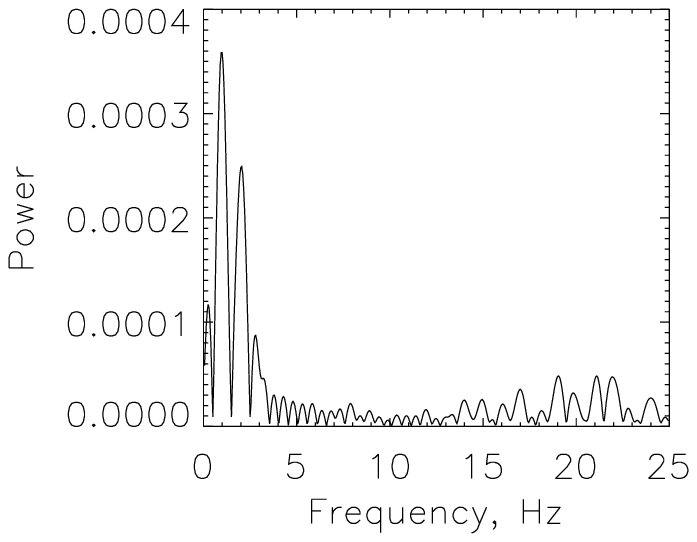}}
	\caption{Sausage mode, high magnetic field. Fourier spectra at a few selected radio frequencies selected in  the low-frequency range. Strong departure from a simple sinusoidal oscillations is clearly visible. }
	\label{sausage_lowfreq_f_fourier}
\end{figure}

\begin{figure}[htp!]
	\centering
	\includegraphics{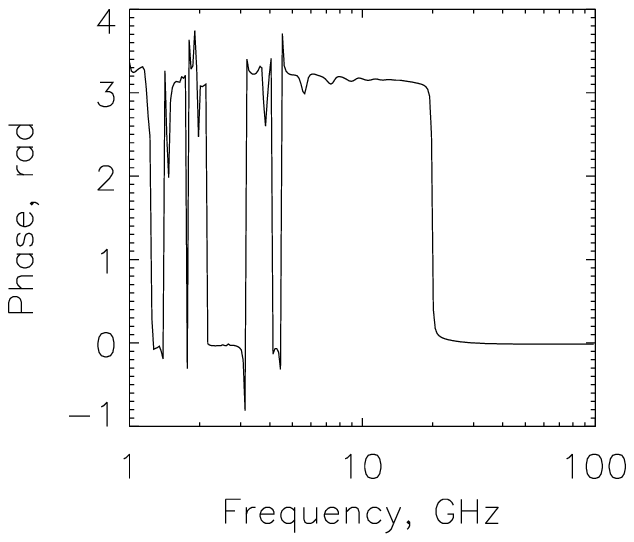}
	\includegraphics{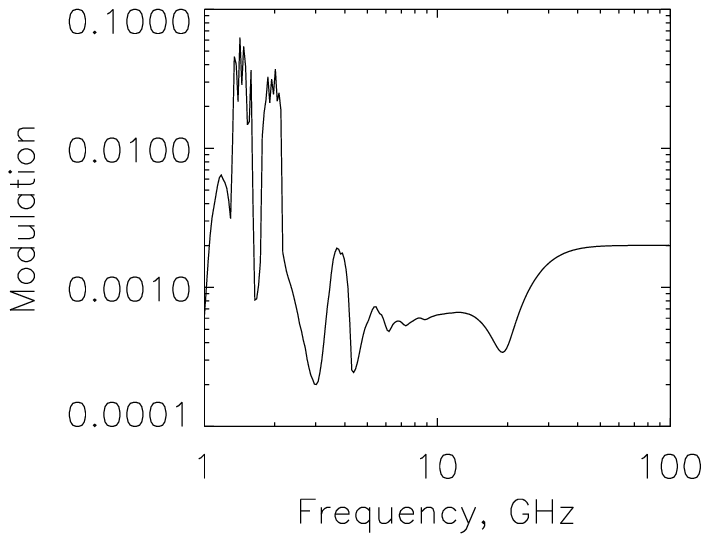}
	\caption{Sausage mode, high magnetic field. Left: Relative phase of the flux density at $\nu=1$~Hz. Right: Full modulation amplitude of the flux density. For higher frequencies, it is similar to the cause of default parameters.  At low frequencies, however, the modulation amplitude fluctuates strongly: it increases noticeably at the frequencies where the gyroharmonics are enhanced.}
	\label{sausage_lowfreq_f_analysis}
\end{figure}

\begin{figure}[htp!]
	\centering
	\includegraphics{figs/sausage/lowfreq/sausage_lowfreq_p_spectrum.eps}
	\includegraphics{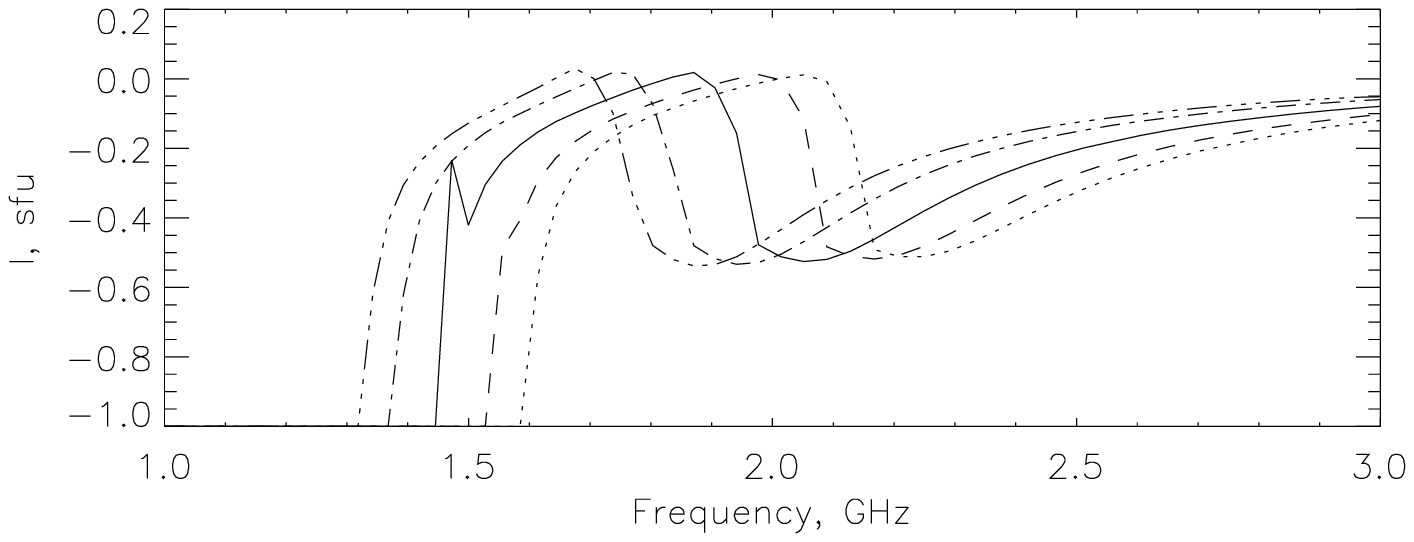}
	\includegraphics{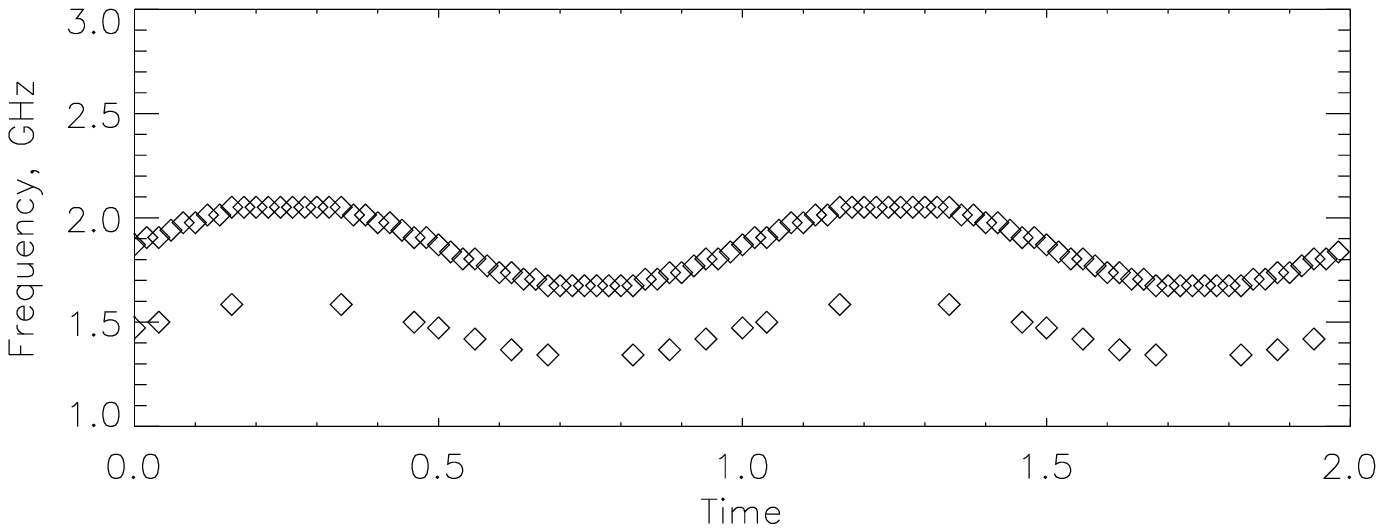}
	\caption{Sausage mode, high magnetic field, oscillations of polarization. Same as in Figure~\ref{sausage_lowfreq_f_light} but for the polarization.}
	\label{sausage_lowfreq_p_light}
\end{figure}

\begin{figure}[htp!]
	\centering
	\subfloat[1.6 GHz]{\includegraphics[width=0.5\linewidth]{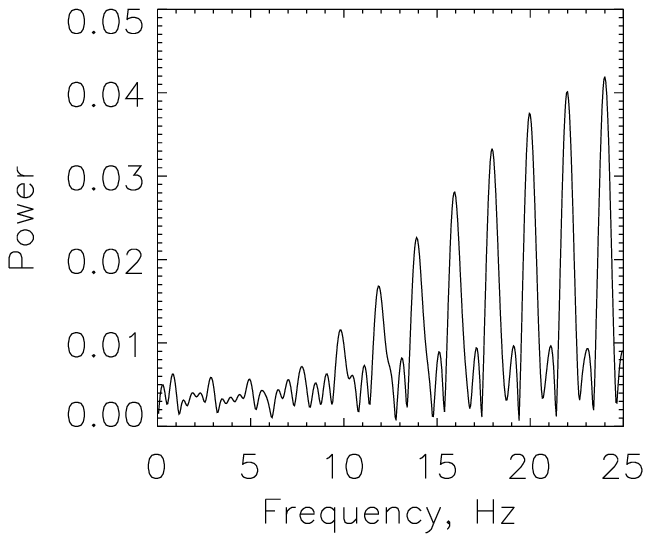}}
	\subfloat[1.7 GHz]{\includegraphics[width=0.5\linewidth]{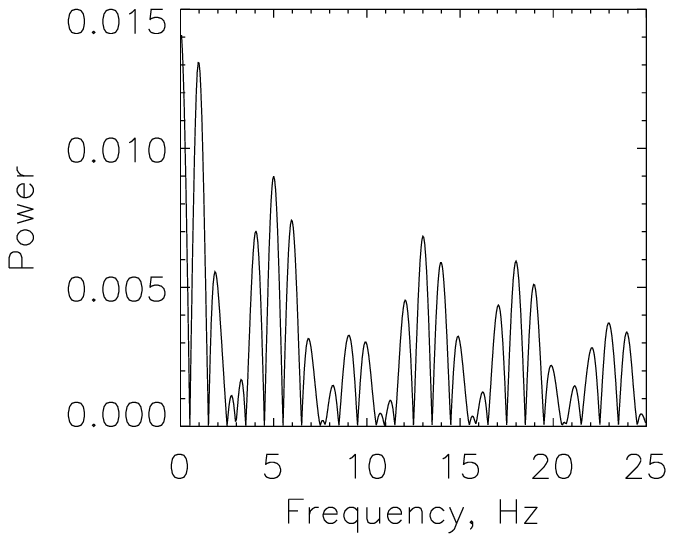}}
	\vspace{-8mm}
	\subfloat[1.9 GHz]{\includegraphics[width=0.5\linewidth]{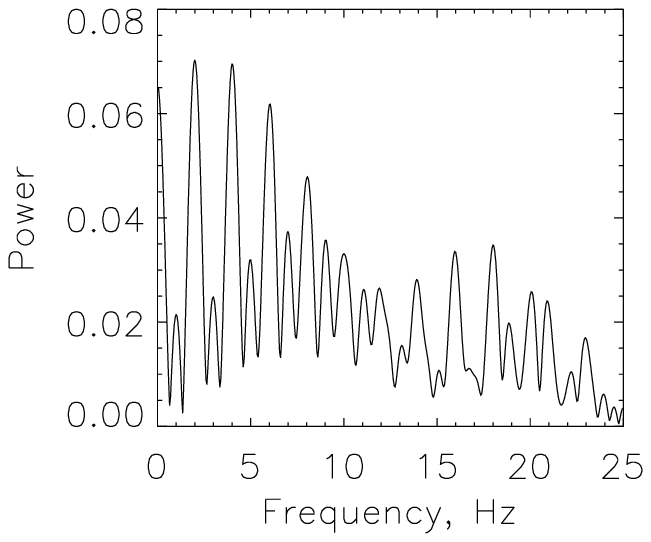}}
	\subfloat[2.1 GHz]{\includegraphics[width=0.5\linewidth]{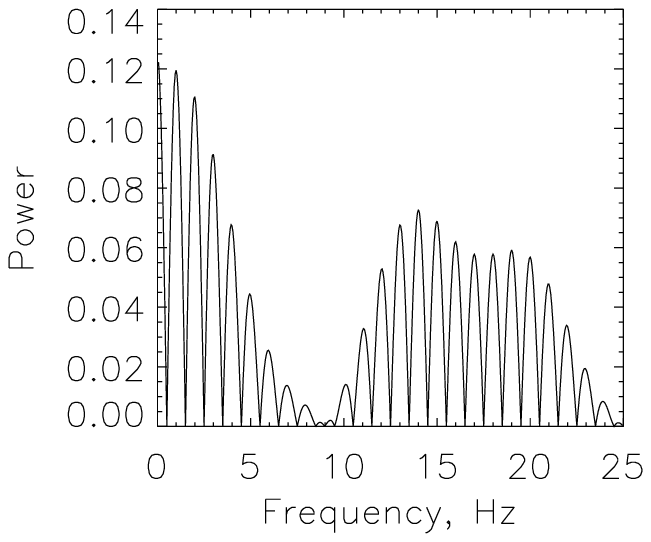}}
	\vspace{-5mm}
	\subfloat[2.3 GHz]{\includegraphics[width=0.5\linewidth]{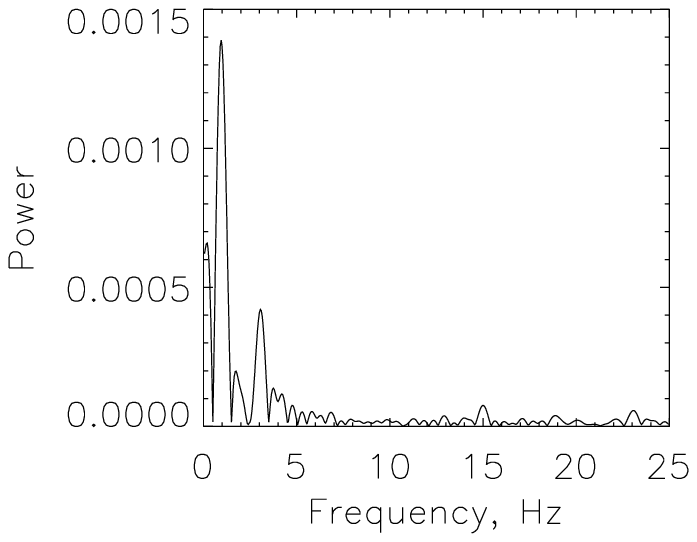}}
	\caption{Sausage mode, high magnetic field, oscillations of polarization. Same as in Figure~\ref{sausage_lowfreq_f_fourier} but for the polarization. }
	\label{sausage_lowfreq_p_fourier}
\end{figure}

\begin{figure}[htp!]
	\centering
	\includegraphics{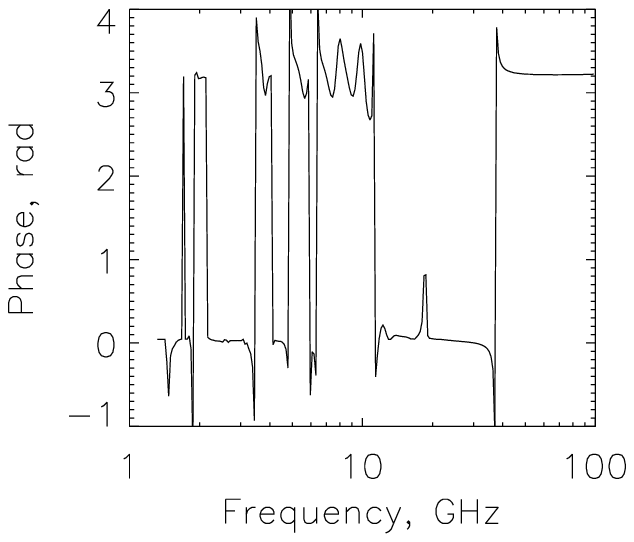}
	\includegraphics{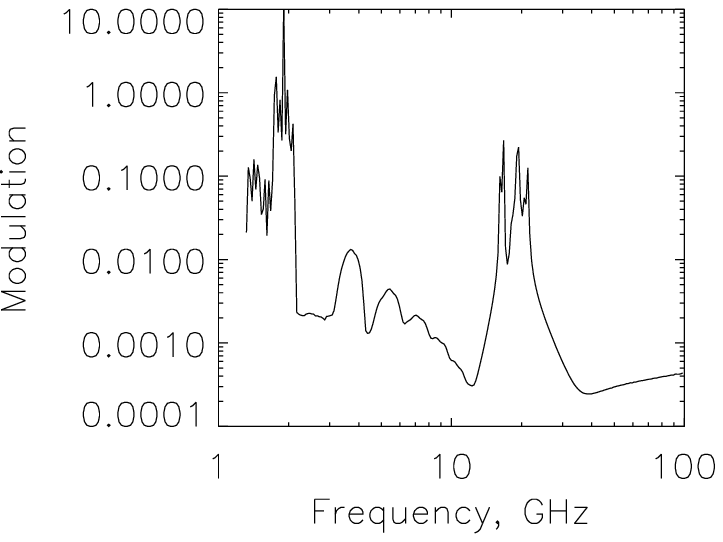}
	\caption{Sausage mode, high magnetic field, oscillations of polarization. Same as in Figure~\ref{sausage_lowfreq_f_analysis}. For higher frequencies,  the modulation amplitude is similar to the cause of default parameters.  At low frequencies, it fluctuates decreasing between gyroharmonics emission frequencies.}
	\label{sausage_lowfreq_p_analysis}
\end{figure}

\begin{figure}[htp!]
	\centering
	\includegraphics{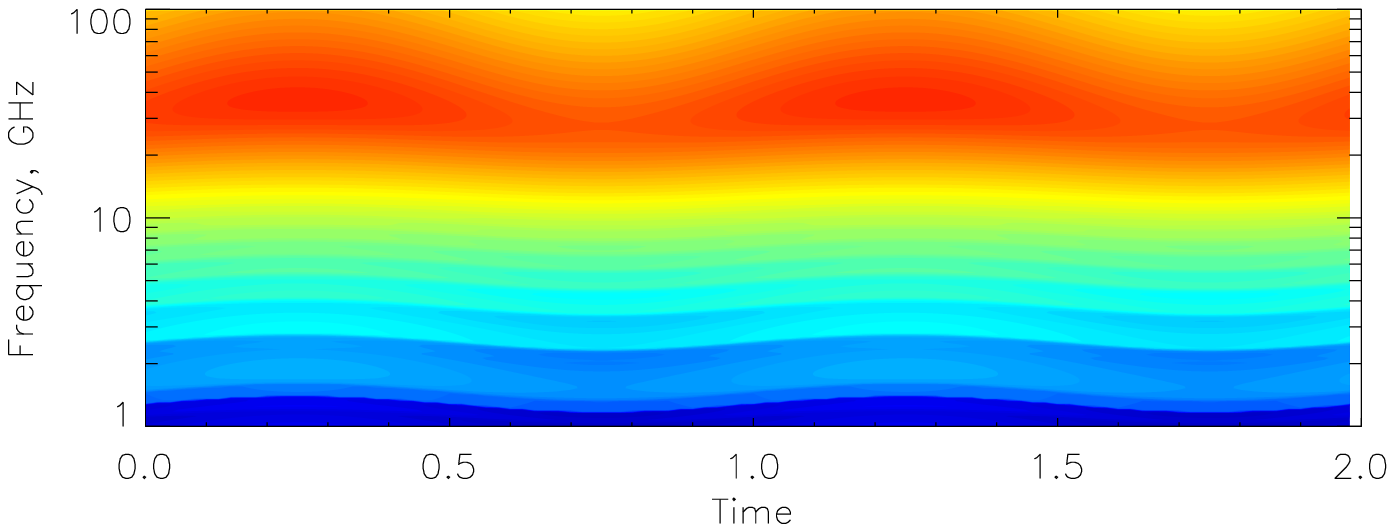}
	\includegraphics{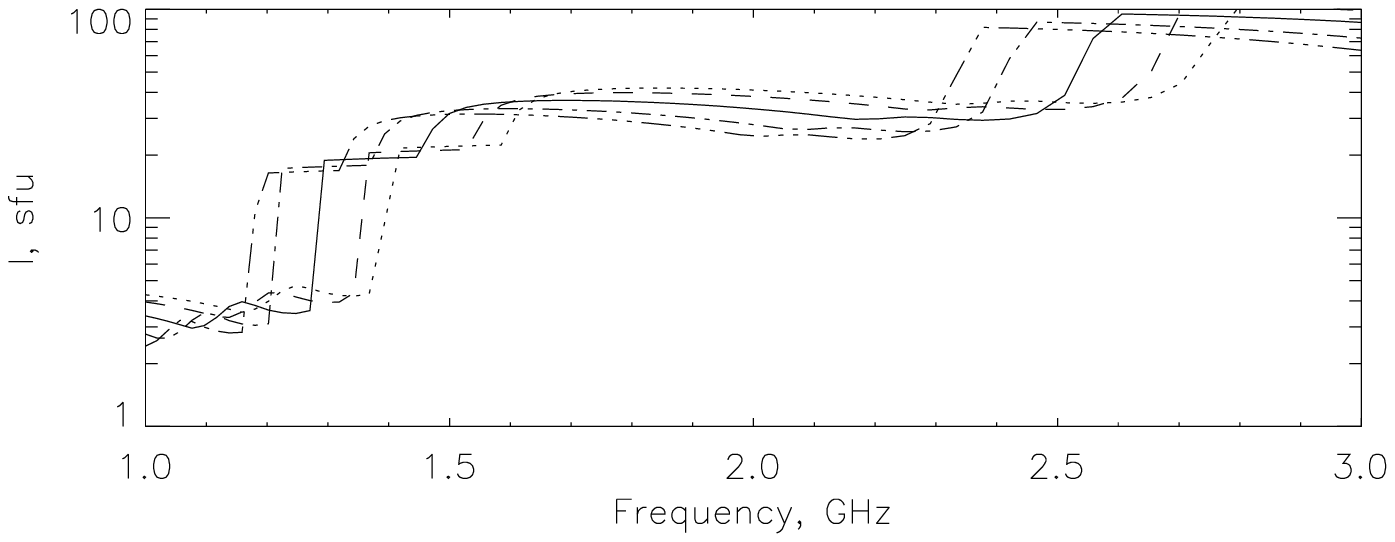}
	\includegraphics{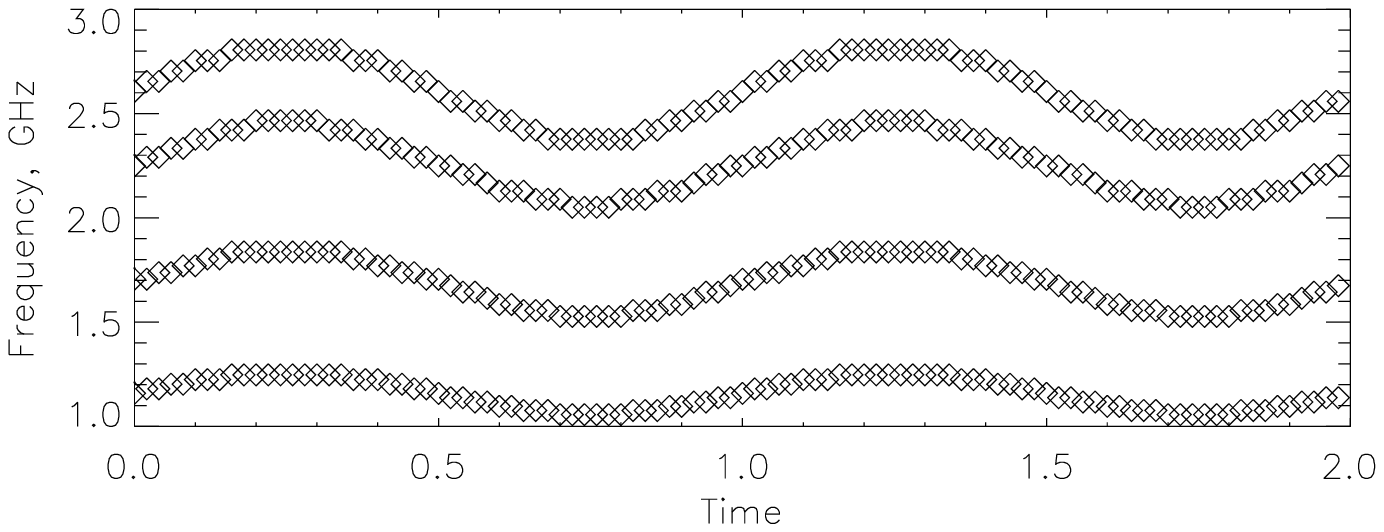}
	\caption{Sausage mode, high magnetic field, oscillations of the flux density. Same as in Figure~\ref{sausage_lowfreq_f_light} but for viewing angle 80$^\circ$.  }
	\label{sausage_theta80_lowfreq_f_light}
\end{figure}

\begin{figure}[htp!]
	\centering
	\subfloat[1.6 GHz]{\includegraphics[width=0.5\linewidth]{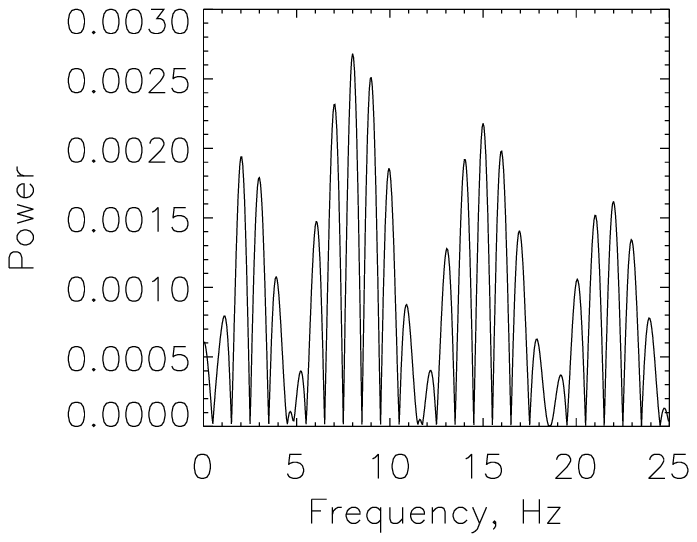}}
	\subfloat[1.7 GHz]{\includegraphics[width=0.5\linewidth]{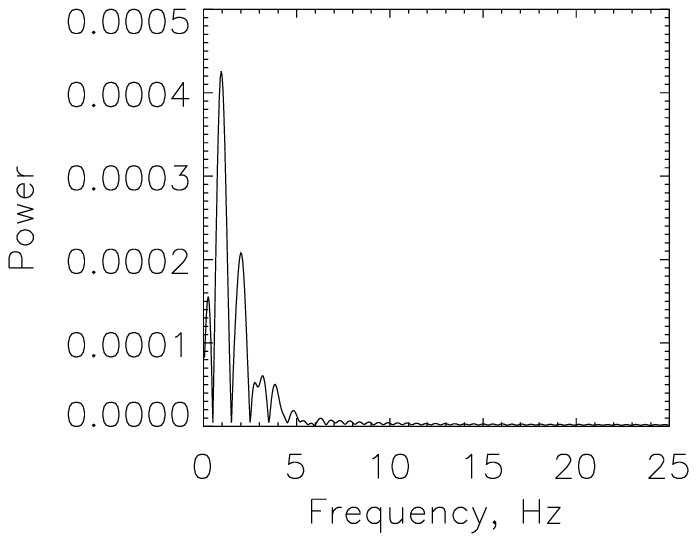}}
	\vspace{-8mm}
	\subfloat[1.9 GHz]{\includegraphics[width=0.5\linewidth]{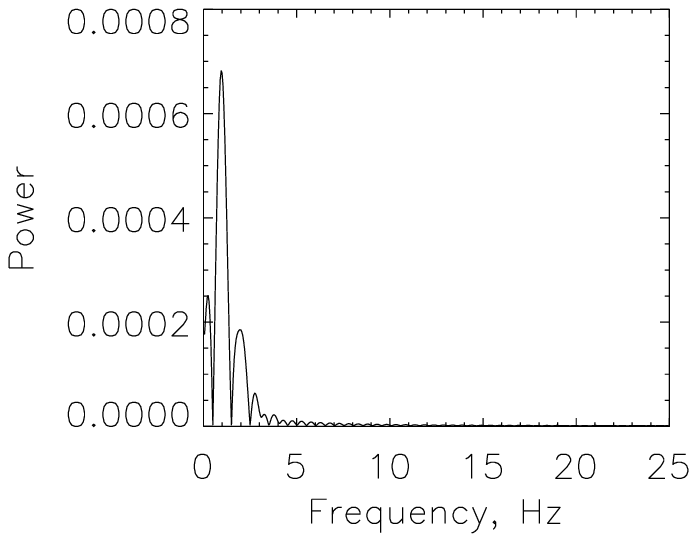}}
	\subfloat[2.1 GHz]{\includegraphics[width=0.5\linewidth]{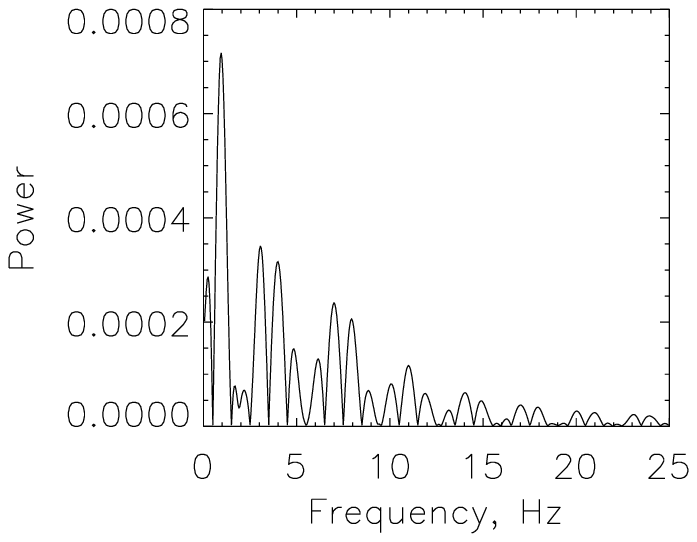}}
	\vspace{-8mm}
	\subfloat[2.3 GHz]{\includegraphics[width=0.5\linewidth]{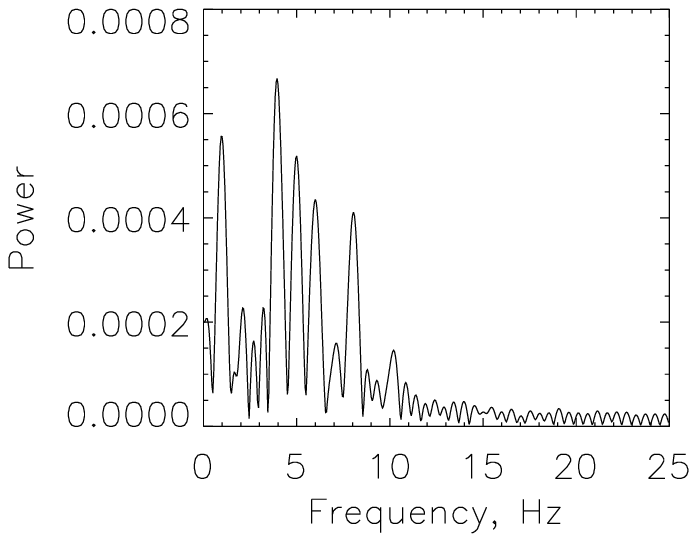}}
	\subfloat[2.5 GHz]{\includegraphics[width=0.5\linewidth]{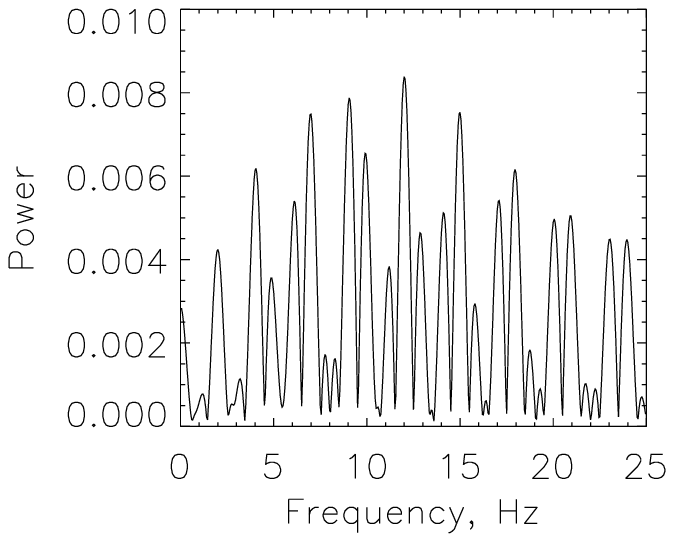}}
	\caption{Sausage mode, high magnetic field, oscillations of the flux density. Same as in Figure~\ref{sausage_lowfreq_f_fourier} but for viewing angle 80$^\circ$. }
	\label{sausage_theta80_lowfreq_f_fourier}
\end{figure}

\begin{figure}[htp!]
	\centering
	\includegraphics{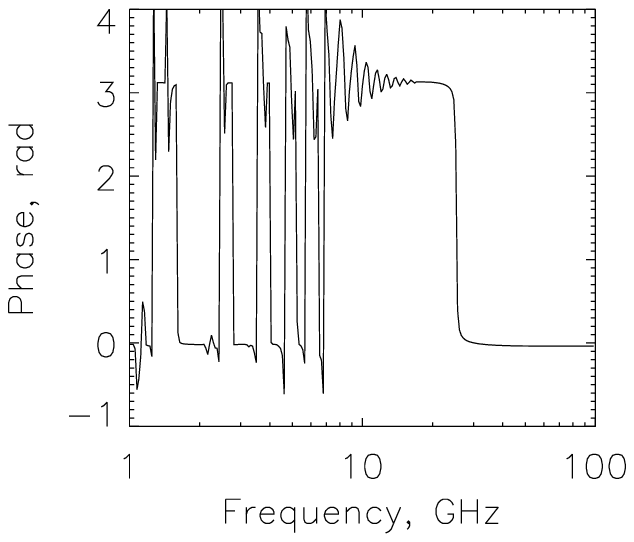}
	\includegraphics{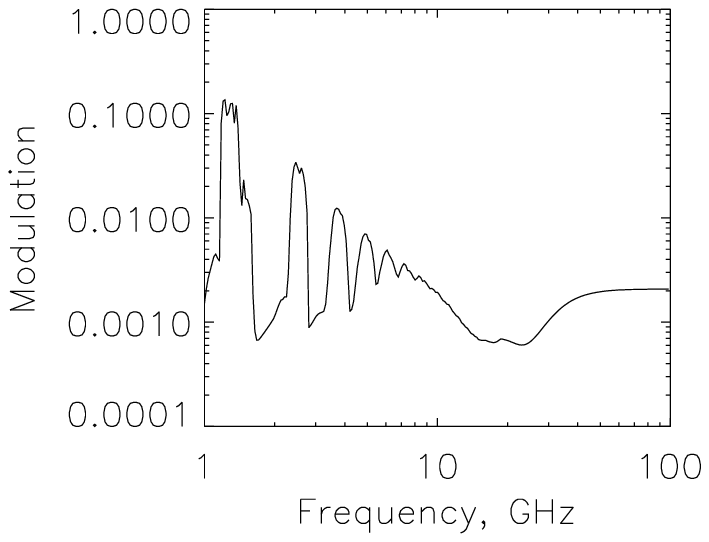}
	\caption{Sausage mode, high magnetic field, oscillations of the flux density. Same as in Figure~\ref{sausage_lowfreq_f_analysis} but for viewing angle 80$^\circ$.}
	\label{sausage_theta80_lowfreq_f_analysis}
\end{figure}

\begin{figure}[htp!]
	\centering
	\includegraphics{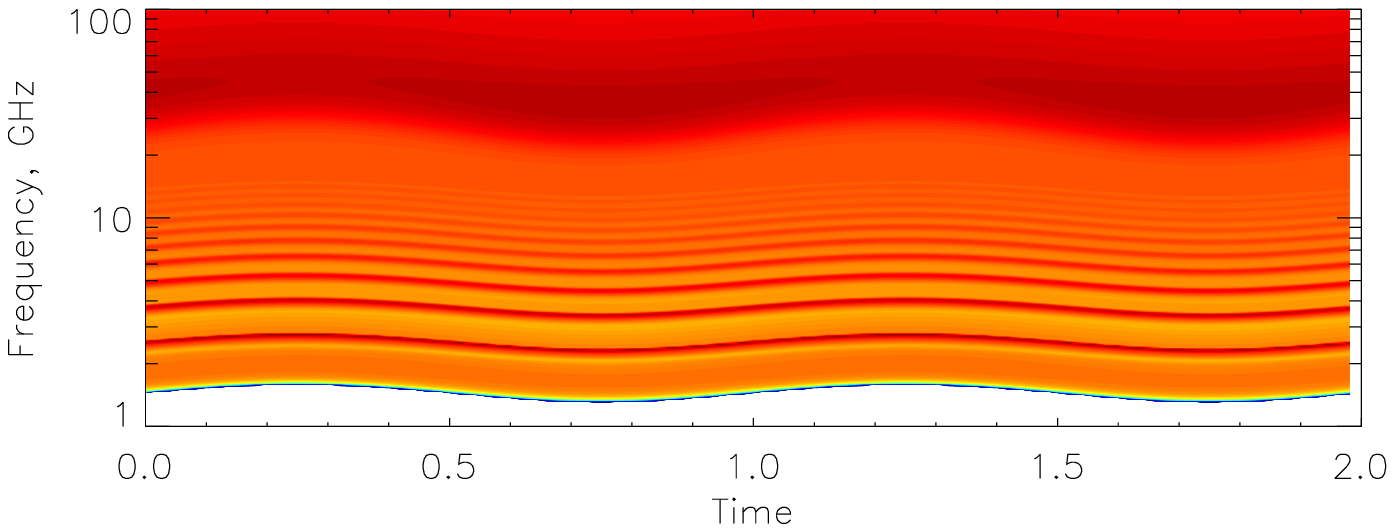}
	\includegraphics{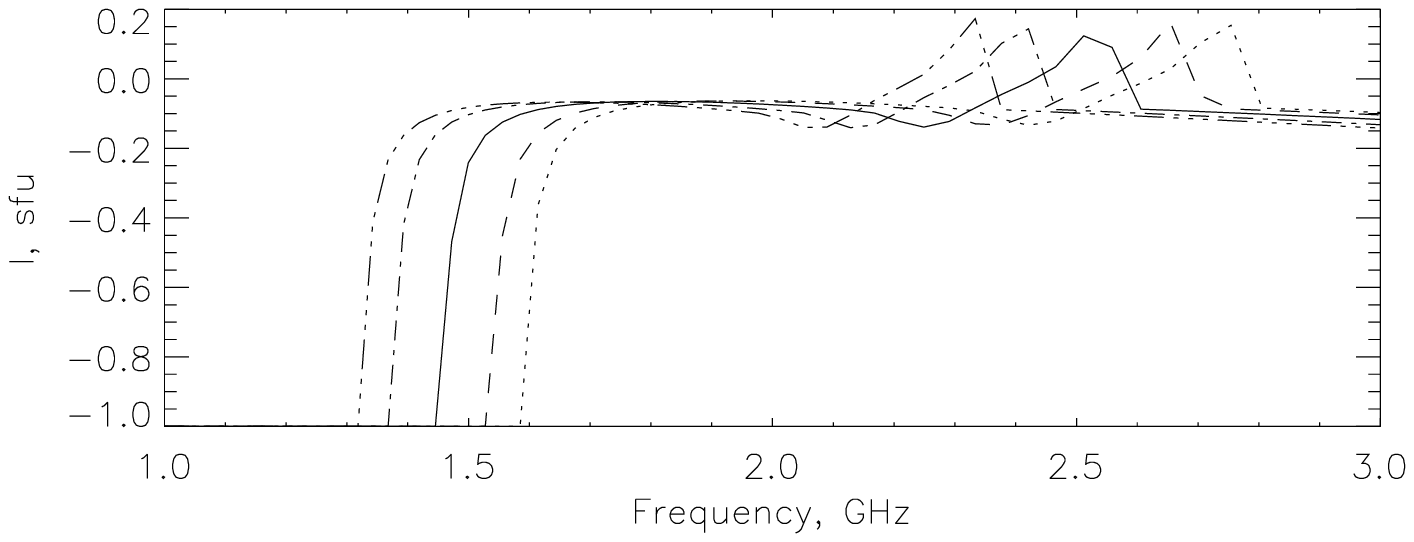}
	\includegraphics{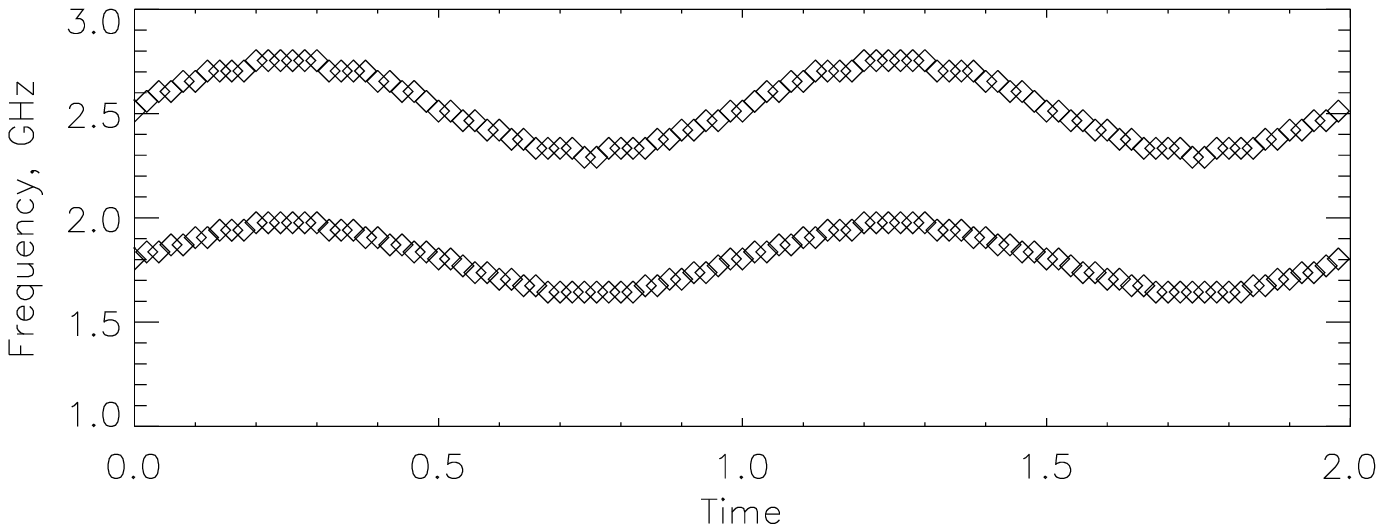}
	\caption{Sausage mode, high magnetic field, oscillations of the polarization. Same as in Figure~\ref{sausage_lowfreq_p_light} but for viewing angle 80$^\circ$.}
	\label{sausage_theta80_lowfreq_p_light}
\end{figure}

\begin{figure}[htp!]
	\centering
	\subfloat[2.1 GHz]{\includegraphics[width=0.5\linewidth]{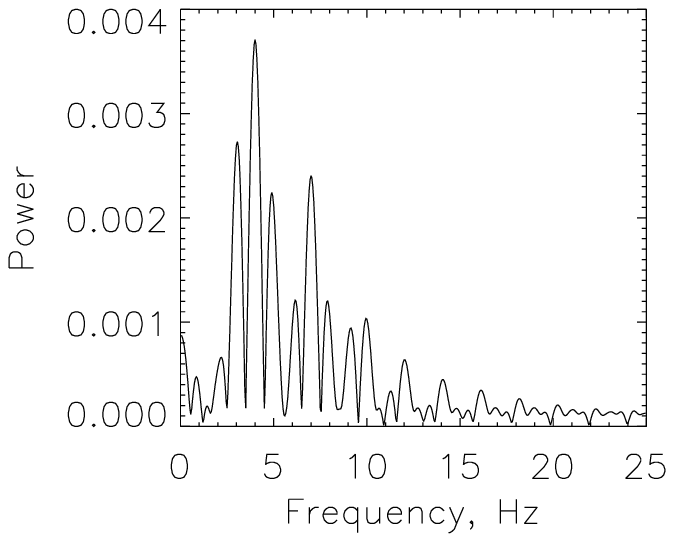}}
	\subfloat[2.3 GHz]{\includegraphics[width=0.5\linewidth]{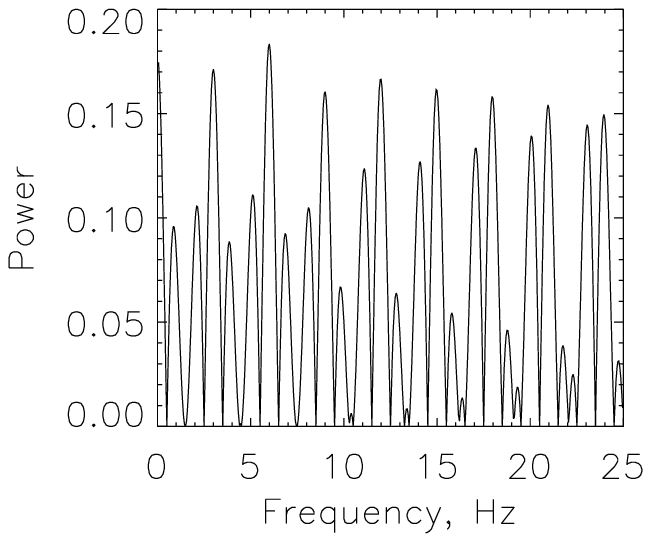}}
	\vspace{-8mm}
	\subfloat[2.5 GHz]{\includegraphics[width=0.5\linewidth]{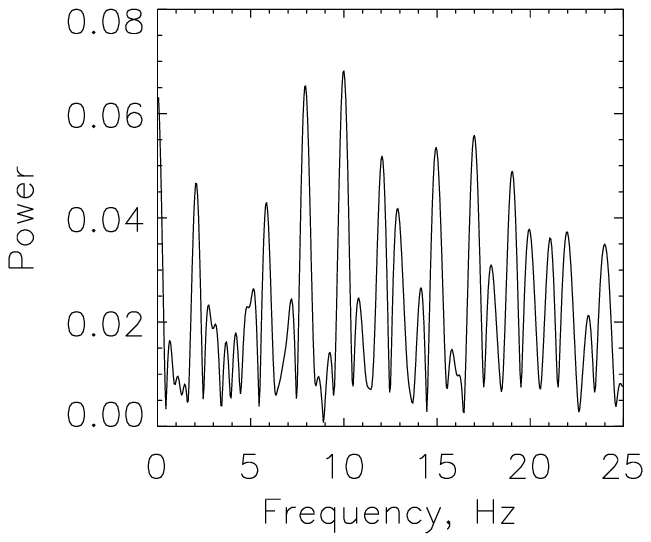}}
	\subfloat[2.7 GHz]{\includegraphics[width=0.5\linewidth]{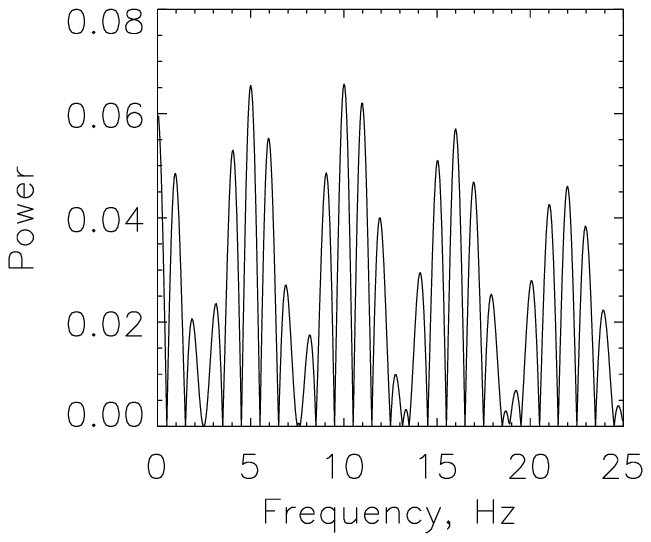}}
	\vspace{-5mm}
	\subfloat[2.9 GHz]{\includegraphics[width=0.5\linewidth]{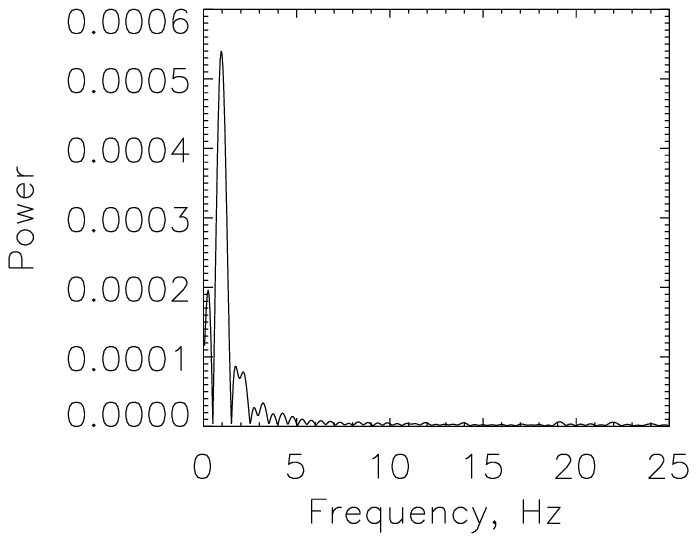}}
	\caption{Sausage mode, high magnetic field, oscillations of the polarization. Same as in Figure~\ref{sausage_lowfreq_p_fourier} but for viewing angle 80$^\circ$. }
	\label{sausage_theta80_lowfreq_p_fourier}
\end{figure}

\begin{figure}[htp!]
	\centering
	\includegraphics{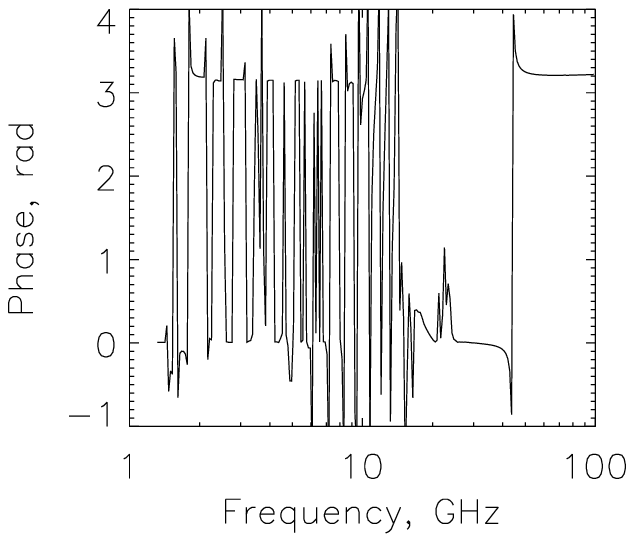}
	\includegraphics{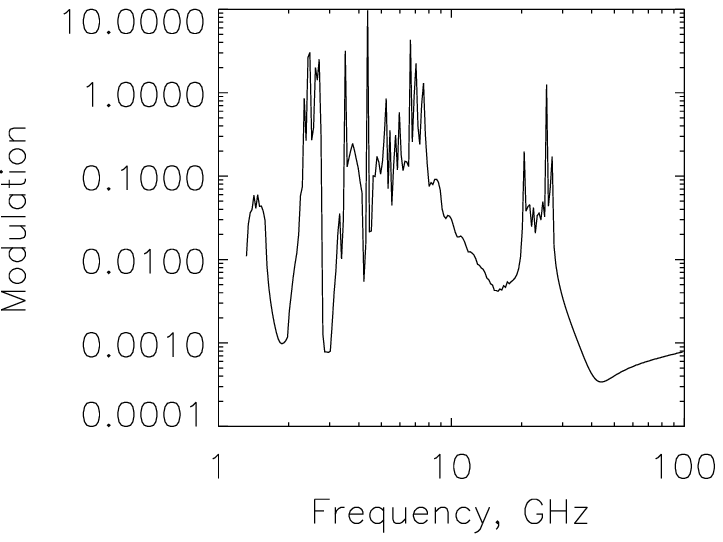}
	\caption{Sausage mode, high magnetic field, oscillations of the polarization. Same as in Figure~\ref{sausage_lowfreq_p_analysis} but for viewing angle 80$^\circ$.}
	\label{sausage_theta80_lowfreq_p_analysis}
\end{figure}

The presence of low-frequency peaks in the emission is an indicator of high magnetic field. In emission from a pulsating source, these peaks oscillate as well, Figure~\ref{sausage_lowfreq_f_light}, accordingly, the Fourier spectra consist of many peaks, Figure~\ref{sausage_lowfreq_f_fourier}. The oscillation phase repeatedly changes around the gyroharmonics and the modulation amplitude fluctuates at the low-frequency spectral range, Figure~\ref{sausage_lowfreq_f_analysis}.
Figures~\ref{sausage_lowfreq_p_light}--\ref{sausage_lowfreq_p_analysis} demonstrate the same parameters for the degree of polarization, while Figures~\ref{sausage_theta80_lowfreq_f_light}--\ref{sausage_theta80_lowfreq_p_analysis} display the same parameters for a different viewing angle, 80$^\circ$.
	
\section{Kink mode}

\begin{figure}[htp!]
	\centering	
	\subfloat[1.5 GHz]{\includegraphics[width=0.3\linewidth]{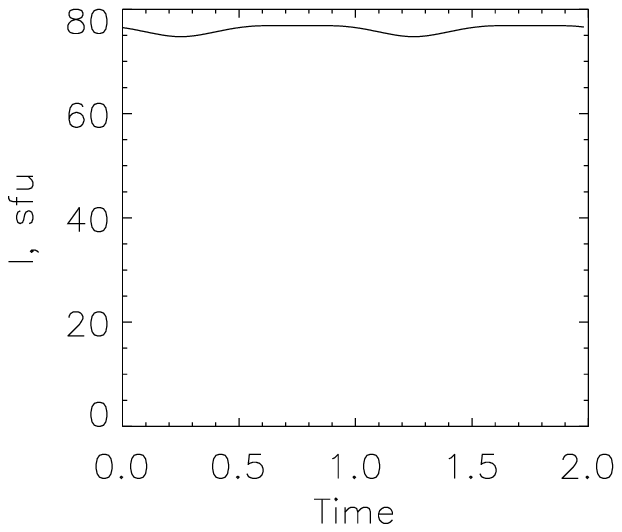}}
	\subfloat[3.1 GHz]{\includegraphics[width=0.3\linewidth]{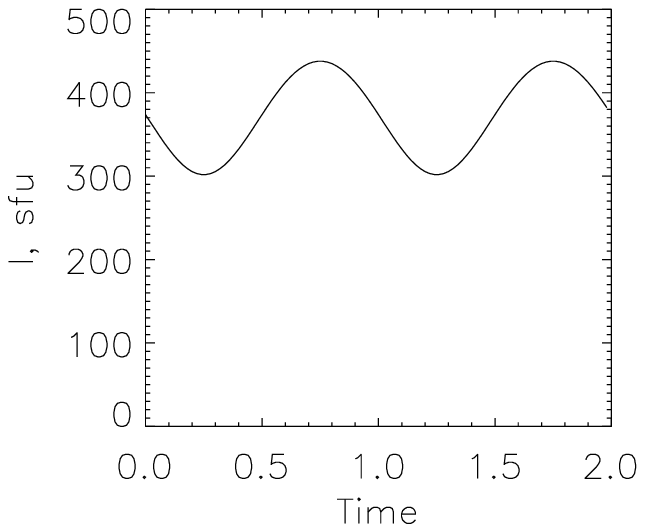}}
	\subfloat[13.8 GHz]{\includegraphics[width=0.3\linewidth]{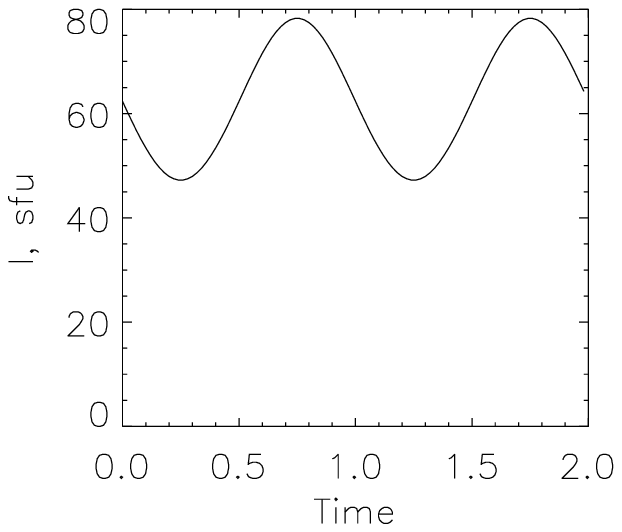}}
	\caption{Kink mode, default parameters, oscillation of the flux density at indicated frequencies.}
	\vspace{-5mm}
	\label{kink_default_osc_f}
\end{figure}

\begin{figure}[hbp!]
	\centering
	\subfloat[1.5 GHz]{\includegraphics[width=0.3\textwidth]{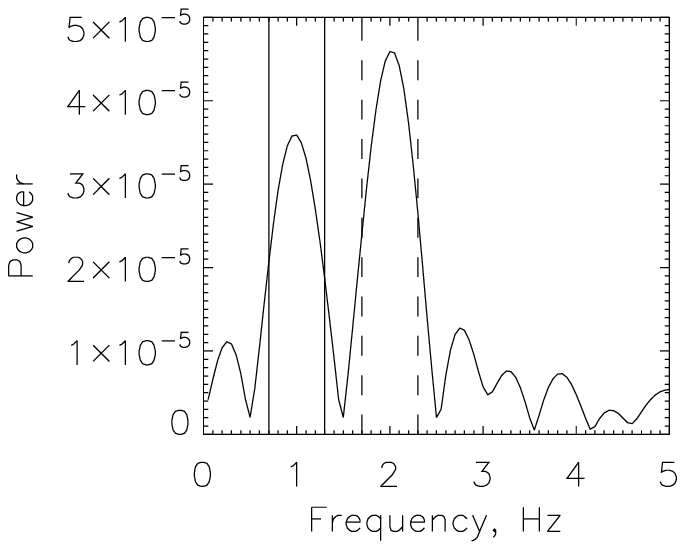}}
	\subfloat[3.1 GHz]{\includegraphics[width=0.3\textwidth]{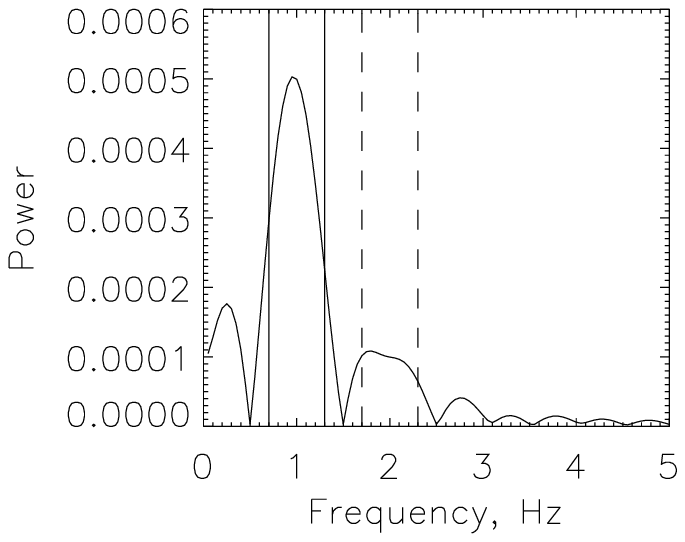}}
	\subfloat[13.8 GHz]{\includegraphics[width=0.3\textwidth]{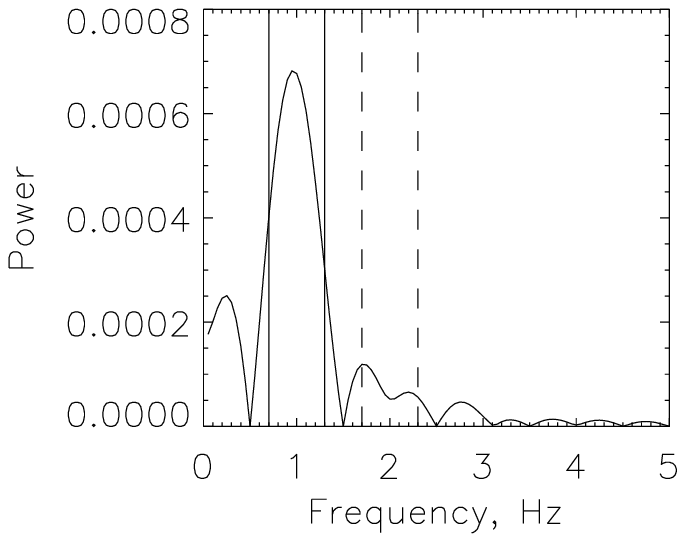}}
	\vspace{-5mm}
	\subfloat[Phase]{\includegraphics[width=0.3\textwidth]{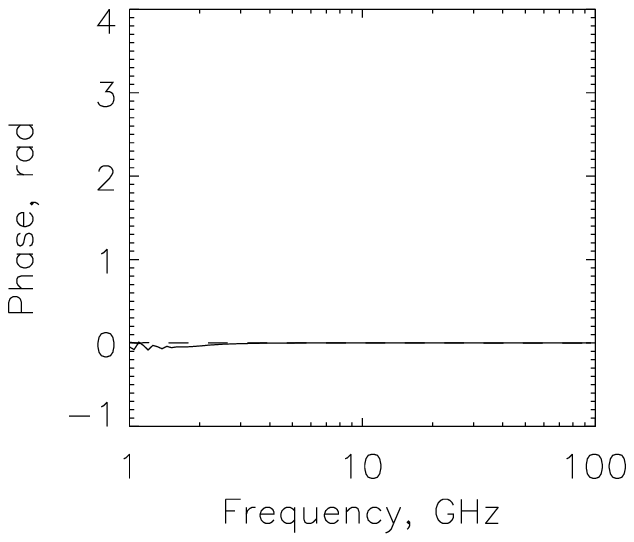}}
	\subfloat[Partial modulation]{\includegraphics[width=0.3\textwidth]{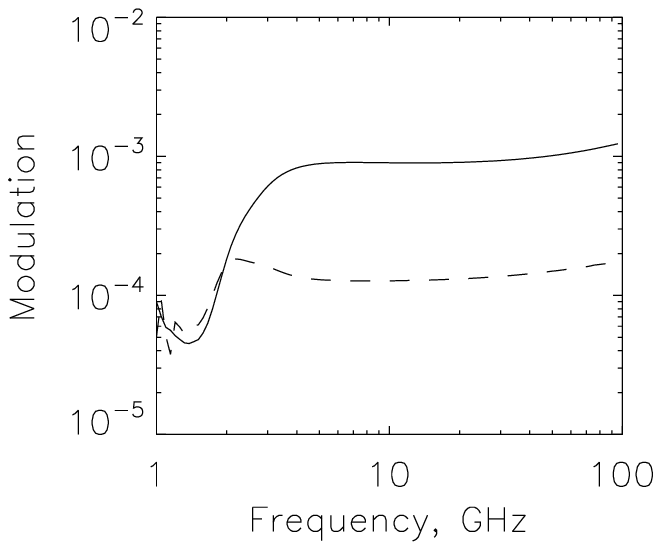}}
	\subfloat[Full modulation]{\includegraphics[width=0.3\textwidth]{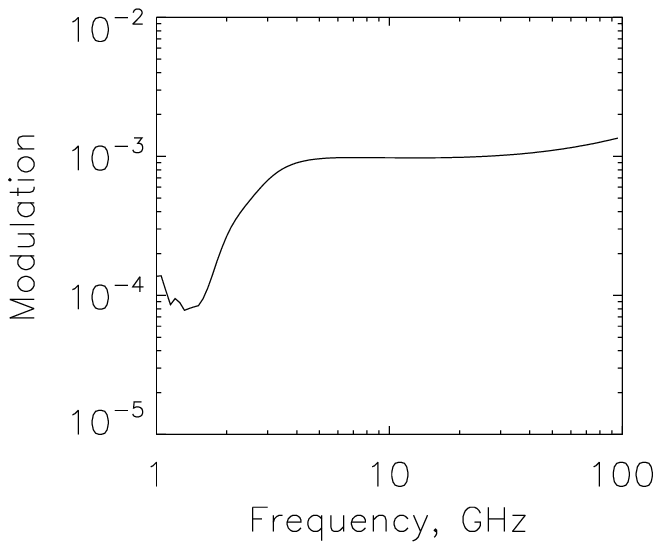}}
	\caption{Kink mode, default parameters, oscillations of the flux density. {\bf(a)-(c)} Fourier spectra for the emission frequencies indicated, vertical lines mark off the  peak regions considered for partial modulation. {\bf(d)} Relative phase of flux, solid line is the fundamental harmonics (1~Hz), dashed line is the second harmonics  (2~Hz); note the  phase constancy across the entire spectral range. {\bf(e)} Partial modulation amplitude for the indicated range of Fourier coefficients for two harmonics: the solid (dashed) line represents integration between the solid (dashed) vertical lines in panels {\bf(a)-(c)}.  {\bf(f)} Full modulation amplitude for the oscillating component; it decreases toward the optically thick part of the spectrum in contrast to sausage mode, which displays a minimum modulation around the spectrum peak frequency.}
	\label{kink_default_osc_f_analysis}

\end{figure}

\begin{figure}[htp!]
	\centering	
	\subfloat[1.5 GHz]{\includegraphics[width=0.3\linewidth]{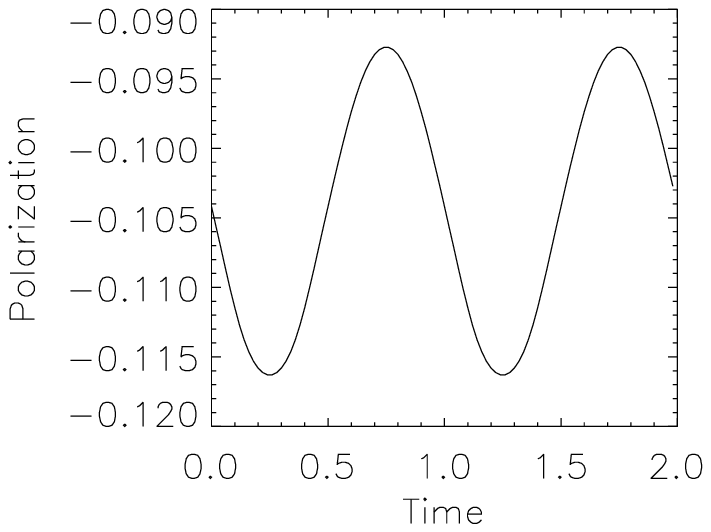}}
	\subfloat[3.1 GHz]{\includegraphics[width=0.3\linewidth]{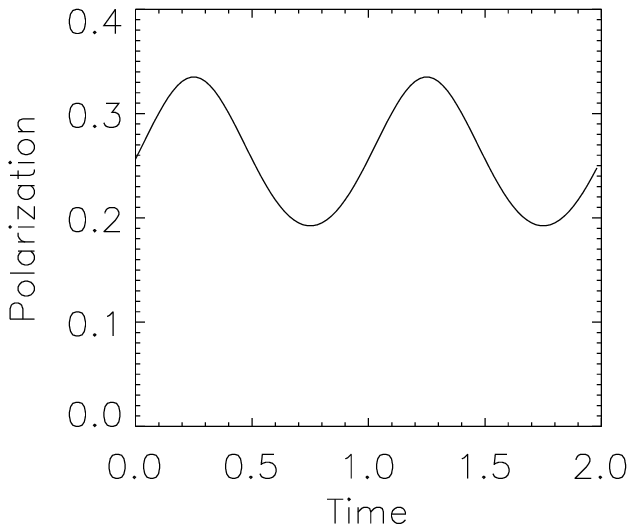}}
	\subfloat[13.8 GHz]{\includegraphics[width=0.3\linewidth]{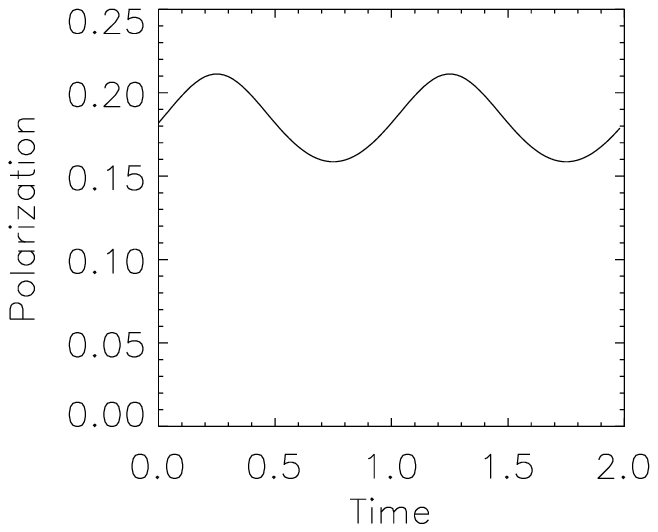}}
	\caption{Kink Mode, default parameters, oscillation of polarization at indicated frequencies. Note, the polarization is $\pi$ out of phase with the flux density at high frequencies, while in phase at low (optically thick) frequencies, sf Figure~\ref{kink_default_osc_f}.}
	\vspace{-5mm}
	\label{kink_default_osc_p}
\end{figure}

\begin{figure}[hbp!]
	\centering
	\subfloat[1.5 GHz]{\includegraphics[width=0.3\textwidth]{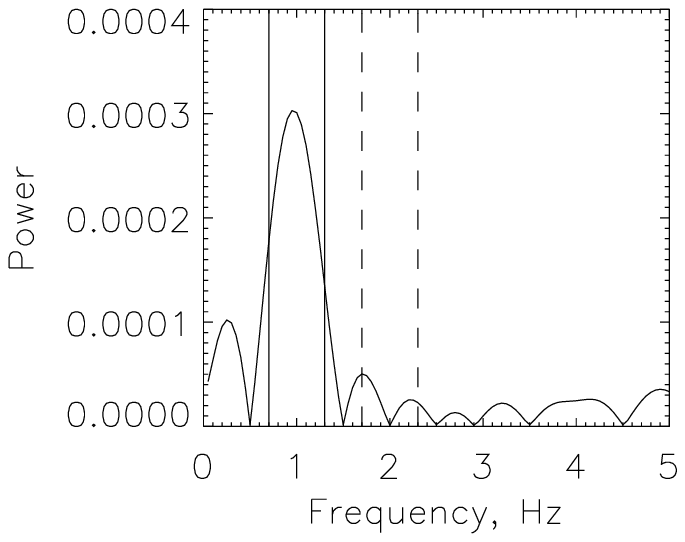}}
	\subfloat[3.1 GHz]{\includegraphics[width=0.3\textwidth]{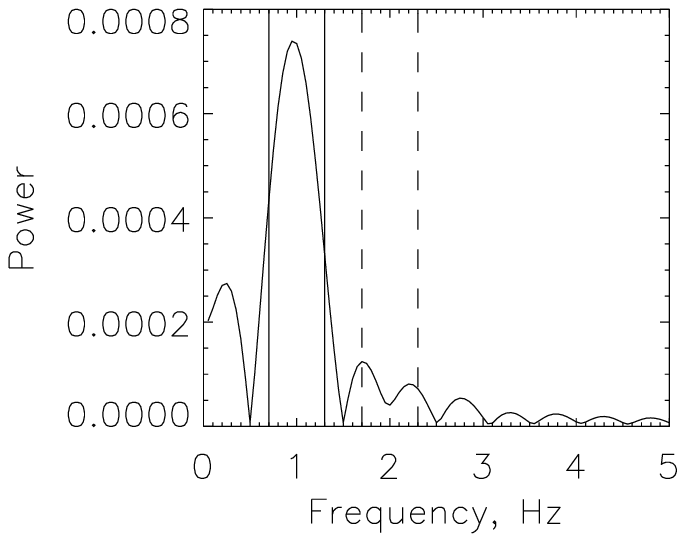}}
	\subfloat[13.8 GHz]{\includegraphics[width=0.3\textwidth]{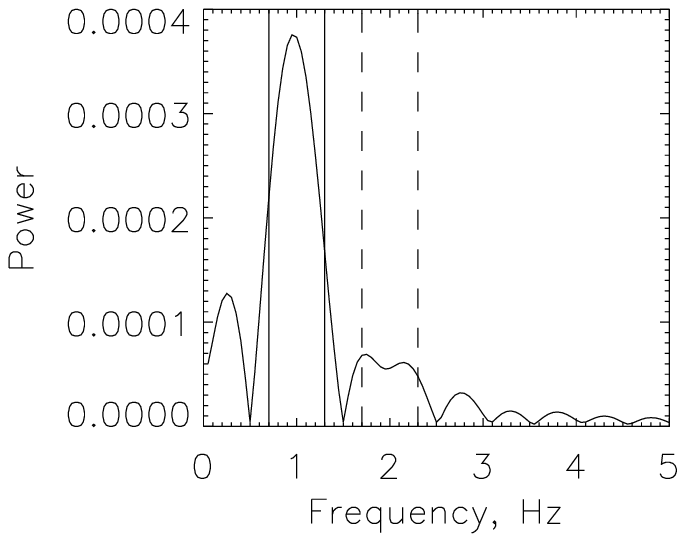}}
	
	\subfloat[Phase]{\includegraphics[width=0.3\textwidth]{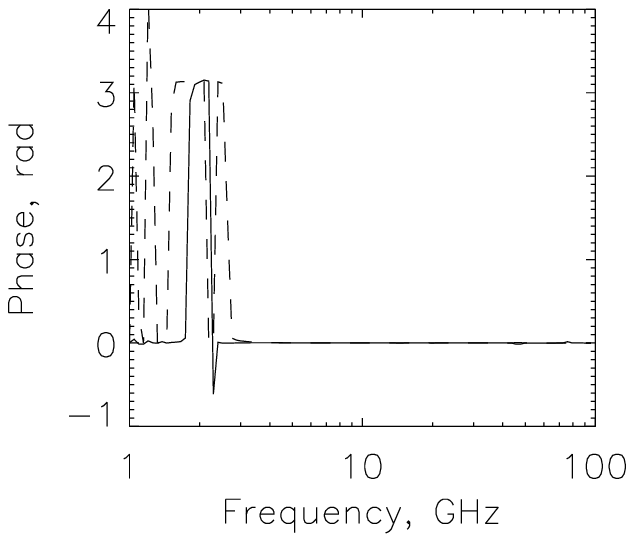}}
	\subfloat[Partial modulation]{\includegraphics[width=0.3\textwidth]{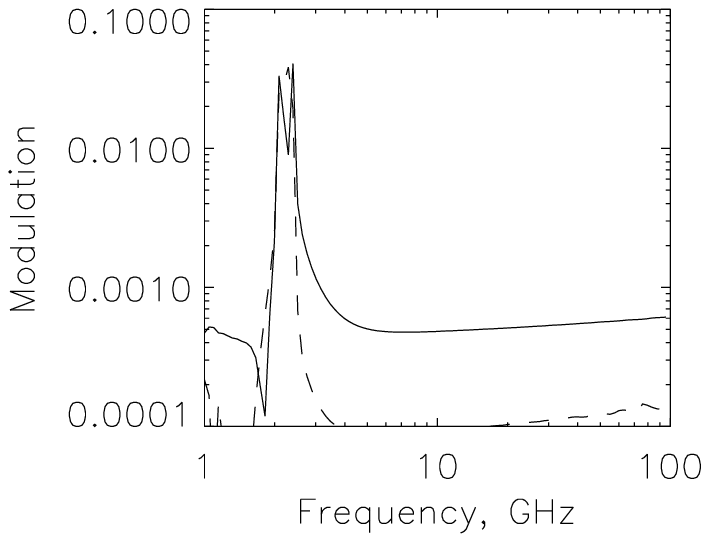}}
	\subfloat[Full modulation]{\includegraphics[width=0.3\textwidth]{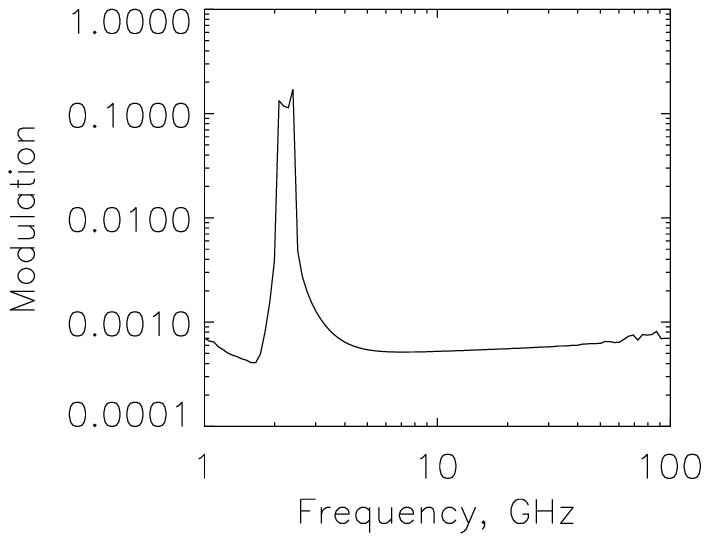}}
	\caption{Kink mode, default parameters. Same as in Figure~\ref{kink_default_osc_p_analysis} but for  oscillations of polarization. }
	\label{kink_default_osc_p_analysis}
\end{figure}

The \emph{kink mode}, default parameters, Figures~\ref{kink_default_osc_f}--\ref{kink_default_osc_p_analysis} is distinguishable from the sausage mode.  Indeed, in the optically thin regime at high frequencies, the flux and polarization are $\pi$ out of phase with each other, see Figures~\ref{kink_default_osc_f} and \ref{kink_default_osc_p}, the oscillation phase does not change with frequency, and the modulation amplitude curve has a distinctly different shape decreasing at lower frequencies (at the optically thick region), Figure~\ref{kink_default_osc_f_analysis}.

\begin{figure}[htp!]
	\centering	
	\subfloat[1.5 GHz]{\includegraphics[width=0.3\linewidth]{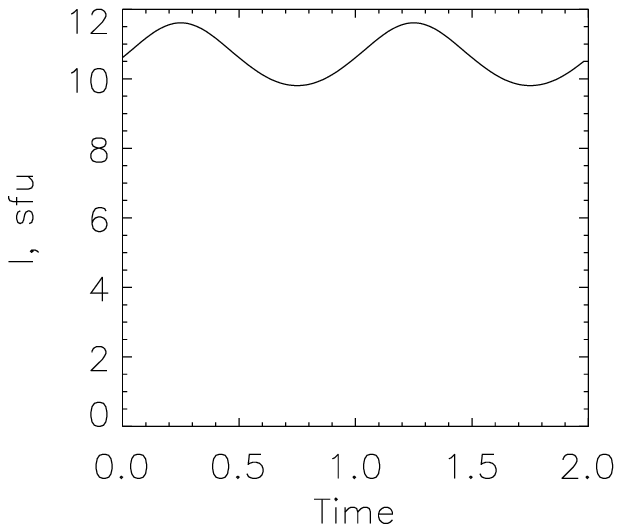}}
	\subfloat[2.2 GHz]{\includegraphics[width=0.3\linewidth]{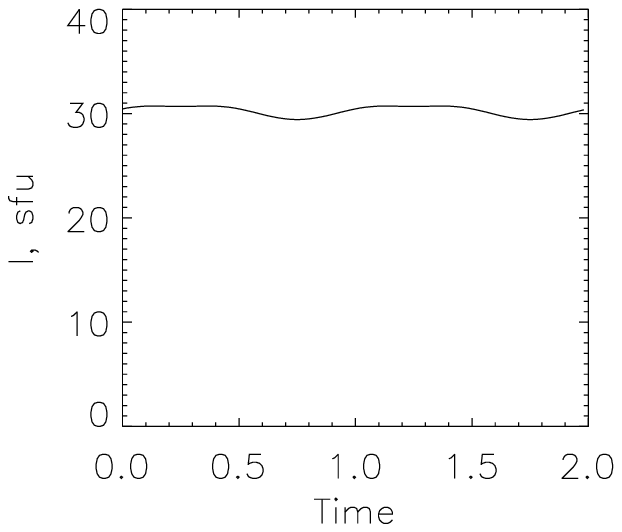}}
	\subfloat[13.8 GHz]{\includegraphics[width=0.3\linewidth]{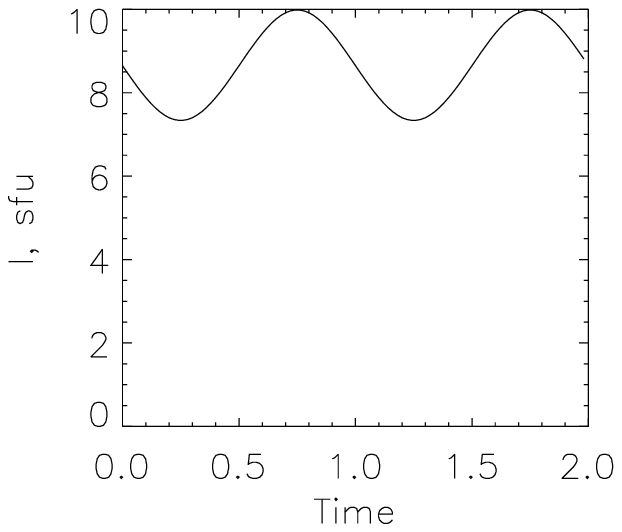}}
	\caption{Kink mode,  single pixel view, oscillation of flux at indicated frequencies.}
	\vspace{-5mm}
	\label{kink_pixel_osc_f}
\end{figure}

\begin{figure}[hbp!]
	\centering
	\subfloat[1.5 GHz]{\includegraphics[width=0.3\textwidth]{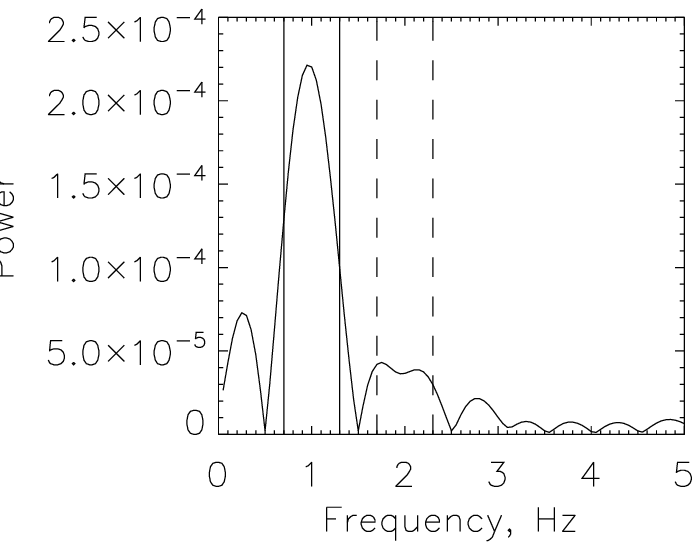}}
	\subfloat[2.2 GHz]{\includegraphics[width=0.3\textwidth]{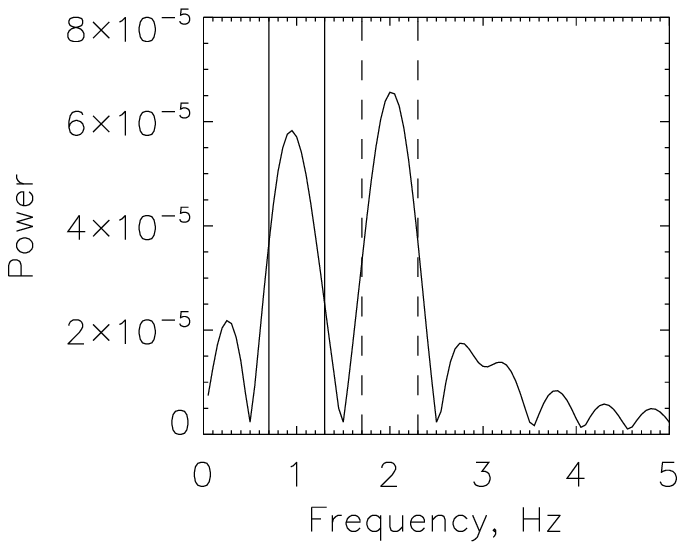}}
	\subfloat[13.8 GHz]{\includegraphics[width=0.3\textwidth]{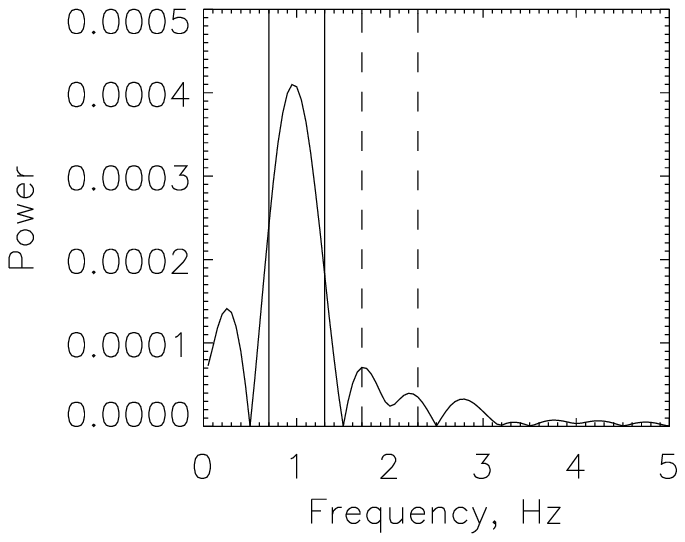}}
	\vspace{-5mm}
	\subfloat[Phase]{\includegraphics[width=0.3\textwidth]{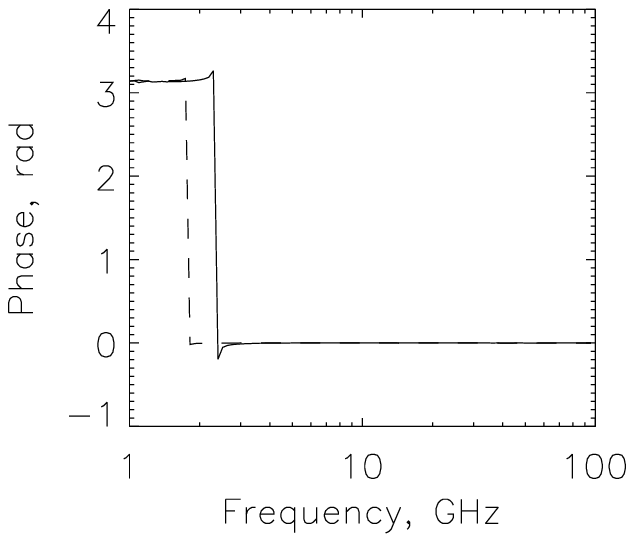}}
	\subfloat[Partial modulation]{\includegraphics[width=0.3\textwidth]{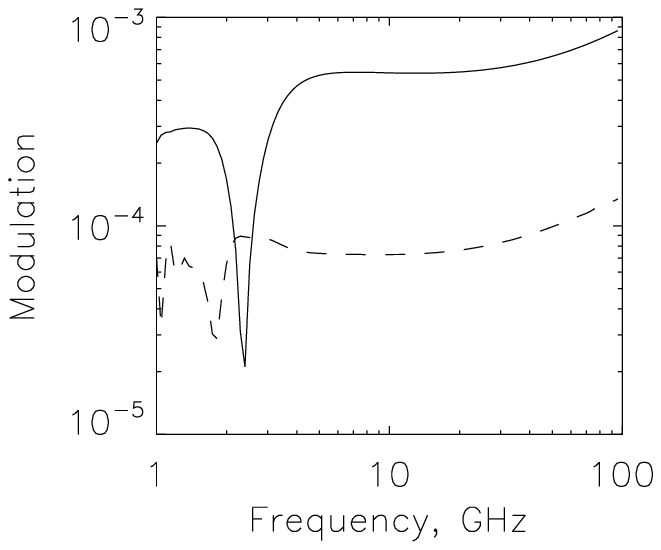}}
	\subfloat[Full modulation]{\includegraphics[width=0.3\textwidth]{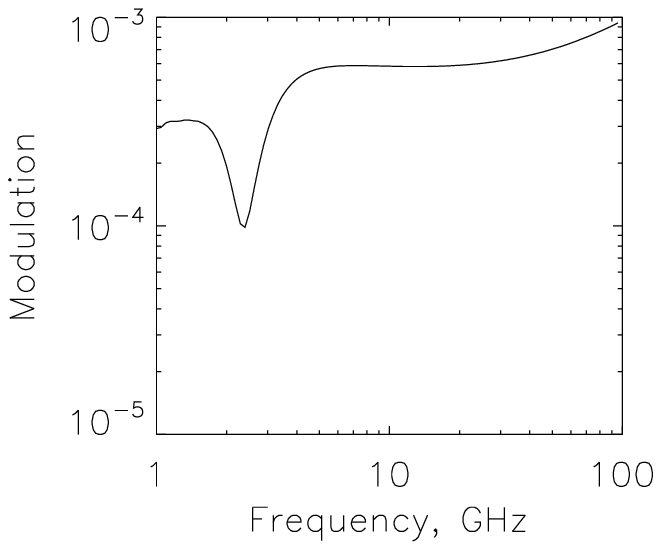}}
	\caption{Kink mode, single pixel view, oscillations of flux. {\bf(a)-(f)} Same as in Figure~\ref{kink_default_osc_f_analysis}. Note that unlike Figure~\ref{kink_default_osc_f_analysis}, here the phase experiences the $\pi$ shift, panel {\bf(d)},  and modulation amplitude displays a curve with a deep minimum, panels {\bf(e, f)}, around the spectrum peak frequency}
	\label{kink_pixel_osc_f_analysis}

\end{figure}

\begin{figure}[htp!]
	\centering	
	\subfloat[1.5 GHz]{\includegraphics[width=0.3\linewidth]{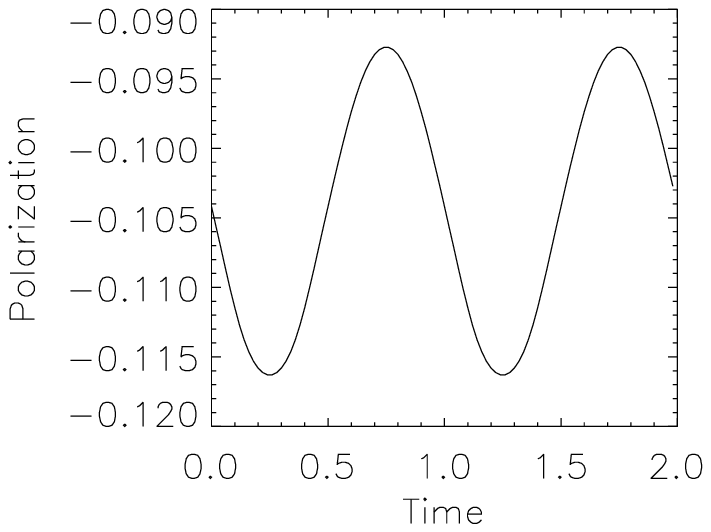}}
	\subfloat[2.2 GHz]{\includegraphics[width=0.3\linewidth]{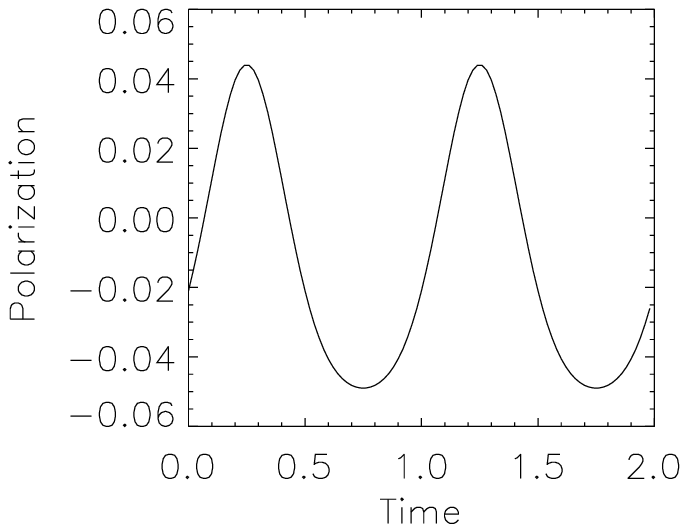}}
	\subfloat[13.8 GHz]{\includegraphics[width=0.3\linewidth]{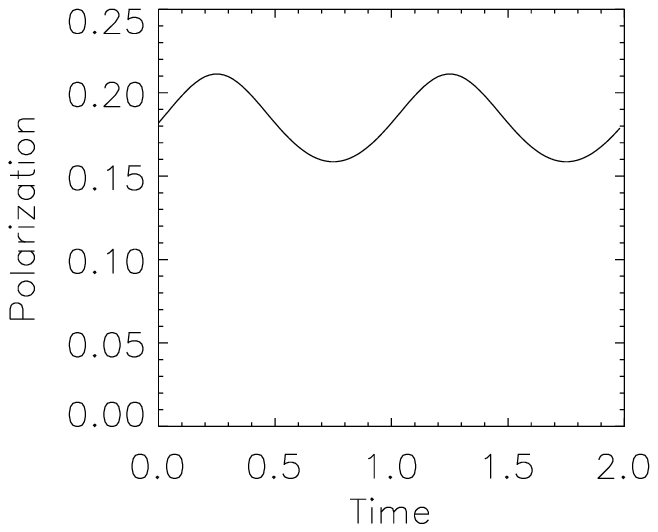}}
	\caption{Kink Mode, single pixel view, oscillation of polarization at indicated frequencies. Now at any frequency (thin or thick) the polarization is $\pi$ out of phase relative to the flux density at the same frequency. }
	\vspace{-5mm}
	\label{kink_pixel_osc_p}
\end{figure}

\begin{figure}[hbp!]
	\centering
	\subfloat[1.5 GHz]{\includegraphics[width=0.3\textwidth]{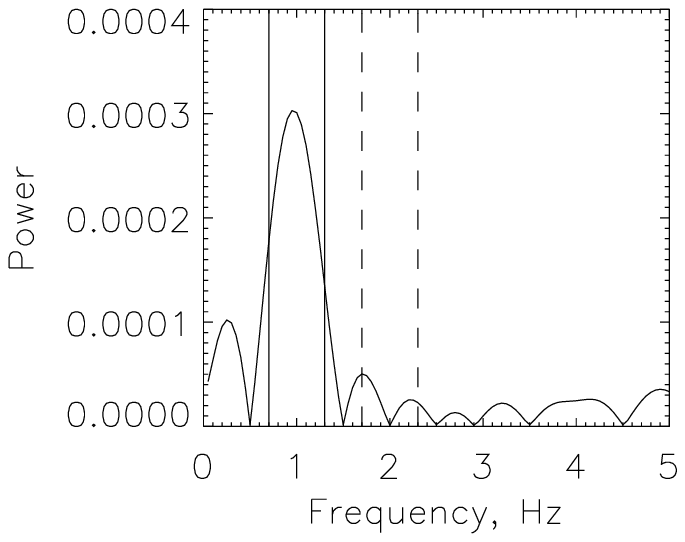}}
	\subfloat[2.2 GHz]{\includegraphics[width=0.3\textwidth]{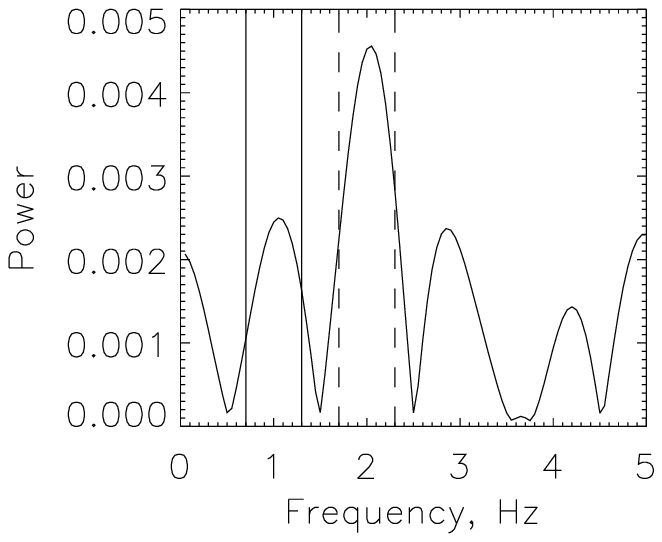}}
	\subfloat[13.8 GHz]{\includegraphics[width=0.3\textwidth]{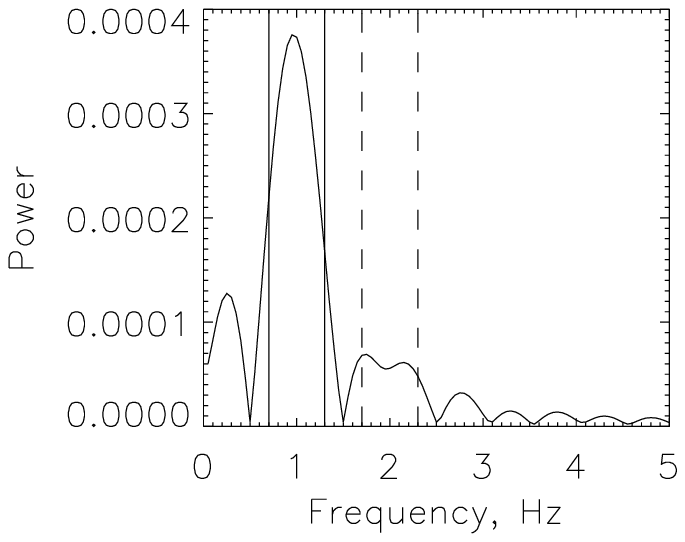}}
	
	\subfloat[Phase]{\includegraphics[width=0.3\textwidth]{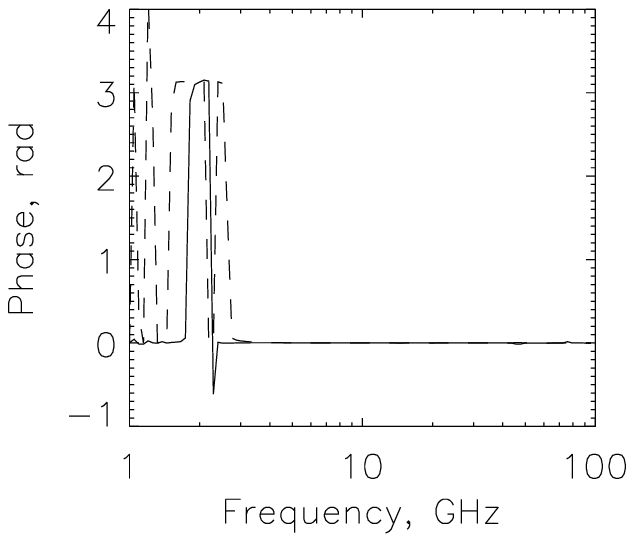}}
	\subfloat[Partial modulation]{\includegraphics[width=0.3\textwidth]{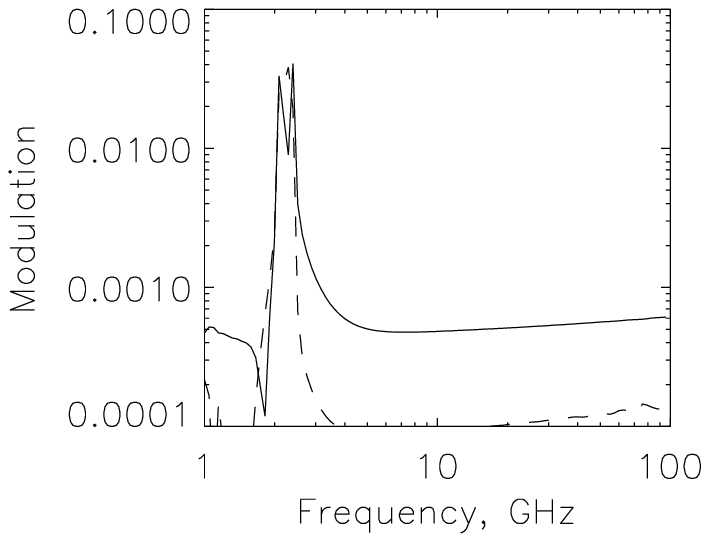}}
	\subfloat[Full modulation]{\includegraphics[width=0.3\textwidth]{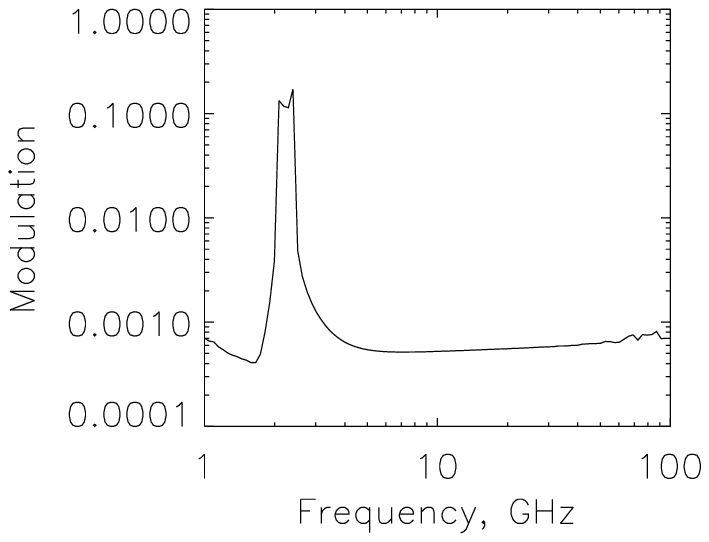}}
	\caption{Kink mode, single pixel view, oscillations of polarization. {\bf(a)-(c)}Fourier spectrum for the emission frequency indicated. Dashed lines mark off peaks considered for partial modulation. {\bf(d)} Relative phase of flux. Solid line is the fundamental harmonic, dashed line is second harmonic.  {\bf(e)} Partial modulation amplitude for the indicated range of Fourier coefficients, for the first two harmonics. {\bf(f)} Full modulation amplitude.}
	\label{kink_pixel_osc_p_analysis}
\end{figure}

The single pixel view is significantly different from the total power data because no visible source area oscillations are involved in the QPPs from a single pixel, Figures~\ref{kink_pixel_osc_f}--\ref{kink_pixel_osc_p_analysis}. Here the phase experiences the $\pi$ shift around the spectrum peak frequency and the modulation amplitude forms a curve with a clear minimum around the spectrum peak. These differences between the total power and single pixel view results could be highly valuable to distinguish the kink mode in the imaging spectroscopy observations.
	
\begin{figure}[htp!]
	\centering	
	\subfloat[1.5 GHz]{\includegraphics[width=0.3\linewidth]{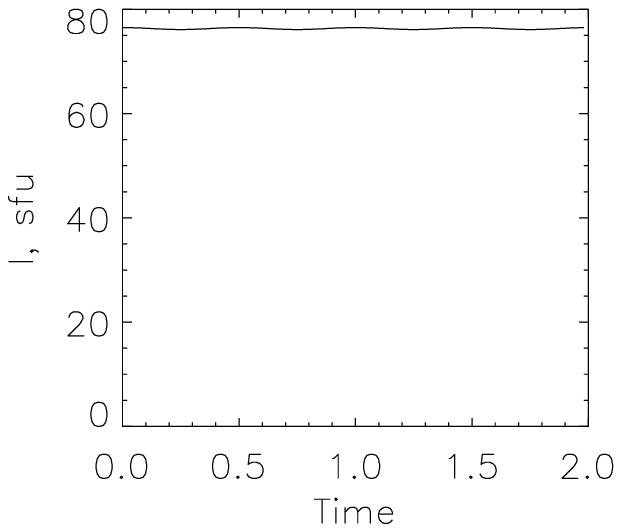}}
	\subfloat[3.1 GHz]{\includegraphics[width=0.3\linewidth]{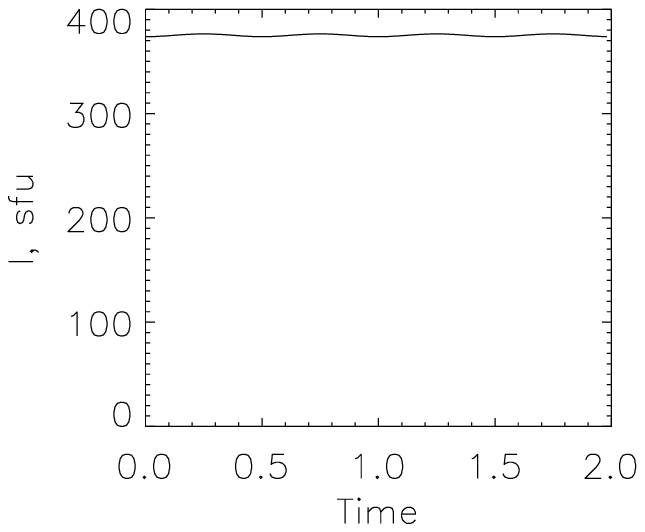}}
	\subfloat[13.8 GHz]{\includegraphics[width=0.3\linewidth]{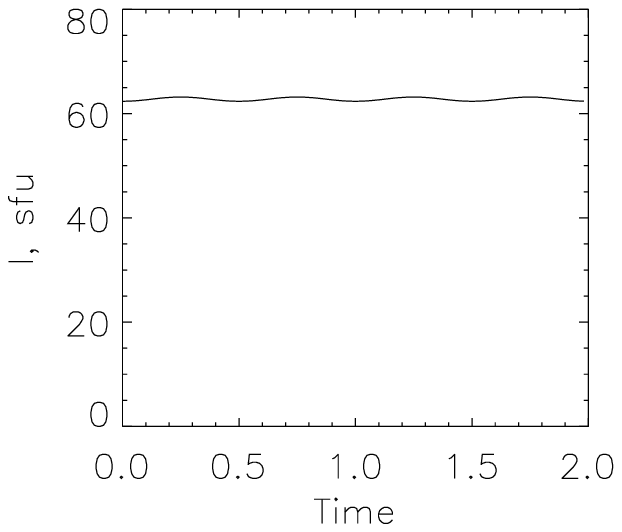}}
	\caption{Kink mode,  no line-of-sight oscillations, oscillation of the flux density at indicated frequencies.}
	\vspace{-5mm}
	\label{kink_noaz_osc_f}
\end{figure}

\begin{figure}[hbp!]
	\centering
	\subfloat[1.5 GHz]{\includegraphics[width=0.3\textwidth]{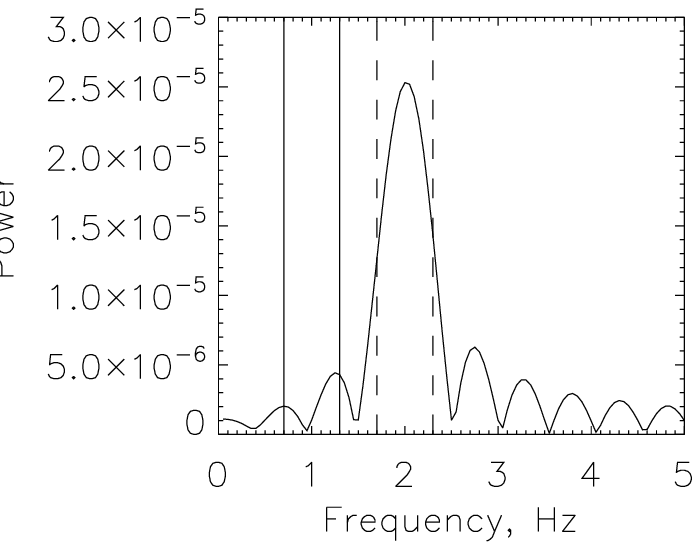}}
	\subfloat[3.1 GHz]{\includegraphics[width=0.3\textwidth]{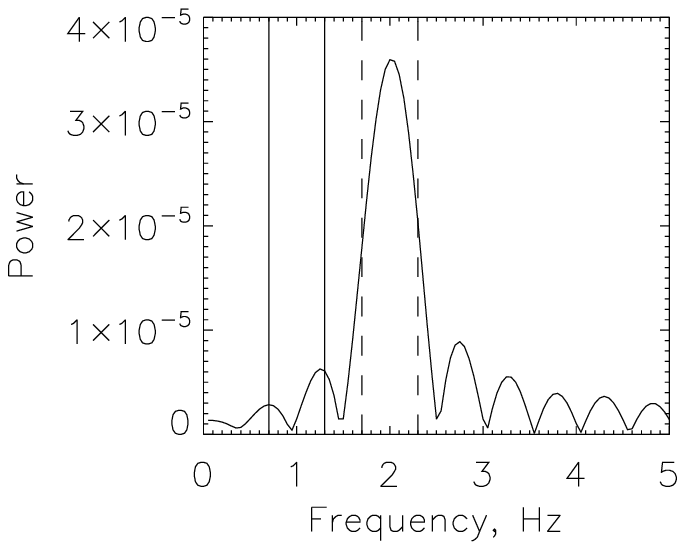}}
	\subfloat[13.8 GHz]{\includegraphics[width=0.3\textwidth]{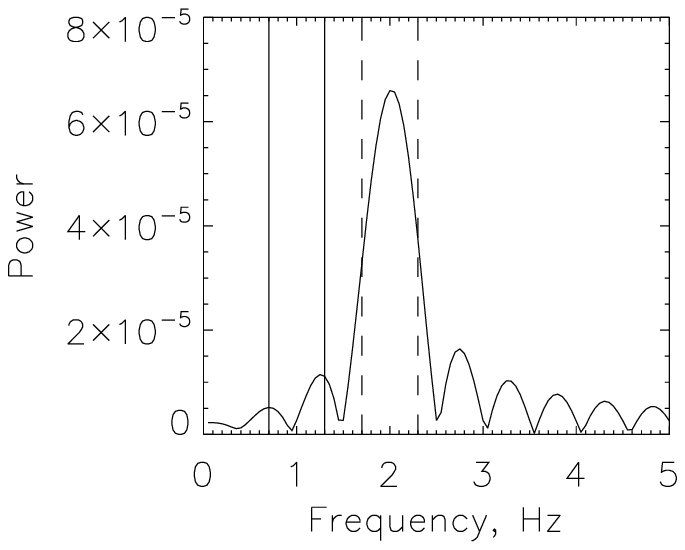}}
	\vspace{-5mm}
	\subfloat[Phase]{\includegraphics[width=0.3\textwidth]{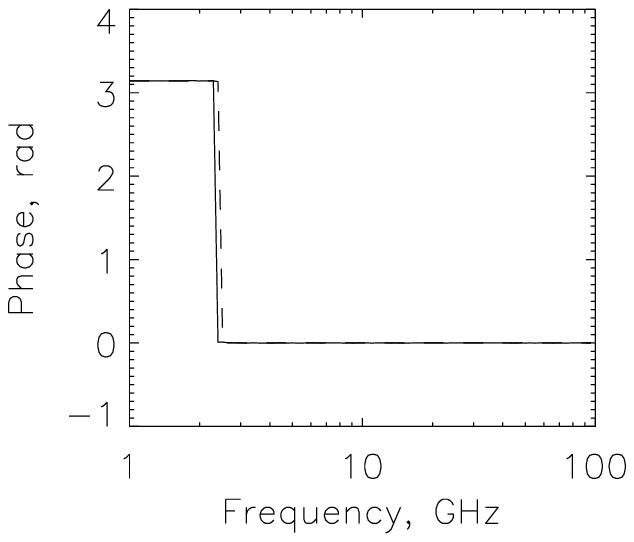}}
	\subfloat[Partial modulation]{\includegraphics[width=0.3\textwidth]{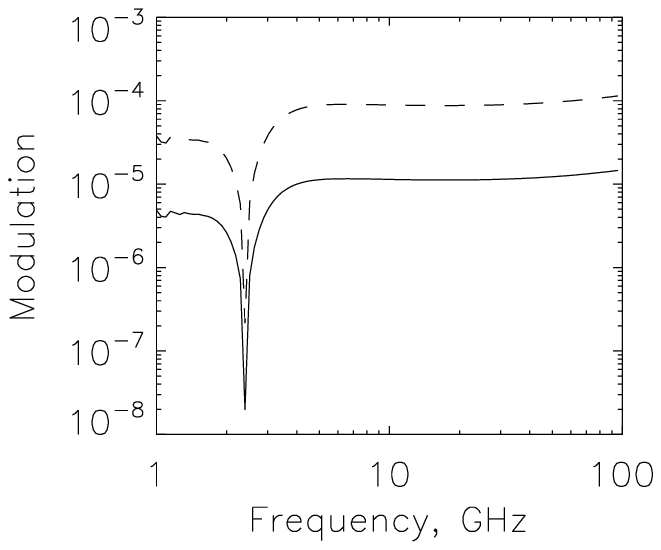}}
	\subfloat[Full modulation]{\includegraphics[width=0.3\textwidth]{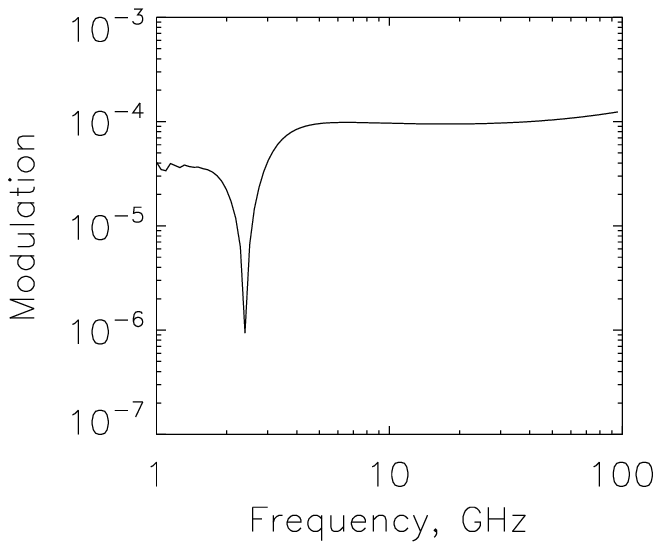}}
	\caption{Kink mode, no line-of-sight oscillations, oscillations of flux. {\bf(a)-(f)} Same as in Figure~\ref{kink_default_osc_f_analysis}. Note enhanced second harmonic oscillations at all frequencies and overall small modulation amplitude as it is a second-order effect in this case.}
	\label{kink_noaz_osc_f_analysis}

\end{figure}

\begin{figure}[htp!]
	\centering	
	\subfloat[1.5 GHz]{\includegraphics[width=0.3\linewidth]{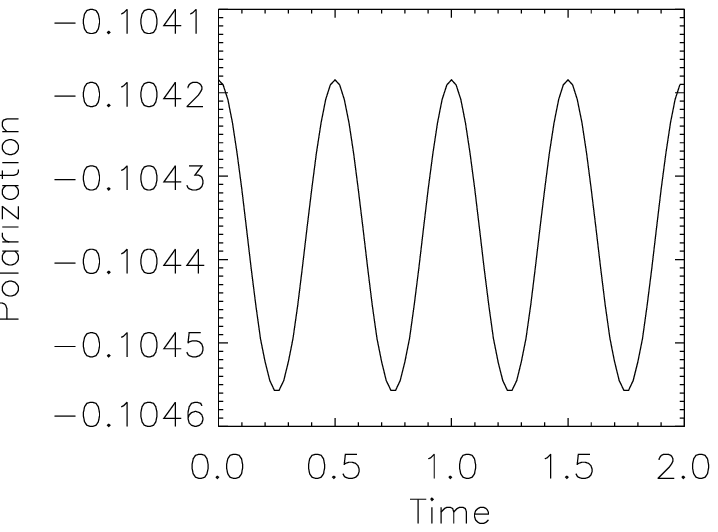}}
	\subfloat[3.1 GHz]{\includegraphics[width=0.3\linewidth]{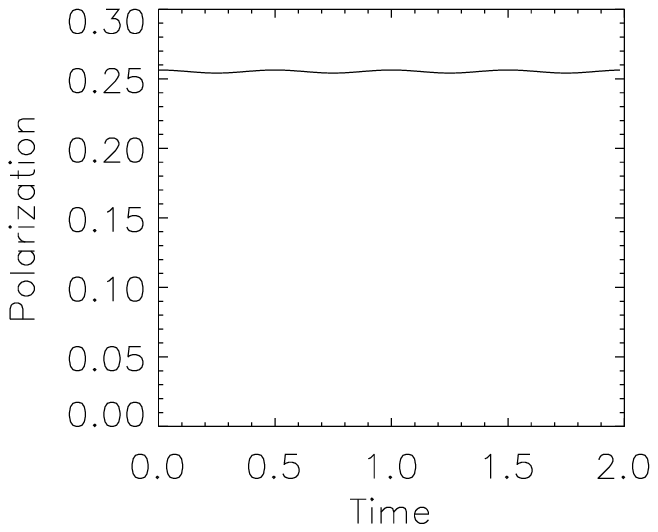}}
	\subfloat[13.8 GHz]{\includegraphics[width=0.3\linewidth]{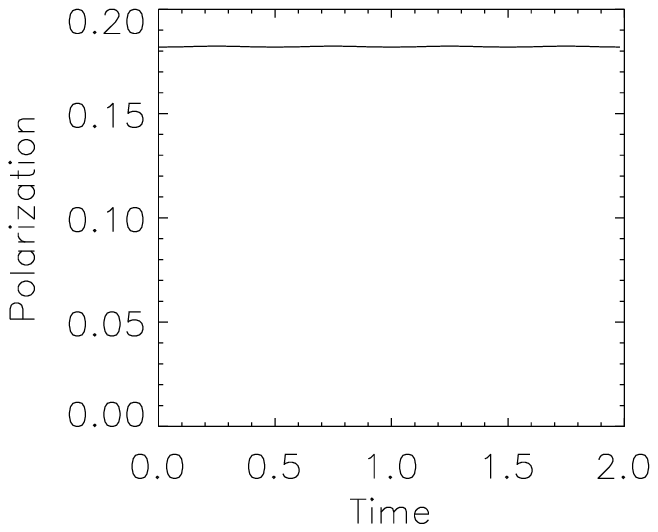}}
	\caption{Kink Mode, no line-of-sight oscillations, oscillation of polarization at indicated frequencies.}
	\vspace{-5mm}
	\label{kink_noaz_osc_p}
\end{figure}

\begin{figure}[hbp!]
	\centering
	\subfloat[1.5 GHz]{\includegraphics[width=0.3\textwidth]{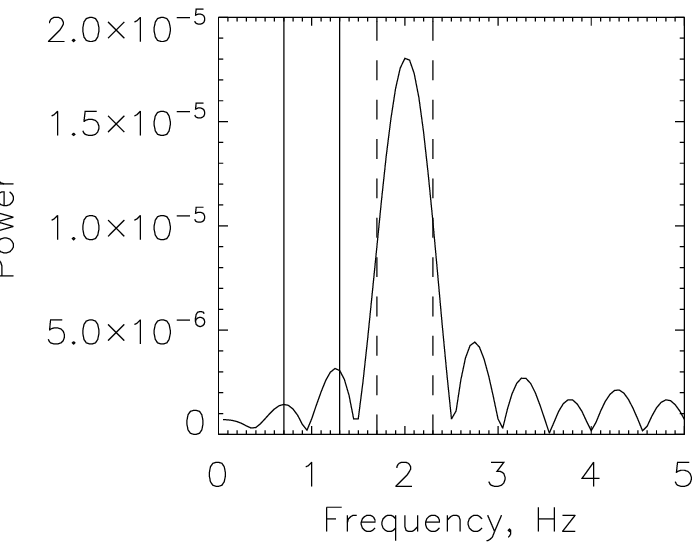}}
	\subfloat[3.1 GHz]{\includegraphics[width=0.3\textwidth]{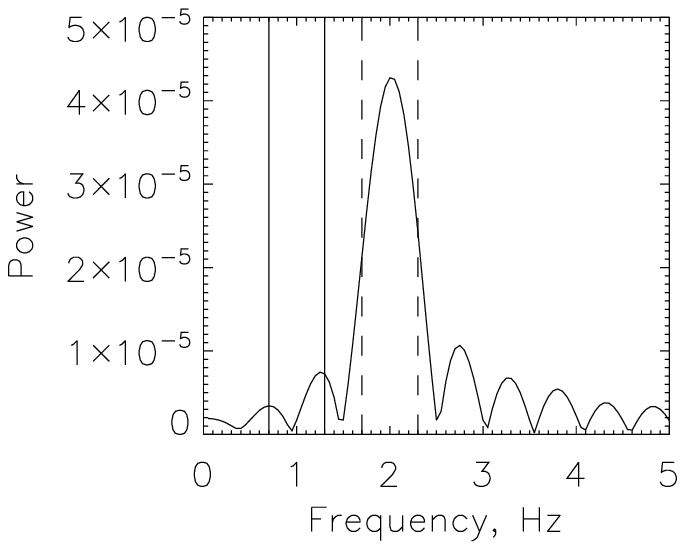}}
	\subfloat[13.8 GHz]{\includegraphics[width=0.3\textwidth]{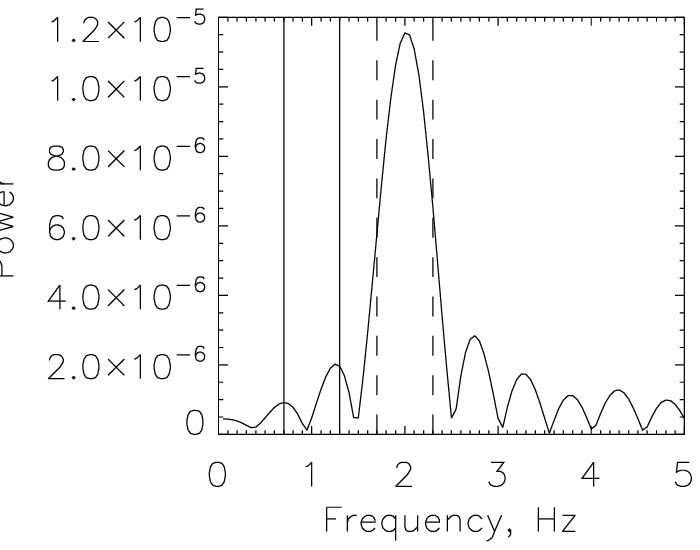}}
	
	\subfloat[Phase]{\includegraphics[width=0.3\textwidth]{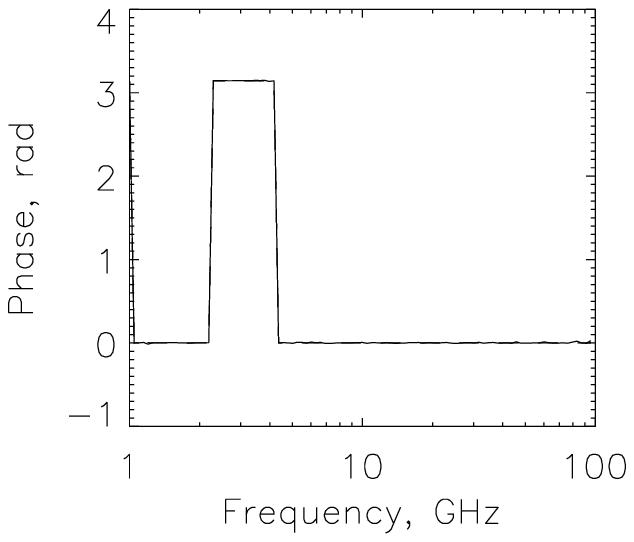}}
	\subfloat[Partial modulation]{\includegraphics[width=0.3\textwidth]{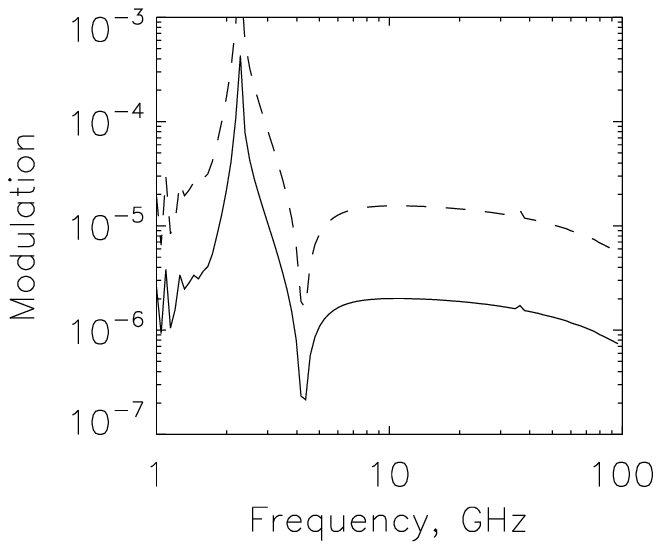}}
	\subfloat[Full modulation]{\includegraphics[width=0.3\textwidth]{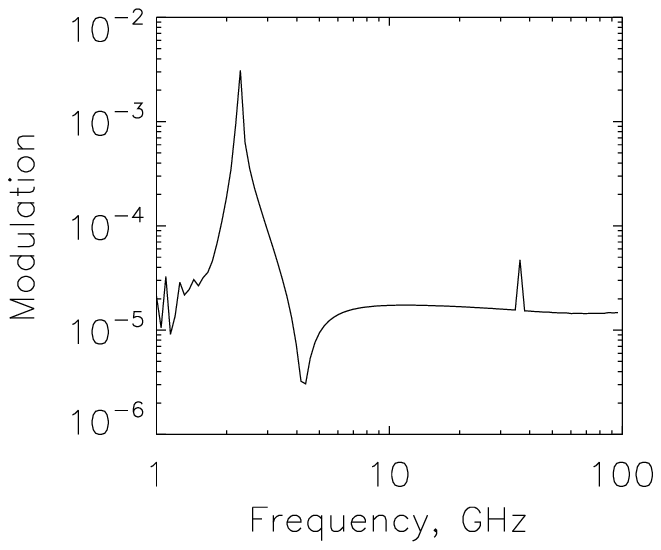}}
	\caption{Kink mode, no line-of-sight oscillations, oscillations of polarization. {\bf(a)-(f)}Same as in Figure~\ref{kink_default_osc_p_analysis}.}
	\label{kink_noaz_osc_p_analysis}
\end{figure}

Figures~\ref{kink_noaz_osc_f}--\ref{kink_noaz_osc_p_analysis} display the results for the kink mode oscillations in the plane transverse to the line of sight. In this case the effect is second order over the magnetic field perturbation. Accordingly, the second harmonics oscillation dominates the Fourier spectrum and the modulation amplitudes are very small.
	
\section{Torsional mode}

\begin{figure}[htp!]
	\centering	
	\subfloat[1.5 GHz]{\includegraphics[width=0.3\textwidth]{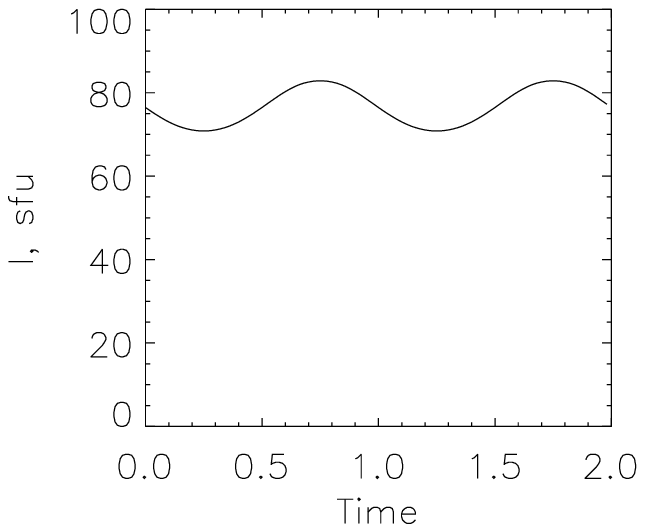}}
	\subfloat[2.2 GHz]{\includegraphics[width=0.3\textwidth]{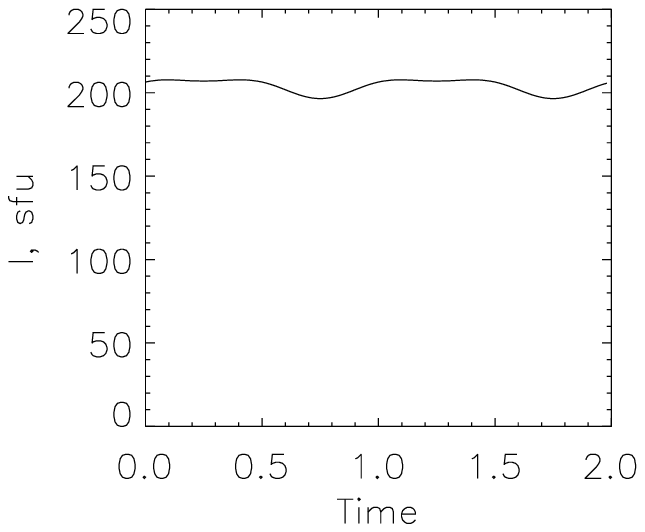}}
	\subfloat[13.8 GHz]{\includegraphics[width=0.3\textwidth]{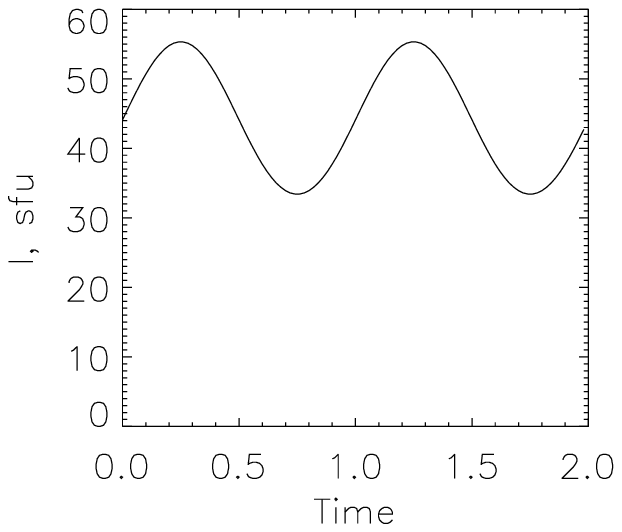}}
	\caption{Torsional mode, default parameters, oscillation of the flux density at the indicated frequencies.}
	\vspace{-5mm}
	\label{torsional_default_osc_f}
\end{figure}

\begin{figure}[hbp!]
	\centering
	\subfloat[1.5 GHz]{\includegraphics[width=0.3\textwidth]{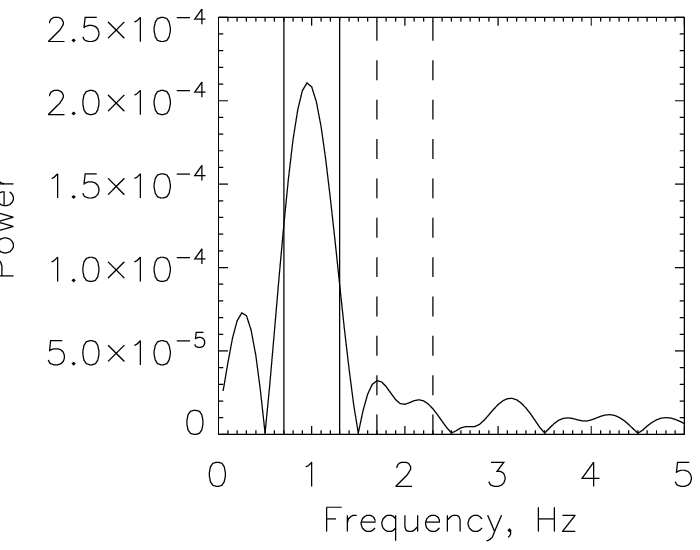}}
	\subfloat[2.2 GHz]{\includegraphics[width=0.3\textwidth]{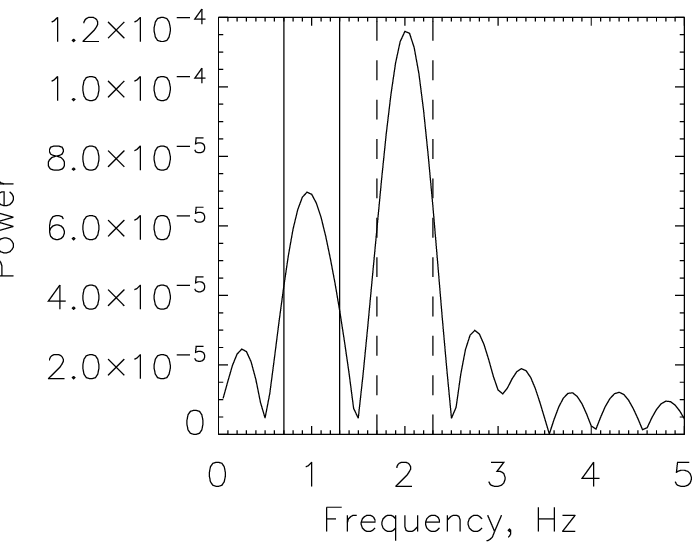}}
	\subfloat[13.8 GHz]{\includegraphics[width=0.3\textwidth]{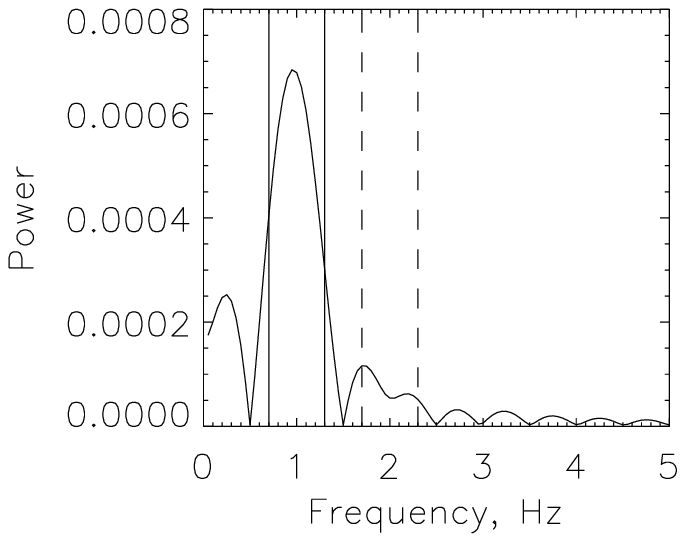}}
	
	\subfloat[Phase]{\includegraphics[width=0.3\textwidth]{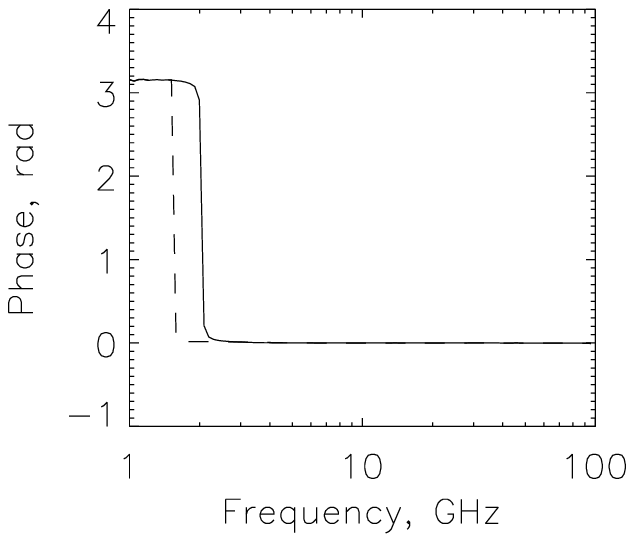}}
	\subfloat[Partial modulation]{\includegraphics[width=0.3\textwidth]{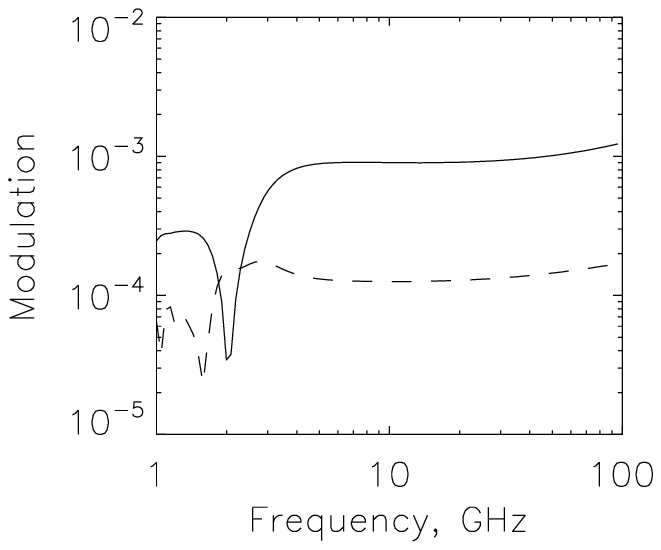}}
	\subfloat[Full modulation]{\includegraphics[width=0.3\textwidth]{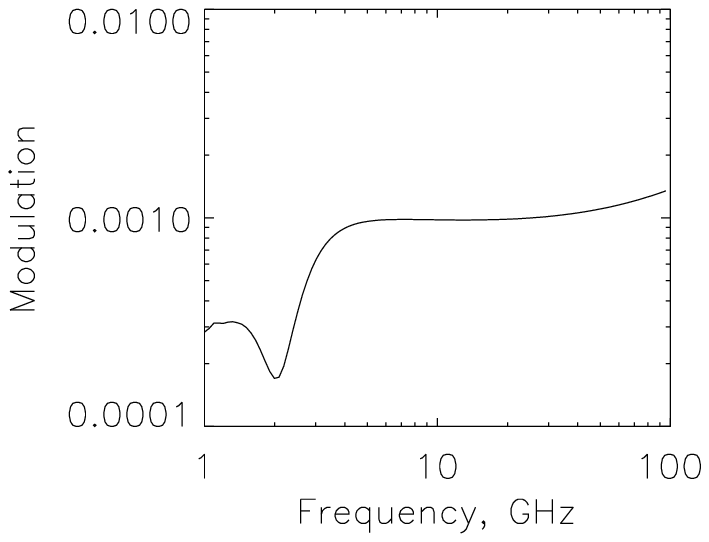}}
	\caption{Torsional mode, default parameters, oscillations of the flux density. {\bf(a)-(f)} Same as in Figure~\ref{kink_default_osc_f_analysis}.  }
	\label{torsional_default_osc_f_analysis}
\end{figure}

\begin{figure}[htp!]
	\centering	
	\centering	
	\subfloat[1.5 GHz]{\includegraphics[width=0.3\textwidth]{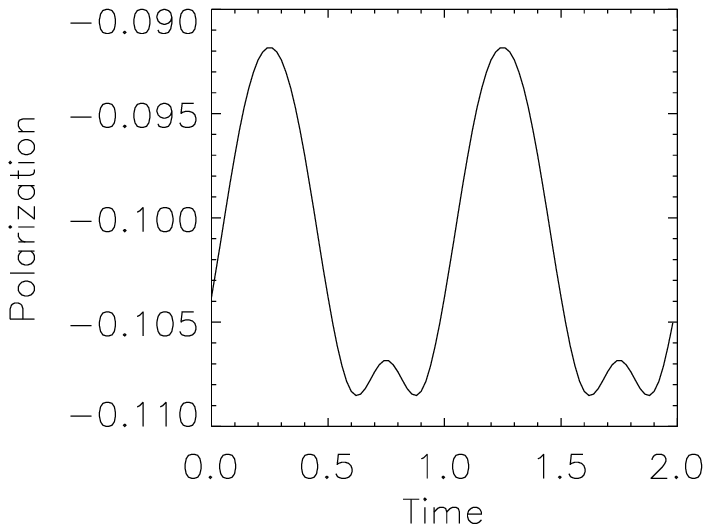}}
	\subfloat[2.2 GHz]{\includegraphics[width=0.3\textwidth]{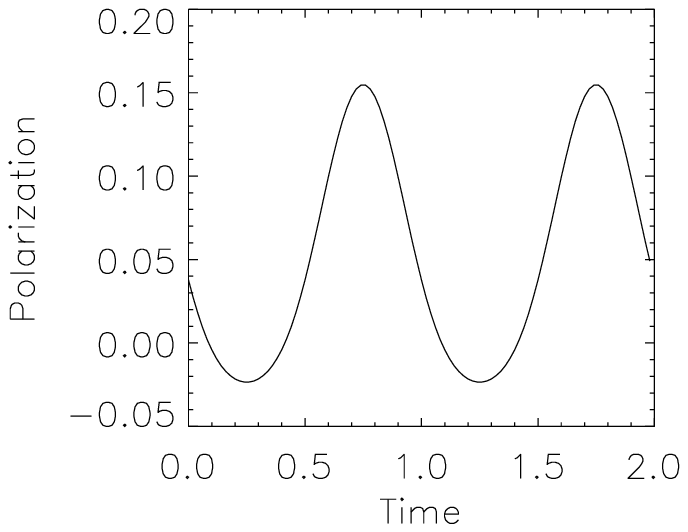}}
	\subfloat[13.8 GHz]{\includegraphics[width=0.3\textwidth]{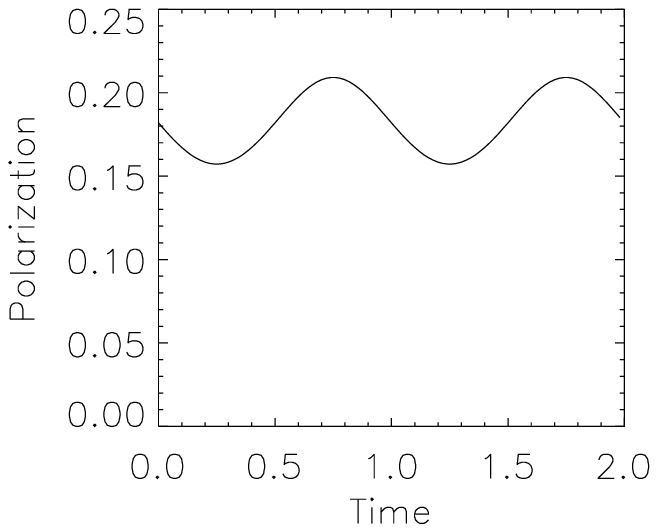}}
	\caption{Torsional mode, default parameters, oscillation of polarization at the indicated frequencies.}
	\vspace{-5mm}
	\label{torsional_default_osc_p}
\end{figure}

\begin{figure}[hbp!]
	\centering
	\subfloat[1.5 GHz P]{\includegraphics[width=0.3\textwidth]{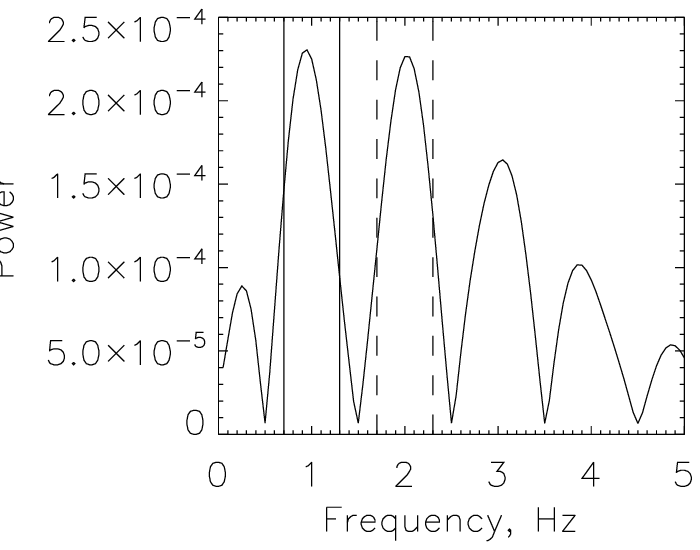}}
	\subfloat[2.2 GHz P]{\includegraphics[width=0.3\textwidth]{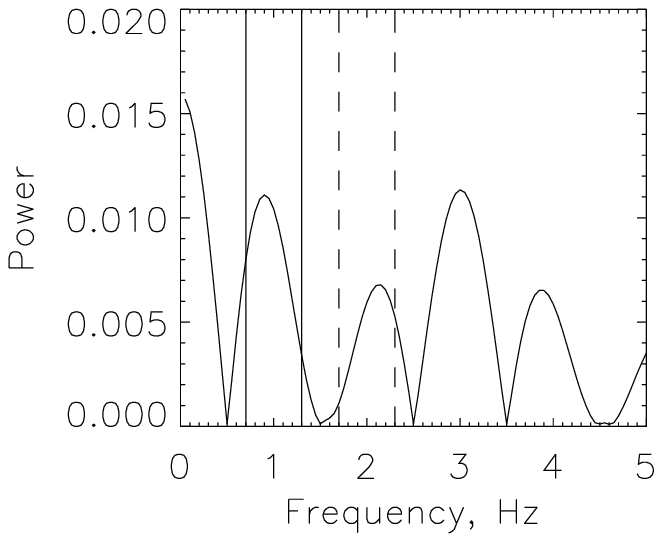}}
	\subfloat[13.8 GHz P]{\includegraphics[width=0.3\textwidth]{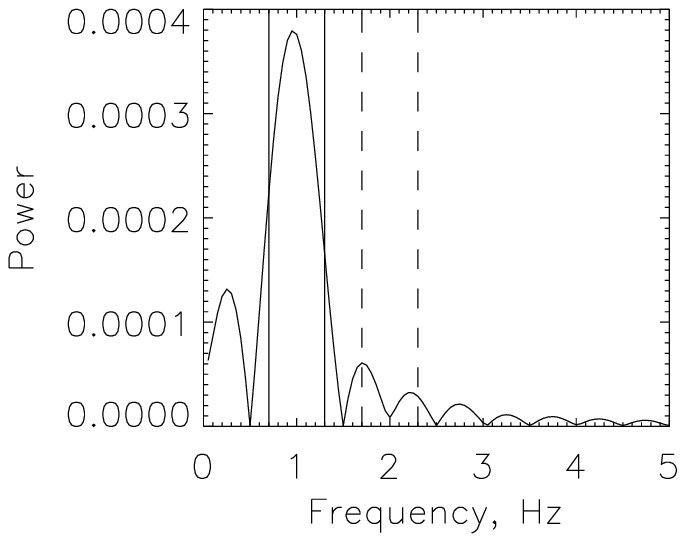}}
	
	\subfloat[Phase]{\includegraphics[width=0.3\textwidth]{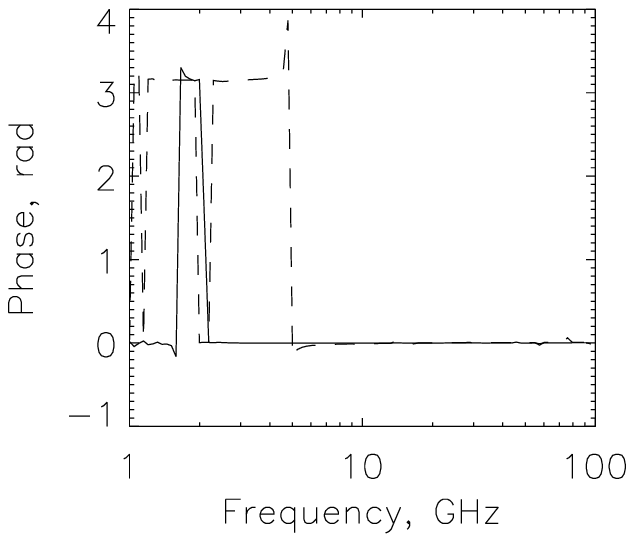}}
	\subfloat[Partial modulation]{\includegraphics[width=0.3\textwidth]{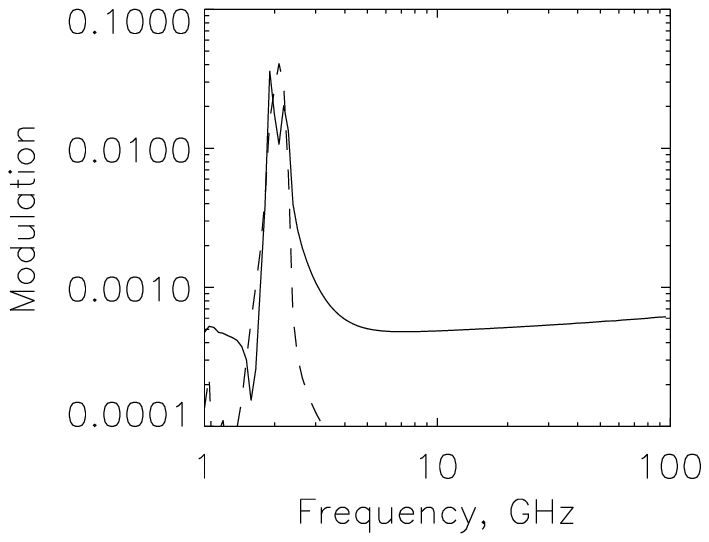}}
	\subfloat[Full modulation]{\includegraphics[width=0.3\textwidth]{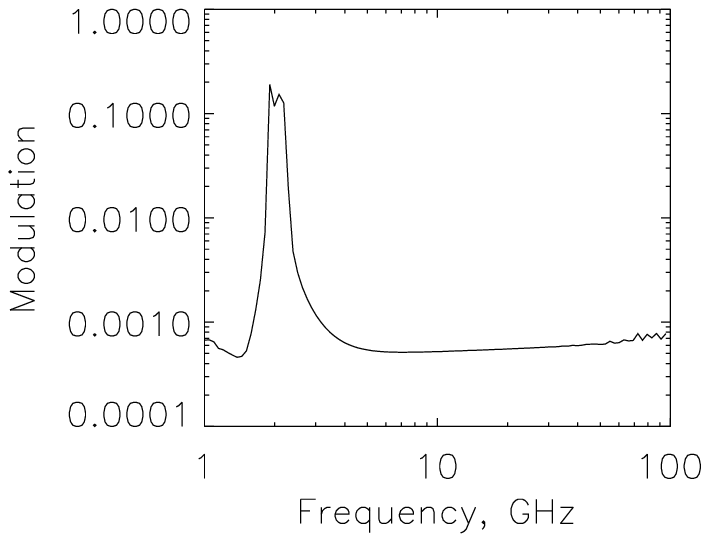}}
	\caption{Torsional mode, default parameters, oscillations of polarization. {\bf(a)-(f)} Same as in Figure~\ref{kink_default_osc_p_analysis}.}
	\label{torsional_default_osc_p_analysis}
\end{figure}

\begin{figure}[hbp!]
	\centering
	\subfloat[1.5 GHz]{\includegraphics[width=0.3\textwidth]{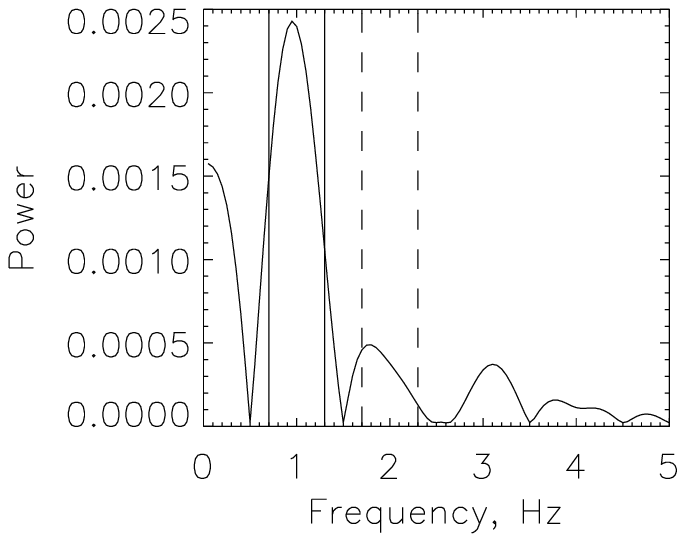}}
	\subfloat[6.6 GHz]{\includegraphics[width=0.3\textwidth]{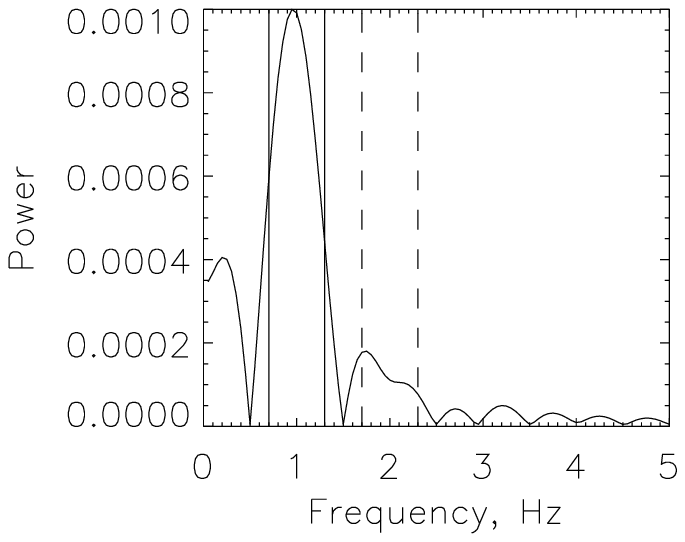}}
	\subfloat[13.8 GHz]{\includegraphics[width=0.3\textwidth]{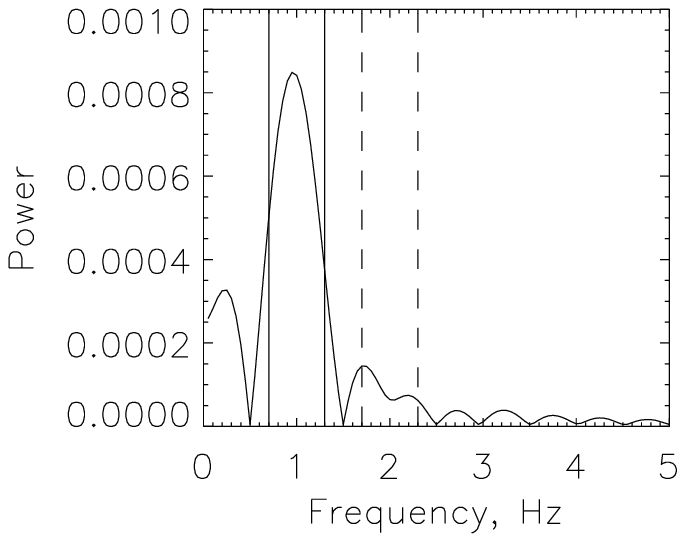}}
	
	\subfloat[Phase]{\includegraphics[width=0.3\textwidth]{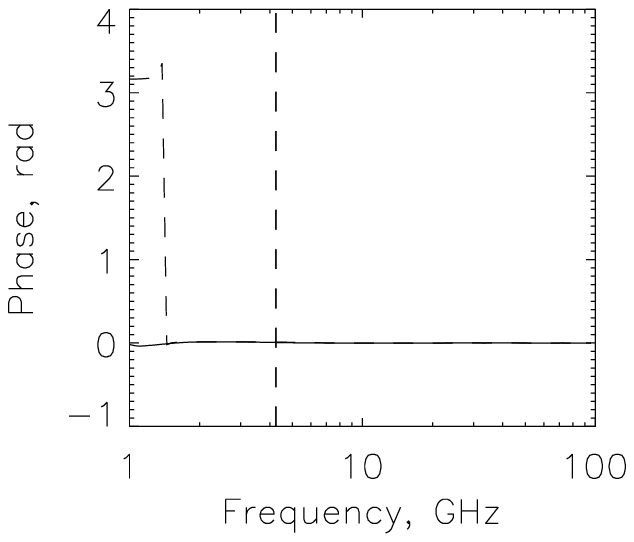}}
	\subfloat[Partial modulation]{\includegraphics[width=0.3\textwidth]{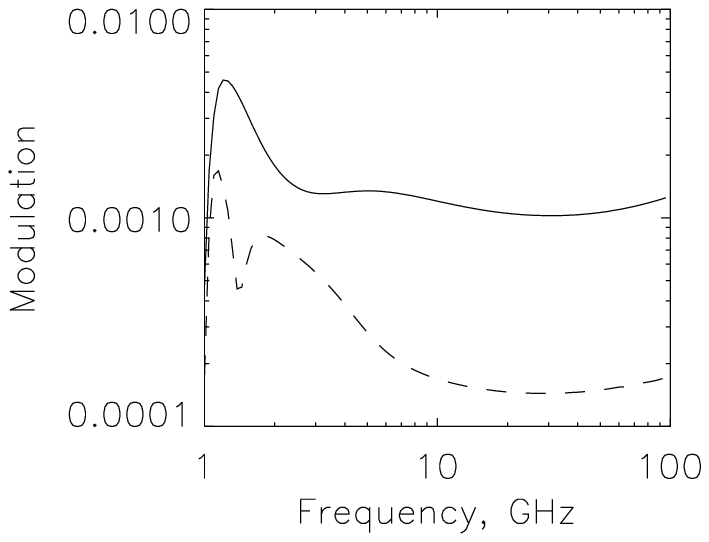}}
	\subfloat[Full modulation]{\includegraphics[width=0.3\textwidth]{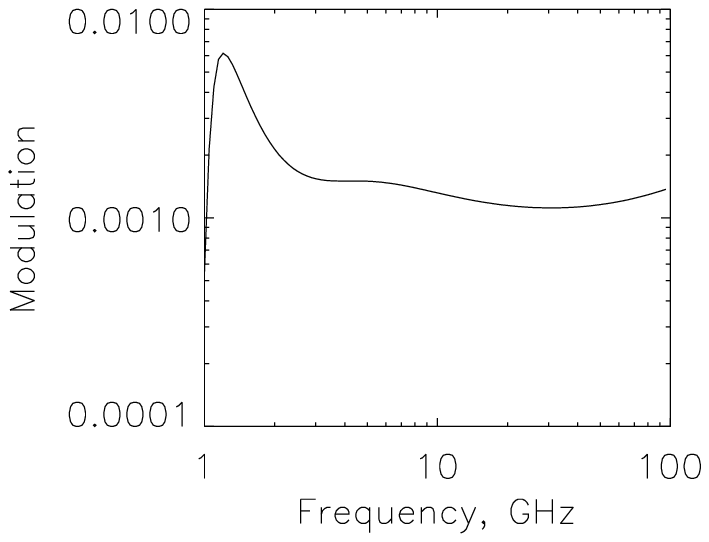}}
	
	\subfloat[1.5 GHz]{\includegraphics[width=0.3\textwidth]{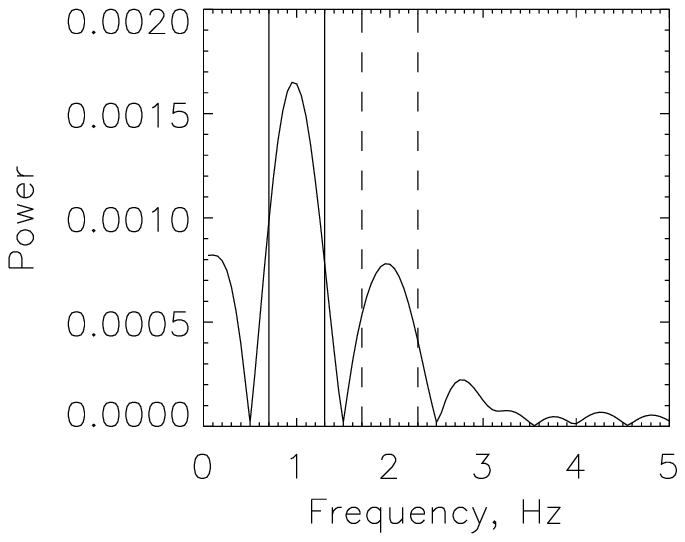}}
	\subfloat[6.6 GHz]{\includegraphics[width=0.3\textwidth]{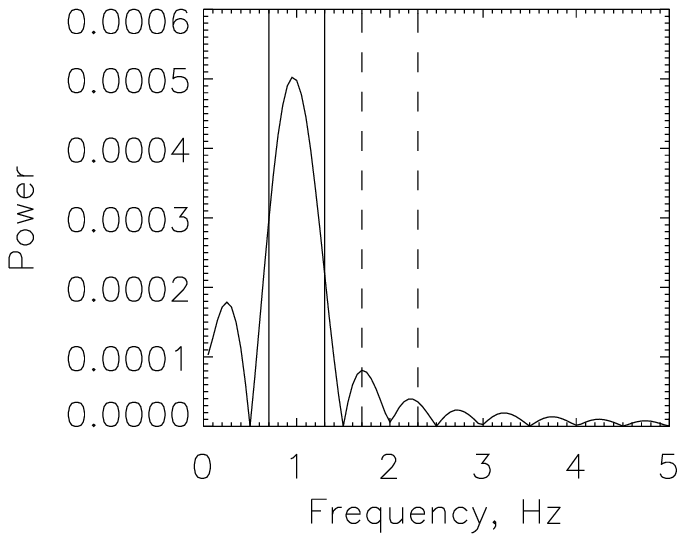}}
	\subfloat[13.8 GHz]{\includegraphics[width=0.3\textwidth]{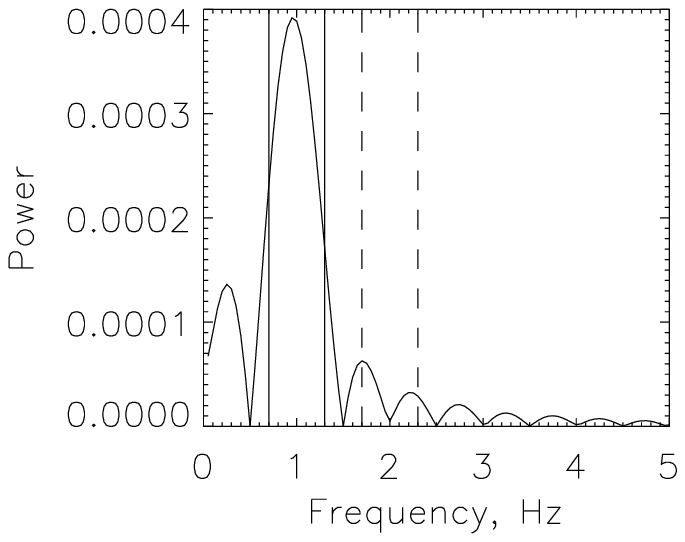}}
	
	\subfloat[Phase]{\includegraphics[width=0.3\textwidth]{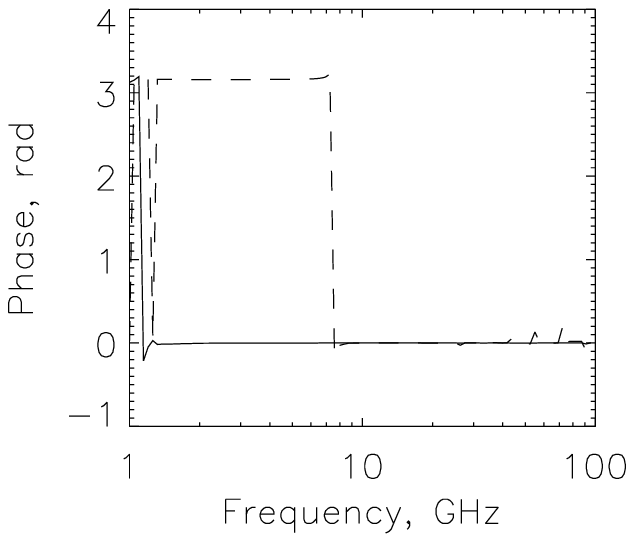}}
	\subfloat[Partial modulation]{\includegraphics[width=0.3\textwidth]{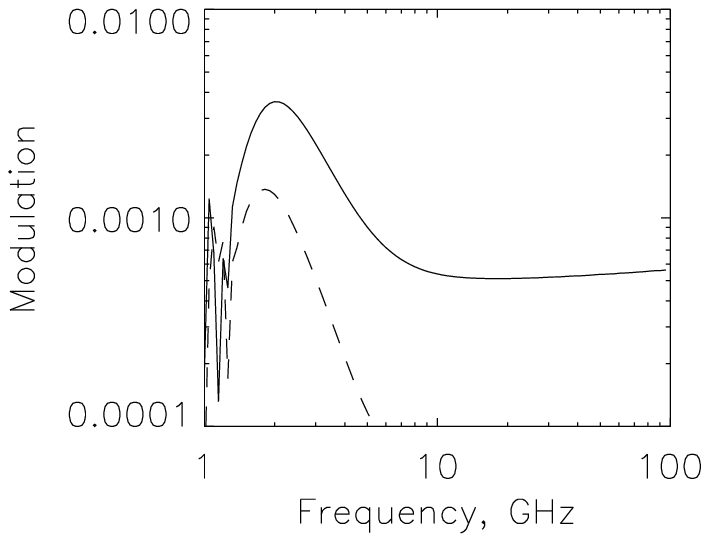}}
	\subfloat[Full modulation]{\includegraphics[width=0.3\textwidth]{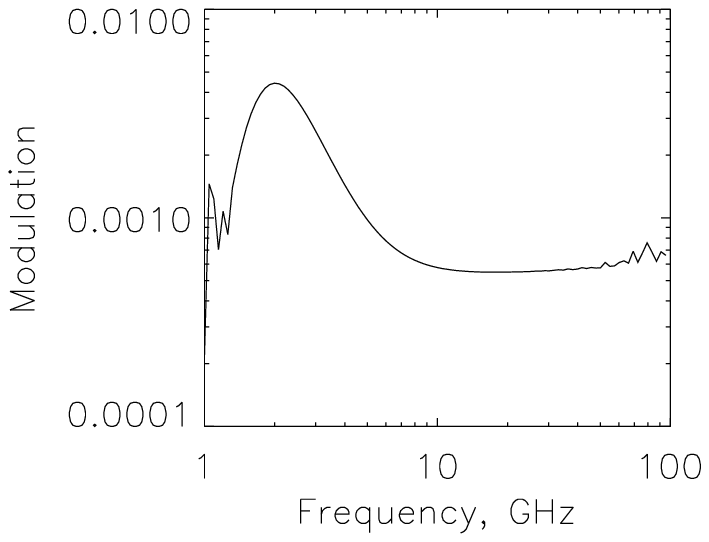}}

	\caption{Torsional mode, Razin effect, oscillations of flux and polarization. {\bf(a)-(l)} Same as in Figure~\ref{sausage_pixel_all}.}
	\label{torsional_razin_all}
\end{figure}

\begin{figure}[htp!]
	\centering
	\includegraphics{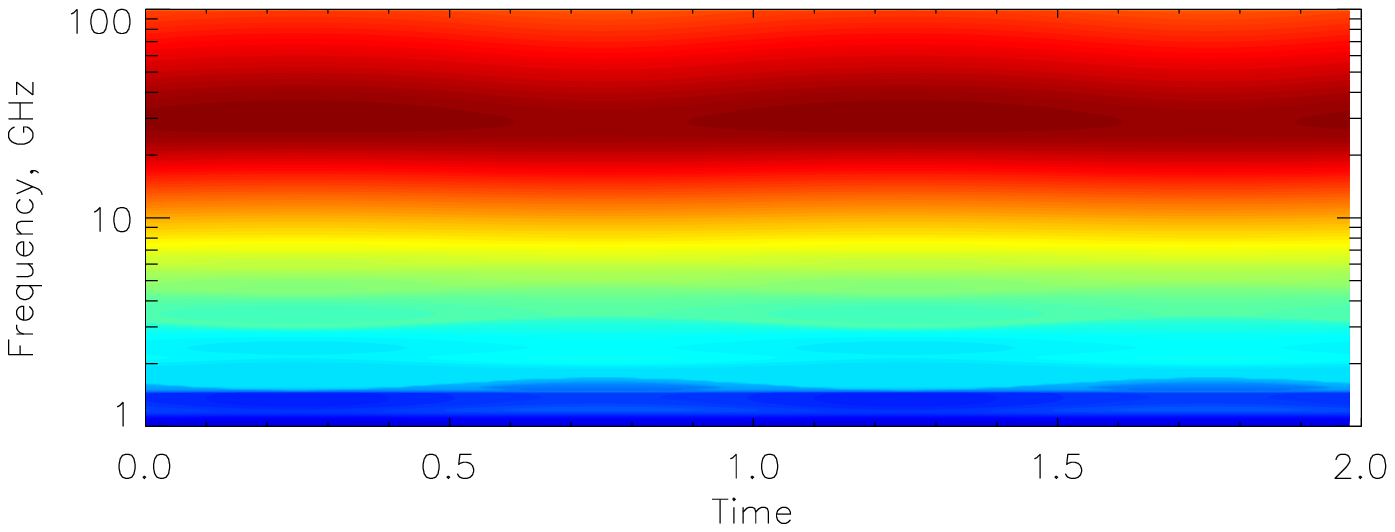}
	\includegraphics{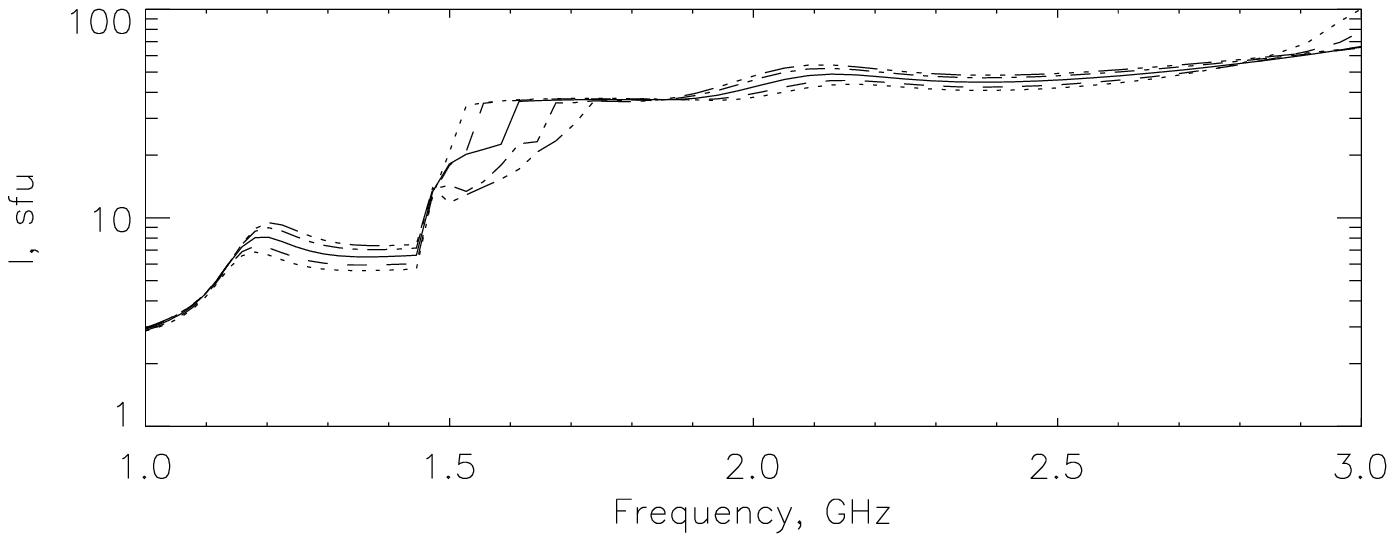}
	\includegraphics{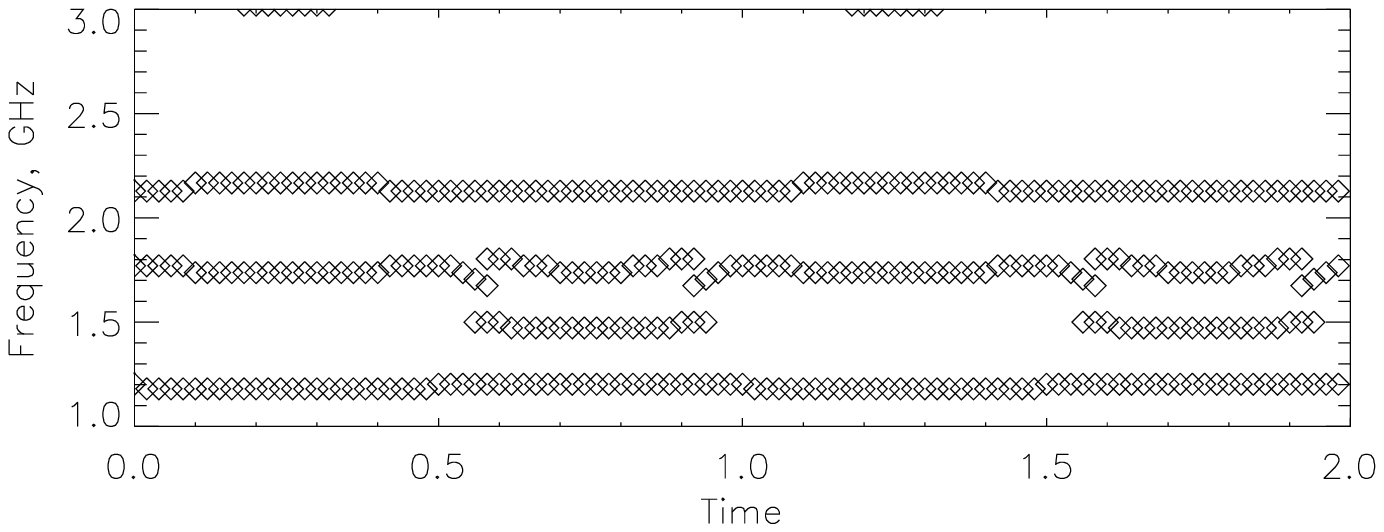}
	\caption{Torsional mode,  high magnetic field, oscillations of the flux density. Same as in Figure~\ref{sausage_lowfreq_f_light}.}
	\label{torsional_lowfreq_f_light}
\end{figure}

\begin{figure}[htp!]
	\centering
	\subfloat[1.6 GHz]{\includegraphics[width=0.5\linewidth]{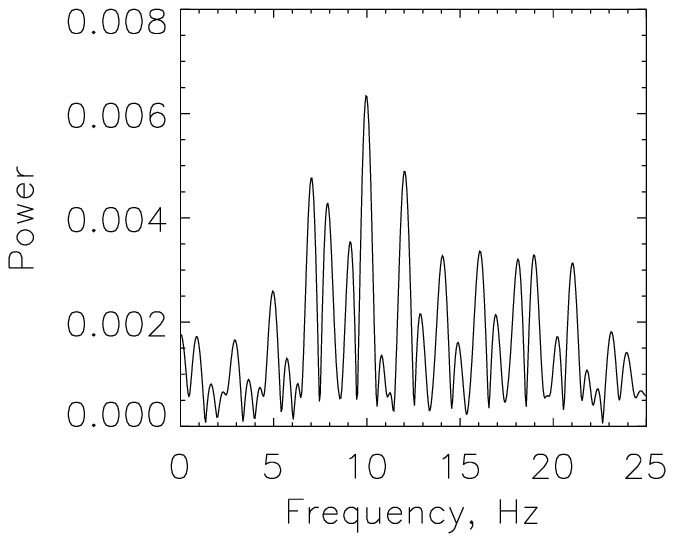}}
	\subfloat[1.7 GHz]{\includegraphics[width=0.5\linewidth]{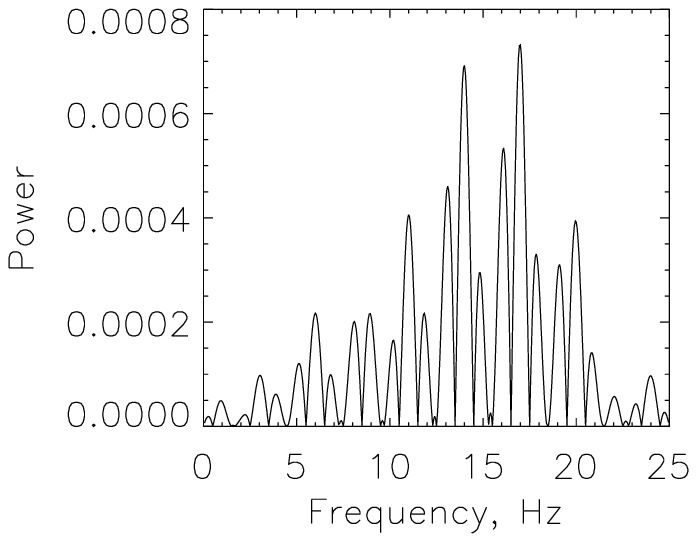}}
	\vspace{-8mm}
	\subfloat[1.9 GHz]{\includegraphics[width=0.5\linewidth]{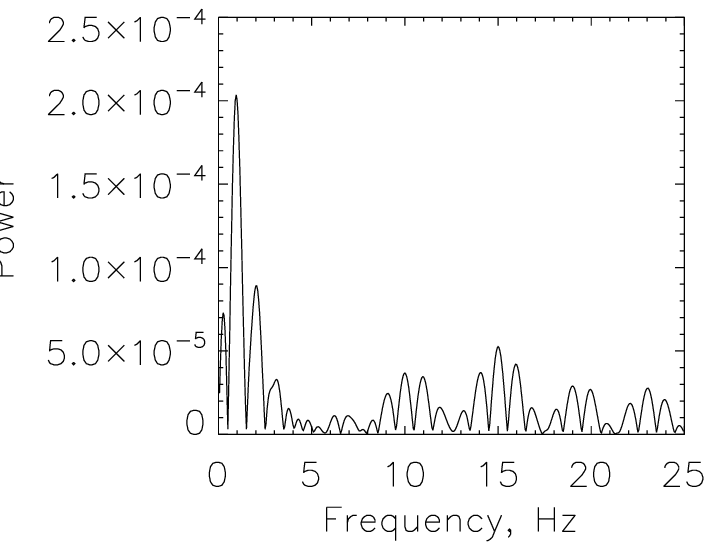}}
	\subfloat[2.1 GHz]{\includegraphics[width=0.5\linewidth]{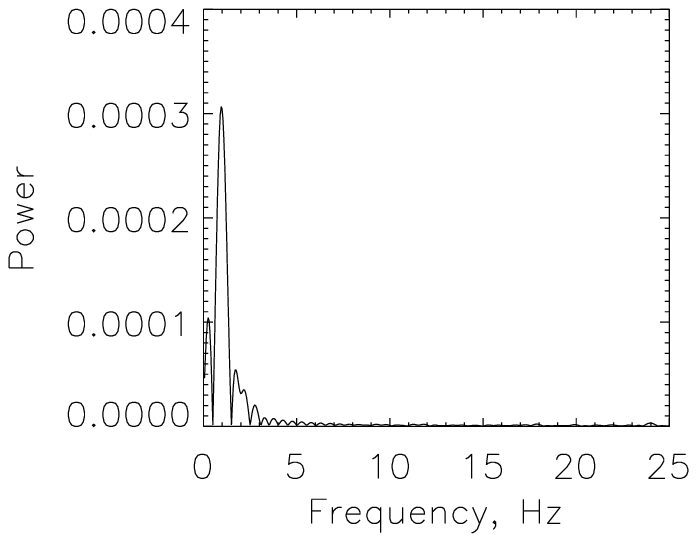}}
	\vspace{-8mm}
	\subfloat[2.3 GHz]{\includegraphics[width=0.5\linewidth]{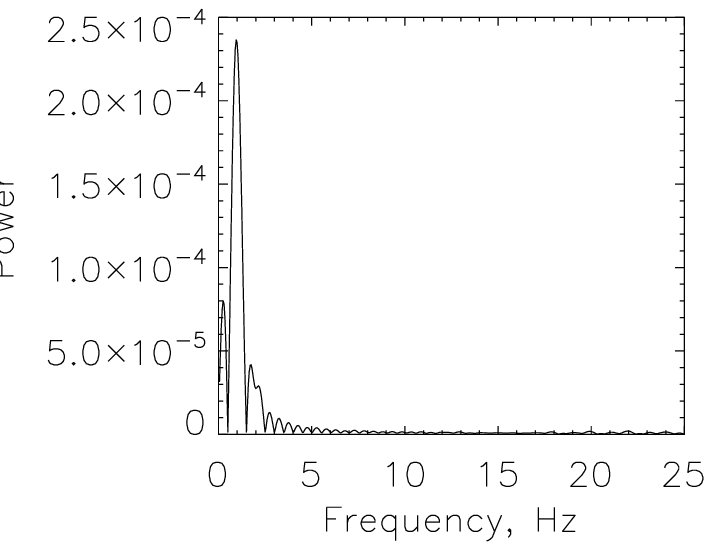}}
	\subfloat[2.5 GHz]{\includegraphics[width=0.5\linewidth]{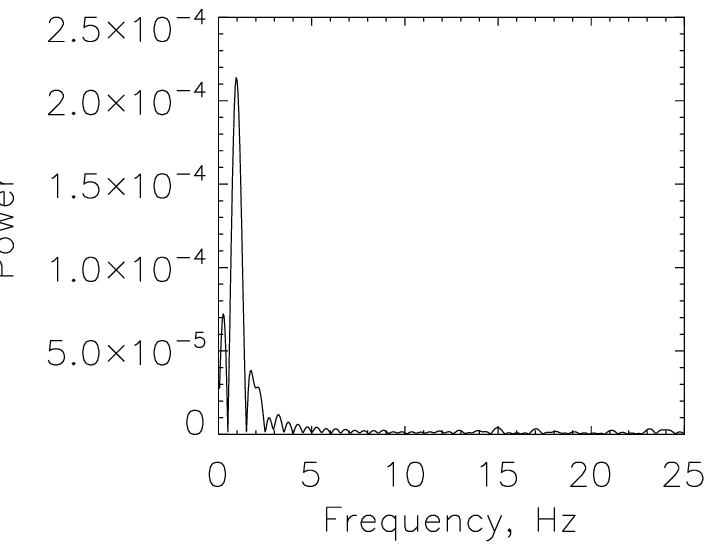}}
	\caption{Torsional mode,  high magnetic field, oscillations of the flux density. Same as in Figure~\ref{sausage_lowfreq_f_fourier}. }
	\label{torsional_lowfreq_f_fourier}
\end{figure}

\begin{figure}[htp!]
	\centering
	\includegraphics{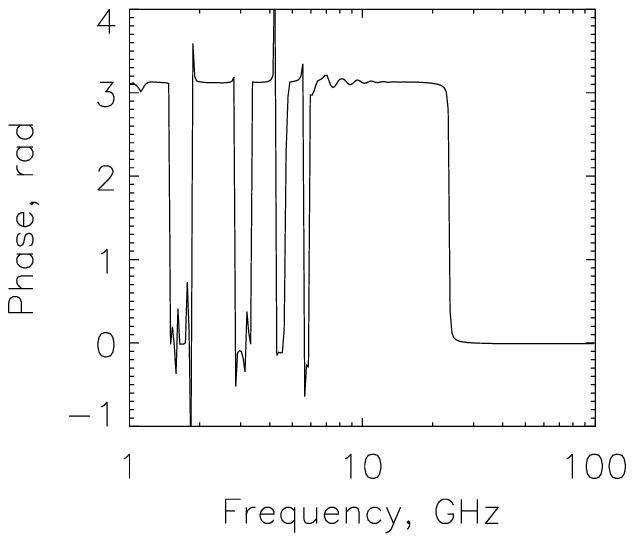}
	\includegraphics{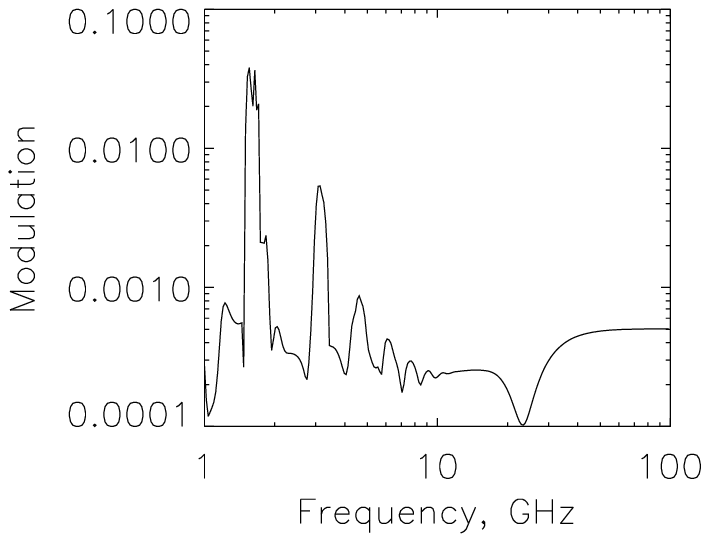}
	\caption{Torsional mode,  high magnetic field, oscillations of the flux density. Same as in Figure~\ref{sausage_lowfreq_f_analysis}.}
	\label{torsional_lowfreq_f_analysis}
\end{figure}

\begin{figure}[htp!]
	\centering
	\includegraphics{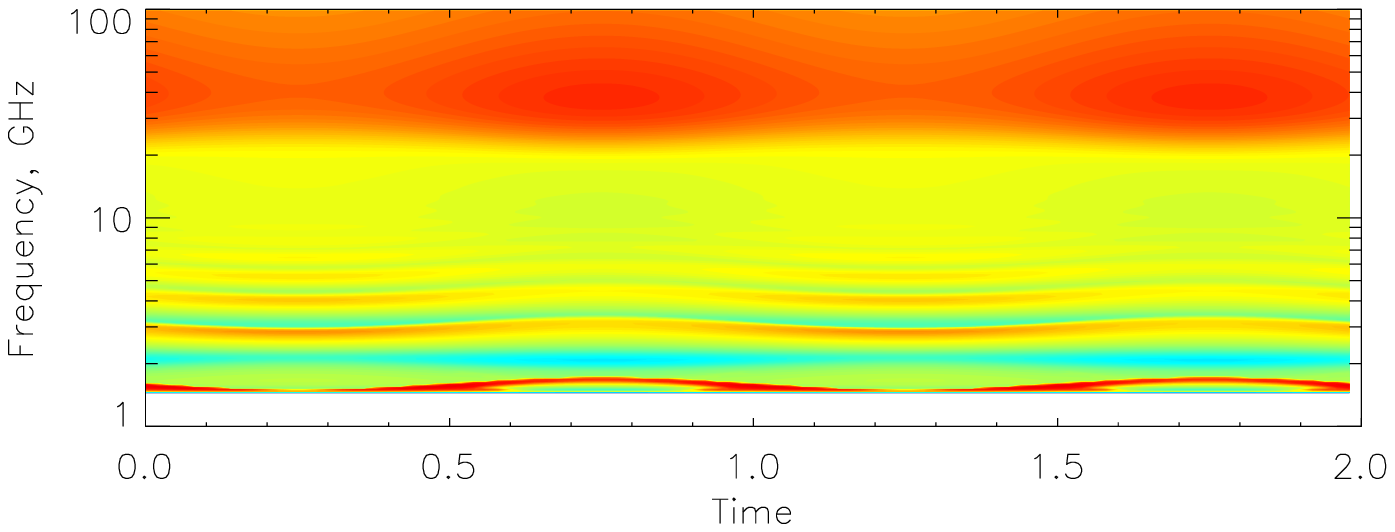}
	\includegraphics{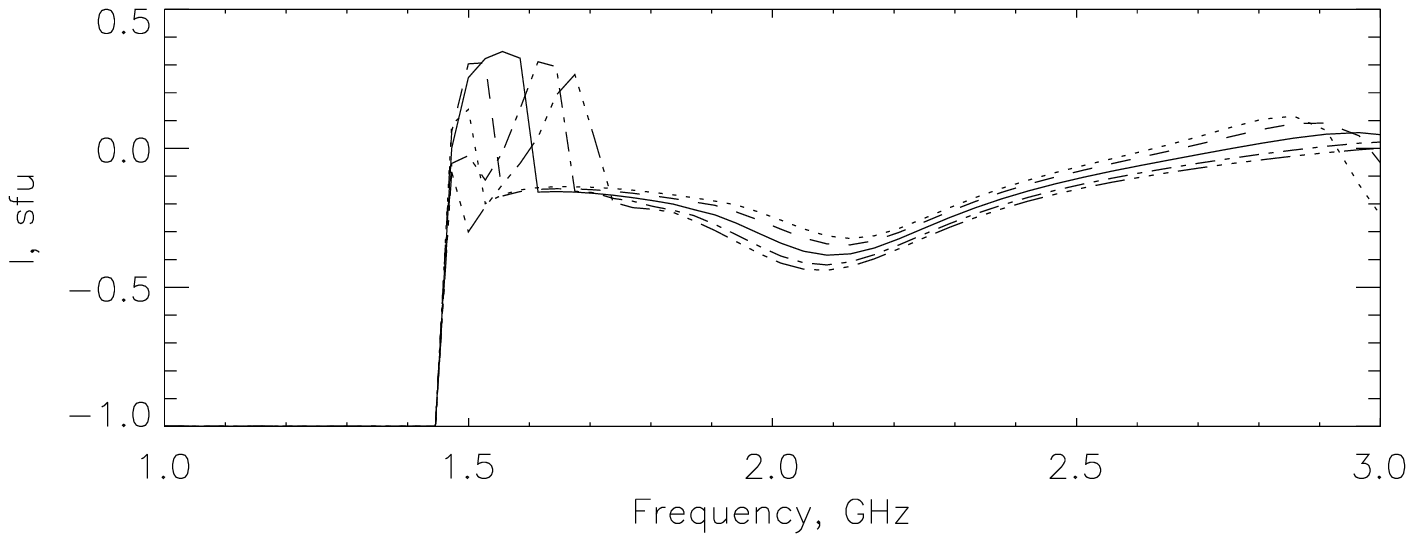}
	\includegraphics{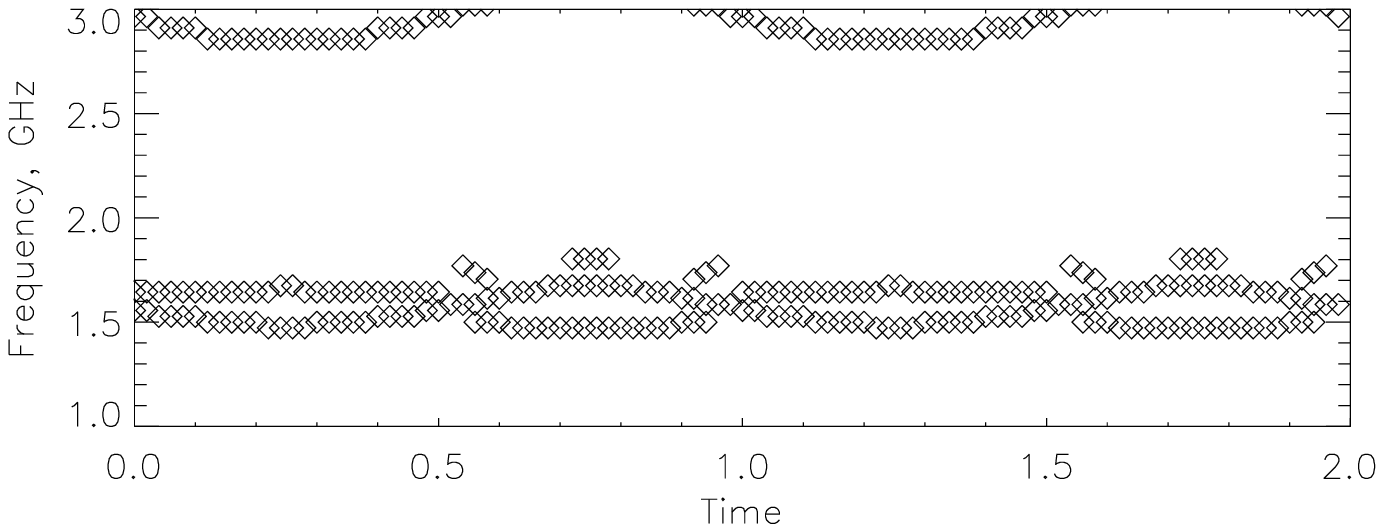}
	\caption{Torsional mode,  high magnetic field, oscillations of polarization. Same as in Figure~\ref{sausage_lowfreq_p_light}.}
	\label{torsional_lowfreq_p_light}
\end{figure}

\begin{figure}[htp!]
	\centering
	\subfloat[1.7 GHz]{\includegraphics[width=0.5\linewidth]{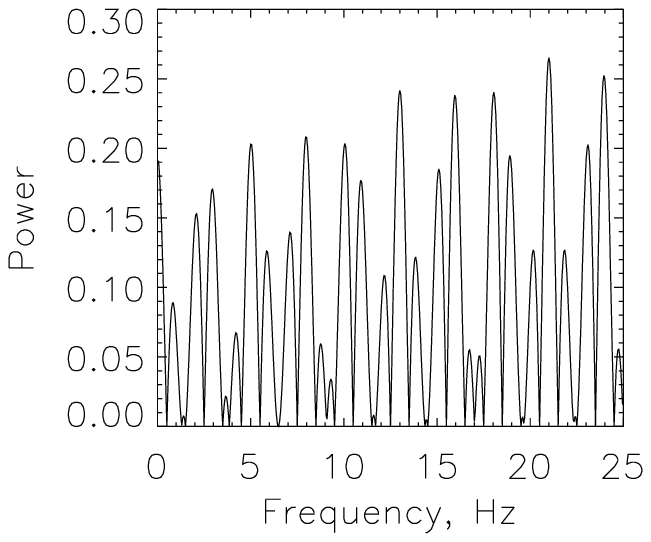}}
	\subfloat[1.9 GHz]{\includegraphics[width=0.5\linewidth]{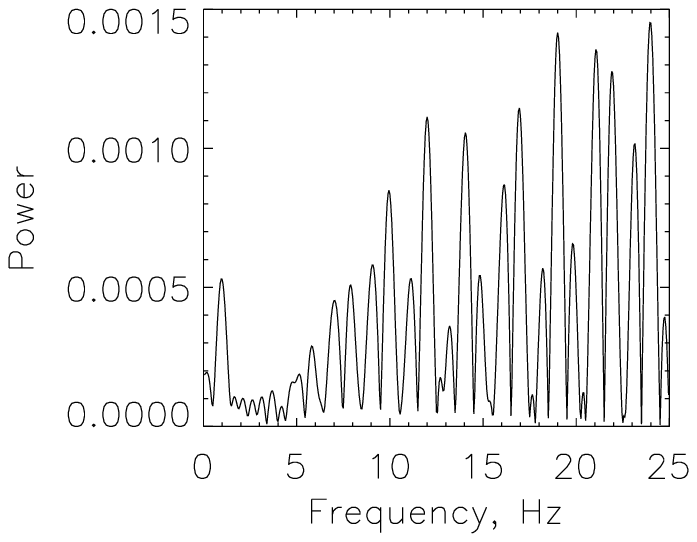}}
	\vspace{-8mm}
    \subfloat[2.1 GHz]{\includegraphics[width=0.5\linewidth]{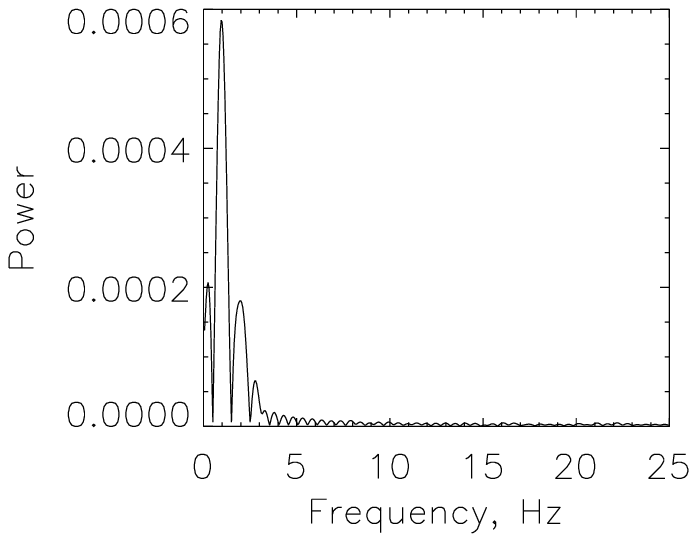}}
	\subfloat[2.3 GHz]{\includegraphics[width=0.5\linewidth]{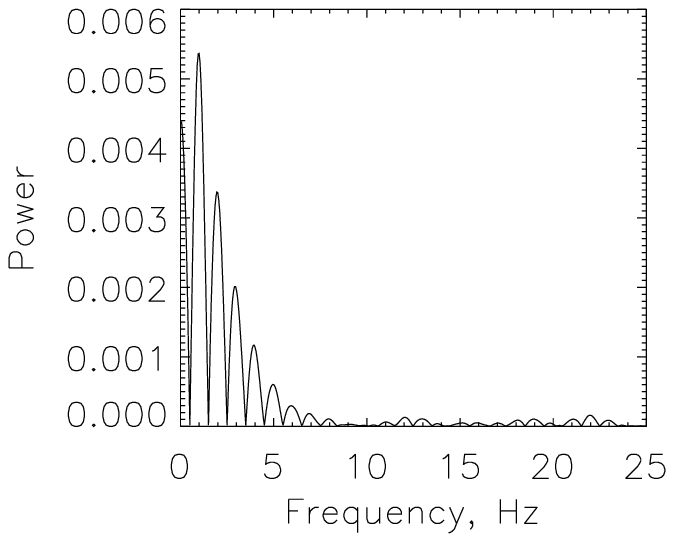}}
	\caption{Torsional mode,  high magnetic field, oscillations of polarization. Same as in Figure~\ref{sausage_lowfreq_f_fourier}. }
	\label{torsional_lowfreq_p_fourier}
\end{figure}

\begin{figure}[htp!]
	\centering
	\includegraphics{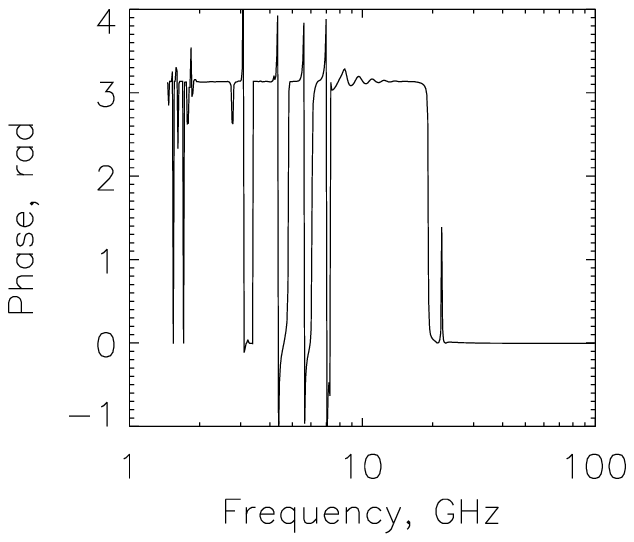}
	\includegraphics{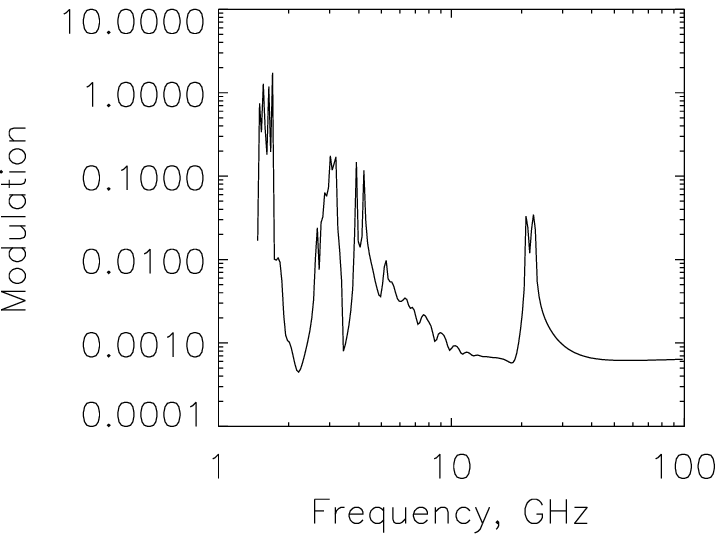}
	\caption{Torsional mode,  high magnetic field, oscillations of polarization. Same as in Figure~\ref{sausage_lowfreq_f_analysis}.}
	\label{torsional_lowfreq_p_analysis}
\end{figure}

The torsional mode oscillations are  to first order the oscillations of the viewing angle with no source visible area or depth oscillations. As such, the torsional mode produces results similar to the kink mode line-of-sight oscillations for the single pixel view and for the Razin-effect parameter regime, see Figures~\ref{torsional_default_osc_f}--\ref{torsional_lowfreq_p_analysis}.




\clearpage

\clearpage

\clearpage

\clearpage

\clearpage

\clearpage

\clearpage

\clearpage


\begin{figure}[htp!]
	\centering
	\includegraphics{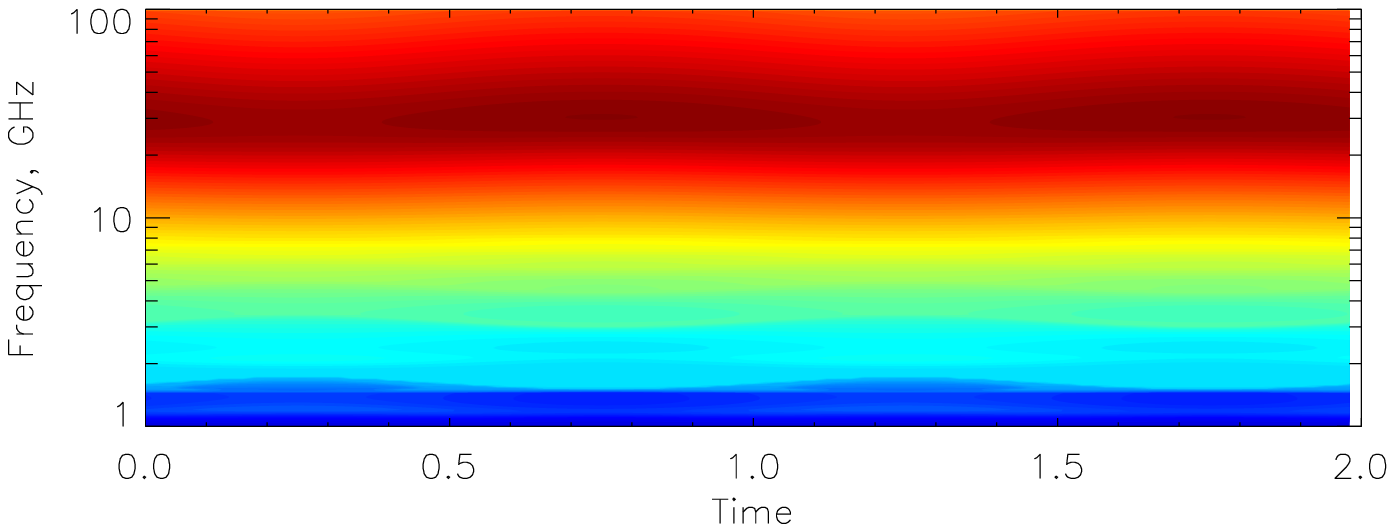}
	\includegraphics{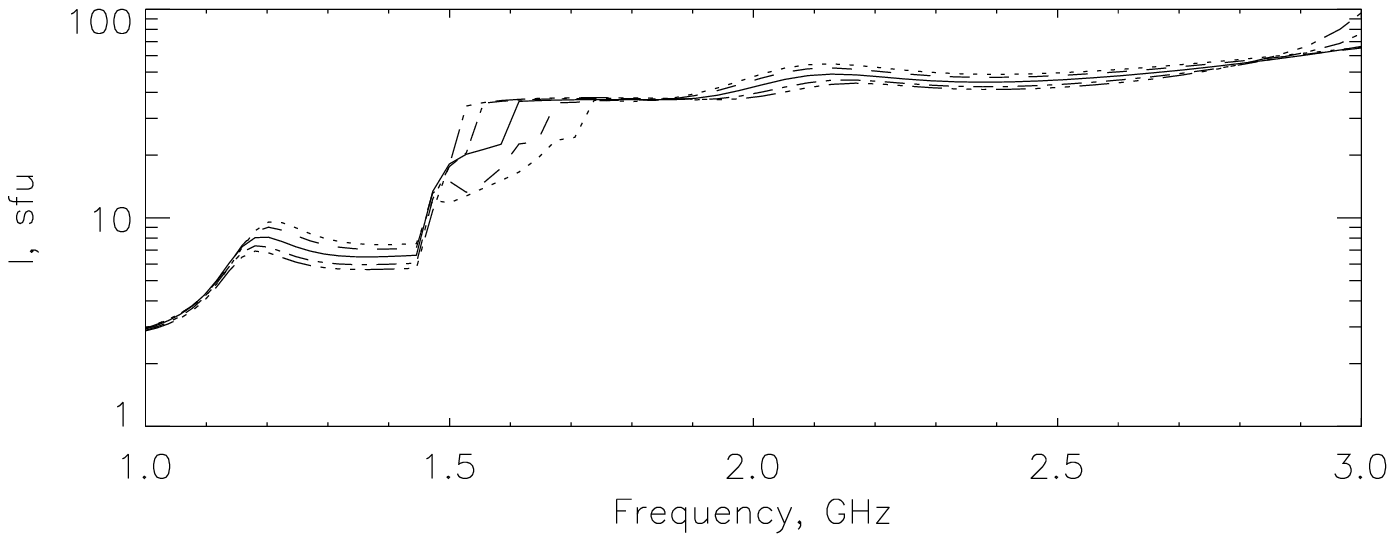}
	\includegraphics{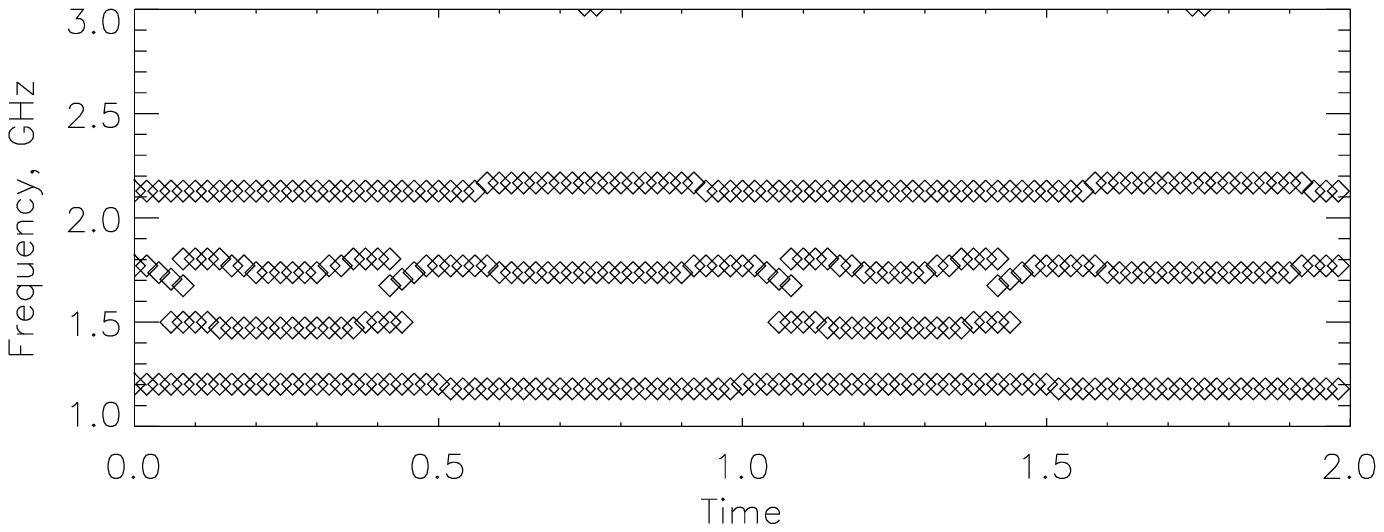}
	\caption{Kink mode, high magnetic field. Same as in Figure~\ref{sausage_lowfreq_f_light}. }
	\label{kink_lowfreq_f_light}
\end{figure}
\begin{figure}[htp!]
	\centering
	\subfloat[1.6 GHz]{\includegraphics[width=0.5\linewidth]{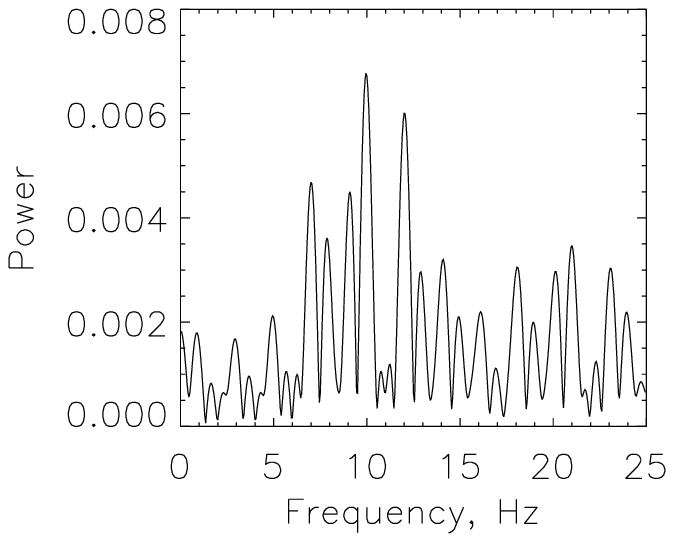}}
	\subfloat[1.7 GHz]{\includegraphics[width=0.5\linewidth]{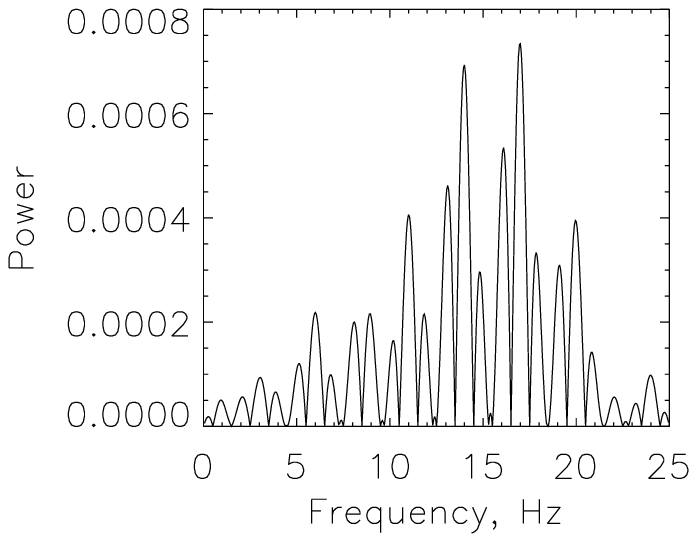}}
	\vspace{-8mm}
	\subfloat[1.9 GHz]{\includegraphics[width=0.5\linewidth]{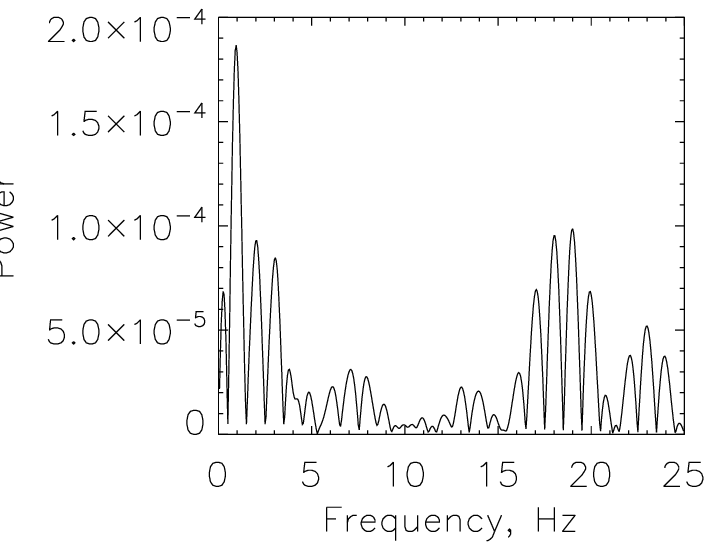}}
	\subfloat[2.1 GHz]{\includegraphics[width=0.5\linewidth]{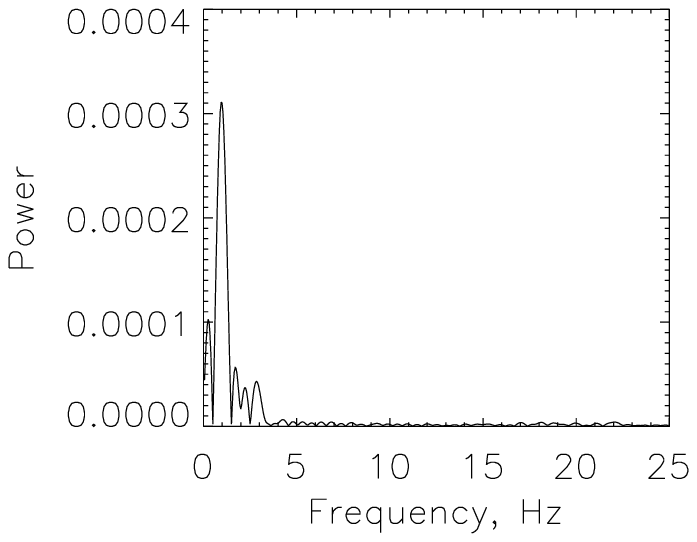}}
	\vspace{-8mm}
	\subfloat[2.3 GHz]{\includegraphics[width=0.5\linewidth]{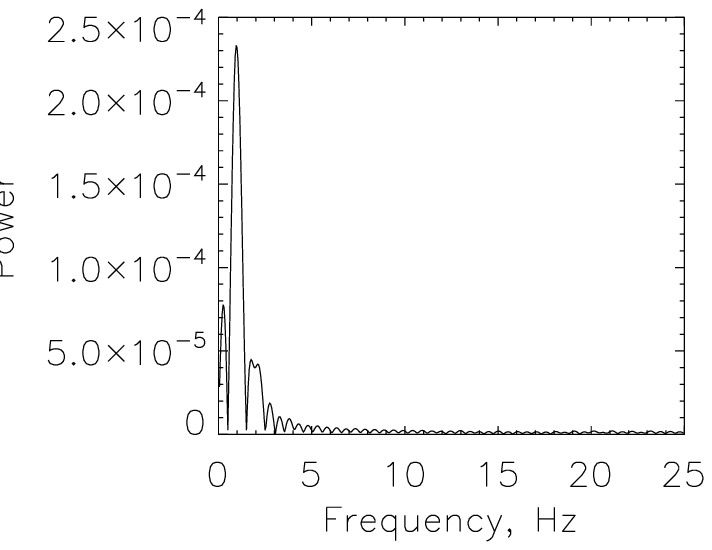}}
	\subfloat[2.5 GHz]{\includegraphics[width=0.5\linewidth]{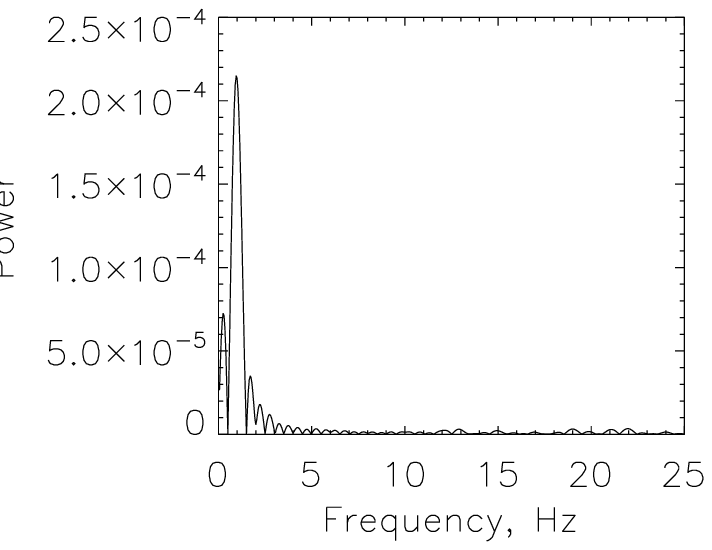}}
	\caption{Kink mode, low-frequency peaks in the flux density. Same as in Figure~\ref{sausage_lowfreq_f_fourier}. }
	\label{kink_lowfreq_f_fourier}
\end{figure}

\begin{figure}[htp!]
	\centering
	\includegraphics{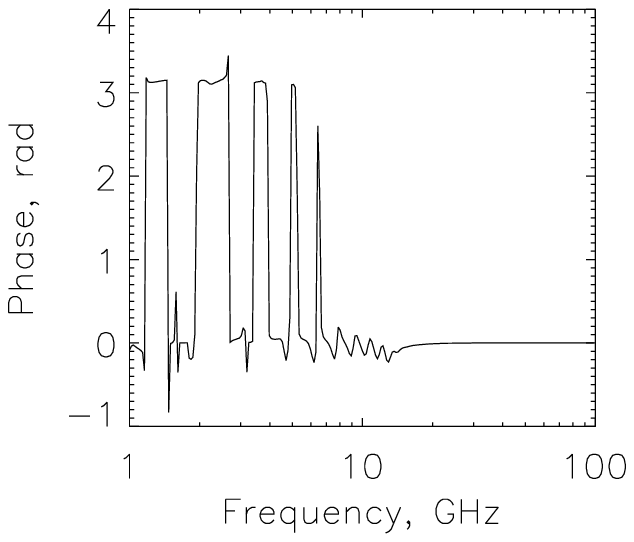}
	\includegraphics{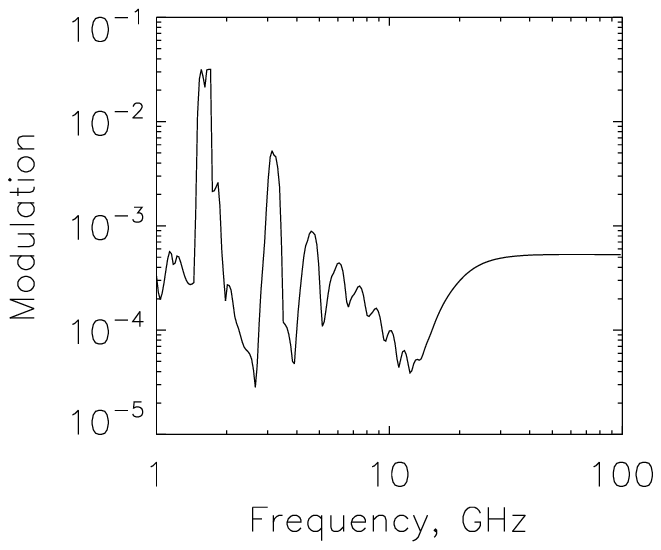}
	\caption{Kink mode, low-frequency peaks in the flux density. Left: Relative phase of flux. Right: Full modulation amplitude of flux. Note oscillating behavior of the modulation amplitude at low frequencies.}
	\label{kink_lowfreq_f_analysis}
\end{figure}

\begin{figure}[htp!]
	\centering
	\includegraphics{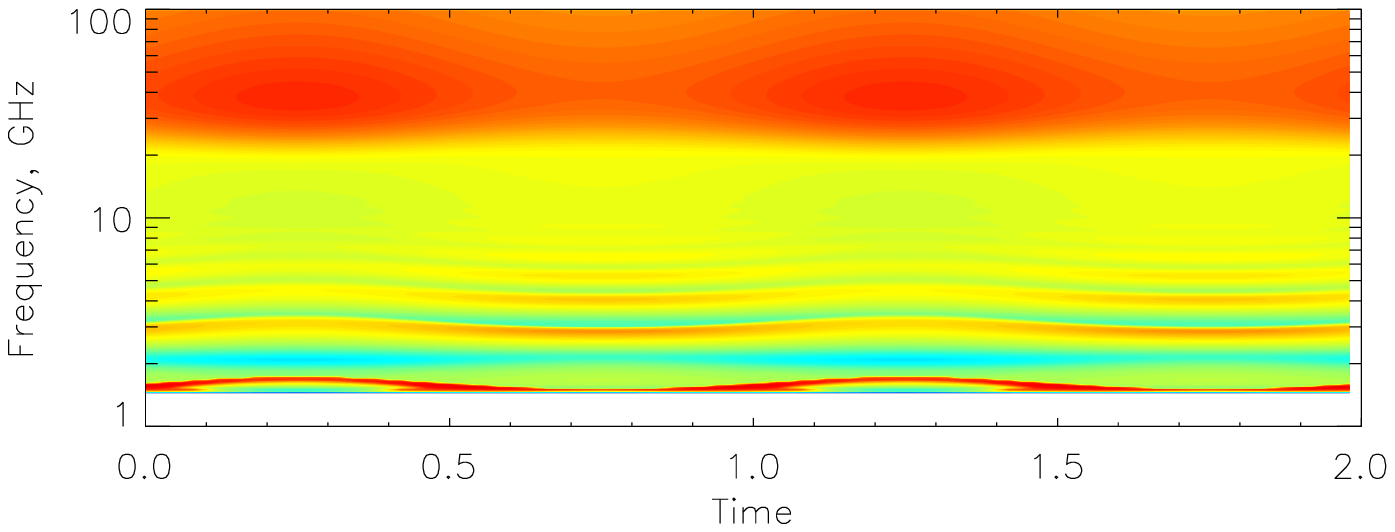}
	\includegraphics{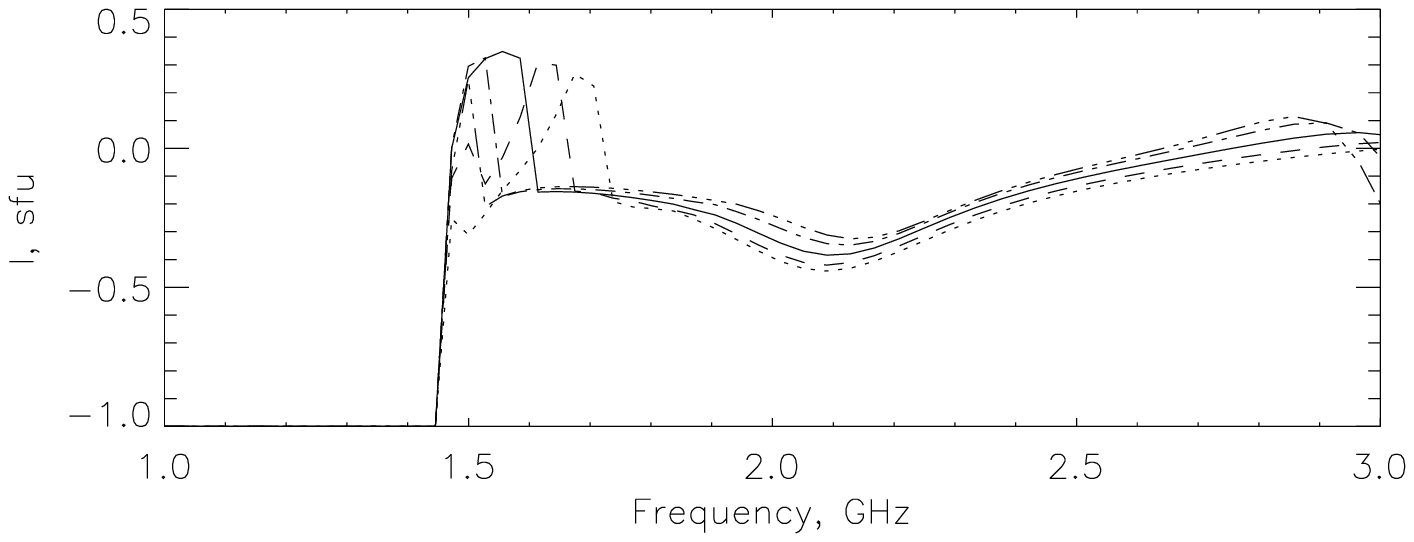}
	\includegraphics{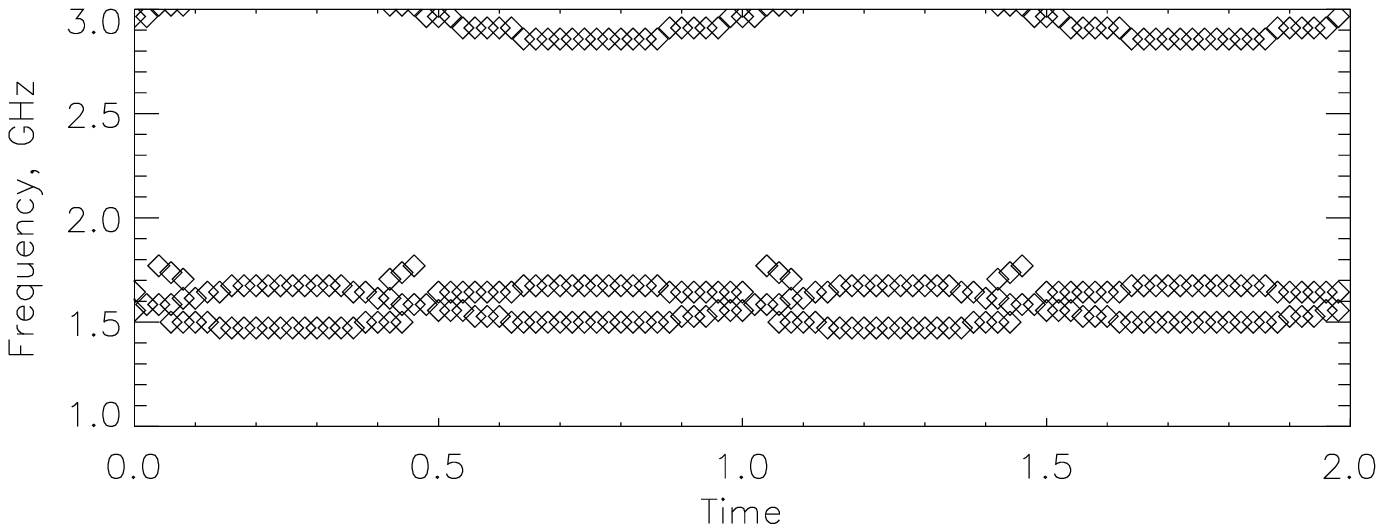}
	\caption{Kink mode, high magnetic field, oscillations of polarization. Same as in Figure~\ref{sausage_lowfreq_p_light}.}
	\label{kink_lowfreq_p_light}
\end{figure}
\clearpage

\begin{figure}[htp!]
	\centering
	\subfloat[1.7 GHz]{\includegraphics[width=0.5\linewidth]{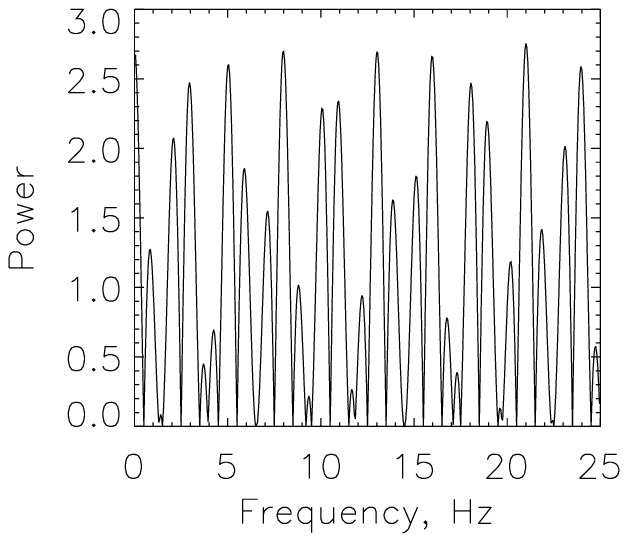}}
	\subfloat[1.9 GHz]{\includegraphics[width=0.5\linewidth]{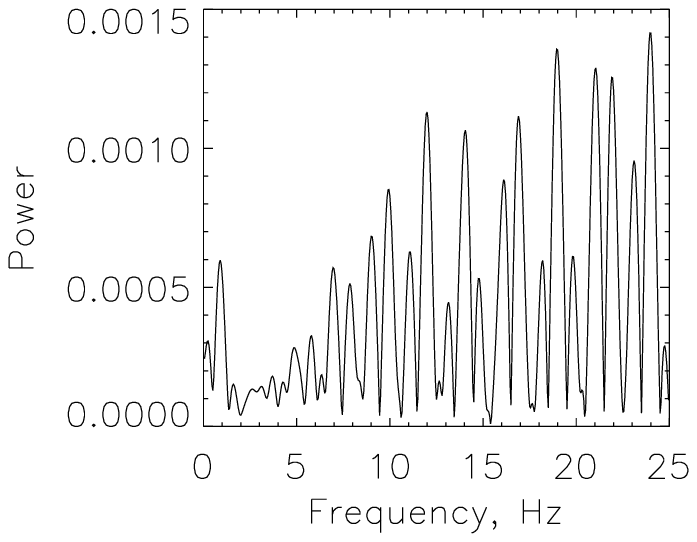}}
    \vspace{-8mm}
	\subfloat[2.1 GHz]{\includegraphics[width=0.5\linewidth]{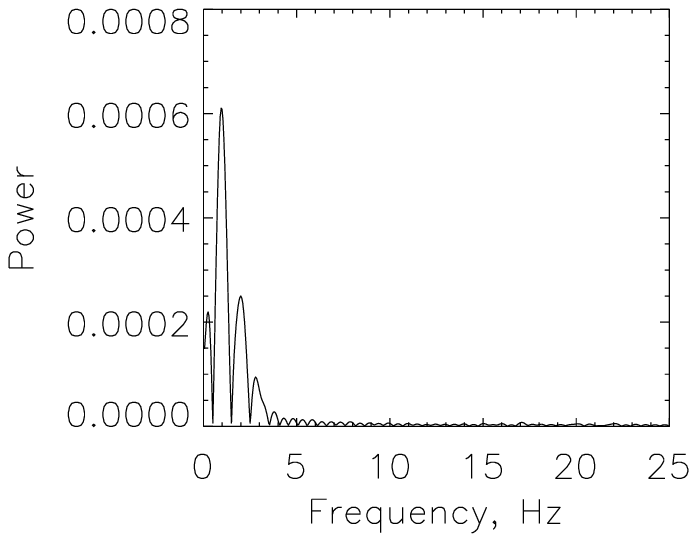}}
	\subfloat[2.3 GHz]{\includegraphics[width=0.5\linewidth]{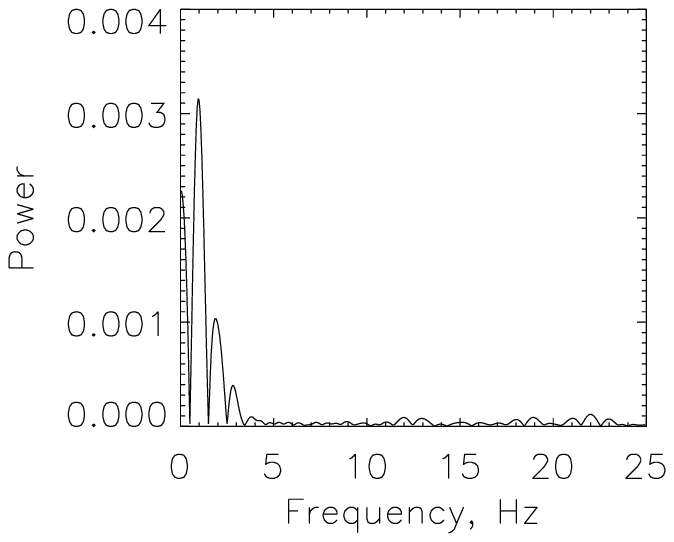}}
	\caption{Kink mode, high magnetic field, oscillations of polarization. Same as in Figure~\ref{sausage_lowfreq_p_fourier}. }
	\label{kink_lowfreq_p_fourier}
\end{figure}

\begin{figure}[htp!]
	\centering
	\includegraphics{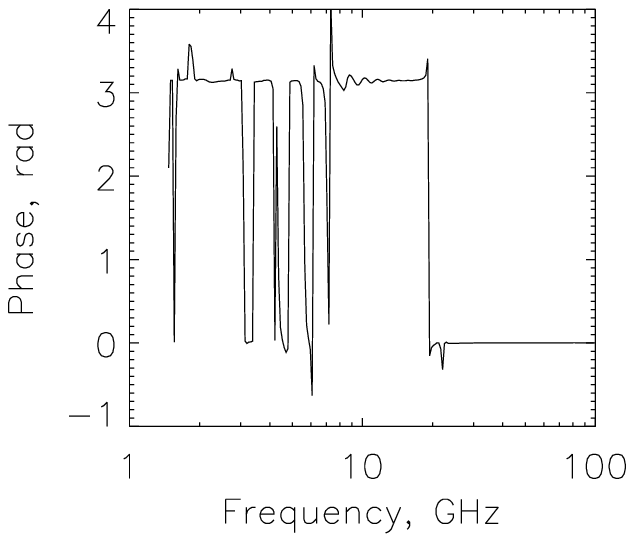}
	\includegraphics{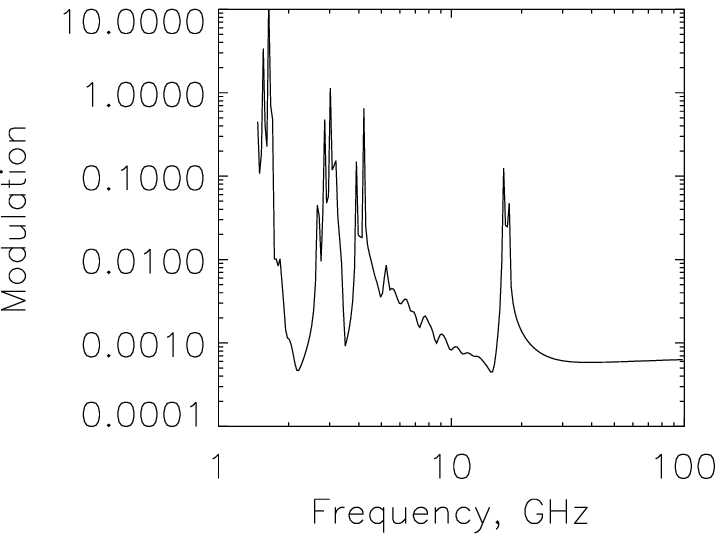}
	\caption{Kink mode, high magnetic field, oscillations of polarization. Same as in Figure~\ref{sausage_lowfreq_p_analysis}.}
	\label{kink_lowfreq_p_analysis}
\end{figure}

\begin{figure}[htp!]
	\centering
	\includegraphics{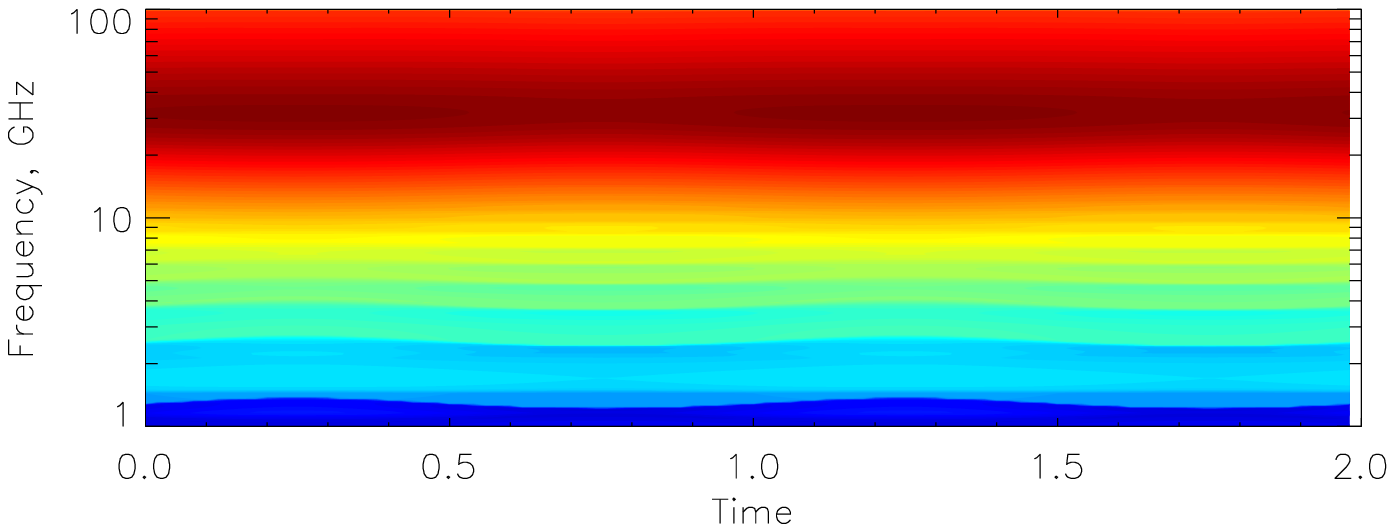}
	\includegraphics{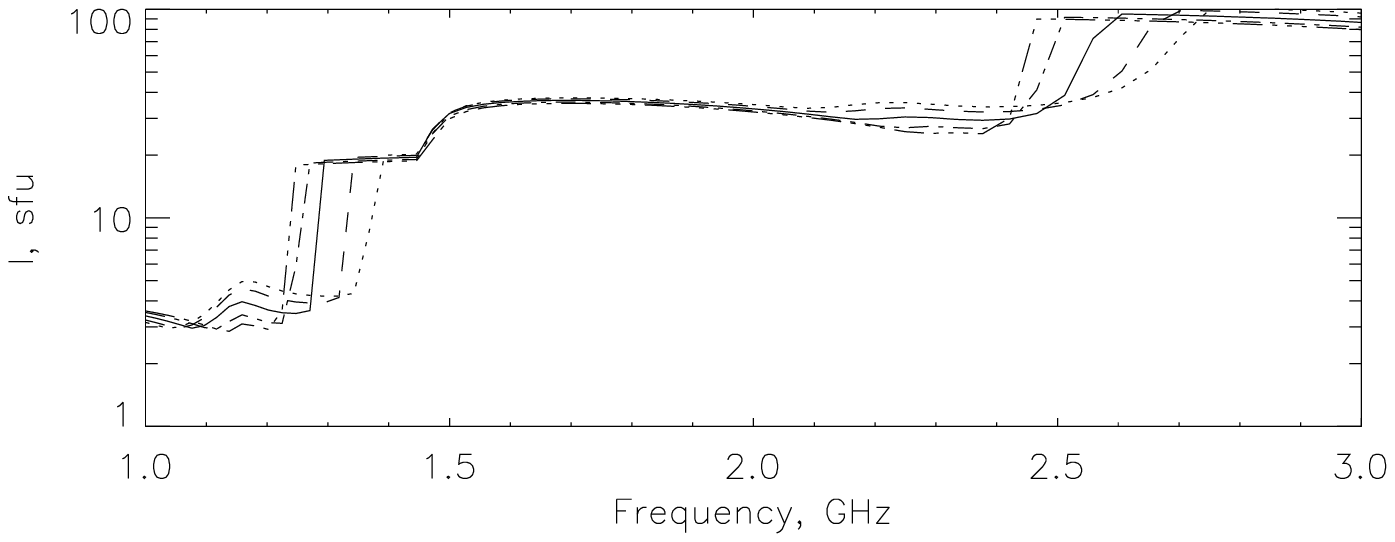}
	\includegraphics{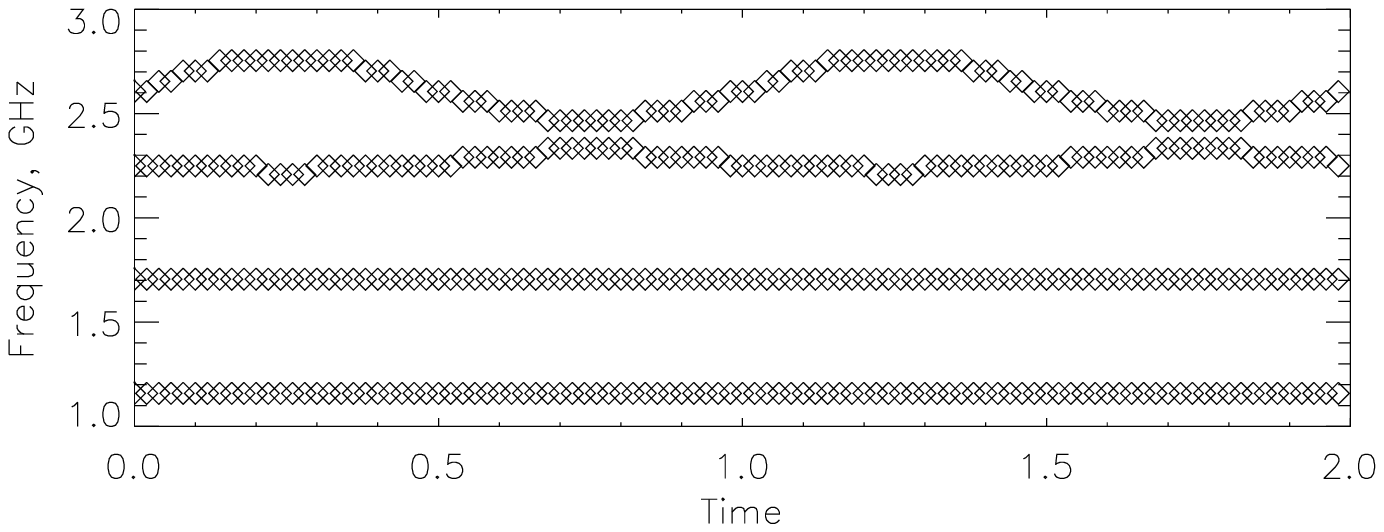}
	\caption{Kink mode, high magnetic field, oscillations of the flux density, viewing angle 80$^\circ$. Same as in Figure~\ref{sausage_lowfreq_f_light}.}
	\label{kink_theta80_lowfreq_f_light}
\end{figure}
\clearpage

\begin{figure}[htp!]
	\centering
	\subfloat[1.6 GHz]{\includegraphics[width=0.5\linewidth]{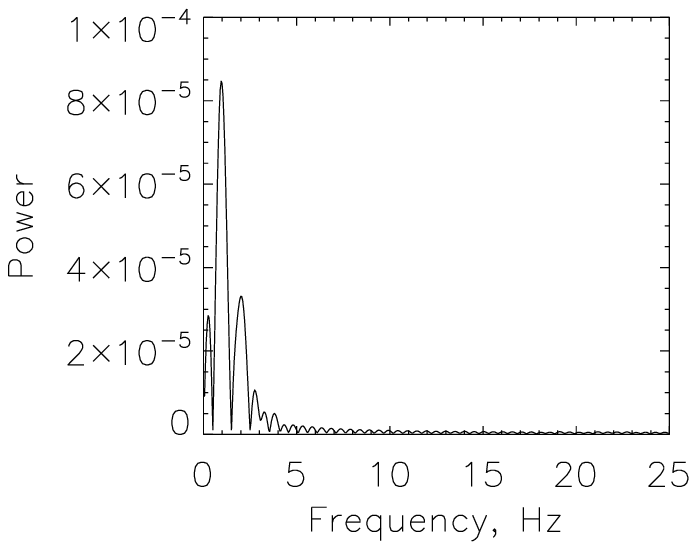}}
	\subfloat[1.7 GHz]{\includegraphics[width=0.5\linewidth]{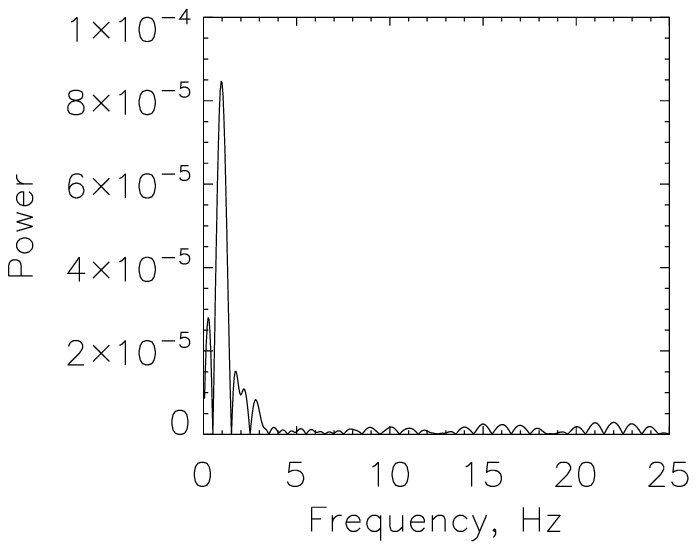}}
	\vspace{-8mm}
	\subfloat[1.9 GHz]{\includegraphics[width=0.5\linewidth]{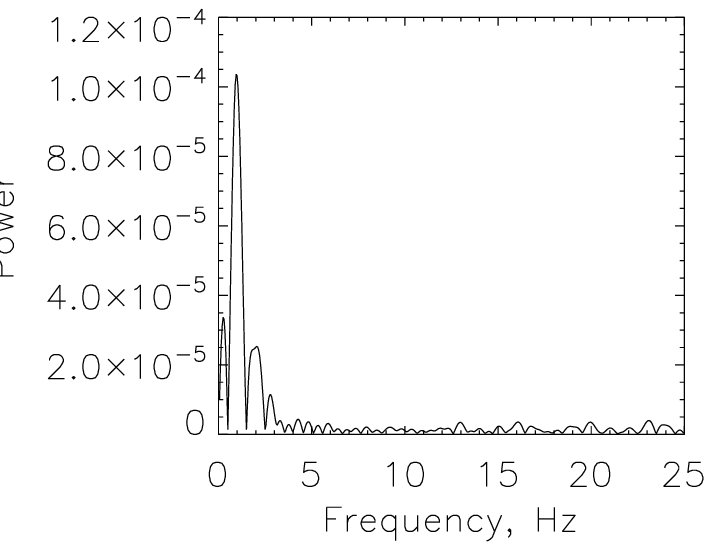}}
	\subfloat[2.1 GHz]{\includegraphics[width=0.5\linewidth]{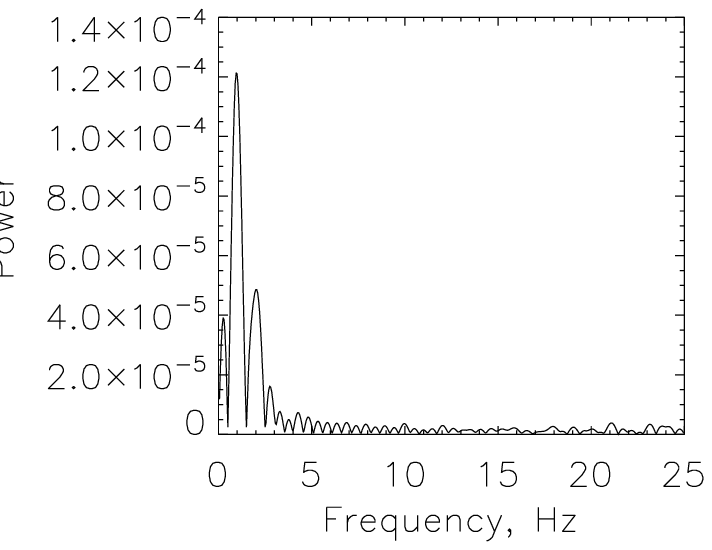}}
	\vspace{-8mm}
	\subfloat[2.3 GHz]{\includegraphics[width=0.5\linewidth]{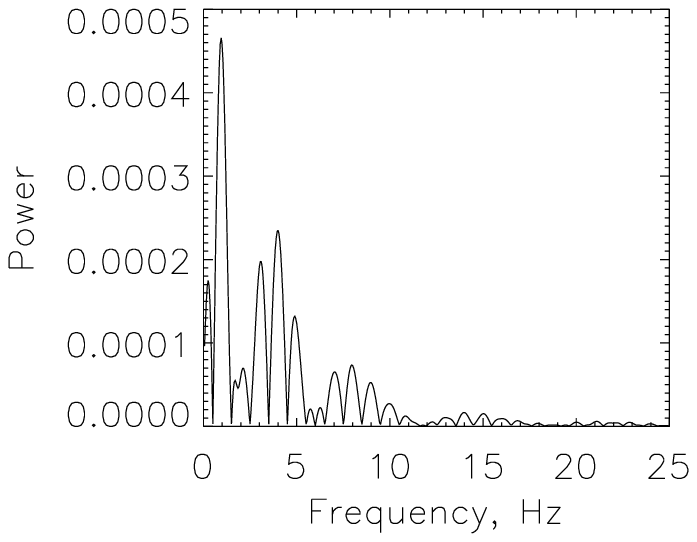}}
	\subfloat[2.5 GHz]{\includegraphics[width=0.5\linewidth]{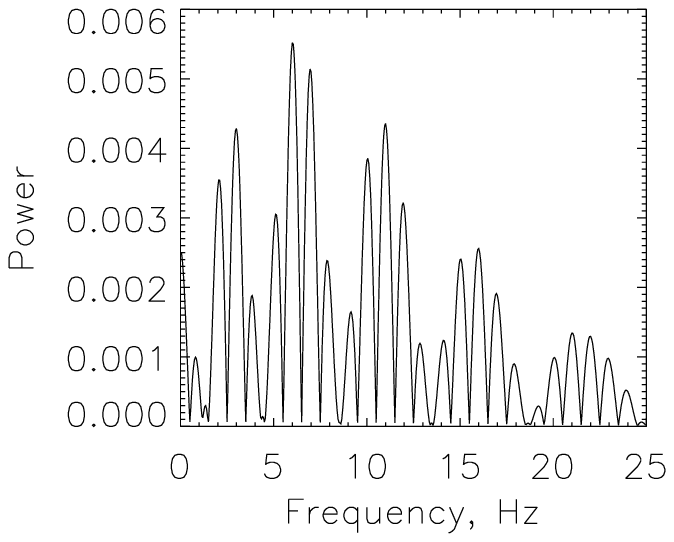}}
	\caption{Kink mode, high magnetic field, oscillations of the flux density, viewing angle 80$^\circ$. Same as in Figure~\ref{sausage_lowfreq_f_fourier}.}
	\label{kink_theta80_lowfreq_f_fourier}
\end{figure}

\begin{figure}[htp!]
	\centering
	\includegraphics{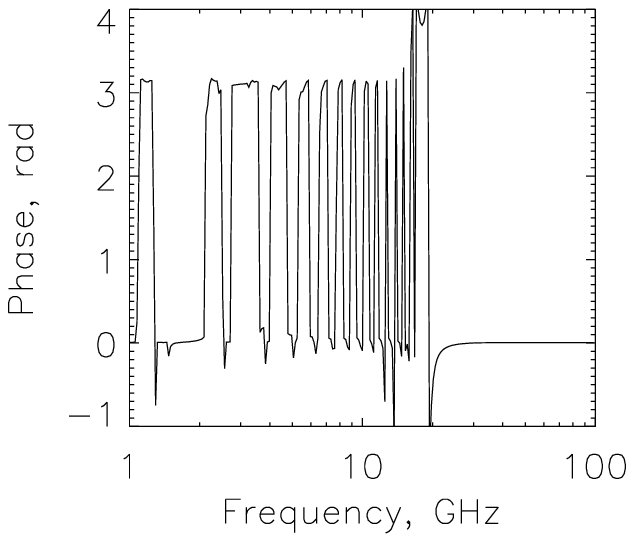}
	\includegraphics{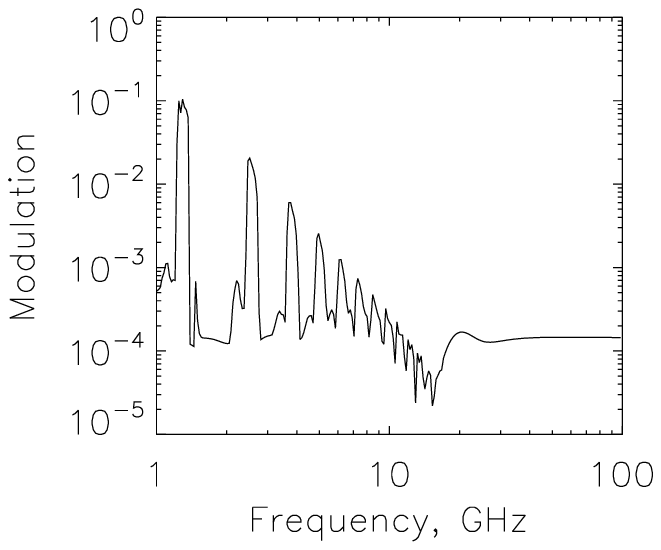}
	\caption{Kink mode, high magnetic field, oscillations of the flux density, viewing angle 80$^\circ$. Same as in Figure~\ref{sausage_lowfreq_f_analysis}. Note prominent oscillating behavior of the modulation amplitude at low frequencies.}
	\label{kink_theta80_lowfreq_f_analysis}
\end{figure}

\begin{figure}[htp!]
	\centering
	\includegraphics{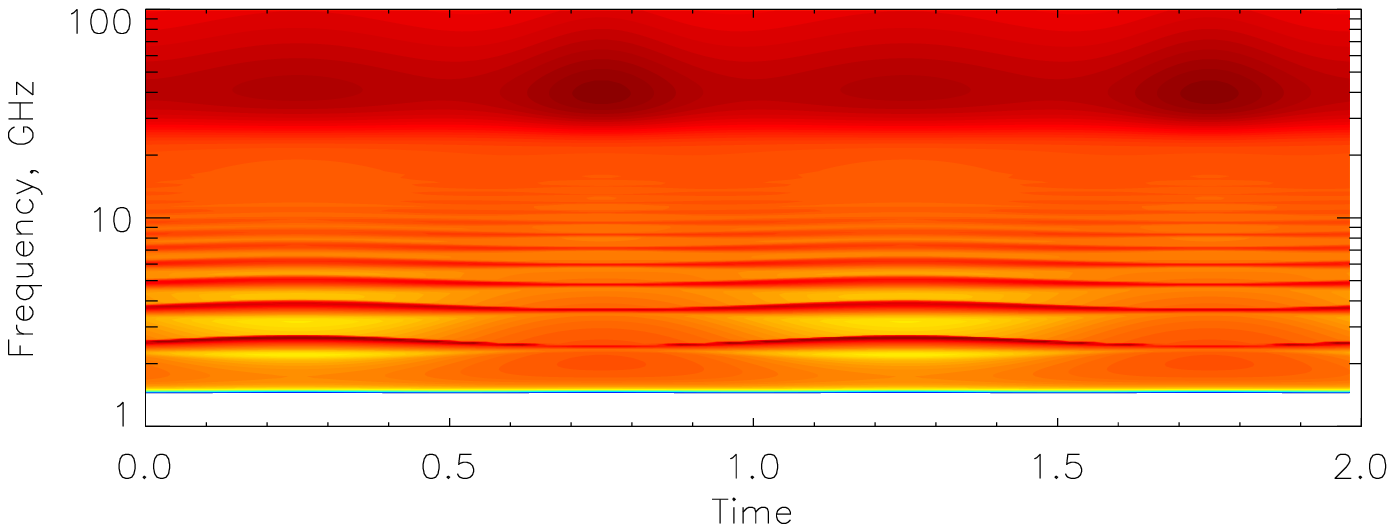}
	\includegraphics{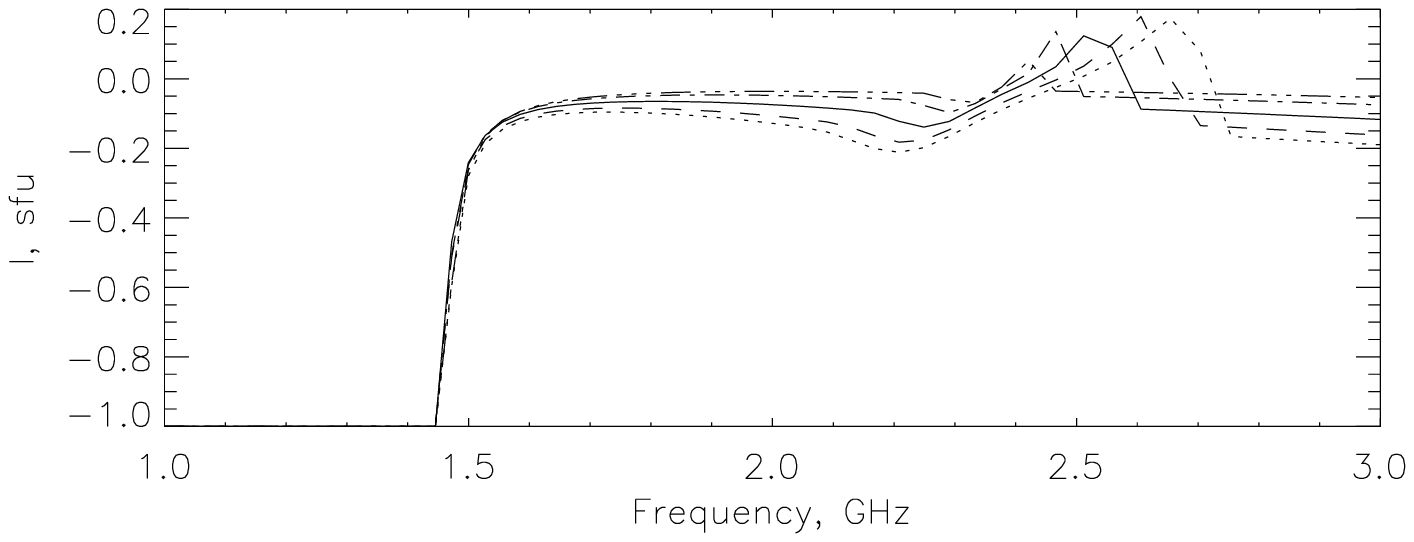}
	\includegraphics{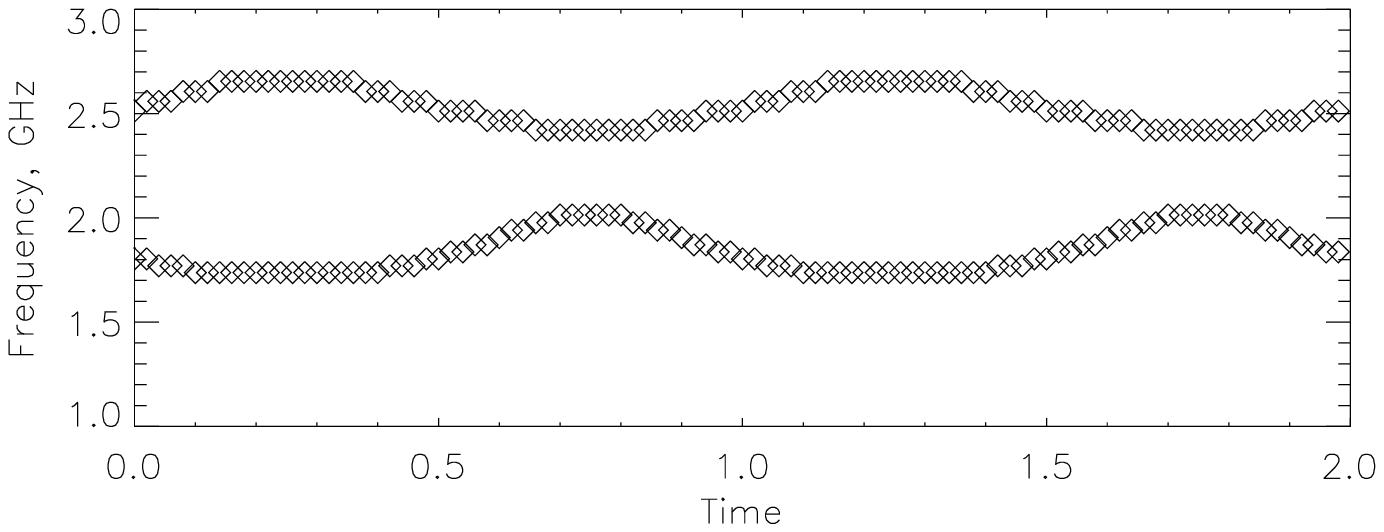}
	\caption{Kink mode, high magnetic field, oscillations of polarization, viewing angle 80$^\circ$. Same as in Figure~\ref{sausage_lowfreq_p_light}. }
	\label{kink_theta80_lowfreq_p_light}
\end{figure}
\clearpage

\begin{figure}[htp!]
	\centering
	\subfloat[1.6 GHz]{\includegraphics[width=0.5\linewidth]{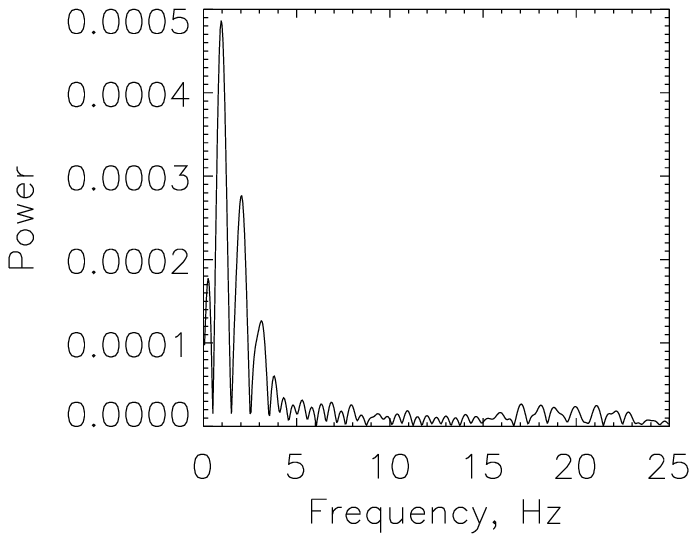}}
	\subfloat[1.7 GHz]{\includegraphics[width=0.5\linewidth]{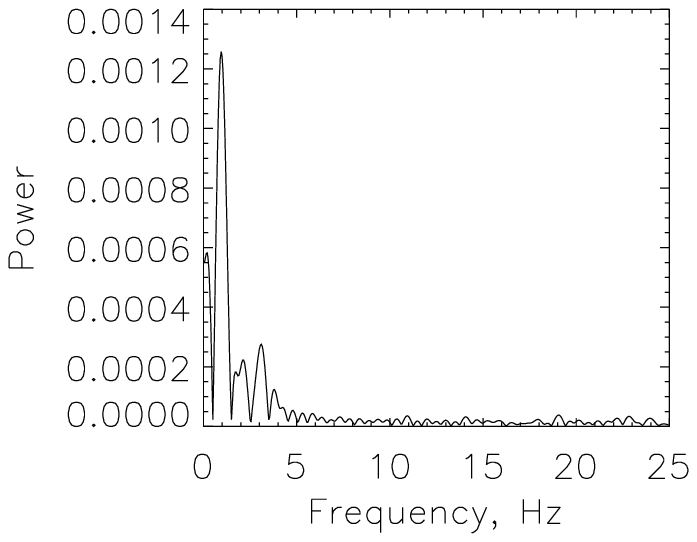}}
	\vspace{-8mm}
	\subfloat[1.9 GHz]{\includegraphics[width=0.5\linewidth]{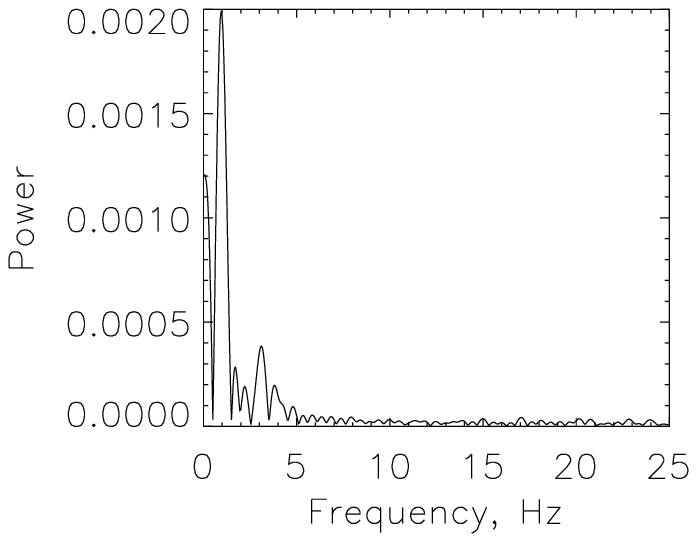}}
	\subfloat[2.1 GHz]{\includegraphics[width=0.5\linewidth]{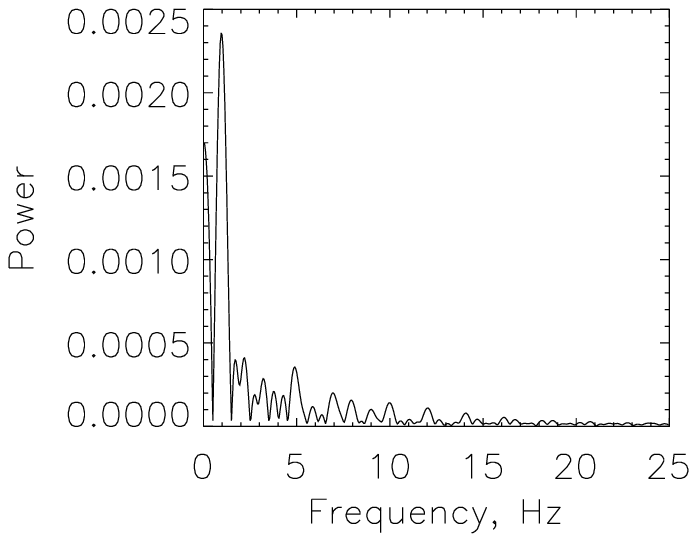}}
	\vspace{-5mm}
	\subfloat[2.3 GHz]{\includegraphics[width=0.5\linewidth]{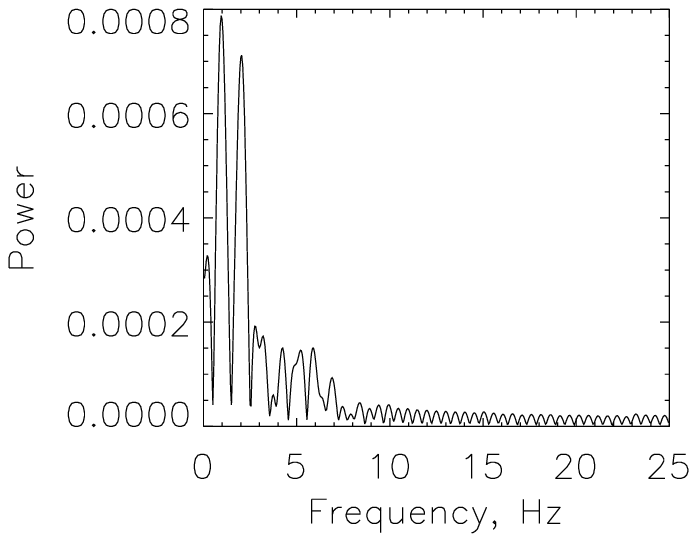}}
	\caption{Kink mode,  high magnetic field, oscillations of polarization, viewing angle 80$^\circ$. Same as in Figure~\ref{sausage_lowfreq_p_fourier}. }
	\label{kink_theta80_lowfreq_p_fourier}
\end{figure}

\begin{figure}[htp!]
	\centering
	\includegraphics{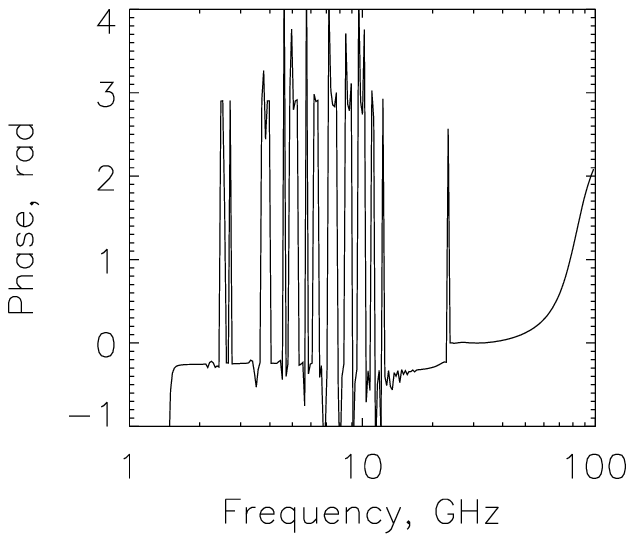}
	\includegraphics{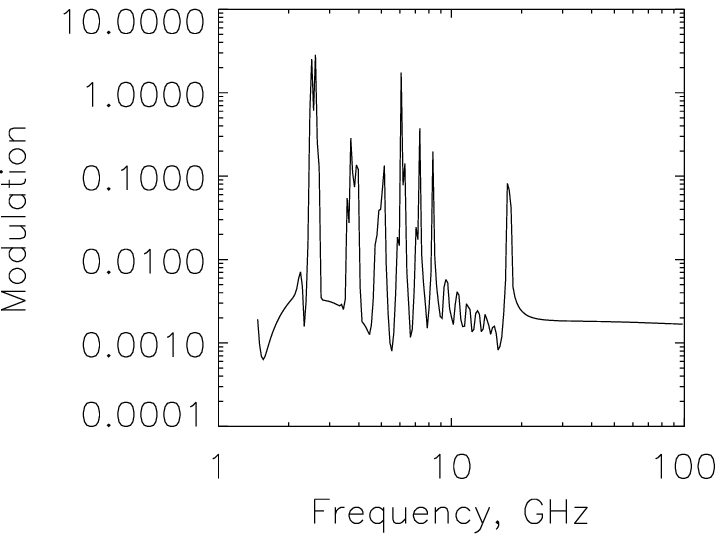}
	\caption{Kink mode,  high magnetic field, oscillations of polarization, viewing angle 80$^\circ$. Same as in Figure~\ref{sausage_lowfreq_p_analysis}.}
	\label{kink_theta80_lowfreq_p_analysis}
\end{figure}


\clearpage

\clearpage

\clearpage


\clearpage
\clearpage

\clearpage

\end{document}